\documentclass[reprint,amsmath,amssymb,aps,prl]{article}

\usepackage{graphicx}% Include figure files
\usepackage{dcolumn}% Align table columns on decimal point
\usepackage{bm}% bold math

\usepackage[utf8]{inputenc}
\usepackage[T1]{fontenc}
\usepackage{mathptmx}

\usepackage[german,english]{babel}
\usepackage{fontenc}
\usepackage{amssymb}
\usepackage{subcaption}
\usepackage{graphicx}
\usepackage{color}
\usepackage{units}
\usepackage{hyperref}
\usepackage{tabularx}
\usepackage{booktabs}
\usepackage[table]{xcolor}
\usepackage{listings}

\hypersetup{colorlinks,%
           citecolor=blue,%
          filecolor=blue,%
           linkcolor=blue,%
           urlcolor=blue,%
       pdfborder=0 0 0}
\usepackage[version=3]{mhchem} % Formula subscripts using \ce{}
\usepackage{chemfig} %chemical reactions
\usepackage[utf8]{inputenc}
\usepackage{csquotes}
\usepackage[exponent-product = \cdot]{siunitx}
\usepackage{todonotes}
\usepackage{xspace}
\usepackage{placeins}
\usepackage{grffile} %for directly loading plots which have a dot in the filename
\newcommand{\Rend}[0]{R_\mathrm{e}}
\newcommand{\Req}[0]{R_\mathrm{eq}}
\newcommand{\Rmax}[0]{R_\mathrm{max}}
\newcommand*{\res}{^{\mathrm{res}}}
\newcommand*{\str}{_\mathrm{str}}

\newcommand{\kT}[0]{k_\mathrm{B}T}
\newcommand{\LB}[0]{\lambda_\mathrm{B}}
\newcommand{\cs}[0]{c_\mathrm{salt}^\mathrm{res}}

\newcommand{\Rout}[0]{R_\mathrm{out}}
\newcommand{\Veq}[0]{V_\mathrm{eq}}
\newcommand*{\Vchain}{V_\mathrm{chain}}
\newcommand{\Pres}[0]{P_\mathrm{res}}
\newcommand{\Pin}[0]{P_\mathrm{gel}}
\newcommand{\Pcap}[0]{P_\mathrm{cap}}
\newcommand{\Pside}[0]{P_\mathrm{side}}

\newcommand{\muresex}[0]{\mu^\mathrm{res,ex}}

\newcommand\epsilonr[0]{\varepsilon_\mathrm{r}}
\newcommand*{\etal}{{et~al.}\xspace}

\begin{document}

\title{Cell Model Approaches for Predicting the Swelling and Mechanical
  Properties of Polyelectrolyte Gels}

\author{Jonas Landsgesell\thanks{Institute for Computational Physics, University of Stuttgart, D-70569 Stuttgart, Germany} \and Christian Holm\thanks{Institute for Computational Physics, University of Stuttgart, D-70569 Stuttgart, Germany} }

\begin{abstract}
  We present two successive mean-field approximations for describing
  the mechanical properties and the swelling equilibrium of
  polyelectrolyte gels in contact with a salt solution.  The first
  mean-field approximation reduces the many-chain problem of a gel to
  a corresponding single chain problem. The second mean-field step
  integrates out the degrees of freedom of the flexible chain and the
  ions. It replaces the particle-based description of the
  polyelectrolyte with suitable charge distributions and an effective
  elasticity term. These simplifications result in a computationally
  very efficient Poisson-Boltzmann cell-gel description. Despite their
  simplicity, the single chain cell-gel model shows excellent and the
  PB model very good agreement with explicit molecular dynamics
  simulations of the reference periodic monodisperse network model for
  varying chain length, polymer charge fraction, and external
  reservoir salt concentrations. Comparisons of our models to the
  Katchalsky model reveal that our approach is superior for strongly
  charged chains and can also predict the bulk moduli more
  accurately. We further discuss chain length polydispersity effects,
  investigate changes in the solvent permittivity, and demonstrate the
  robustness of our approach to parameter variations coming from
  several modeling assumptions.
\end{abstract}
\maketitle
%

%\tableofcontents
%\newpage
\section{Polyelectrolyte Gels}
Polyelectrolyte gels consist of crosslinked charged polymers. These
gels can swell tremendously when placed in an aqueous environment and
their reversible uptake of water is exploited in numerous applications
such as superabsorbers, desalination agents
\cite{hoepfner13b,richter17a},
%~\cite{zohuriaanmehr08a},
cosmetics, pharmaceuticals~\cite{peppas00a,jia09a,jagurgrodzinski09a},
osmotic engines \cite{arens17a}, or in
% meenach09a,tamura11a,sannino03a,zavan09a},
agriculture~\cite{zohuriaanmehr10a,kazanskii92a}.

Optimizing polyelectrolyte gels to their respective application
requires gel models which can yield accurate quantitative predictions
of the swelling equilibria and their mechanical properties, a task
that goes beyond known analytical approaches
\cite{flory43a,katchalsky55a,ricka84a,khokhlov93a,rubinstein96a,claudio09a,longo11a,quesadaperez11b,jha11a,liu15a}.
Using the predicted pressure-volume curves the desalination
performance of the gels as a function of, e.g., charge fraction or
chain length can be assessed. Likewise, osmotic engines (gels that can
translate differences in chemical potential to mechanical work) are
another area for applications that benefit from quantitative
predictions in order to optimize the engine efficiency prior to gel
synthesis.

% Where are the open problems
Macroscopic polyelectrolyte gels with monodisperse chain lengths have
been simulated using molecular dynamics (MD) simulations with periodic
boundary conditions (PBCs) (cf.~\textit{periodic gel
  model})~\cite{schneider02a,yan03a,edgecombe04a,yin05a,mann05a,mann06a,yin08a,mann11a,quesadaperez11a,kosovan13a,kosovan15a,rud17a}.
%A unit gel section is then connected periodically to yield an infinite gel
%without boundaries. 
These coarse-grained simulations provide predictions about
mechanical and swelling properties of macroscopic gels and revealed
microscopic insights about validities of various analytical
predictions.  Similar coarse-graining approaches already proved to be
useful for microgels\cite{claudio09a,colla14a,denton16a} as well.

Nevertheless, some features of real gels like polydispersity are hard
to include into such models. In order to represent polydispersity
faithfully, one would have to simulate a huge volume element with many
chains of different length and a sufficiently large number of
realizations.
%First experiments which systematically investigated the effect of polydispersity were performed in the 1980s by Mark \etal using bimodal polymer networks\cite{mark84a,mark85a}. 
%Early theoretical work on chain length heterogeneity in polydisperse polymer networks was performed by Higgs \etal \cite{higgs88a} who introduced an electric resistor analogy. Schimmel \etal \cite{schimmel91a} investigated the elastic modulus of polydisperse networks in experiments and observed a lowering with increasing broadness of the chain length distribution. The explanation given by Schimmel \etal is that spacial clustering of crosslinks reduces the effective number of crosslinks and therefore the elastic modulus of polydisperse gels is reduced compared to monodisperse gels.
%Fist computer simulations for crosslinked polymer networks were performed by Everaers and Kremer in 1996~\cite{everaers96a}.
Huge particle-based simulations with more than $~10^5$ monomers have
been performed for investigating polydispersity in uncharged networks
by \cite{gavrilov14a} and \cite{svaneborg05a}
%280000 monomers simulated by Svaneborg
%They investigated polydispersity in simulations where the crosslinking process was explicitly taken into account. 

To our knowledge, the only simulation study that treats polydispersity
in charged polymer gels is that of \cite{edgecombe07a}.  Due to the
computational cost of simulations with explicit charges Edgecombe
\etal were only able to simulate gels with $~10^2$ monomers
~\cite{edgecombe07a}. %16 chains a 20 monomers, in total 320 monomers
% Since their simulation setup is so small
% they cannot realize a representative Flory-Schulz distribution in
% their simulation box.
With this setup the polyelectrolyte chains are highly correlated since
the small unit cell is periodically repeated. We compare the results obtained by Edgecombe
qualitatively to a simple extension of or our models.

In a previous letter\cite{landsgesell19a} we introduced two new
mean-field gel models to which we refer to as the \enquote{single chain
cell-gel model} (CGM = cell-gel model), and the \enquote{Poisson-Boltzmann (PB) CGM} in the
following.  We demonstrated the general applicability of the CGMs, and
give a simple extension of the PB CGM to treat also weak (pH-sensitive) polyelectrolyte gels.

\begin{figure}
\centering
\includegraphics[width=0.7\textwidth]{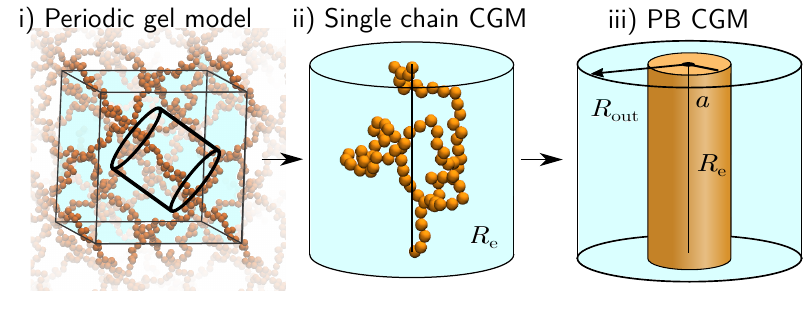}
\caption{\label{fig:schematic}A schematic of the i)
  macroscopic gel; ii) single chain CGM; and iii) PB CGM of a
  macroscopic gel, which are in equilibrium with a salt reservoir. The symbols are introduced in the text.}
\end{figure}

For a quick recapitulation we illustrate in Figure \ref{fig:schematic}
various levels of gel models: Starting from the periodic gel model i)
we see in ii) the single chain CGM and iii) displays the second
mean-field PB CGM~\cite{landsgesell19b}.  In our models, the polymer
chains are characterized by the number of monomers $N$ per chain, and
the charge fraction per monomer $f\in[0,1]$. The reservoir, on the
other hand, is characterized by the salt concentration $\cs$,
which is related to the reservoir concentrations of positive and
negative ions $\cs=c_+\res=c_-\res$ ensuring
electroneutrality.

Two important features of all models i--iii) that deserve to be highlighted
are the following two equilibrium conditions
\cite{khokhlov93a,barrat92a}: First, the chemical potentials
for all reservoir particle species $i$ are
constant (in the following, we use species $i\in \{+, -\}$). Therefore, the chemical potentials (of each species $i$) in the gel ($\mu_i^\mathrm{gel}$) and in the reservoir ($\mu_i\res$) are equal: %\footnote{In the case of no particle reservoir coupled to the cell the equilibrium condition is instead that all counterions are included in the cell}
\begin{equation}
\mu_i^\mathrm{gel}=\mu_i\res.
\end{equation}
Second, the pressure of the gel ($\Pin$) has to balance the pressure that is
exerted on the system by the
reservoir ($\Pres$):%\footnote{In the case of no salt bath coupled to the cell the equilibrium condition is instead that the pressure of the system is zero}
\begin{equation}
  \Pin(\Veq)=\Pres.
  \label{eq:pressure}
\end{equation}
In all models we approximate the reservoir pressure by the pressure
of an ideal gas $\Pres \approx \kT \sum_i c\res_i$ (unless specified
otherwise). The gel, while in contact with a reservoir, is simulated
at different volumes and the equilibrium volume $\Veq$ is determined
by eq. $\eqref{eq:pressure}$. In mechanical and chemical equilibrium the
end-to-end distance is the equilibrium chain extension $\Req$.

The outline of the paper is as follows: First, we describe the
periodic gel simulations which provide the benchmark data for our
comparison. We then present in detail the single chain CGM and the PB
CGM. We further compare the swelling equilibria and the elastic moduli
of these models to the Katchalsky approach\cite{kosovan15a} and
demonstrate the superiority of our new models. Additionally, we
introduce polydispersity in our CGMs and investigate the effects that
solvents of different polarity have on swelling. 
%Finally, we also
%demonstrate that our CGMs are insensitive to model parameters like
%the \enquote{diamond parameter} $A$ or the exact polymer charge
%distribution in the PB CGM.

\section{Periodic Gel Model}
Before describing the mean-field models, we briefly present the
periodic gel simulations serving as a benchmark to evaluate the
quality of our modeling approaches. The MD simulations are used to
model periodically connected tetrafunctional strong gels on a
coarse-grained level. In the periodic gel MD simulations we deal with
crosslinked bead-spring polyelectrolytes in an implicit solvent. An
example snapshot of a periodic gel can be seen in figure
\ref{fig:schematic} i). All MD simulations presented in this paper
have been performed using the MD simulation package
ESPResSo~\cite{weik19a} . 
We obtained the periodic gel benchmark data form simulations performed explicitly for this paper. 
The simulations are performed in full analogy to the setup described by Ko{\v{s}}ovan\cite{kosovan15a}.
In contrast to our previous publication\cite{landsgesell19b} we did not use data from {Ko{\v{s}}ovan} et al~\cite{kosovan15a} 
since the grand canonical scheme was flawed. For
clarity, we give a short recap on the simulation setup here: We simulate a unit cell of a tetrafunctional gel. 
The number of monomers per chain in our simulation varies: $N\in \{40,64,80\}$. 
The unit cell contains 16 chains and 8 nodes, i.e. in total $16N+8$ gel monomers.
All particles interact via a Weeks-Chandler-Andersen (WCA)
potential~\cite{weeks71a} with the effective particle size
$\sigma=0.355\unit{nm}$ and strength of the interaction
$\epsilon=1.0\,\kT$. Bonds connect the neighbouring monomers via
FENE bonds~\cite{kremer90a}, with the typical Kremer-Grest
parameters~\cite{grest86a} $k_{\mathrm{FENE}}=30.0\,\kT/\sigma^2$ and
$\Rmax=1.5\sigma$.  Temperature is accounted for via a
Langevin thermostat (at room temperature approx. $\unit[300]{K}$), while electrostatic interactions are calculated
via the P3M method~\cite{deserno98a} tuned to an accuracy in the
electrostatic force of at least $10^{-3} \kT/\sigma$~\cite{deserno98b}. In our primitive model the relative
electrostatic permittivity $\epsilonr$ is chosen to be that of pure
water at room temperature, i.e.
$\epsilonr\approx 80$. The Bjerrum length which
corresponds to this choice is that of water at room temperature,
i.e. $\LB=\unit[7.1]{}$ \AA (unless otherwise noted).
%In a real gel the relative permittivity might vary spatially, which is not taken into account here.
The simulation is performed in a semi-grand canonical
ensemble\cite{hill86a} %p. 342
where salt and counterions are allowed to enter and leave the gel.
% while the gel particles are not allowed to be exchanged with the reservoir. 
%As macroscopic gels are electroneutral some of the ions may not leave the gel\cite{claudio09a}. 
The exchange of salt ions is performed by inserting pairs of salt ions.
The acceptance probability for the insertion of one positive and one negative ion is\cite{frenkel02b}:
\begin{equation}
\text{acc}_\text{insertion}=\min \left(1, V^2 c_+\res c_-\res \exp(\beta \muresex_s) \frac{1}{(N_-+1)(N_+ +1)} \exp(-\beta \Delta E_\mathrm{pot}) \right),
\end{equation}
with $\beta=\frac{1}{\kT}$ and $\Delta E_\mathrm{pot}$ the potential energy change due to the Monte Carlo move. $N_i$ is the number of ions of species $i$ in the simulation volume $V$ and $c_i\res$ is the concentration of species $i$ in the reservoir. 
The semi-grand
canonical simulation makes use of the pair excess chemical potential in the reservoir $\muresex_s=\muresex_+ + \muresex_-$
for inserting a positive ($i=+$) and a negative ion ($i=-$) at the same time.
The pair excess chemical potential is calculated in independent simulations of the reservoir (at different but fixed $\cs$) via the Widom insertion method\cite{frenkel02b}.
The adapted formula for determining the pair excess chemical potential for inserting a positive and a negative ion at the same time is:
\begin{equation}
\label{eq:widom}
\muresex_s=-\kT \ln \left( \frac{1}{V^2} \int_V \int_V d^3r_{N+1} d^3 r_{N+2} \langle \exp(-\beta \Delta E_\mathrm{pot}) \rangle_N \right),
\end{equation}
where $\langle \cdot \rangle_N$ is the canonical ensemble average over the N-particle configuration space\cite{frenkel02b} and where $\Delta E_\mathrm{pot}$ is the potential energy change due to the insertion of a positive and a negative ion. The simulated system is electroneutral at any time. As explained by \cite{frenkel02b} the two integrals in equation \eqref{eq:widom} are calculated via brute force sampling and inserting the ions at random positions inside the box.

This semi-grand canonical procedure ensures electrostatic neutrality as well
as equality of the chemical potentials. % $\mu^\mathrm{gel}_{i}=\mu^\mathrm{res}_{i}$ of species $i$ 
%In order to find the mechanical equilibrium with the reservoir the gel is simulated at different volumes. 
For each simulation the volume and ensemble averaged virial pressure inside the
gel is recorded yielding a pressure-volume ($PV$) curve (compare figure \ref{fig:result
  PV curve}). The Coulomb part of the pressure is calculated as
described by \cite{essmann95a}.
%The pressure in the reservoir is, to a good approximation, given by the
%ideal gas expression $\Pres$. 

\section{Single Chain Models}
\label{sec:single chain models}

In contrast to the previously discussed expensive MD simulations of periodic gels, we proposed a reduction
of the many-chain problem to a single chain problem in our recent publication \cite{landsgesell19b}. 
In this section we describe common features of the (particle-based) single chain CGM and the PB
CGM.

Using simple geometric considerations it is found that for a
fully stretched tetrafunctional monodisperse gel the volume $\Vchain$, which is associated to
a single chain, depends
on the end-to-end distance $\Rend$ via the
relation~\cite{kosovan15a}
\begin{equation}
\label{eq:chain volume}
\Vchain =\Rend^3/A.
\end{equation} 
The denominator has the value $A=\sqrt{27}/4$ for a diamond lattice \cite{kosovan15a}.
Whenever referring to the volume in the single chain models, we mean the volume $\Vchain$ per chain.
For a not
completely stretched gel, the denominator $A$ is non-trivially
depending on the end-to-end distance\cite{kosovan15a}. For simplicity,
we neglect this dependence and assume for the rest of the paper that
$A$ is a constant (we discuss this point in the supporting information).
% Via using this affine compression factor $A=\sqrt{27}/4$ we match
% the volume per chain in the the cell model to the volume per chain
% in a tetrafunctional gel under affine compression.\footnote{Note
% that simulations show that the compression of a gel is not always
% truly affine\cite{kosovan15a} Maybe also because the bond length
% changes for the used simulation parameters in the paper by Richter
% and {Ko{\v{s}}ovan}. We also checked for other prefactors A,
% e.g. A=1.8 and obtained only small deviating results (~ 2\%)
% regarding the swelling equilibrium.}

Relation $\eqref{eq:chain volume}$ connects the end-to-end distance
and the volume which is available for a chain: on compression the
end-to-end distance decreases since the ends of the polymer chains are
brought closer together — at the same time also the volume which is
available per chain shrinks. 

Having fixed the volume which is accessible to a single chain in a
periodic gel, we still need to decide on the geometry of the
volume: In a macroscopic polyelectrolyte gel the monomers on one chain
experience forces which originate from the electrostatic environment
imposed by the other chains and from the elastic response
to the crosslinking. Emulating this environment we
choose a cylindrical volume and connect the chain ends
through the cylinder top and bottom by applying PBCs.

Due to restrictions in the single chain CGM and application of PBCs (see section
\ref{sec:MD single chain}) the cylinder height needs to be
$L(\Rend)=\Rend+b$ with the bond length $b \approx 0.966 \sigma \ll \Rend$. 
Here $b$ takes a value typical for the Kremer-Grest FENE
potential\cite{kremer90a}.
In order to use results obtained for an affine compression of the cylinder (see section \ref{sec:pressure affine compression}), and since $b$ is typically much smaller than $\Rend$, we approximate the volume available per chain via:
\begin{equation}
\label{eq:chain volume approx}
\Vchain\approx L(\Rend)^3/A.
\end{equation} 

%For less strong FENE bonds the bond length is not constant anymore and depends on the overall prescribed end-to-end distance. Therefore a end-to-end dependent bond length would have to be prescribed when wanting to compare to simulations of polymers with weak FENE bonds.

From the cylinder height and the cell volume \eqref{eq:chain volume approx} the outer cell radius $\Rout$ is given by:
\begin{equation}
\Rout=\sqrt{\frac{\Vchain(\Rend)}{L(\Rend) \pi}}=\sqrt{\frac{L(\Rend)^3}{L(\Rend) \pi A} }=\frac{L(\Rend)}{ \sqrt{\pi A} },
\end{equation} 
as depicted in figure \ref{fig:schematic}. Since we consider only affine compressions
 we keep the aspect ratio $\alpha=\Rout/L$ constant in the single chain CGM and the PB CGM.
\subsection{Pressures in an Affinely Compressed Cylinder}
\label{sec:pressure affine compression}
%For an affine compression both the cap position and the radial boundary move during a volume change:

Applying the chain rule for the volume derivative of the free energy of the system $F_\mathrm{gel}$
(at fixed aspect ratio $\alpha$) we find the pressure in the
cylindrical cell geometry as~\cite{antypov06a}:
\begin{align}
\Pin&=-\frac{\partial F_\mathrm{gel}}{\partial \Vchain}\Bigg|_\alpha=-\frac{\partial F_\mathrm{gel}}{\partial \Rout} \frac{\partial \Rout}{\partial \Vchain}\Bigg|_\alpha -\frac{\partial F_\mathrm{gel}}{\partial L} \frac{\partial L} {\partial \Vchain} \Bigg|_\alpha\\
%&=f_R \frac{\partial R}{\partial \Vchain}\Bigg|_\alpha +f_z \frac{\partial L}{\partial \Vchain}\Bigg|_\alpha \\
%&=P_R \cdot (2\pi R L(\Rend) ) \underbrace{\frac{\partial R}{\partial \Vchain}\Bigg|_\alpha}_{\overset{...}{=}\frac{R}{3\Vchain} } +P_z \cdot (\pi R^2) \underbrace{ \frac{\partial L} {\partial \Vchain}\Bigg|_\alpha}_{\overset{...}{=} \frac{L(\Rend)}{3\Vchain} }\\
&=\frac{2}{3}\Pside+\frac{1}{3}\Pcap.
\label{eq:pressure affine compression}
\end{align}
This result allows to express the pressure of the
cylindrical system via the pressure contribution $\Pcap$ from the top and a
pressure contribution $\Pside$ from the side of the cylinder.

%In order to make use of this result for affine compressions in the
%single chain CGM (where due to PBCs the
%cylinder height needs to be $L=\Rend+b$) we employ from equation
%\eqref{eq:chain volume approx} the approximate volume available per
%chain.

\subsection{Particle-Based Single Chain CGM}
\label{sec:MD single chain}
Using the above described geometrical considerations, we introduce the
first simplification for solving the many-chain problem of a periodic gel: the particle-based single
chain CGM depicted in figure \ref{fig:schematic} ii). In the remainder of the text we call this model \enquote{single
chain CGM}.
%In the particle-based single chain CGM simulations we take samples from a suitable
%single-chain partition function.
Our single chain CGM consists of a single polyelectrolyte
chain, confined to a cylinder of height $L(\Rend)$ and radius\footnote{Note that in the MD simulation the outer cell radius is
  increased by one $\sigma$ because the WCA wall of the
  outer cylinder interacts with all particles with a WCA interaction
  with cutoff $2^{1/6}\sigma$ which corresponds to good approximation
  to a reduction of the ion available volume of about one $\sigma$ in radial direction. This
  technicality is not further noted in the description of the MD
  simulations.} $\Rout$.
The first and last monomers of the single chain are 
bonded through the periodic boundary conditions and fixed to the cylinder center but are free to move along the cylinder axis. 
Applying PBCs mimics the electrostatic interactions in the gel where one end of the chain sees the beginning of the next chain, and where the fixed endpoints of the chain correspond to crosslinks in the gel. 
A single chain
consists of $N$ monomers. The contour
length of the chain is $\Rmax=(N-1)b$, where $b$ is the average bond length. The particles in the particle-based single chain CGM interact with the same interactions and parameters as in the periodic gel model (FENE bonds, WCA potential, Langevin thermostat, P3M, semi-grand canonical ensemble).
%As in the periodic gel simulations we connect monomers via FENE bonds with Kremer-Grest parameters~\cite{kremer90a}.
% \footnote{In the periodic gel there are $16 \mathrm{MPC}+8$ monomers contained in a box which contains 16 chains: the additional 8 monomers are the 8 nodes which are connecting the 16 chains. Therefore in the single chain CGM a typical chain would have $N=\frac{16 \mathrm{MPC}+8}{16}=\mathrm{MPC}+0.5\approx \mathrm{MPC}$ monomers.}
%We employ a Langevin thermostat~\cite{frenkel02b} to simulate the chain
%at room temperature $\approx \unit[300]{K}$. Particles interact via WCA
%interactions~\cite{weeks71a}. 
To compute the electrostatic forces and energies we
use the P3M algorithm, with 3D periodic boundary conditions,
tuned to an absolute accuracy~\cite{deserno98a} in the electrostatic force of at least $10^{-3} \frac{\kT}{\sigma}$. 
We compared the obtained forces of P3M with the more accurate MMM1D algorithm\cite{arnold05b} designed for 1D periodic systems. 
We observed only small deviations compared to the P3M forces — this is especially true when using a gap in the x-y directions to separate the infinite cylinders. 
%In order to couple a salt reservoir to the
%simulation box we employ the semi-grand canonical ensemble via grandcanonically\cite{frenkel02b}
%inserting salt ion pairs (in the same way as it was previously done in the periodic gel model). 

%\subsubsection{Pressure in single chain CGM}
In order to obtain a prediction for the swelling equilibrium we need to
calculate the pressure inside the gel with the cap contribution $P_\mathrm{cap}$ and the side contribution $P_\mathrm{side}$ (see equation \eqref{eq:pressure affine compression}).
The caps of the cylinder are only imaginary and are not explicitly modeled.
In the particle-based single chain CGM, 
the cap pressure is given by the ensemble average of the $(z,z)$ component of the instantaneous pressure
tensor $\Pcap:=\langle \Pi_{(z,z)} \rangle$:
%This is allowed since the system is homogeneous in the z direction \footnote{Forces due to constraints are not included into the virial pressure (this is good). They shall not be included in the viral.}.

\begin{align*}
\Pi_{(z,z)} = \frac{\sum_{i} {m_{i}v_{i}^{(z)}v_{i}^{(z)}}}{\Vchain} + \frac{\sum_{j>i}{f_{ij}^{(z)}r_{ij}^{(z)}}}{\Vchain} +\Pi^\text{Coulomb, P3M}_{(z,z)}.
\end{align*}

Here $\Vchain$ is the effective available
  volume, $m_{i}$ is the mass, $\vec{v}_i$ the velocity of particle
$i$ (and the superscript $(z)$ denotes the z-component). $\vec{r}_{i,j}$ is the connection vector between particles $i$ and $j$. $\vec{f}_{i,j}$ is the pair force (excluding the electrostatic force) between particles $i$ and $j$.
The last term is the zz-component of the instantaneous Coulomb pressure tensor\cite{essmann95a} and accounts for the electrostatic interactions.
%\footnote{
%We do not exclude any charged ions from the pressure calculations. Then the Coulomb part of the pressure tensor has two contributions:
%\begin{align}
%P^\text{Coulomb, P3M}_{(z,z)} &=P^\text{Coulomb, P3M, dir}_{(z,z)} + P^\text{Coulomb, P3M, rec}_{(z,z)}
%\end{align}
%The first contribution is:
%\begin{equation}
%\small{
% P^\text{Coulomb, P3M, dir}_{(z,z)}= \frac{1}{4\pi \epsilon_0 \epsilonr} \frac{1}{2V} \sum_{\vec{n}}^* \sum_{i,j=1}^N q_i q_j \left( \frac{ \mathrm{erfc}(\beta |\vec{r}_j-\vec{r}_i+\vec{n}|)}{|\vec{r}_j-\vec{r}_i+\vec{n}|^3} +\frac{2\beta \pi^{-1/2} \exp(-(\beta |\vec{r}_j-\vec{r}_i+\vec{n}|)^2)}{|\vec{r}_j-\vec{r}_i+\vec{n}|^2} \right) ((\vec{r}_j-\vec{r}_i+\vec{n})_z)^2}
%\end{equation}
%and the second contribution is: 
%\begin{equation}
%P^\text{Coulomb, P3M, rec}_{(z,z)}= \frac{1}{4\pi \epsilon_0 \epsilonr} \frac{1}{2 \pi V} \sum_{\vec{k} \neq \vec{0}} \frac{\exp(-\pi^2 \vec{k}^2/\beta^2)}{\vec{k}^2} |S(\vec{k})|^2 \cdot (1-2\frac{1+\pi^2\vec{k}^2/\beta^2}{\vec{k}^2} \vec{k}_z \vec{k}_z).
%\end{equation}
%$\beta$ is the P3M splitting parameter , $S(\vec{k})$ is the spacial Fourier transform of the charge density $\rho(\vec{r})=\sum_i q_i \delta(\vec{r}-\vec{r}_i)$, $\vec{k}$ are the wave vectors.}.
Please note that the particle mass is irrelevant to the calculation of the ensemble averaged kinetic pressure tensor component $\langle \Pi_{\mathrm{kin}, (z,z)} \rangle:=\langle\frac{\sum_{i} {m_{i}\vec{v}_{i}^{(z)} \vec{v}_{i}^{(z)}}}{\Vchain}\rangle$. The reason for this is that the velocities are Maxwell-Boltzmann distributed: The probability density to find a velocity vector $\vec{v}_i$ for particle $i$ is $p(\vec{v}_i) = \left(\frac{m_i}{2\pi\kT}\right)^{3/2} \exp(-\beta \frac{m_i}{2} \vec{v}_i^2)$. The probability density to find a certain component (e.g. the $z$ component) is $p({v}_i^{(z)}) = \left(\frac{m_i}{2\pi\kT}\right)^{1/2} \exp(-\beta \frac{m_i}{2} {v_i^{(z)}}^2)$. The ensemble average is, therefore, independent of the particle mass:
\begin{equation}
\langle \Pi_{\mathrm{kin}, (z,z)} \rangle = \frac{1}{\Vchain} \sum_{i} {\int_{v_i^{(z)}=-\infty}^\infty m_{i}{v}_{i}^{(z)} {v}_{i}^{(z)}}p(v_i^{(z)}) dv_i^{(z)} = \frac{\sum_j N_\mathrm{j} \kT}{\Vchain} ,
\end{equation}
where the first sum runs over all particles $i$ and the last sum runs over all species $j$.
This result is necessary in order to recover the limiting case of an ideal gas (which has no interactions) and for which the result of the isotropic pressure is well known: $P_\mathrm{ideal}=\langle \Pi_{\mathrm{kin}, (z,z)} \rangle =\frac{\sum_j N_j \kT}{V}$. 
Therefore, we chose the arbitrary particle mass $m_i=1$ simulation unit.

The cylinder boundaries are the side walls. To constraint the particles in the cylinder all particles are repelled from the wall via a WCA interaction.
The contribution of the pressure $\Pside$ acting on the side
walls is directly measured as
the average normal force on the outer wall per area.
The side pressure mainly arises due to the pressure that the mobile ions
exert onto the wall. %The average electric field vanishes at the outer wall and the polymer chain rarely interacts with it. 
Therefore, the side pressure could have also been
obtained using the contact value theorem for the cylindrical cell
model\cite{wennerstrom82a}. However, the contact theorem does not account for a possible interaction of the polymer chain with the
wall and is therefore potentially less accurate. 
%Therefore, we have chosen to directly measure the normal forces on the wall, which does not rely on the assumption that the polymer chain does not interact with the side walls.

In order to increase the accuracy of our description in the particle
based models (namely the single chain CGM and the periodic gel
model), we include the excess pressure in the pressure of the reservoir (for
Bjerrum lengths $\LB>2\sigma$, i.e. $\epsilonr<80$):
$\Pres=\kT \sum_i c\res_i +P^\mathrm{ex, res}$ (measured via independent
MD simulations of the reservoir). In this case the excess
pressure significantly lowers the pressure in the reservoir compared
to the ideal gas pressure, due to the attractive nature of the electrostatic interaction.

Next, we present the second mean-field model which further simplifies the single chain CGM.

\subsection{Poisson-Boltzmann CGM}
%We could account for excluded volume effects and pressures in our continuum PB model by employing a modification of the modified PB (m2PB) equation presented by Borukhov~\cite{borukhov1997steric}. %%I did that but swelling equilibria changed only by 2%. It s not worth to bother the reader with its derivation here.

The PB model uses a density based description of the polymer charges and the
surrounding ions instead of a particle-based description. As before, the polymer and the ion densities are enclosed in a cylinder having an external radius
$\Rout$ and a height of $\Rend+b$ as shown in
figure~\ref{fig:schematic} iii). The electrostatic environment in a gel is mimicked via periodically replicating the system along the z-axis. Therefore, we calculate the electrostatic interaction in the
gel via solving the electrostatic problem for an infinite
rod with prescribed charge density. In contrast to the original rod cell model\cite{alfrey51a, fuoss51a} our rod is penetrable to the ions. The maximal radial
extension of the polymer density is denoted by the radius $a$ in
figure~\ref{fig:schematic} iii).

In the PB model the distribution of the monomers is, in principle,
arbitrary with the following two restrictions: 1) the maximal
extension $a$ needs to be smaller than the cell radius $\Rout$ and 2)
the monomer distribution should be as realistic and as close as possible to that of
a charged polymer bead-spring chain while
maintaining the simplicity of the model. To fulfill restrictions 1) and 2) we use data from our previous
single chain CGM simulations. From this data we extract a fitting curve for the average
distance $\langle r \rangle_\text{MD}$ of the chain monomers from the cylinder axis as a function of $\Rend$:
\begin{equation}
\langle r \rangle_\text{MD}(\Rend) = N \sigma \left( C_1 \left(\frac{\Rend }{N \sigma}\right)^2 +
  C_2 \frac{\Rend }{N \sigma} + C_3 \right),
\end{equation}  
with $C_1=-0.17$, $C_2=0.14$ and $C_3=0.03$ (fit to single chain CGM
data for $N \in \{39,79,300\}$,
$\epsilonr=80$, for a fully
charged chain without added salt). %\footnote{The curves barely change (within fluctuations around the mean value) when changing salt concentration from $0 mol/l$ to $0.2 mol/l$ or when changing the charge fraction $f$ from $0.125$ to 1).}
%fit of a_\text{MD} happens in \url{/tikhome/jlandsgesell/PB/MD/force_extension/chain_salt_charged_confinement_excluded_volume_stiff_FENE_stress_tensor/plot_master_curve_average_distance_from_end_to_end_vector.gp} for different chain lengths at charge fraction 1 and salt concentration 0.1 mol/l. We find that the charge fraction dependence and salt bath concentration dependence is weak within fluctuations around the mean!

This average distance of the monomers $\langle r \rangle_\text{MD}$ from the end-to-end vector is
imposed on the monomer densities occurring in the PB model.  To
investigate the influence of the chosen monomer densities we
investigate two distributions i) a rectangular distribution and in the supporting information a
more realistic approximately Gaussian distribution with compact
support ii).
The probability density to find a monomer in a given distance $r$ from the cylinder axis
has, in the case of a rectangular distribution, the form
\begin{equation}
p(\vec{r})=\mathcal{N}_1 H(-(r-a)),
\end{equation}
where $H(x)$ denotes the Heaviside function, and $\mathcal{N}_1$ a
normalization such that $p(\vec{r})$ is a probability density. The
normalization condition for $p(\vec{r})$ gives
$\mathcal{N}_1=\frac{1}{\pi a^2 L}$.  The parameter $a$ is chosen such
that the average distance from the end-to-end vector
$\langle r\rangle =\int_{\Vchain} p(\vec{r}) r\,\mathrm{d}V$ is equal to the
average distance from the end-to-end vector which was observed in
simulations of the single chain CGM:
$\langle r \rangle \overset{!}{=}\langle r
\rangle_\text{MD}$. Therefore, we obtain for
the parameter $a=\frac{3}{2}\langle r \rangle_\text{MD}$.

The above probability density to find a monomer $p(\vec{r})$, together with the charge fraction $f$ and the
number of monomers $N$ in the chain, imply a corresponding charge
density $\rho_f(\vec{r})= - N f e_0 p(\vec{r})$, where $e_0$ is the absolute value of the elementary charge.
%\footnote{
%Typically one encounters the choice of $\rho_f(\vec{r})=\frac{Q}{2\pi a L} \delta(r-a)$ which represents a rod which carries its charge on the surface and where ions are prohibited to enter the volume of the rod. 
%}
%. For the delta distributed charge distribution we prohibit the counterions from entering the rod at $r<a$\footnote{In literature the rod is typically modeled to be impenetrable which means that the mobile ions are excluded from entering the space where $r<a$ -- see for example~\cite{naji06b,fuoss51a}}.
This charge density is the key input to the Poisson-Boltzmann equation
which describes the electrostatic interaction in the system:
\begin{align}
\nabla^2\psi= -\frac{1}{\epsilonr\epsilon_0} \left(q_+ c_+(\vec{r}) + q_- c_-(\vec{r}) +\rho_f(\vec{r}) \right),
\label{eq:PB}
\end{align}
where $\psi$ is the total electric potential, and $\epsilonr$ is
the relative permittivity.
densities via $\rho_i(\vec{r})=c_i(\vec{r}) q_i$) are given by
standard Poisson-Boltzmann theory\cite{markovich16a-pre} (with the choice of the reservoir
potential $\psi\res=0$):
\begin{equation}
c_i(r)=c_i\res e^\frac{-q_i \psi(x,y,z)}{\kT},
\end{equation}
where $c_i\res=\cs$ are the ion densities in the reservoir of species $i$, where $i$ is either $+$ or $-$.

Since we model macroscopic gels we impose charge neutrality, hence, there is no flux of electric field through the surface of the
cell. The two boundary conditions which we employ for solving the PB equation \eqref{eq:PB} are, therefore, that the radial electric field $E_r(r)$ is zero at $r=0$ and $r=\Rout$. 
%\footnote{Applying a third incompatible boundary condition like e.g. setting the electric potential $\psi(\Rout)=0$ would be wrong and electroneutrality of the system would be violated. Setting the electric potential to the value in the reservoir would also violate the Donnan potential as discussed in the supplemental information.} 
%This is a direct result from applying Maxwells equations and Gauss' law: ($\vec{E}_\bot=(-\nabla \psi)_\bot(R) =\vec{0}$) at R is zero since this ensures charge neutrality. %(Gauss law yields $Q(cell)=\int_{\partial \Vchain} d\vec{S} \vec{E}=0$ $\implies \vec{E}_\bot=\vec{0}$)

%\subsubsection{Pressures in the Cell Model}
%In the salt free case we only needed to search for the pressure of the cell system itself becoming zero, which is equivalent (via $p=-\frac{\partial F}{\partial \Vchain}$) to searching the free energy minimum of the cell with respect to volume changes.
In the PB model the pressure has two contributions 1) the combined
Maxwell and kinetic pressure and 2) the stretching pressure. For the
standard PB theory the first contribution is given by\cite{trizac97a}:
\begin{align}
\Pside^\text{ions}&=\kT \sum_i c_i(R)\\
\Pcap^\text{ions}&=\kT \sum_i \langle c_i \rangle_z +\frac{\epsilon_0\epsilonr}{2} \langle E_r^2 \rangle_z,
\end{align}
%The result for $P_z$ looks similar to the one from Antypov in~\cite{antypov2006osmotic}. However Antypov missed the electrostatic term in $p_L$ due to an oversimplified derivation of the pressure!
%In the case of the modified Poisson-Boltzmann equation\cite{borukhov1997steric} where one introduces the entropy of a lattice gas the pressure tensor is given by Trizac in~\cite{trizac1999long}. Essentially the local ideal gas pressure is replaced by the local lattice gas pressure.
where $E_r=-\partial_r \psi(r)$ is the radial component of the
electric field, $\epsilon_0$ is the vacuum permittivity, and where
$\langle \mathcal{A}(r) \rangle_z=\frac{2\pi}{\pi \Rout^2}
\int_0^{\Rout} r \mathcal{A}(r) \mathrm{d}r$ denotes the average taken over the cap.
The side pressure is given by $\kT$ times the sum of the ion densities at the outer cylinder wall. This result is in agreement with the contact value theorem\cite{wennerstrom82a}. We want to note that the densities at the outer cylinder wall do not have to coincide with the densities in the reservoir.
The second contribution is the stretching pressure which only acts on
the cap. The stretching behavior of linear macromolecules under confinement is a delicate problem on its own and we refer the reader to further literature\cite{birshtein91a}.
The simple stretching term which we use is motivated in the following. The stretching pressure is
given as a volume derivative of a stretching free energy $F\str$:
\begin{align}
\Pcap^\text{str}&=-\frac{\partial F\str(\Rend)}{\partial \Vchain}\\%:=-\frac{d F\str(\Rend)}{d \Vchain}\bigg|_{\{N\},T}\\%=-\frac{d F\str}{d \Rend}\bigg|_{\{N\},T} \frac{d \Rend}{d \Vchain}\\
%&=-\left(\underbrace{\frac{\partial F\str}{\partial a}}_{=0} \frac{\partial a}{\partial \Rend} +\underbrace{\frac{\partial F\str}{\partial R}}_{=0} \frac{\partial R}{\partial \Rend} +\underbrace{\frac{\partial F\str}{\partial \Rend}}_{\text{force extension curve}}\right)\Bigg|_{\{N\},T}\frac{d \Rend}{d \Vchain}\\
&=-\frac{\partial F\str}{\partial L}\underbrace{\frac{\partial L}{\partial \Vchain}}_{1/(\pi \Rout^2)}=-\frac{\partial F\str}{\partial \Rend}\underbrace{\frac{\partial \Rend}{\partial L}}_{1} \frac{1}{\pi \Rout^2}.
\end{align}
Since the stretching free energy needs to contain a confinement
contribution $F_\mathrm{conf}$ and a tensile chain contribution $F_\mathrm{chain}$ it consists of two
terms:
$F\str=F_\mathrm{chain}+F_\mathrm{conf}$.  %The added confinement contribution allows to predict gel sizes for neutral gels or when the gel barely carries charges (in the limit of $f\to 0$).
%Although the Flory-Rehner ansatz is widely used, treating all free energy contributions as uncoupled and additive is known to be an approximation only.
Therefore one obtains:
\begin{equation}
\Pcap^\text{str}=-\frac{1}{\pi \Rout^2} \left( \frac{\partial F_\mathrm{chain}}{\partial \Rend}+ \frac{\partial F_\mathrm{conf}}{\partial \Rend} \right).
\end{equation}
The first term is given by the force extension curve of a freely
jointed chain
$-\frac{\partial F_\mathrm{chain}}{\partial \Rend}=\frac{\kT}{b}
\mathcal{L}^{-1}\left(\frac{\Rend}{\Rmax}\right)$ (with the
inverse Langevin function $\mathcal{L}^{-1}$ or a numerical
approximation\cite{cohen91a} and the maximal contour length\cite{rubinstein03a}
$\Rmax=(N-1)b$). %\footnote{The assumed polymer model here is arbitrary. Instead of the freely jointed chain model we could use the worm like chain extension curve and make use of a persistence length which depends on the electrostatic interactions which are present.} and identifying the force $f_\text{str}=\frac{\partial F_\text{str}}{\partial \Rend}$.

For the confinement free energy we use the result for a
free ideal polymer:
$F_\text{conf}=B\kT
\frac{Nb^2}{\Rend^2}$~\cite[p. 115]{rubinstein03a}.  We choose the
proportionality constant $B$ such that the stretching force is zero at
the equilibrium extension:
$\frac{\partial F_\text{str}}{\partial
  \Rend}\bigg|_{R_0}+\frac{\partial F_\text{conf}}{\partial
  \Rend}\bigg|_{R_0}=0$ yielding
$B=\frac{R_0^3}{2 Nb^3}\mathcal{L}^{-1}(\frac{R_0}{R_\text{max}})$.
%Alternatively we can obtain a similar force acting against compression using the constant $\langle |\vec{R}_E| \rangle$ (length) ensemble\cite{neumann01a-pre}. The $\Rend$ dependence for the pushing pressure is however different.
The final expression for the stretching pressure is:
\begin{equation}
\Pcap^\text{str}=\frac{1}{\pi \Rout^2} \frac{\kT}{b}\left( \frac{R_0^3}{\Rend^3} \mathcal{L}^{-1}\left(\frac{R_0}{\Rmax}\right) -\mathcal{L}^{-1}\left(\frac{\Rend}{\Rmax}\right) \right)
\end{equation}
The equilibrium extension $R_0$ is a free parameter, and we use the
relation $R_0=1.2 b N^{0.588}$ which was found for a
system of
neutral chains interacting via a WCA potential~\cite{kosovan15a}.

In total the cap pressure is given by
$\Pcap=\Pcap^\mathrm{str}+\Pcap^\text{ions}$, and together with
eq. \eqref{eq:pressure affine compression} one obtains the pressure in
the gel model $\Pin$.  The PB model is expected to work well in
aqueous solutions if there are no multivalent ions, high charge
densities (e.g. at high compression of the gel) or high ionic
concentrations present~\cite{andelman95a,deserno00a}. Problems in PB
theories arise due to neglecting ionic and excluded volume
correlations. Also the simplistic stretching pressure
$\Pcap^\text{str}$ (which is independent of the charge fraction and
the salt concentration) adds a source of inaccuracy.  These
limitations are not present in our single chain CGM.

%PB is only derived for low electrostatic coupling parameters $\Xi \to 0$\cite{naji06b} $\Rend >10 \sigma$ seems to be OK, then $\Xi \approx <10$.
%plot electrostatic coupling parameter
%p 2*pi*2**2*N*1/(2*pi*sqrt(x*(N-1)*1.03))

%%%%%%%%%%%%%%%%%%%%%%%%%%%%%%%%%%%%%
\section{Results}
%%%%%%%%%%%%%%%%%%%%%%%%%%%%%%%%%%%%%

%%%%%%%%%%%%%%%%%%%%%%%%%%%%%%%%%%%%%
\subsection{Pressure-Volume Curves}
\begin{figure}
  \begin{subfigure}{0.45\textwidth}
  \includegraphics[width=\columnwidth]{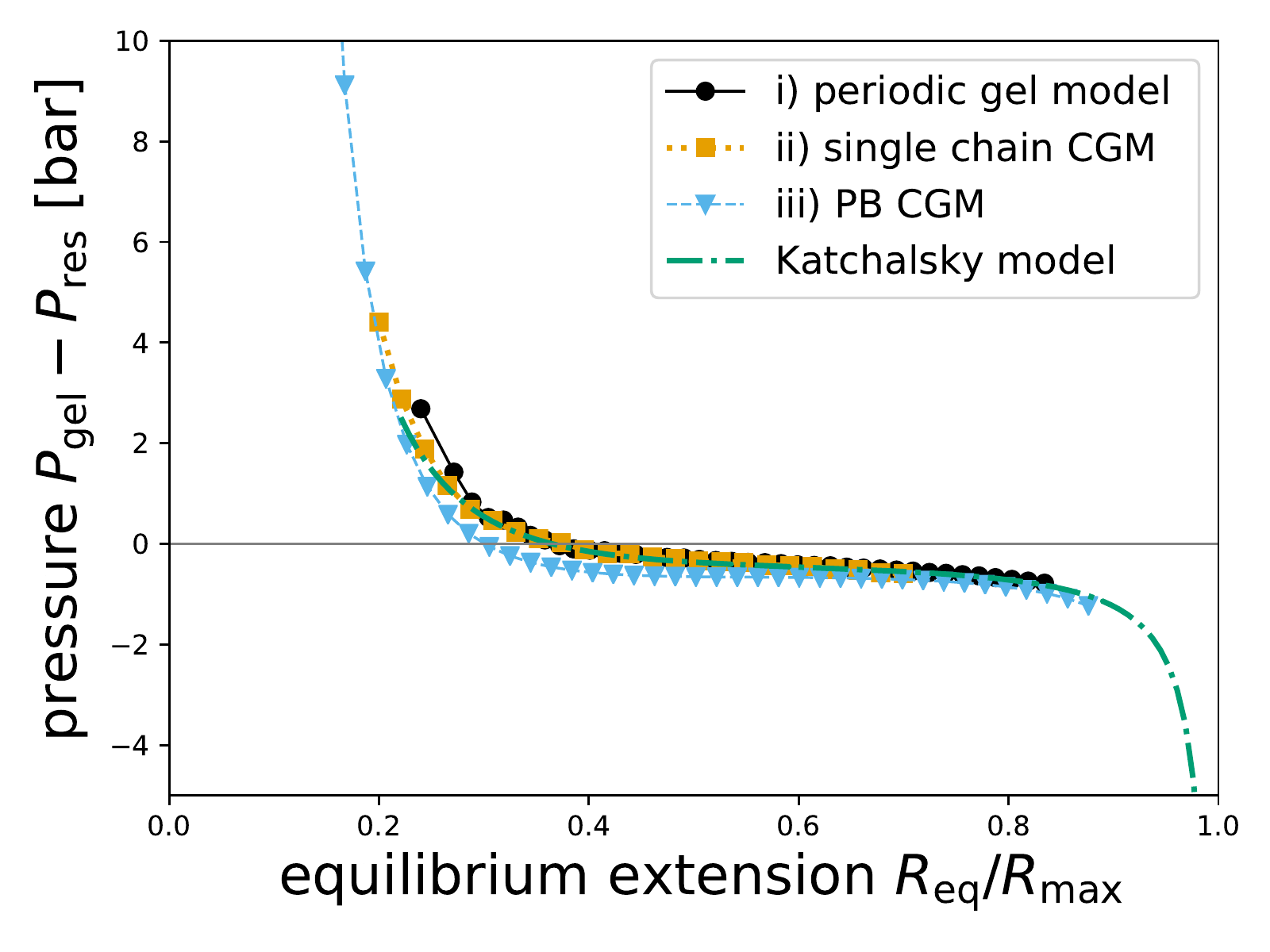}
  \caption{}
  \end{subfigure}
  \begin{subfigure}{0.45\textwidth}
  \includegraphics[width=\columnwidth]{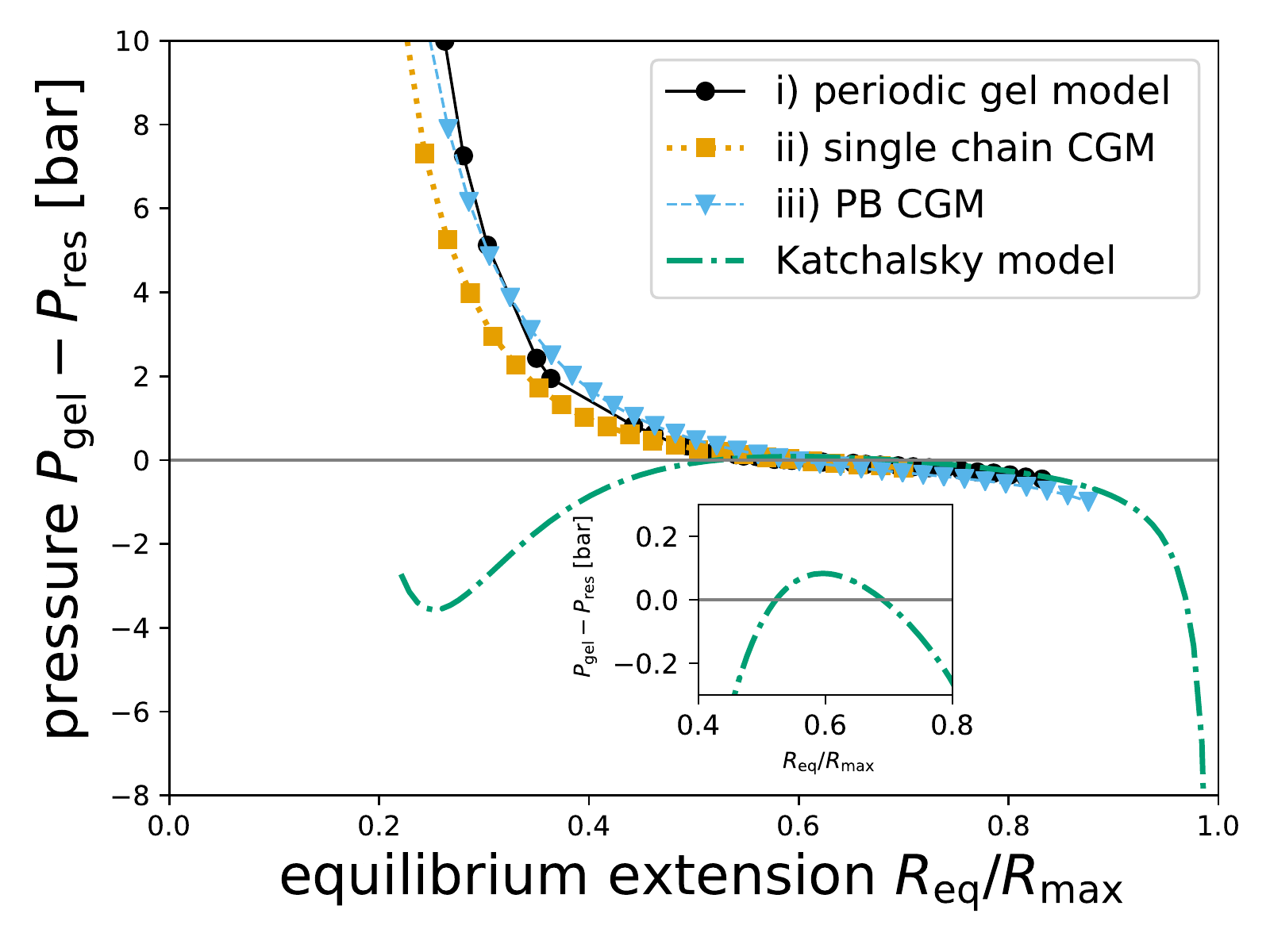}
  \caption{}
  \end{subfigure}
  \caption{$P(\Rend)$ curves for a) $f$=0.125 and b) $f$=1, both for
    $N$=64, $\cs$=\unit[0.01]{mol/l}, $\LB=2\sigma$
    (corresponds to $\epsilonr=80$). The inset in figure b) shows the
    Katchalsky model data from figure b) enlarged. In the inset two
    zero crossings are visible.  The abscissa is the equilibrium
    end-to-end distance divided by the maximum elongation of the chain
    and therefore parameterizes the volume per chain via
    eq. \eqref{eq:chain volume approx}.  As the end-to-end distance
    can be easily converted to a volume via equation \eqref{eq:chain
      volume approx} we refer to the $P(\Rend)$ as $PV$ curve.  $PV$
    curves for other parameters can be found in the supplemental
    information.}
\label{fig:result PV curve}
\end{figure}

%%%%%%%%%%%%%%%%%%%%%%%%%%%%%%%%%%%%%
% \FloatBarrier
All models which were described in the previous sections give pressure-volume curves (similar to stress-strain curves), see figure \ref{fig:result PV curve}. 
In addition, we included $PV$ curves from
the Katchalsky model which is a simple free energy model, for details please consult the papers authored by \cite{kosovan15a} and \cite{katchalsky55a}.
Compared to the Katchalsky model, our new models work for all charge
fractions. At high charge fractions ($f >0.5$ ~\cite{kosovan15a}) the
Katchalsky model produces non-physical $PV$ curves which is obvious
due to: a) the Katchalsky $PV$ curves have multiple zero crossings
(see figure \ref{fig:result PV curve}) and b) the Katchalsky $PV$
curves do not have a form compatible with the $PV$ curves of the
periodic gel model. At high charge fractions the Katchalsky model
fails due to the used electrostatic energy functional which is derived
from a linearization of the PB
equation\cite{katchalsky55a}. Therefore, the this model is only of
limited use for the prediction of desalination energy costs to gels
with low charge fractions\cite{richter17a}. In contrast to the
Katchalsky model, our new PB CGM gives monotonic $PV$ curves even for
highly charged gels ($f>0.5$). All $PV$ curves used in this work can
be found in the supplemental information. Additionally, our PB CGM
also shows better agreement with the periodic gel model than the
self-consistent field theory presented in Ref.~\cite[Figure
2]{rud17a}. Hence, the $PV$ curves could be used for an improved
prediction of the energy costs for desalinating sea water using highly
charged gels\cite{richter17a}.

From figure \ref{fig:result PV curve} the swelling equilibrium for
the corresponding gel can be obtained by identifying the zero
crossing (where the reservoir pressure and the pressure in the gel
balance).  The dependence of the swelling equilibria on experimental
conditions like the salt concentration or gel parameters like the
charge fraction of the monomers are examined in the next section.

%\FloatBarrier
%%%%%%%%%%%%%%%%%%%%%%%%%%%%%%%%%%%%%
\subsection{Swelling Equilibria}
%%%%%%%%%%%%%%%%%%%%%%%%%%%%%%%%%%%%%

\begin{figure}
%plots originate from scripts in /tikhome/jlandsgesell/phd/own_papers/2018-PB_gel_short/figures/data
   \begin{subfigure}{0.45\textwidth}
  \includegraphics[width=\columnwidth]{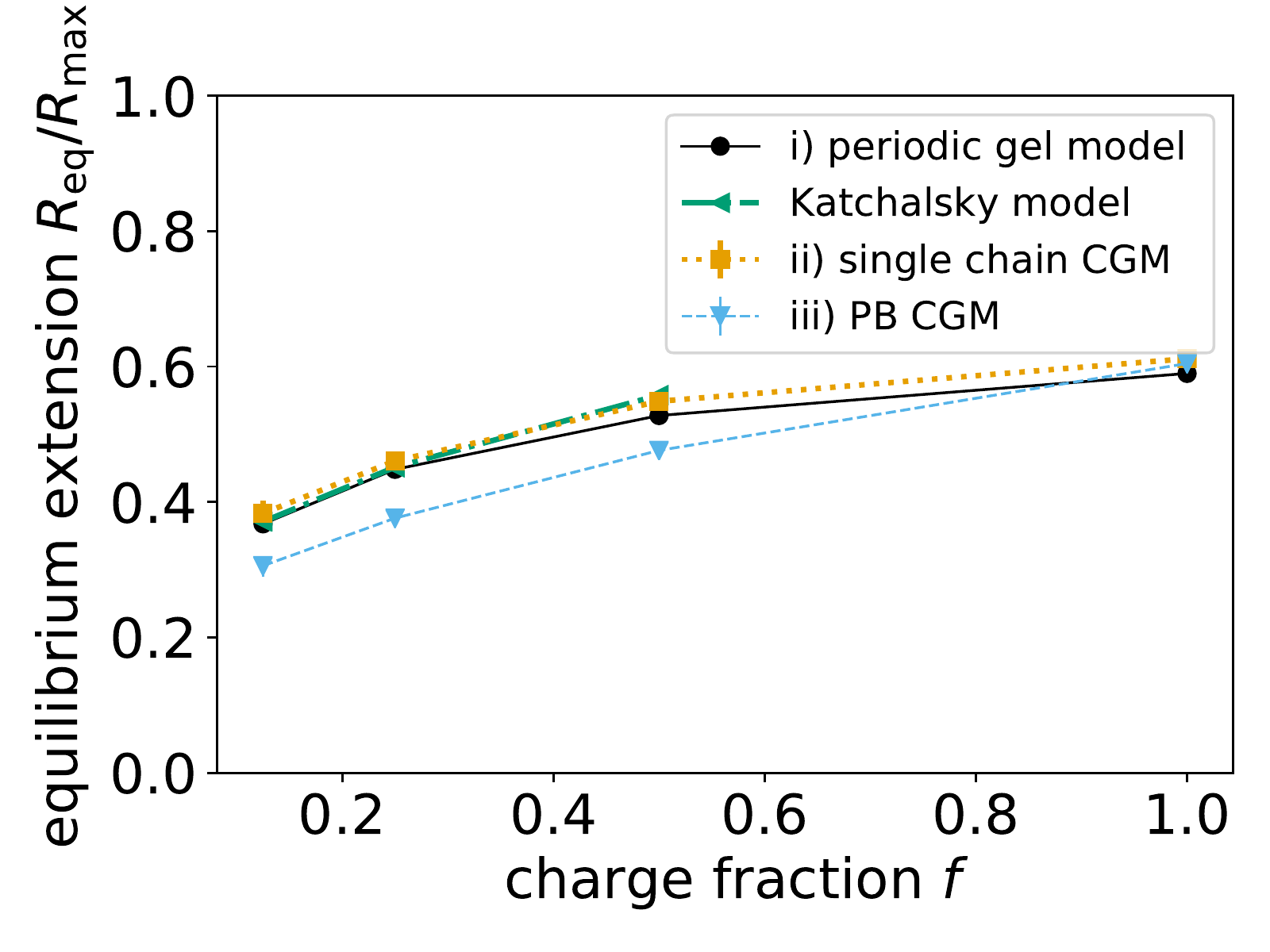}
  \caption{\label{fig:charge fraction}}
  \end{subfigure}
 \begin{subfigure}{0.45\textwidth}
  \includegraphics[width=\columnwidth]{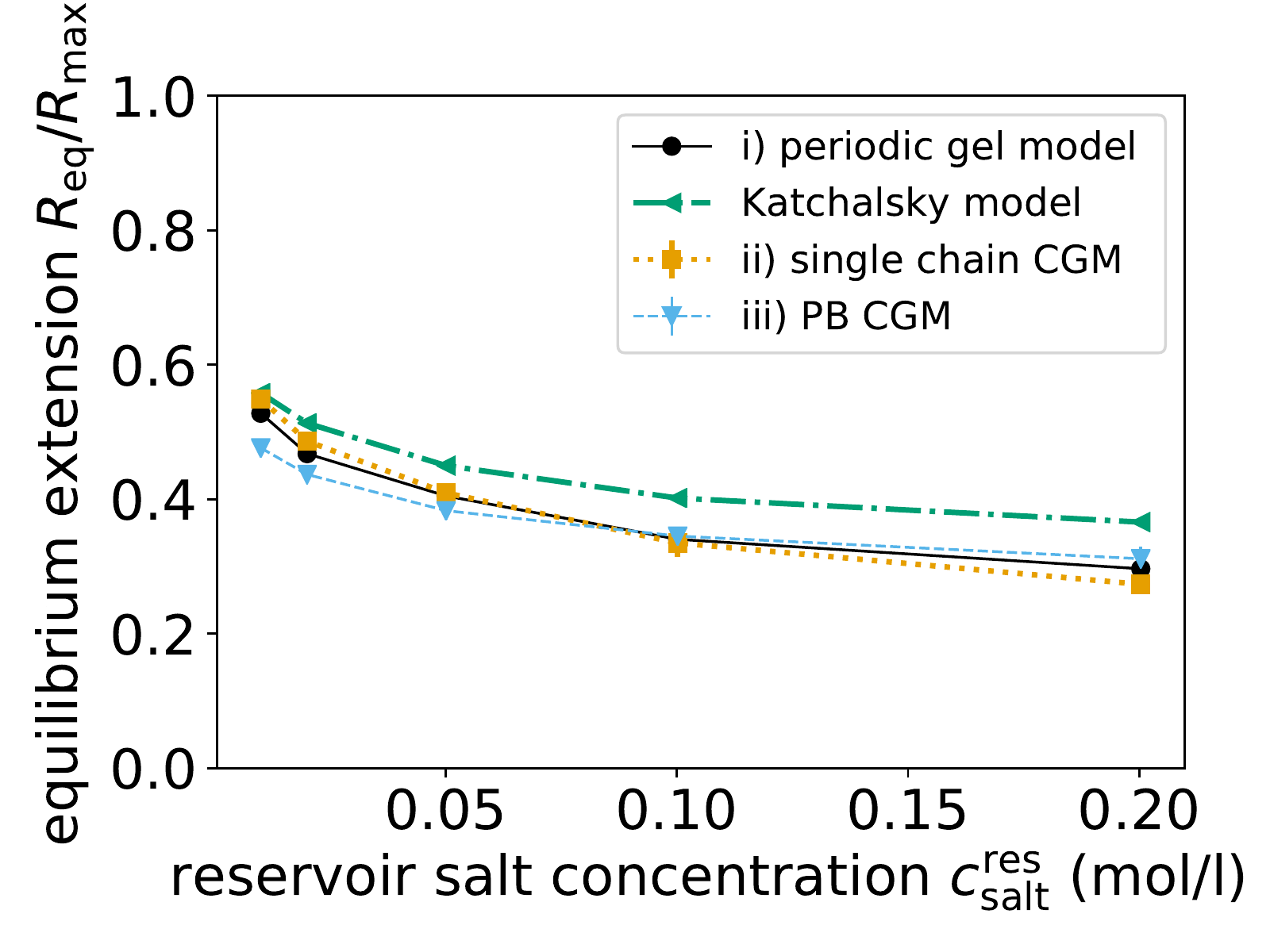}
    \caption{\label{fig:salt}}
  \end{subfigure}
  \caption{Comparison of swelling
    predictions of the periodic gel model i), the two new CGMs ii-iii) and the Katchalsky model for a polymer of chain length $N=64$ in aqueous solution ($\epsilonr=80$).  The
    equilibrium swelling $\Req$ as a) a function of the charge
    fraction $f$ for
    $\cs=\SI{0.01}{\mole\per\liter}$
%    b) the monomers $N$ in the polymer segments between cross links for $\cs=\SI{0.02}{\mole\per\liter}$ and $f=1$; 
    and b) as a function of the reservoir salt concentration $\cs$ for
    $f=0.5$. The equilibrium volume is determined by locating the
    volume where $\Pin(\Veq)$ and $\Pres$ are equal by using a linear
    interpolation. The error bar is the width of the interval at which
    the intersection happens. The errorbar is typically smaller than
    symbol size.  Note that in the first %(two plots)
    plot the points for the Katchalsky model do not cover the whole
    plot range. This is due to the fact that the Katchalsky model
    fails for these parameters~\cite{kosovan15a}.  }
\label{fig: swelling equilibria}
\end{figure}

In figure \ref{fig: swelling equilibria} the scaled equilibrium
extension $\Req/\Rmax$ is shown as a function of the salt
concentration in the reservoir ($\cs$) and the charge fraction ($f$).
In agreement with
literature\cite{schneider02a,edgecombe04a,longo11a,mann04a,mann05a,mann06a,mann11a,quesadaperez11a,quesadaperez12a,quesadaperez12b,kosovan13a,kosovan15a}
we find that the gel swells a) more with increased charge fraction $f$
and b) less with higher salt concentration in the reservoir.

In both subfigures of figure \ref{fig: swelling equilibria} we observe
that the single chain CGM and the periodic gel model agree very
well. Therefore, the usage of the single-chain CGM saves us an order
of magnitude in CPU time without sacrificing the accuracy of the
predictions of $\Req$. The PB CGM shows deviations from out reference data whose
size depends on the chosen parameters, but in general the trend is
reproduced. We find that the PB CGM also works for gels which are
highly charged (where the Katchalsky model fails). In agreement with
\cite{kosovan15a} we find, that the Katchalsky model offers very good
gel swelling predictions for gels with low charge fraction.  The exact
quantification of how big the error of the different models is,
however, a difficult problem. This arises from two facts:

\begin{itemize}
    \item The PB model predictions can only be compared to a finite set of
      reference data which are computationally expensive to generate.
    \item The high dimensionality of the parameter space: predictions
      for gel swelling $\Req$ are made for different chain lengths,
      charge fractions, salt concentrations and dielectric
      permittivities. We observe that the PB and the Katchalsky model
      show differing suitability in predicting the swelling
      equilibrium in various parts of the parameter space.
\end{itemize}

For completeness, we include the swelling equilibria which were
determined for all parameter combinations in the supplemental
information. This data reveals that the PB CGM predicts too low
swelling equilibria at low salt concentrations (compared to the
periodic gel MD data) and too large swelling equilibria for highly
charged gels $f \to 1$.  In the following we want to explicitly name
some simplifying assumptions made in the PB model which could cause
these devations:

 \begin{itemize}
 \item The polymer charge density $\rho_f(\vec{r})$ in the PB model
   is, for simplicity, not dependent on the salt concentration or the
   charge fraction which are imposed in the model.
 \item The stretching contribution $\Pcap^\text{str}$ is independent
   of the salt concentration and the charge fraction $f$. For a
   charged polymer, we would expect a different stiffness depending on
   a) the salt concentration and b) the charge fraction. At low salt
   concentrations or high charge fractions the stiffness (or
   persistence length) should increase, resulting in a higher
   extension. Therefore, the force-extension of a polyelectrolyte
   should favor more stretched states at low salt concentrations or
   high charge fractions.
  \end{itemize}
%Since this observation depends on the electrostatic interaction, we think that in our simple PB CGM there is some fortunate cancellation of errors between the simplified stretching term and the description of the electrostatic pressure contributions.
 As mentioned above, the Katchalsky model does not provide valid
 predictions for high charge fractions which can be seen in the
 missing points in figure \ref{fig:charge fraction}.  Another tendency
 which can be see for some parameter combinations (e.g. in figure
 \ref{fig:salt}) is that the Katchalsky model exhibits bigger
 deviations from the periodic gel model for higher salt
 concentrations. This could be related to the fact that the
 Debye-H\"uckel approximation works only well for low ion
 concentrations (below $0.01$ mol/l for 1:1
 electrolytes\cite{hummel02a} when used for predicting the ionic
 activity coefficient). Another possible reason could be that the
 elastic pressure contribution in the Katchalsky model is independent
 of charge fraction and salt concentration (similar to the PB model
 above).

\begin{figure}
  \centering
  \includegraphics[width=0.7\textwidth]{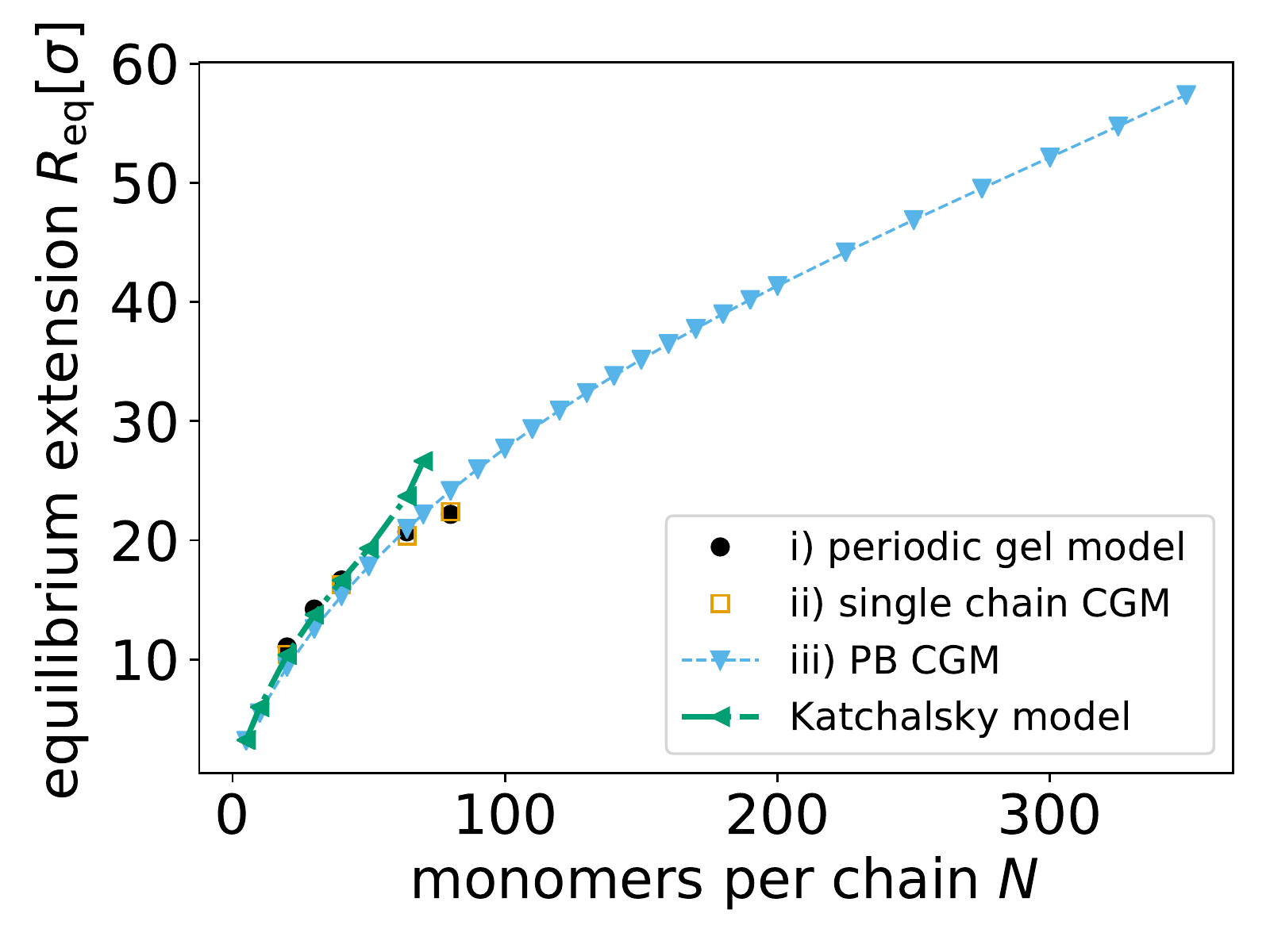}\\
%plots originate from scripts in /tikhome/jlandsgesell/phd/own_papers/2018-PB_gel_short/figures/data
  \caption{\label{fig:Req_N}Predictions
    for the equilibrium end-to-end distance of monodisperse gels of
    different chain length $N$ for $f=0.5$, $\epsilonr=80$ and
    $\cs=\unit[0.1]{mol/l}$.}
\end{figure}

Our new PB CGM also allows to obtain predictions for the
equilibrium extension of long chains in a gel which were previously too expensive to
simulate. For one set of parameters we show this prediction in figure \ref{fig:Req_N} where
the equilibrium end-to-end distance is plotted as a function of the
chain length $N$. Additionally, we also show the predictions of the single chain CGM, the periodic gel model and the Katchalsky model.
As one would expect the equilibrium end-to-end
distance increases with chain length. The exact results, in
figure \ref{fig:Req_N}, are important for the later treatment of
polydisperse gels (as explained in section \ref{sec:polydisperse
  gels}).
At this point we want to note that the Katchalsky model fails for chain lengths $N \gtrsim 80$ (at $f=0.5$) and already shows significant deviations to the periodic gel model at $N=80$. 
As reported by Richter\cite[Figure 70.]{richter17b} the electrostatic pressure contributions of the Katchalsky model are too negative for $f=0.5$ compared to the periodic gel model. These negative pressure contributions also introduce multiple zero-crossings in the $PV$ curve of the Katchalsky model for $N \gtrsim 80$ where the model fails.

\begin{figure}
\includegraphics[width=0.7\textwidth]{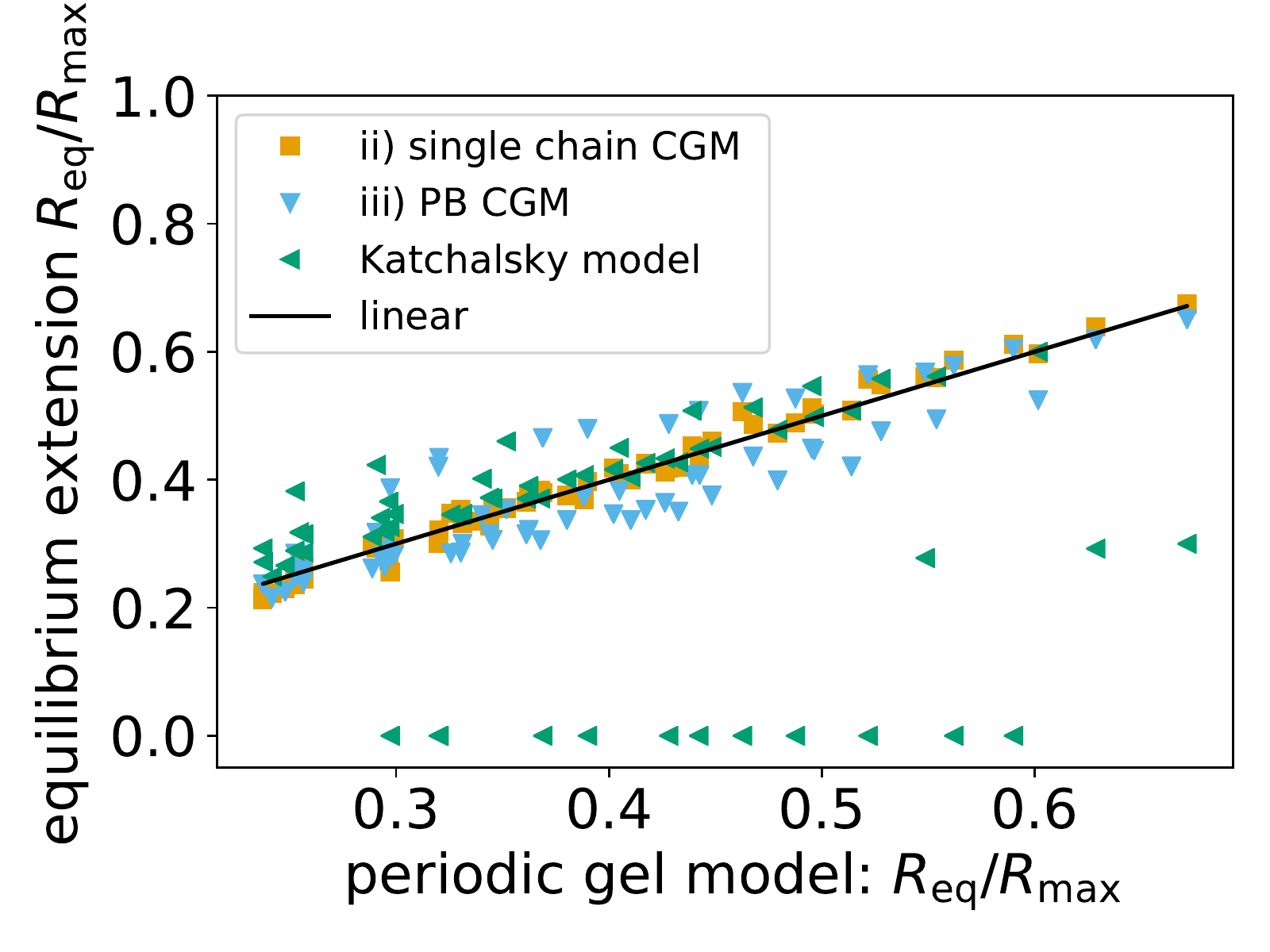}
\centering
\caption{\label{fig:correlation}The swelling equilibria
  of the single chain CGM and the PB CGM compared to the more detailed periodic gel
  model.  The results are presented for a wide set of
  parameter combinations with $\epsilonr=80$, $N \in \{40,64, 80\}$, $f \in \{0.125,0.25,0.5,1\}$ and
  \mbox{$\cs \in \{0.01,0.02,0.05, 0.1, 0.2\}$
    \si{\mole\per\liter}}. The linear function has the form
  $y(x)=x$ and is referred to as \enquote{ideal (prediction) line}.  Note that the Katchalsky model produces outliers at
  high charge fractions which go hand in hand with high degrees of
  swelling in the periodic gel model. In order to show the failure of the Katchalsky model 
  we included data points for which the Katchalsky model failed to give a prediction. For illustration purposes we assign those data points to $\Req/\Rmax=0$. We want to note that our comparison includes more charge fractions below $f=0.5$ than above. This might lead to the wrong impression that the Katchalsky model works in most of the charge fraction although it does not work for $f>0.5$. }
\end{figure}

The overall agreement between the two new models and the expensive
periodic gel simulations is evaluated in figure
\ref{fig:correlation} which is a parametric plot with the periodic gel data on the abscissa and the data of the other models on the ordinate. A straight line with slope one would indicate
perfect agreement with the periodic gel simulations (this \enquote{ideal line} is indicated with the label \enquote{linear} in figure \ref{fig:correlation}). The single chain CGM  fits the periodic gel data very good in the whole parameter space and therefore lies close to the \enquote{ideal line} (a fitted line $y(x)=mx$ through the single chain CGM data has the slope $m=1.01$ and a coefficient of determination $R^2=0.998$). The PB CGM in general has a similar trend as the periodic gel data but has deviations to the periodic gel data (a fitted line through the PB CGM data has slope $m=0.968$ and $R^2=0.98$). As outlined above, the data points where the swelling is below the \enquote{ideal line} are in tendency data at low salt concentration. PB CGM data which are above the \enquote{ideal line} are in tendency data at high charge fraction $f$.
In contrast to the Katchalsky model our new models can be applied even at high charge fractions and high salt concentrations. The Katchalsky model data above the ideal line are due to deviations at high salt concentrations, while Katchalsky model data below the ideal line are due to deviations at high charge fraction. A fitted line to the Katchalsky model data including the outliers (excluding the outliers at $f=1$) has a slope $m=0.74$ ($m=1.07$) and a coefficient of determination $R^2=0.72$ ($R^2=0.992$).

As we see in figure \ref{fig:correlation} there seems to be \enquote{scattering} around the predicted periodic gel data which serve as benchmark.
The \enquote{scattering} of the different model data around the periodic gel data in figure \ref{fig:correlation} is due to the models not working perfectly and projecting the results obtained in a three-dimensional parameter space: $\{\cs, N, f\}$ onto a one dimensional abscissa. While there is no \enquote{scattering} (in the sense of a strongly non-monotonic behavior) of the data around the ideal line for a single parameter set (e.g. $\Req(\{\cs\}, N=\mathrm{fix}, f=\mathrm{fix})/\Rmax$) there is apparent \enquote{scattering} when plotting two data sets (e.g. $\Req(\{\cs\}, N=40, f=0.125)/\Rmax$ and $\Req(\{\cs\}, N=80, f=1)/\Rmax$, see supplemental infromation) from separate regions of the parameter space together in figure \ref{fig:correlation}.
This \enquote{scattering} does not mean that predictions of the models vary in a strongly non-monotonic when only varying one parameter.

Deviations between the simplified models and the periodic gel model are either due to simplified descriptions of the interactions (as discussed above) or at least in part due to the fact that the parameter $A$ is not a constant but rather a function of the end-to-end distance\cite{kosovan15a}. This non-affine behavior exists in the periodic gel model and probably in real polymer networks\cite{broedersz14a}.
Therefore, a refined theory would also take into account that the compression of a gel is not affine and would deal with $A(\Rend)$. In the supporting information we investigate the effect of changing $A$ on the predicted end-to-end distance in the PB CGM and find, however, that the influence is small.
%Another reason for the observed deviations may also be the neglect of correlations between the different chains in the single chain models. In the periodic gel model the space charge density around a node is increased due to four chains meeting. This is a problem which is, however, not straightforward to model in the single chain CGMs. 

%%%%%%%%%%%%%%%%%%%%%%%%%%%%%%%%%%%%%
\subsection{Bulk Modulus}
%%%%%%%%%%%%%%%%%%%%%%%%%%%%%%%%%%%%%

Mechanical properties are important, for example, when evaluating the energetic costs for desalination with
polymer gels\cite{richter17a}. 
One mechanical property of interest is the isothermal bulk modulus $K$ of a gel as a measure for the mechanical strength:
\begin{equation}
K=-V \frac{\partial P_\mathrm{gel}}{\partial V} \approx -V \frac{\Delta P_\mathrm{gel}}{\Delta V}.
\label{eq:bulk modulus}
\end{equation}
In the case of the single chain CGMs we use $\Vchain$ instead of $V$.
The volume derivative of the pressure curve is approximately
calculated via the finite difference quotient of the points which are
at the intersection of the reservoir and system pressure.

The scaling behavior of the shear modulus is connected to the scaling
of the bulk modulus $K$ via\footnote{We note that our single
  chain CGM and the PB CGM have a Poisson ratio
  $\nu=-\frac{\Delta \Rout}{\Rout}\frac{L}{\Delta L} =-1$ and
  therefore the two models itself are auxetic ($\nu<0$) -
  i.e. stretching the chain enlarges the volume in the dimension
  perpendicular to the applied force. However, real gels are not
  auxetic materials. It is, therefore, important to remember that the
  single chain CGM and the PB CGM are models for a gel
  under isotropic compression: Compressing the gel reduces the volume
  which is available per chain and reduces the end-to-end distance of
  the chains in the gel.} 
the Poisson ratio\cite{demtroder04b} $\nu$:
\begin{equation}
K= \frac{2(1+\nu)}{3(1-2\nu)}G.
\end{equation}

The scaling analysis by Barrat et al.\cite{barrat92a} (which is also
based on the pressure balance of the osmotic and the elastic pressure using
the ideal Donnan equilibrium) predicts that in swelling
equilibrium the shear modulus of a polymer gel is given by
$G=\kT \left( \frac{c_m}{N}\right) \left(\frac{\Req^2}{N
    b^2}\right)$, where $c_m=N/\Req^3$ is the monomer concentration in
the gel in equilibrium\cite{barrat92a}, $N$ the number of monomers per chain, and $b$
the size of the monomer. The equilibrium end-to-end distance scales
with the number of monomers per chain
$\Req \propto N^{\nu_\text{gel}}$, where $\nu_\text{gel}$ is a
Flory exponent. Note that the Flory exponent of a free chain and
the Flory exponent $\nu_\text{gel}$ do not agree in general
(compare Barrat et al.\cite{barrat92a}).  Using this we obtain
$G\propto N^{-(\nu_\text{gel}+1)}$.  In salt free solution we further
have the following relation for the equilibrium end-to-end distance in
a gel\cite{barrat92a} $\Req \propto N b$. Therefore, we expect
$K\propto G \propto 1/N^2$. For a gel in contact with
a saline solution and in a good solvent we expect\cite{barrat92a}
$\Req \propto b N^{3/5}$ which therefore alters the scaling prediction
$K\propto G\propto N^{-8/5}=N^{-1.6}$.

\begin{figure}
    \centering
  \begin{subfigure}[t]{0.45\textwidth}
  \includegraphics[width=1.\columnwidth]{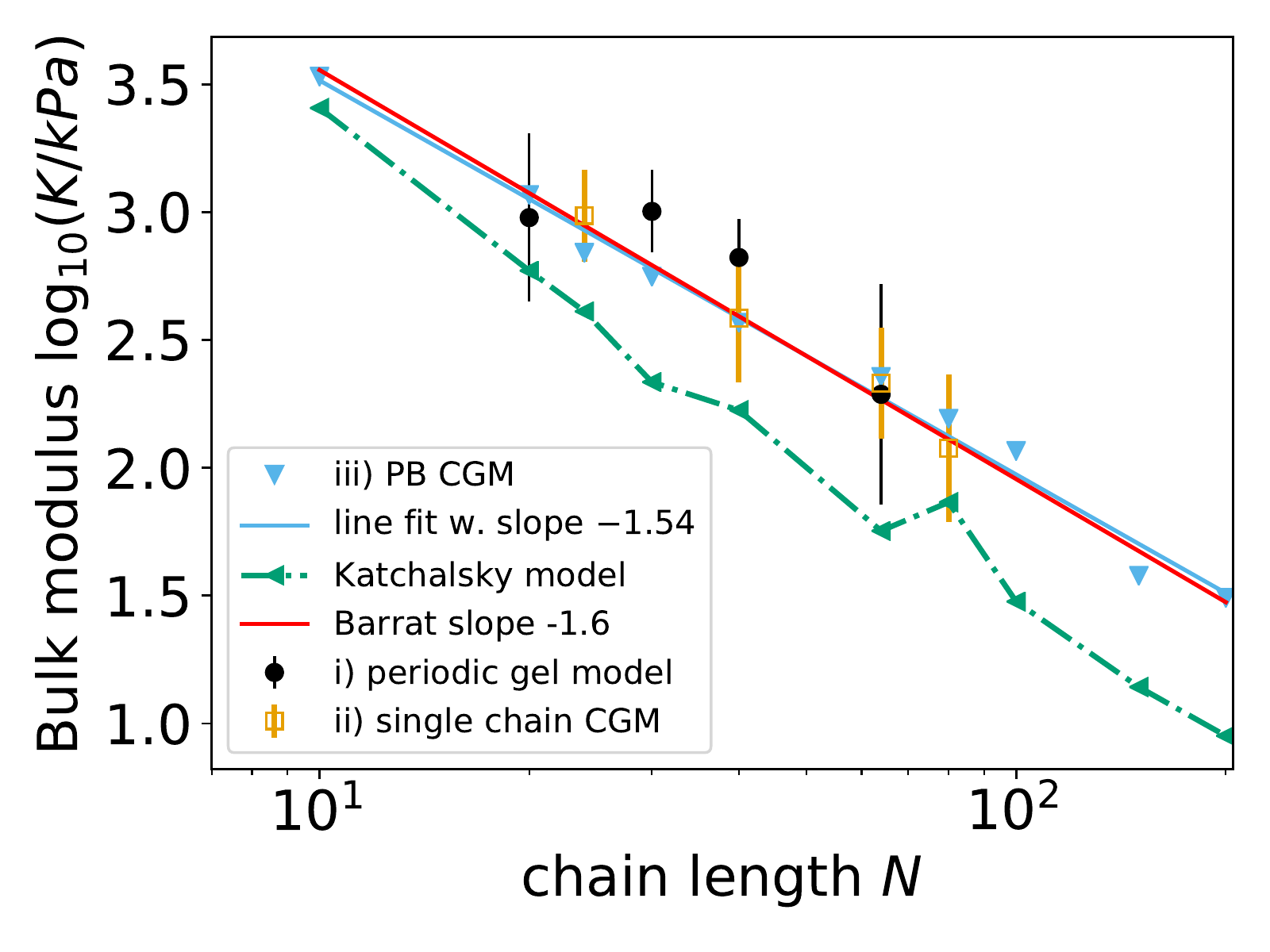}\\
%plots originate from scripts in /tikhome/jlandsgesell/phd/own_papers/2018-PB_gel_short/figures/data
%analysis for katchalsky is taken from /tikhome/jlandsgesell/phd/own_papers/paper_draft_poisson_boltzmann_cell_under_tension/Katchalsky_Model/analyze_pressures_and_get_bulk_modulus.py
  \caption{\label{fig:bulk moduli 0125}}
  \end{subfigure}\hfill% 
  
  \begin{subfigure}[t]{0.45\textwidth}
  \includegraphics[width=1\columnwidth]{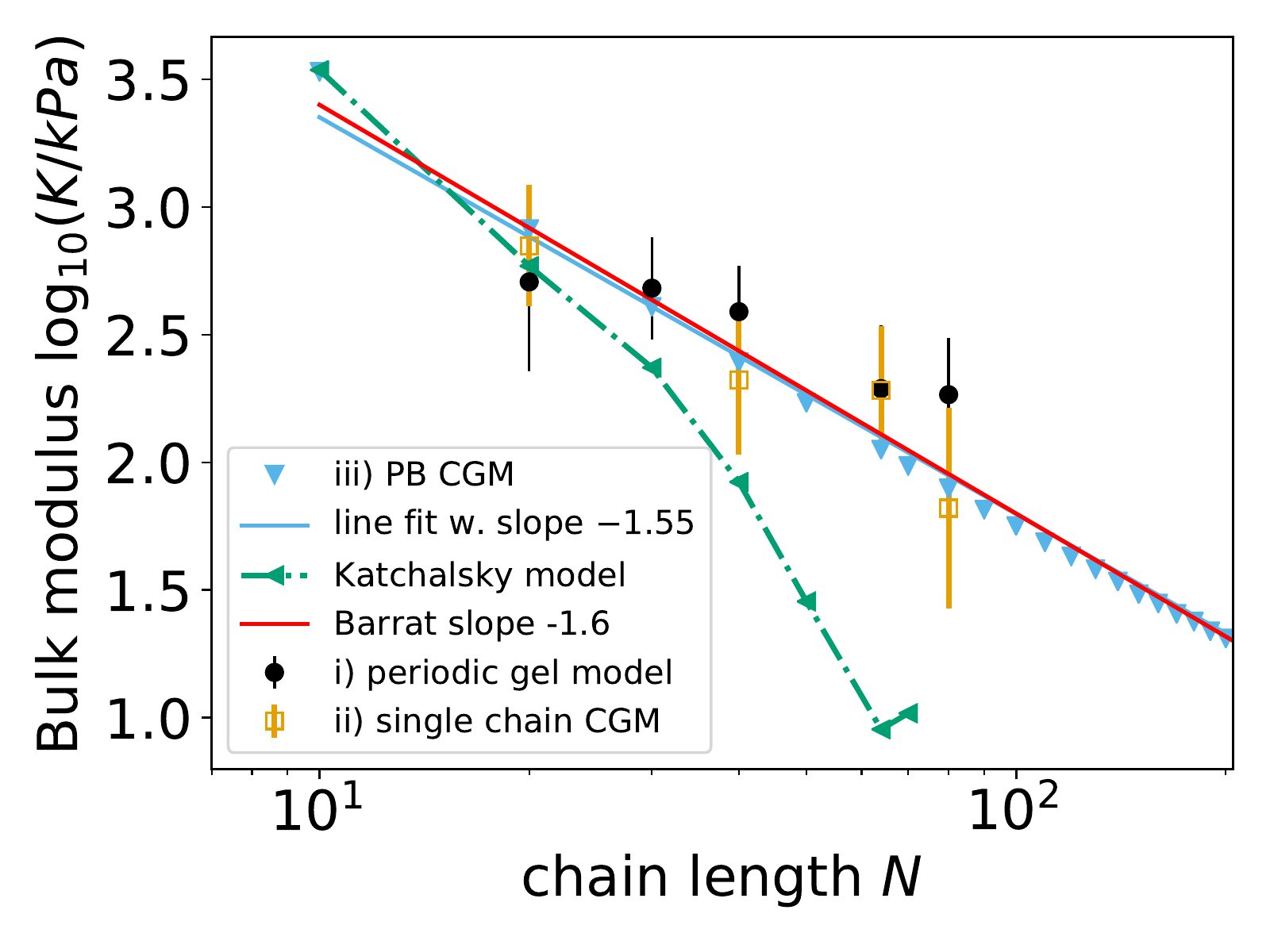}
  \caption{\label{fig:bulk moduli 05}}
  \end{subfigure}
  
  \begin{subfigure}[t]{0.45\textwidth}
  \includegraphics[width=1\columnwidth]{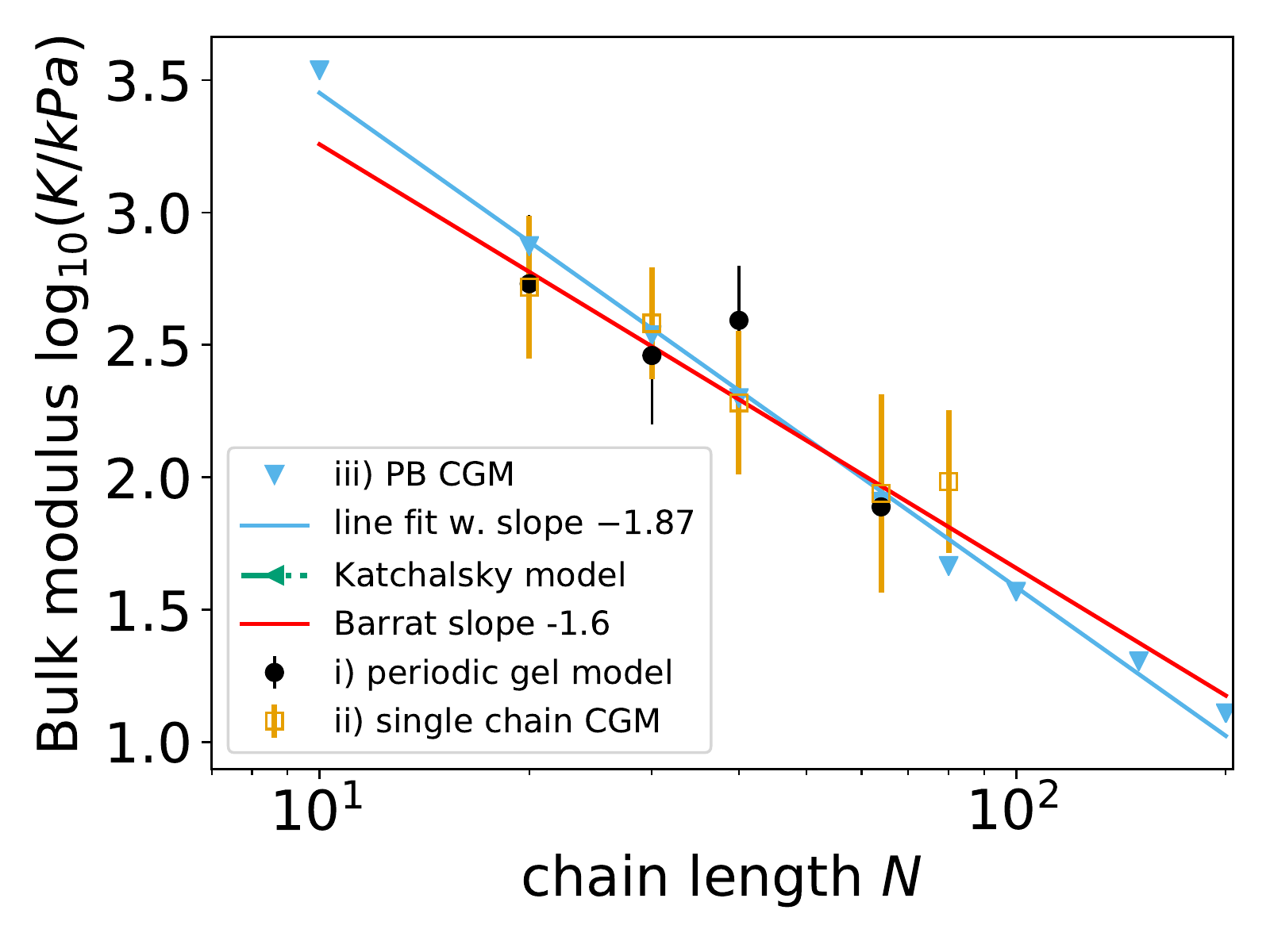}\\
  \caption{\label{fig:bulk moduli 1}}
  \end{subfigure}
  
\caption{\label{fig:bulk moduli}Predictions of periodic gel model i), the two single chain CGMs
  ii--iii) and the Katchalsky model for the bulk moduli $K$ of
  monodisperse gels of different chain length $N$ for a) $f=0.125$, b) $f=0.5$ and c) $f=1$ at
  $\epsilonr=80$ and $\cs=\unit[0.1]{mol/l}$. For $f=1$ the Katchalsky model fails and therefore there are no data points.}
\end{figure}

The bulk modulus obtained by the PB CGM is displayed in figure \ref{fig:bulk moduli} together with
values obtained from the single chain CGM, the periodic gel
model, and the Katchalsky model. For the two particle-based models the $PV$-curve and the errors in the pressure are recorded during the simulation. 
The resulting error in the bulk modulus is then calculated according to error propagation in the volume $V$ and the slope $\partial P_\mathrm{gel}/\partial V$.
The used formula is:
\begin{equation}
\Delta K=\left|\left(\frac{\partial P_\mathrm{gel}}{\partial V}\right)_{\mathrm{eq}} \right|\Delta V_\mathrm{eq}+V_\mathrm{eq} \Delta \left(\frac{\partial P_\mathrm{gel}}{\partial V} \right)_\mathrm{eq},
\end{equation}
where the symbol $\Delta$ denotes that the error margin is positive.
The error margins $\Delta V_\mathrm{eq}$ and $ \Delta \left(\frac{\partial P_\mathrm{gel}}{\partial V} \right)_\mathrm{eq}$ are determined using the error bars of the pressure next to the equilibrium point. 

In our PB CGM we find scaling exponents $K \propto N^{-1.54\pm 0.06}$ (for $f=0.125$), $K \propto N^{-1.55\pm 0.03}$ (for
$f$=0.5) or $K \propto N^{-1.87\pm 0.06}$ (for
$f$=1) via fitting\footnote{The errors
($\pm$) are the standard deviations of the slope obtained from the
square root of the corresponding entry in the covariance matrix of the
fit. For fitting a line in the log-log plot the Levenberg-Marquardt
algorithm was used.} to data $N > 5$.
The PB CGM data for the bulk modulus do not follow a perfect power law as predicted by Barrat\cite{barrat92a}, there are deviations at very small chain lengths and large chain lengths. 
Therefore, the scaling exponents for the $N$ dependence of the bulk modulus should be taken with care.
We observe that the scaling exponents of the bulk modulus with $N$ are close
to the scaling prediction $K\propto G \propto N^{-1.6}$ when the gels carry a low charge fraction.
%The following numbers are outdated. However, they do not provide additional insight, therefore removing them
%This behavior corresponds to a Flory exponent of
%$\nu_\text{gel}=1.08\pm 0.1$ for $f=1$,
%$\nu_\text{gel}=0.72\pm 0.06$ for $f=0.5$ or
%$\nu_\text{gel}=0.72\pm 0.09$ for $f=0.125$ (using
%$K\propto N^{-(\nu_\text{gel}+1)}$ ) compared to the expected behavior
%of $\nu_\text{gel}=3/5=0.6$.  
We also find that the PB CGM, the
single chain CGM, and the periodic gel model agree within error bars. 
The Katchalsky model, on the other hand, shows significant deviations to the periodic gel model, it deviates from the particle-based model predictions both at charge fraction $f$=0.125 and $f$=0.5 or even fails at $f=1$ (where we have no Katchalsky model prediction). For low charge fractions the slope of the Katchalsky data is still compatible with Barrat's scaling prediction while for charge fraction $f=0.5$ the deviation to Barrats scaling prediction and the periodic gel data is already significant. 
We conclude that our new
models provide an improved description of the mechanical properties of gels at intermediate or high charge fractions ($f\geq 0.5$) when compared to the Katchalsky model.

%The PB CGM bulk modulus of chains of length $N=5$ deviate significantly from the best fit power law (for all charge fractions) and at chain lengths $N>200$. 
We also want to note that the error bars on the bulk modulus for the particle-based models (obtained via propagation of error) are big.  Due to this fact, we do not fit scaling exponents to the particle-based model data. It seems that in figure \ref{fig:bulk moduli} b) the periodic gel model would have a different best fit line than the PB CGM. One possible reason for this could be that the PB CGM uses a stretching pressure which is derived from an ideal model, neglecting a possible salt dependence or charge fraction dependence of this pressure contribution. A chain length dependent correction to the ideal behaviour would also change the bulk modulus predicted with equation \eqref{eq:bulk modulus}.

\begin{figure}
  \includegraphics[width=0.7\columnwidth]{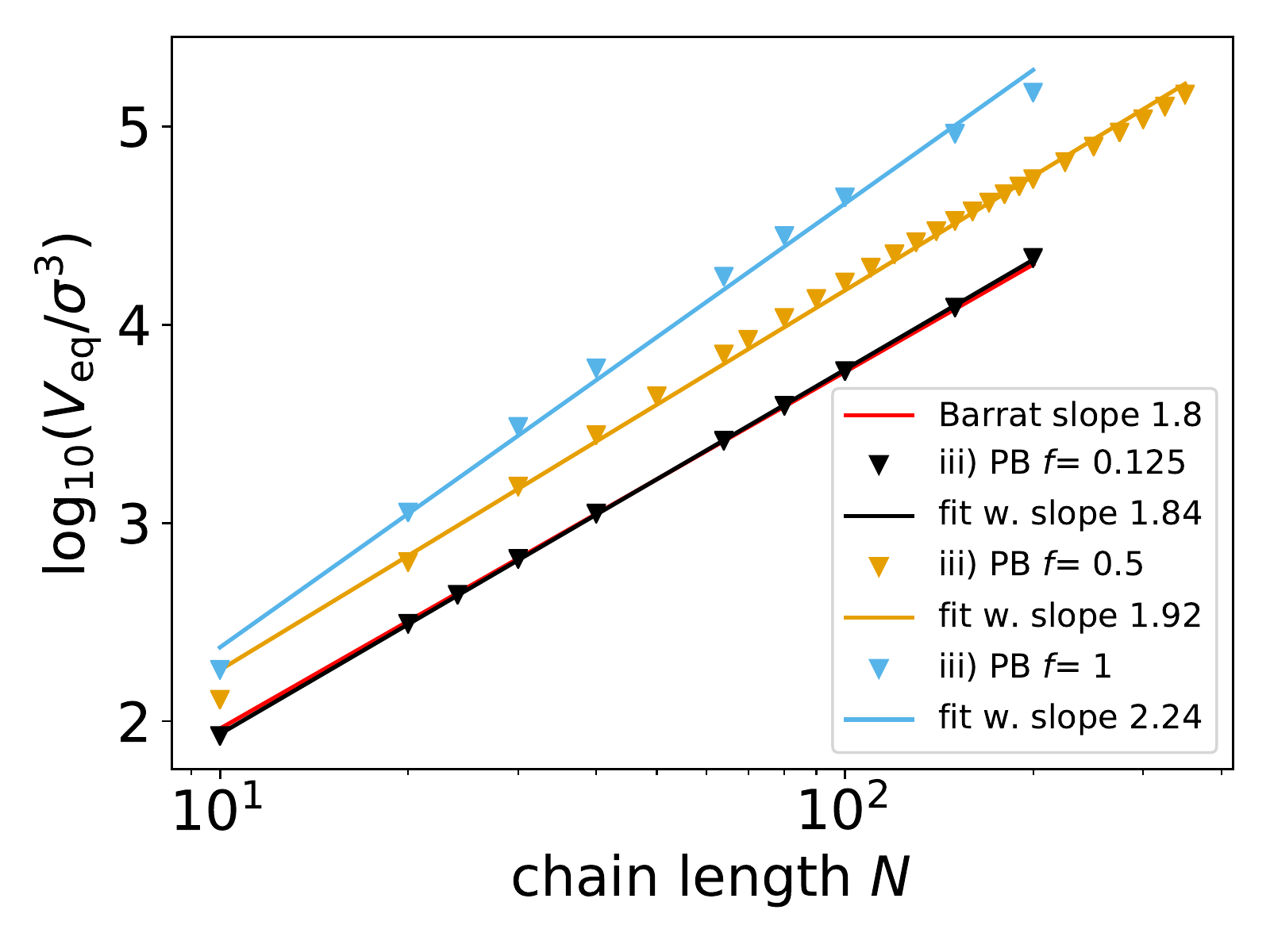}\\
%plots originate from scripts in /tikhome/jlandsgesell/phd/own_papers/2018-PB_gel_short/figures/data
  \caption{\label{fig:gel volume scaling} PB CGM
    predictions for the chain length ($N$) dependence of the volume per chain of monodisperse gels at $\epsilonr=80$ and $\cs=\unit[0.1]{mol/l}$. The different charge fractions $f$ are encoded via color as indicated in the legend. The volume is calculated according to eq. \eqref{eq:chain volume approx} using the equilibrium end-to-end distance. The scaling prediction by Barrat is shown in red (with fitted y-intercept to the PB model data for $f=0.125$). We expect the idealized model by Barrat\cite{barrat92a} to work best for low charge fractions.}
\end{figure}

In figure \ref{fig:gel volume scaling} we also display the volume at
swelling equilibrium as a function of $N$. Like Barrat, we assume that the
volume of the gel is proportional to the end-to-end distance
cubed\cite{barrat92a}. We expect the following scaling behavior in swelling equilibrium 
$V_\mathrm{eq}\propto N^3$ (in the salt free case) and
$V_\mathrm{eq} \propto N^{9/5} \approx N^{1.8}$ (in the case of added
salt)\cite{barrat92a}. As one can see in figure \ref{fig:gel volume
  scaling} the PB CGM predicts a scaling of
$\Veq \propto N^{2.24\pm 0.065 }$ (for $f=1$),
$\Veq \propto N^{1.92 \pm 0.03}$ (for $f=0.5$) and
$\Veq \propto N^{1.84\pm 0.01}$ (for $f=0.125$) which is close to
Barrat`s prediction for added the salt ($\Veq \propto N^{9/5}=N^{1.8}$).
In the case of high charge fraction $f=1$ we do not expect the
model by Barrat to work anymore since ions are treated on the ideal level\cite{barrat92a}. Therefore, the
scaling exponent in the PB model is different to the prediction by Barrat\cite{barrat92a}. 
Since $\Veq \propto \Req^3 \propto N^{3\nu_\text{gel}}$, we find \enquote{effective}
Flory exponents $\nu_\text{gel}=0.613 \pm 0.01$ (for $f=0.125$), $\nu_\text{gel}=0.64 \pm 0.01$ (for $f=0.5$) or $\nu_\text{gel}=0.75 \pm 0.03$ (for $f=1$).
For highly charged gels the electrostatic interactions stretch the gel more and the swelling increases (resulting in a higher $\nu_\text{gel}$). 
%These values 
%differ slightly from the ones obtained previously
%using the bulk moduli, but are close when considering error bars.  The
%higher errors in the values of $\nu_\text{gel}$ which are obtained via
%the bulk moduli can partly be explained by the use of the finite
%difference quotient of the $PV$ curve in eq. \eqref{eq:bulk modulus}
%which adds a source of error that depends on the spacing of sampling
%points in the $PV$ curve.

%%%%%%%%%%%%%%%%%%%%%%%%%%%%%%%%%%%%%
\subsection{Influence of the Relative Permittivity}
%%%%%%%%%%%%%%%%%%%%%%%%%%%%%%%%%%%%%

\begin{figure}
  \centering
  \includegraphics[width=0.7\textwidth]{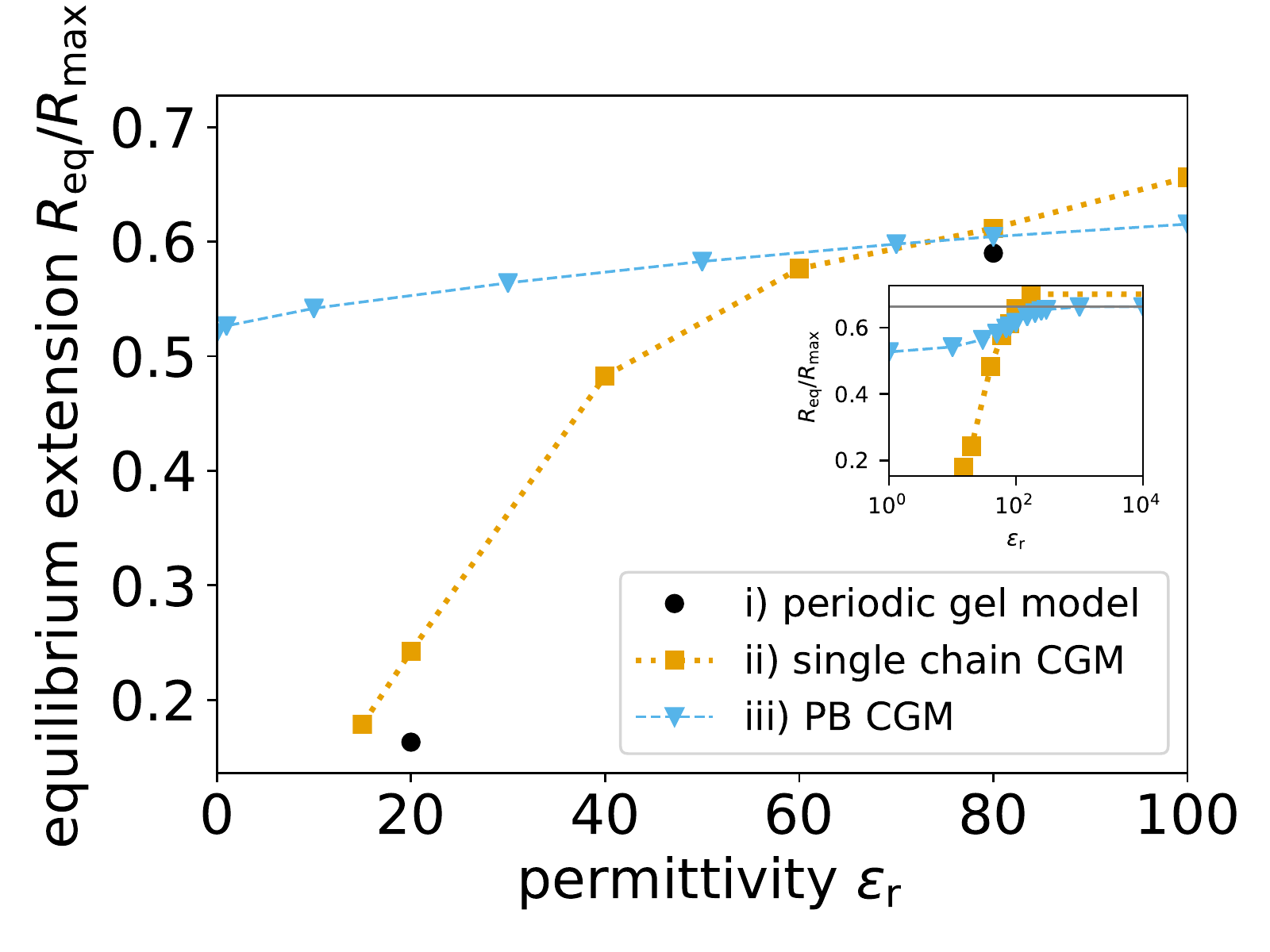}\\
%plots originate from scripts in /tikhome/jlandsgesell/phd/own_papers/2018-PB_gel_short/figures/data
  \caption{\label{fig:Req_div_Rmax_epsilon}PB CGM and single chain CGM predictions for the equilibrium
    end-to-end distance of monodisperse gels for $N=64$, $f=1$ and
    $\cs=\unit[0.01]{mol/l}$ at various different relative
    permittivities $\epsilonr$. Additionally, we confirm the accuracy of the
    single chain CGM simulations by providing two periodic gel data points
    for the swelling equilibrium at $\epsilonr=80$ and
    $\epsilonr=20$. The inset shows the limiting behavior of the
    PB CGM and the single chain CGM for large relative
    permittivities. In this limit salt is partitioned according to the
    ideal Donnan behavior (grey line) for the PB CGM. The data point
    at $\epsilonr=\infty$ for the single chain CGM is
    obtained by switching the electrostatic interactions off. }
\end{figure}

The relative permittivity $\epsilonr$ controls the strength of
the electrostatic interaction in the implicit solvent approach.
Around $\epsilonr\approx 80$ both, the single chain CGM and
the PB CGM, agree well in their prediction for the swelling
equilibrium (see figure
\ref{fig:Req_div_Rmax_epsilon}).  The results for the PB CGM
and the single chain CGM are shown in figure
\ref{fig:Req_div_Rmax_epsilon}.
For lower values of $\epsilonr$ the electrostatic interactions become
so strong that ion correlation effects occur \cite{naji06b}. These
correlation effects cannot be captured by the mean-field
Poisson-Boltzmann approach \cite{naji06b} and produce a stronger
collapse of the gel than predicted by the PB CGM.

The effect of varying $\epsilonr$ was previously investigated (for periodic gel model)
by Schneider et al.~\cite{schneider04a} who
observed a similar trend. We also confirm the accuracy of the single
chain CGM by comparing our data to those of periodic gel simulations at
$\epsilonr=80$ and $\epsilonr=20$.

At larger relative
permittivities, electrostatic interactions diminish. In the limit of $\epsilonr \to \infty$ the
electrostatic interaction energy approaches zero, and the salt is partitioned
according to the ideal Donnan prediction\cite{kosovan15a}:
\begin{equation}
c_\mathrm{salt}^\mathrm{gel}=\sqrt{\left(\frac{fc_m}{2}\right)^2 +{\cs}^2 }-\frac{fc_m}{2}.
\end{equation}
Here $c_\mathrm{salt}^\mathrm{gel}$ is the salt concentration inside the gel, $\cs$ the salt concentration in the reservoir and $c_m$ the concentration of monomers in the gel.

This salt partitioning allows to predict the behavior in
the limit $\epsilonr \to \infty$ since then ions can be treated as an
ideal gas, and the pressure contributions $\Pcap$ and $\Pside$ can be
evaluated easily via the homogeneous densities (the electric field goes to zero for
$\epsilonr \to \infty$). The Donnan partitioning is used for
calculating the cell model pressures and yields the grey line in the
inset in figure \ref{fig:Req_div_Rmax_epsilon}.  In figure
\ref{fig:Req_div_Rmax_epsilon} it is visible that the PB CGM and the
single chain CGM swell slightly different for $\epsilonr \to
\infty$. This difference of roughly 6 \% in $\Rend/\Rmax$
appears due to different stretching pressures in both models and possibly due to the neglect of excluded volume interactions in the PB model.

We want to point out that the strong collapse in solvents with low
dielectric constant cannot be accurately represented by theories which
treat the electrostatic interactions only on an approximate ideal
level (ensuring electroneutrality via Donnan equilibrium) or on a
Debye-H\"uckel level, as is done for example in the approaches by
Katchalsky \cite{katchalsky55a}, Tanaka \cite{ricka84a}, Khokhlov
\cite{khokhlov93a}, or the model by Barrat\cite{barrat92a}.

% \FloatBarrier
%%%%%%%%%%%%%%%%%%%%%%%%%%%%%%%%%%%%%
\subsection{Mass-Based Degree of Swelling (for Monodisperse Gels)}
%%%%%%%%%%%%%%%%%%%%%%%%%%%%%%%%%%%%%
%\FloatBarrier
For better comparison with experimental results it is instructive to
introduce a mapping from the predicted end-to-end distances $\Req$ to the mass the gel has in swelling
equilibrium. This facilitates the comparison of experimental results where
the swelling equilibrium is often reported as a fraction of the mass
of the solvent in the swollen state divided by the mass of the dry
state of the gel $Q_m=m_\text{solvent uptake}/m_\text{dry}$.  This mapping requires some modeling
assumptions. First, the mass of the dry state is determined by
the number of monomers $N$ per chain, the total number of chains $N_\mathrm{chains}$ in the gel and the mass of the monomers $m_0$:
\begin{equation} 
m_\text{dry}=m(N) N_\mathrm{chains},
\end{equation} 
where $m(N)=N m_0$ is the mass of one chain with N monomers. The monomer mass depends on the experimental preparation and we choose $m_0=$\unit[94]{u} so
that the monomers represent sodium acrylate. The second modeling
assumption is that we can measure the mass of the solvent in the
swollen gel via its volume\footnote{Being exact one would need to
  subtract the volume which is occupied by the gel monomers from
  $V(N)$. For swollen gels the excluded volume by the gel is, however,
  negligible.} and an approximate density $\rho_\mathrm{w}\approx \unit[1]{kg/l}$ which is close to that of
pure water:
\begin{equation} 
m_\text{solvent uptake} =\Vchain(N)\rho_\mathrm{w} N_\mathrm{chains},
\end{equation}
where $\Vchain(N)$ is the volume per chain given by eq. \eqref{eq:chain volume approx}. This volume per chain is predicted, e.g., by the single chain CGM or
the PB CGM (compare figure ~\ref{fig:Req_N}).
The volume of the gel is obtained by multiplying $\Vchain$ with the number of chains $N_\mathrm{chains}$ (which cancels in the calculation of $Q_m$).

This simple mapping allows for the prediction of mass-based
degrees of swelling\cite{arens17a}. The results for monodisperse gels
are shown in figure \ref{fig:mass_Q_N} as a black
curve. 
%It is promising that the range of predicted swelling
%equilibria is within the experimentally reported
%range~\cite{}.
In the following, we generalize our mapping of the (mass
based) degree of swelling and introduce polydispersity effects.
%\FloatBarrier
%%%%%%%%%%%%%%%%%%%%%%%%%%%%%%%%%%%%%
\subsection{Chain Length Polydispersity}
%%%%%%%%%%%%%%%%%%%%%%%%%%%%%%%%%%%%%
\label{sec:polydisperse gels}
%\FloatBarrier

So far we have only treated monodisperse macroscopic gels. In reality,
however, most gels are highly
polydisperse\cite{panyukov96b,svaneborg05a,gavrilov14a,tehrani17a}.
Properties of polymeric systems typically depend on the chain length
polydispersity and in many applications this feature may be used to
improve material properties for a given
application\cite[p.10]{strobl07a}.  Nevertheless, polydispersity is
often negelected in theoretical considerations and a monodisperse
molar mass distribution is assumed.

\begin{figure}
  \centering
  \includegraphics[width=0.7\textwidth]{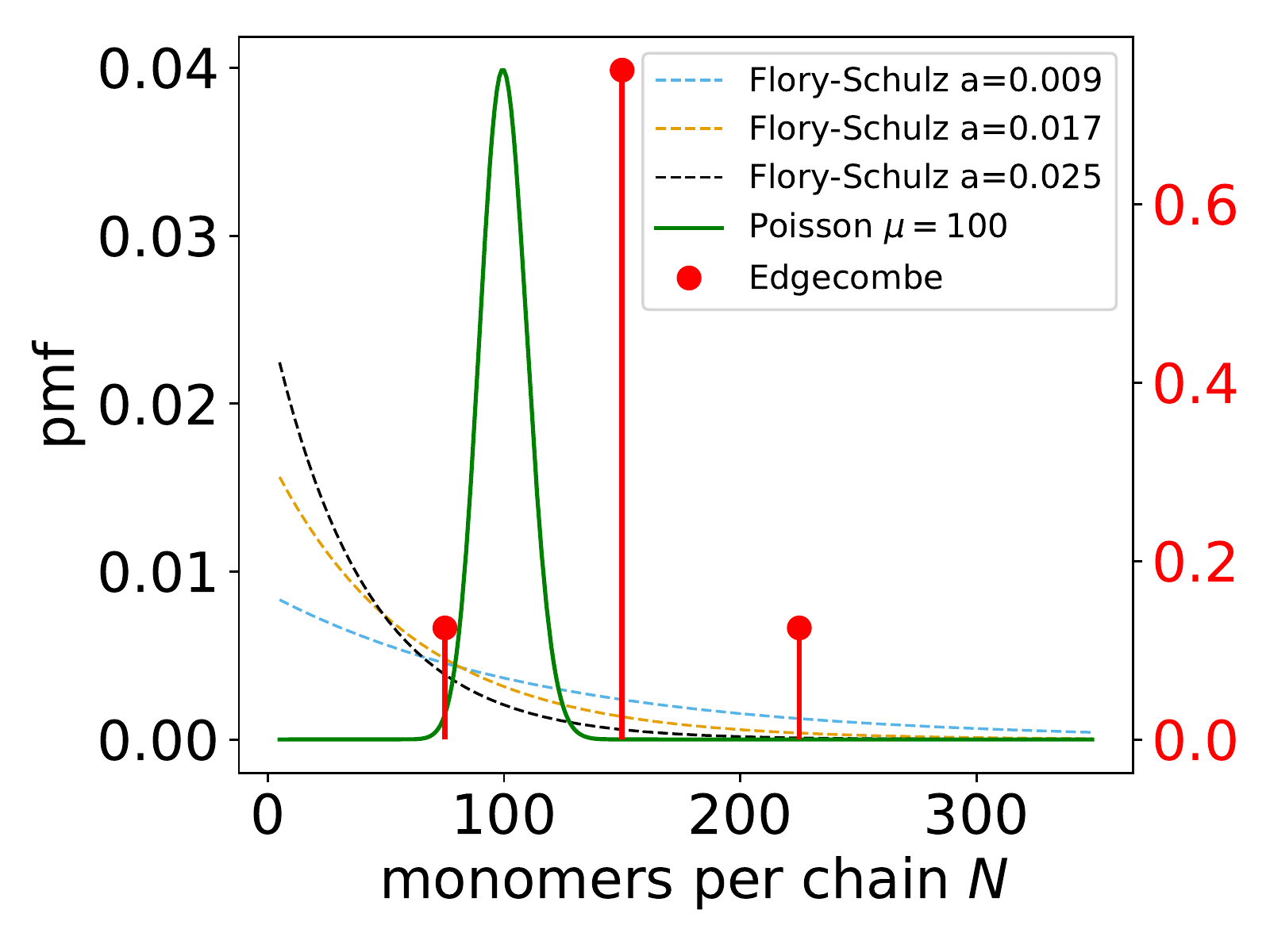}
%plots originate from scripts in /tikhome/jlandsgesell/phd/own_papers/2018-PB_gel_short/figures/data
  \caption{The figure shows the different probability mass functions which we use in this paper: the geometrical distribution (Flory-Schulz), the Poisson distribution and the distribution in accordance with Edgecombe\cite{edgecombe07a}. The Flory-Schulz distribution is shown for different
    parameters $a$. The pmfs are only defined for integer values $N$, the lines
    serves only as a guide to the eye. The three shown values of the
    parameter $a$ are the minimal used $a$, the maximally used $a$ and
    the median of the used values of $a$. The values of $a$ which we
    chose, all guarantee that at least 90 \% of the chains have a chain
    length between $N=5$ and $N=350$ for which we have data
    $\Req(N)$ (see figure \ref{fig:Req_N}). This means we
    choose values of $a$ such that $\sum_{N=5}^{350} p_a(N)>0.9$. Note that the distribution from Edgecome is only nonzero for three values of $N$ and therefore the probability for those values is bigger than usual compared to the other distributions. Therefore, the pmf of Edgecombe has its own ordinate axis in red.}
\label{fig:pmfs}
\end{figure}

Chain length heterogeneity is a parameter for which data for
comparison cannot be easily obtained via MD/MC simulations of periodic charged
gels: A simulation of a heterogeneous gel would require prohibitively time-consuming
simulations with a huge simulation box to realize a representative
distribution of the chain length distribution. Additionally, a strict
comparison to experiments is difficult because the exact chain length
distribution in typical experiments is unknown due to the insufficient
characterization of the topology of real gels. 
We employ here a simple approach where we assume that a polydisperse macroscopic gel can be
described by partitioning the total volume into sub-cells which
contain chains of possibly different length in each cell. 
%We previously used this approach successfully to describe monodisperse gels\cite{landsgesell19b}.
Given that we know the probability mass function (pmf) for the polymer
length in a gel $p(N)$ (see e.g. Panyukov and Rabin\cite{panyukov96b})
we can then easily obtain the degree of swelling for a polydisperse
gel.  In the studies by \cite{gavrilov14a} and \cite{svaneborg05a}
the strand length distribution is exponential or close to exponential
and therefore very broad.  As outlined by \cite{tehrani17a} random
crosslinking %is similar to condensation reactions during bulk polymerization and
gives a Flory-Schulz distributed chain length polydispersity, i.e. a
geometrical distribution or in good approximation an exponential
distribution\cite{tehrani17a}.
%This especially means that the crosslinks are very unevenly distributed along the segments.
%Tehrani \etal recently proposed a model which takes into account polydispersity \cite{tehrani17a,tehrani17b,tehrani17c-pre,tehrani17d-pre,tehrani18a}. 
For simplicity, we assume that the chain length is
distributed\footnote{The monodisperse case for a gel with chain length
  $\tilde{N}$ is simply given by the pmf $p(N)=\delta_{N, \tilde{N}}$,
  where $\delta_{i,j}$ is the Kronecker delta.} according to this
geometric or Flory-Schulz distribution\cite{flory36a,mcnaught97a}:
\begin{equation}
p_a(N)=a (1-a)^{N-1},
\end{equation}
where $0<a<1$ is a fit parameter for a given gel and
$N \in \mathbb{N}$ is the number of monomers per chain. An illustration of the corresponding probability mass function
can be found in figure~\ref{fig:pmfs}.

%For demonstration purposes (without physical motivation) we also investigate a binomial distribution $p_{N_\mathrm{max}}(N) = \binom{N_\mathrm{max}}{N} p^N (1-p)^{N_\mathrm{max}-N}$, where we choose $p$=0.5 for maximal broadness of the distribution.
%For the parameter $N$ we have $N\in{0,...N_\mathrm{max}}$.
For demonstration purposes we also investigate a Poisson distribution $p_\mu(N)=\exp(-\mu) \frac{\mu^N}{N!}$, where $\mu>0$ is the average value $E(N)=\mu$ and the variance $\mathrm{Var}(N)=\mu$. This narrow distribution of chain lengths is for example obtained via chain polymerizations\cite{strobl07a}. 
Finally, we also investigate the distribution that was used by \cite{edgecombe07a}:

\begin{equation}
p_{N_\mathrm{short},N_\mathrm{medium},N_\mathrm{long}}(N)=\begin{cases}
1/8 \text{ if N=} N_\mathrm{short}\\
3/4 \text{ if N=} N_\mathrm{medium}\\
1/8 \text{ if N=} N_\mathrm{long}\\
0 \text{ else.} \\
\end{cases}
\end{equation}
For the different chain lengths which occur in this probability mass function, we have the restriction $\frac{1}{2}(N_\mathrm{short}+N_\mathrm{long})=N_\mathrm{medium}$ \cite{edgecombe07a}. We always choose $N_\mathrm{short}=0.5 N_\mathrm{medium}$ and $N_\mathrm{long}=1.5 N_\mathrm{medium}$ like Edgecombe \etal

Using the geometrical pmf, the Poisson pmf or the Edgecombe pmf we weight the contributions of cells with
different polymer chain lengths in order to predict the
(mass-based) degree of swelling of the gel: The degree of swelling for
a polydisperse gel is given via the following ratio
\begin{equation}
  \langle Q_m \rangle_N=\frac{\langle m_\text{solvent uptake} \rangle_N}{\langle m_\text{dry}\rangle_N}, 
\label{eq:polydisperse degree of swelling}
\end{equation}
where the dry mass of the gel is:
\begin{equation*}
  \langle m_\text{dry} \rangle_N=\sum_{N} p_a(N)m(N) N_\mathrm{chains},
\end{equation*}
where $m(N)=N\cdot m_0$ and $N_\mathrm{chains}$ are as before. The mass of the solvent in the swollen state of the gel is
similarly calculated:
\begin{equation*}
\langle m_\text{solvent uptake} \rangle_N=\sum_{N} p_a(N)\Vchain(N)\rho_\mathrm{w} N_\mathrm{chains},
\end{equation*}
where again $\rho_\mathrm{w}\approx \unit[1]{kg/l}$ and $\Vchain(N)$ is given by eq. \eqref{eq:chain volume approx} (see figure ~\ref{fig:Req_N} for values of $\Req(N)$, for other values of $N$ we
use the linear interpolation in between those points). As before, $N_\mathrm{chains}$ cancels from the calculation of $\langle Q_m \rangle_N$.

\begin{figure}
  \centering
  \includegraphics[width=0.68\textwidth]{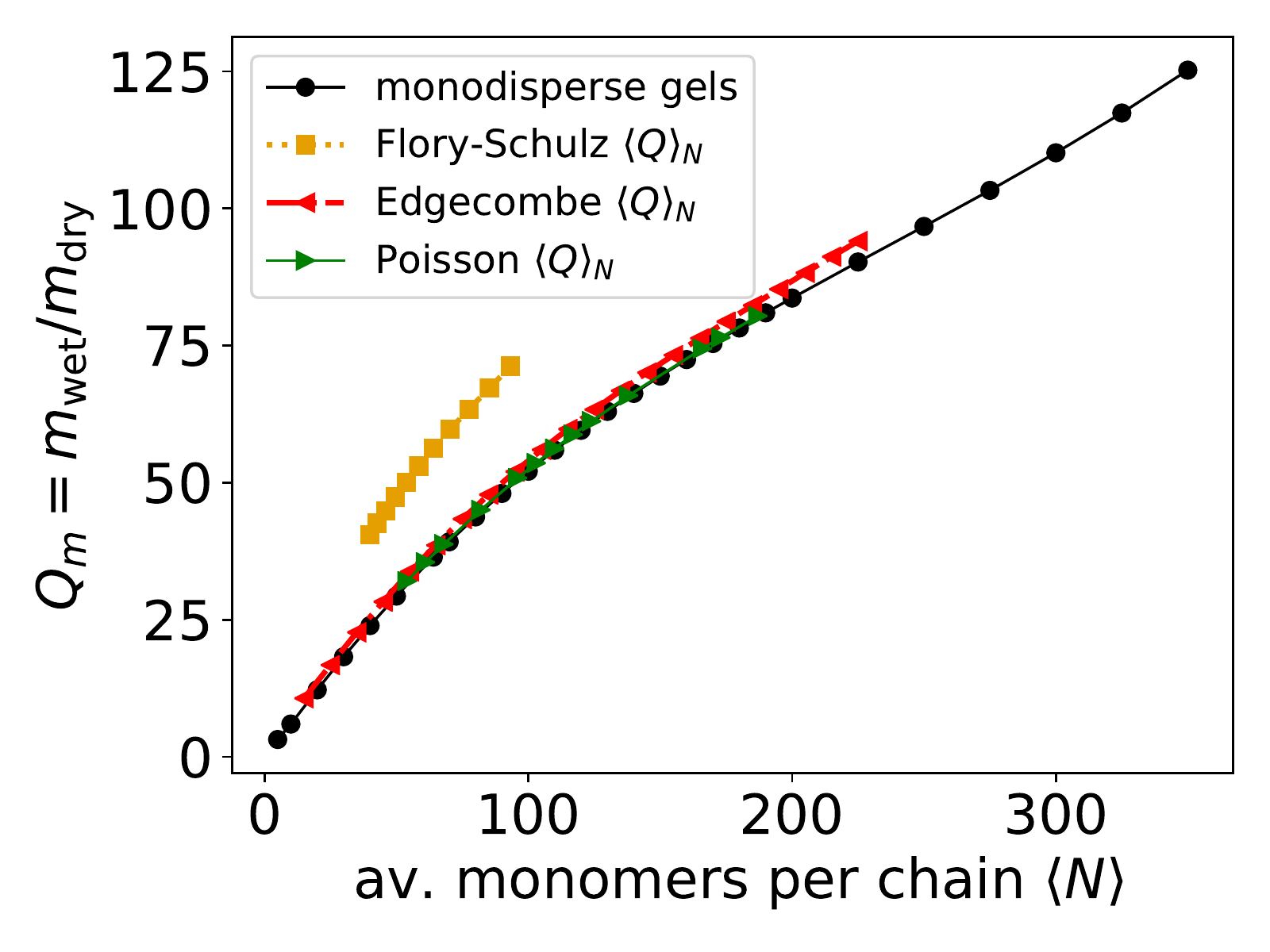}
%plots originate from scripts in /tikhome/jlandsgesell/phd/own_papers/2018-PB_gel_short/figures/data
\caption{PB CGM prediction (for $f=0.5$, $\cs=0.1$
  mol/l, $\epsilonr=80$): The figure shows the (mass-based) degree
  of swelling $Q_m$ as a function of the average chain length $\langle N \rangle$. In black the results for different monodisperse gels are shown.
  For the same charge fraction and reservoir salt concentration, the plot also shows the
  average mass-based degree of swelling $\langle Q_m \rangle_N$ for polydisperse gels (orange, red, green curves)
  plotted over the average chain length $\langle N \rangle_N$. }
\label{fig:mass_Q_N}
\end{figure}

In figure \ref{fig:mass_Q_N} we compare the degree of swelling of monodisperse gels to polydisperse
gels with the same average chain length.
We find that the swelling of polydisperse gels highly depends on the chain length distribution. 
We observe an increased degree of swelling for polydisperse gels which have a geometric chain length distribution. The fact, that the introduced Flory-Schulz polydispersity increases the degree of swelling can be understood as
an effect of the tail of the pmf: gels with
longer chains swell more. Close to no increase in the degree of swelling is observed for a gel with a chain length distribution which is of Poisson type or follows the distribution by Edgecombe\cite{edgecombe07a}.
Since the Poisson distribution is rather sharp it is expected\cite{strobl07a}, and observed in our simple model, that the polydispersity has only little influence on the mass based degree of swelling.
We want to emphasize that \cite{edgecombe07a} reported a decrease in the swelling of his polydisperse gel model compared to the monodisperse gel in contrast to our approach. This difference could be a result of strong correlations in their gel model originating basically from the small unit cell they consider - inter chain correlations are overestimated. Additionally, it is not clear how different chain length heterogeneities like e.g. a Flory-Schulz distribution would impact the results reported by \cite{edgecombe07a}. Representing a broad chain length distribution requires to simulate many chains in a unit cell, hence a huge computational effort.  In contrast to the simulations of Edgecombe, our simplistic model for accounting chain length polydispersity lacks inter-chain correlations which seem to play an important role\cite{broedersz14a}. Measuring the correlations and accounting for them in simulations remains an open task for theorists as well as experimentalists\cite{broedersz14a}.
We conclude that there is a discrepancy between the reported decrease in swelling by Edgecombe \etal \, for polydisperse gels and the data shown in figure \ref{fig:mass_Q_N} for the chain length distribution of Edgecombe in our simple model. Our simplistic model for accounting chain length polydispersity lacks inter-chain correlations (in contrast to the simulations of Edgecombe). Therefore, correlations between the stretching state of different chains in the gel seem to play an important role\cite{broedersz14a}. Correlations and special topological conditions in the gel cannot be taken into account in our simple model of polydispersity since we assume independent chains. 
Measuring the correlations and accounting for them in simulations remains an open task for theorists as well as experimentalists\cite{broedersz14a}. 
Please note that a non-affine deformation of the gel alters the equilibrium volume which is predicted by equation \eqref{eq:chain volume approx}. 
In the case of a non-affine deformation $A$ is a function of the end-to-end distance.
Therefore, non-affine deformations of the gel alter the prediction of the mass-based degree of swelling for monodisperse gels, as well as for polydisperse gels.
%We envision that further coarse graining of our polyelectrolyte network to a spring network with matched spring constants $k_i$ and equilibrium elongations $r_{0,i}$ could give valuable insights in the actual physical behavior of highly polydisperse polyelectrolyte gels. As outlined in the introduction MD simulations of a volume element which truely represents a Flory-Schulz distributed polydisperse polyelectrolyte gel is prohibitively expensive. We imagine that spring mass networks like the ones presented by Kot \etal \cite{kot15a} could be parametrized to represent polydisperse polyelectrolyte networks.
%For an extensive review about polydispersity in crosslinked disordered random spring networks we refer for example to Broedersz \etal \cite{broedersz14a} who also discuss effective medium theories like that of Sheinman \etal \cite{sheinman12a}. 

%\FloatBarrier
%%%%%%%%%%%%%%%%%%%%%%%%%%%%%%%%%%%%%
\section{Conclusion}
%%%%%%%%%%%%%%%%%%%%%%%%%%%%%%%%%%%%%
%Perturbed-Chain Statistical Associating Fluid Theory equation of state is an alternative method for swelling equilibrium

In summary, we have presented two successive mean-field models aimed
at describing gel swelling and the elastic moduli of polyelectrolyte
gels: the single chain CGM and the Poisson-Boltzmann CGM.  We find
that the single chain CGM provides an excellent agreement with the
periodic gel model.  Since it is particle-based, we can use exactly
the same interactions as in the periodic gel model, and hence
also investigate specific ion effects (modeled via different short
range interactions), poor solvent conditions, or the influence of
multivalent ions.  The single chain CGM provides about one order of
magnitude reduction of the computational cost due to the smaller
number of particles that need to be simulated.  The PB CGM and the Katchalsky model
provide several additional orders of magnitude in speed-up compared to the
single chain model. On one hand, this speed-up comes at the cost of
reduced accuracy. On the other hand, computationally cheaper models
allow to screen the the possible parameter space, needed for
optimizing real-world applications, more efficiently.  While the
Katchalsky model fails for charge fractions $f>0.5$, our new PB CGM
still works for highly charged gels.  In addition, both, the PB CGM and
the single chain CGM, can be used to study pH-sensitive
gels\cite{landsgesell19b}, where high charge fractions occur.  The
results for the bulk modulus of the single chain CGM and the PB CGM
are consistent with periodic gel results (within
errorbars).%, while the Katchalsky model again fails to give predictions at high charge fractions.
Additionally, we explore gel swelling as a function of solvent
permittivity. For large relative permittivities the ideal Donnan
prediction is recovered, while for lower relative permittivities
electrostatic correlations lead to the expected deviations between the
PB CGM and the periodic gel model. Nevertheless, our single chain CGM
can still capture these correlation effects correctly.  In addition,
we demonstrated a simplistic approach of introducing chain length
heterogeneity by assuming a mean-field factorization of the gel into
uncorrelated single chain cells.

%We also investigated the
%influence of several model parameters like the diamond parameter $A$
%and the polymer charge distribution in the PB CGM. We showed that
%these parameters have only minor influences on the final results.

\section{Acknowledgments}
Funding from the German Research Foundation (DFG) through grants HO
1108/26-1, 423435431 (HO1108/30-1) as part of FOR 2811, and AR 593/7-1
is gratefully acknowledged. We thank D. Sean, K. Szuttor,
P. Kreissl and J. Zeman for helpful discussions.

\bibliographystyle{unsrt}
\bibliography{icp.bib}

\FloatBarrier

\pagebreak

\section{Supplementary Information}

\subsection{Influence of the Different Imposed Charge Densities in the
  PB CGM}
%%%%%%%%%%%%%%%%%%%%%%%%%%%%%%%%%%%%%
\label{sec: influence of different imposed monomer densities}
%\FloatBarrier

We now investigate the influence of the imposed charge densities
in the PB CGM and exchange the rectangular monomer density with an
approximately Gaussian monomer density with compact support. The
choice of the monomer density affects the charge density $\rho_f$
which is input to the PB equation.

\begin{figure}
\centering
\includegraphics[width=0.7\textwidth]{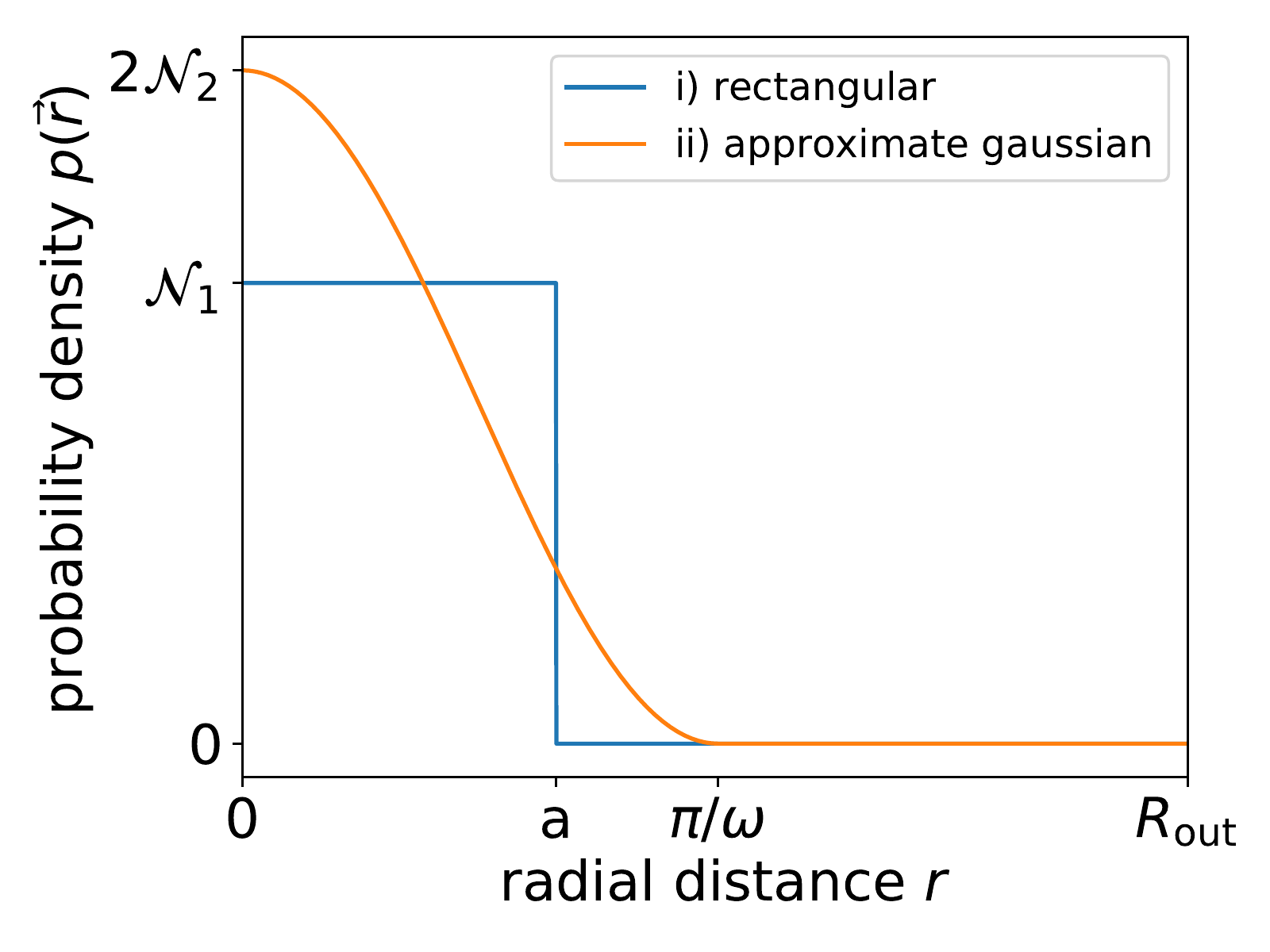}
\caption{Different probability densities to find a monomer in distance $r$ from the end-to-end vector which are used in the PB CGM. The
  figure shows i) the rectangular probability density and
  ii) the approximately Gaussian probability density. }
\label{fig:monomer densities}
\end{figure}

If the probability density to find a monomer in a given distance $r$ from the end-to-end vector is approximately Gaussian, then it is given via the following formula:
%Idea taken from \url{https://www.johndcook.com/blog/2010/04/29/simple-approximation-to-normal-distribution/}
%see innerRodProbabilityFunctionsNormalization.nb
\begin{equation}
p(\vec{r}) = \mathcal{N}_2(1 + \cos(\omega r)) H(-(r-\pi/\omega)).
\end{equation}
$\mathcal{N}_2$ is a normalization constant such that
$p(\vec{r})$ is a probability density
($\int_{\Vchain} \mathrm{d}V p(\vec{r}) =2\pi L \int_0^R \mathrm{d}r r  p(\vec{r}) =1 $). This
normalization criterion yields
$\mathcal{N}_2=\frac{2\omega^2}{-4+\pi^2} \frac{1}{2\pi L}$. The parameter $\omega$ is
chosen such that
$\langle r \rangle \overset{!}{=}\langle r \rangle_\text{MD}$ is
matched to the average distance of the monomers from the end-to-end
vector. This condition yields
$\omega=\frac{2(-6\pi+\pi^3)}{3(-4+\pi^2)\langle r \rangle_\text{MD}}$
which determines the width of the probability density. A
comparison between the approximately Gaussian probability
density and the rectangular probability density to find a monomer in distance $r$ can be seen in
figure \ref{fig:monomer densities}.

Imposing this alternative monomer density and solving the PB CGM
again for the resulting alternative choice of
$\rho_f(\vec{r})=-N f p(\vec{r})$.
We obtain only marginally changed
swelling equilibria (compared to the imposed rectangular monomer
density). This fact is depicted in figure \ref{fig:different monomer
  densities} which barely allows to distinguish the resulting swelling
equilibria.

\begin{figure}
\centering
\includegraphics[width=0.7\textwidth]{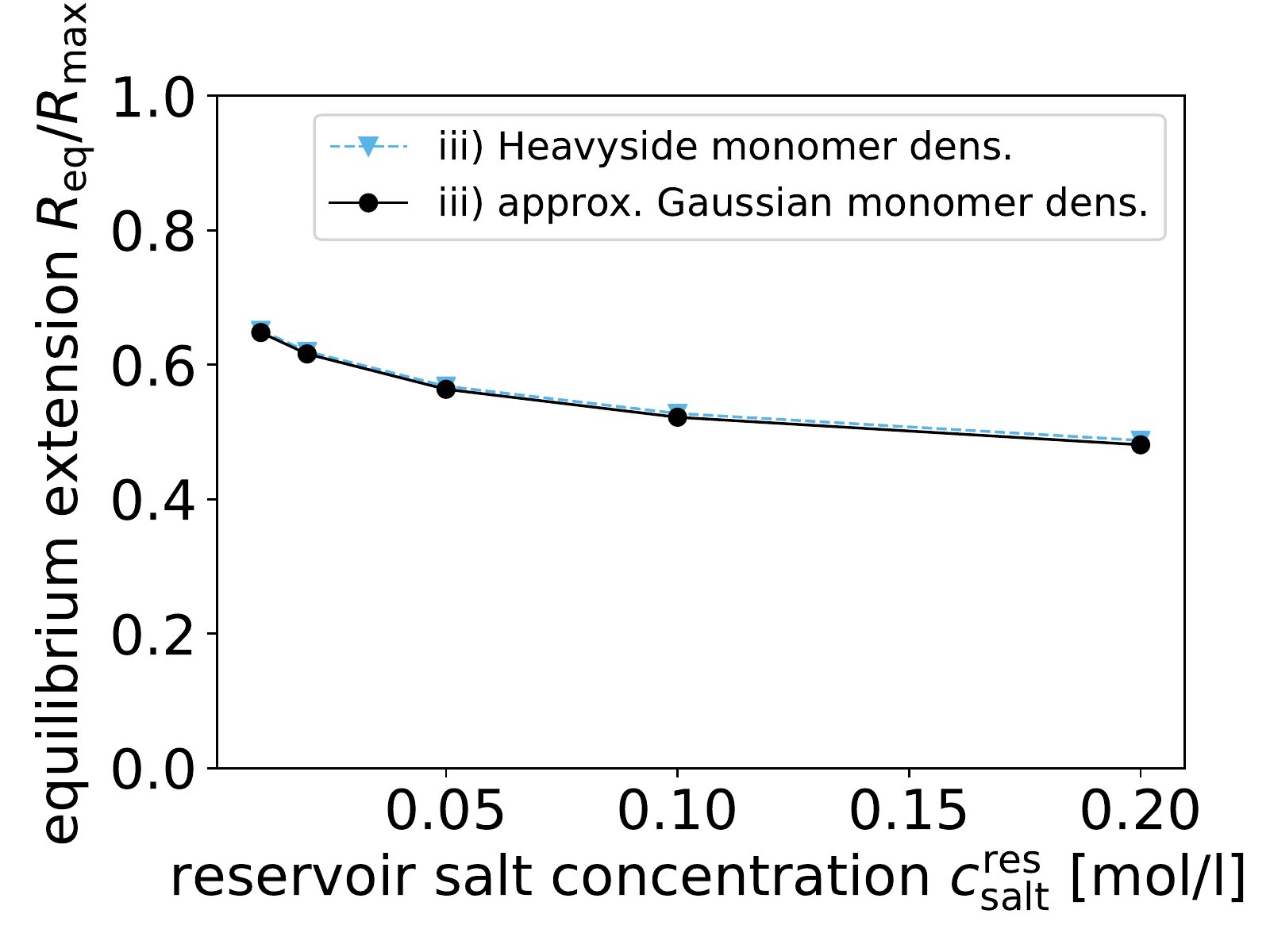}
% plots originate from scripts in
% /tikhome/jlandsgesell/phd/own_papers/2018-PB_gel_short/figures/data
\caption{Comparison for different monomer densities (rectangular or approximate Gaussian density) in the PB model.
The equilibrium swelling length $\Req$ as a function of the reservoir salt concentration $\cs$ for $f=1$, $\epsilonr=80$ and $N= 40$.}\
\label{fig:different monomer densities}
\end{figure}

 \FloatBarrier
%%%%%%%%%%%%%%%%%%%%%%%%%%%%%%%%%%%%%
\subsection{Influence of the Different Imposed Values of $A$}
%%%%%%%%%%%%%%%%%%%%%%%%%%%%%%%%%%%%%
\label{sec: different A}
\FloatBarrier

\begin{figure}
  \centering
  \includegraphics[width=0.7\textwidth]{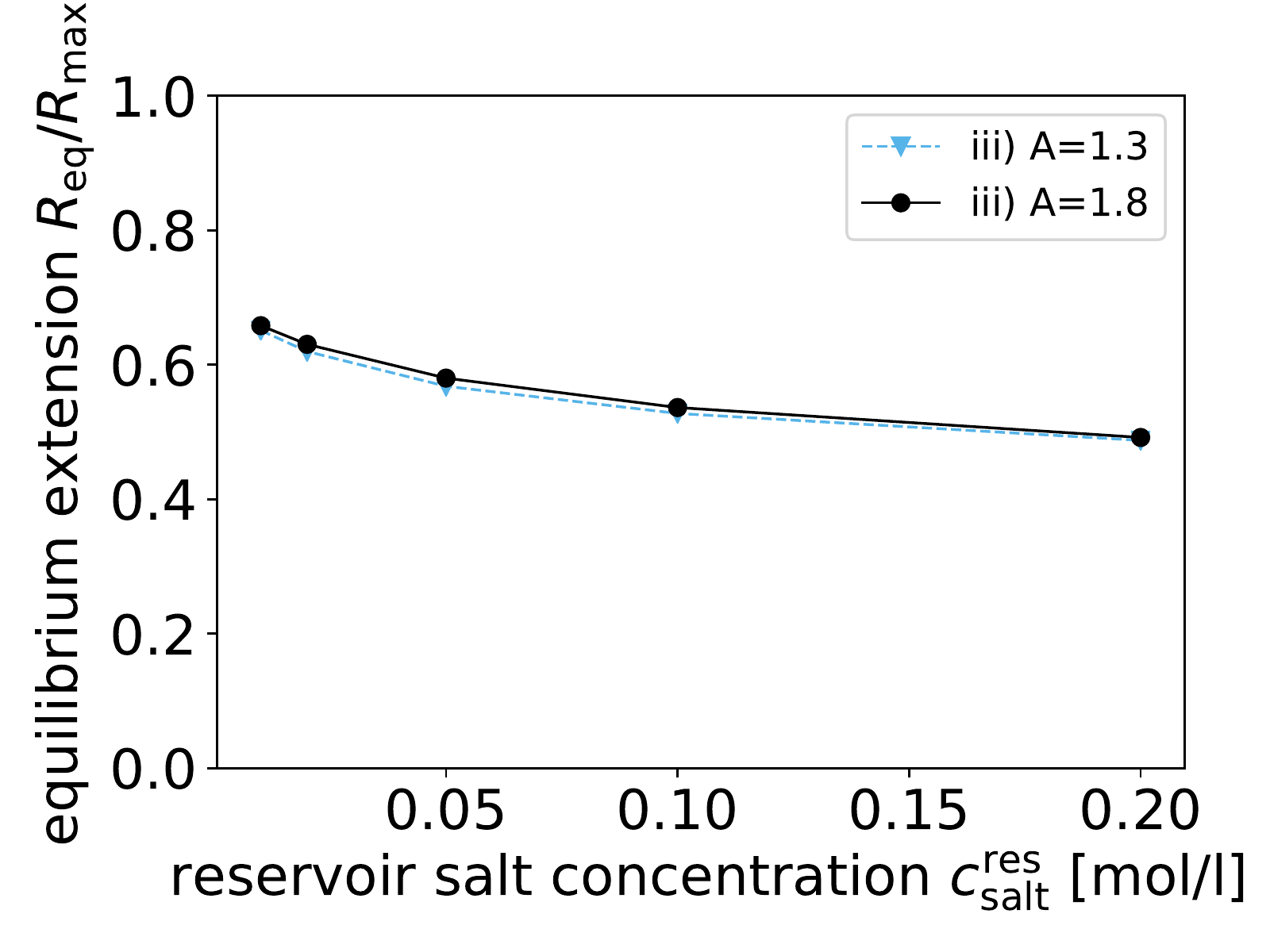}
%plots originate from scripts in /tikhome/jlandsgesell/phd/own_papers/2018-PB_gel_short/figures/data
  \caption{\label{fig:different A}Comparison of
    different values of $A$ in the PB model (for the rectangular
    charge distribution).  The equilibrium swelling length $\Req$ as a
    function of the reservoir salt concentration $\cs$ for $f=1$,
    $\epsilonr=80$ and $N= 40$.}
\end{figure}

{Ko{\v{s}}ovan} et al.~\cite{kosovan15a} investigated the relation
between the end-to-end distance and the volume per chain. They found
(their figure 10) in MD simulations of periodic gels that the ratio
$A=\Rend^3/\Vchain$ varies between roughly 1.8 and 1.5 and
depends on the compression state of the gel, the charge fraction of
the gel and the salt concentration of the reservoir. This fact is in
contrast to our two new models where $A ={\sqrt{27}}/{4}\approx 1.3$
is fixed. The easiest approach to test the influence of $A$ on our
models is to impose a value of $A=1.8$ and compare it to results where
$A$ was chosen to be $\sqrt{27}/4$.  The effect of a change of $A$ on
the predicted swelling equilibria of the PB CGM can be seen in
figure \ref{fig:different A}. We conclude that the value of the end-to-end distance $\Req$, predicted by the PB CGM, is not
sensitive to a change of the exact value of the ratio $A$: The
predicted swelling equilibria barely change on a change of $A$. 
However, the predicted volume per chain is quite sensitive to the value of $A$ through equation \eqref{eq:chain volume approx}:

\begin{equation}
\label{eq:chain volume approx}
\Vchain\approx L(\Rend)^3/A.
\end{equation}

\subsection{P-$\Rend$ curves}
For completeness we show the P-$\Rend$ curves which were used in this paper for determining the swelling equilibria in the following figures.

%generate with: for f in $(ls *.pdf); do echo "\\includegraphics[width=0.32\\linewidth]{$f}" >> log.txt; done;
\begin{figure}
\includegraphics[width=0.32\linewidth]{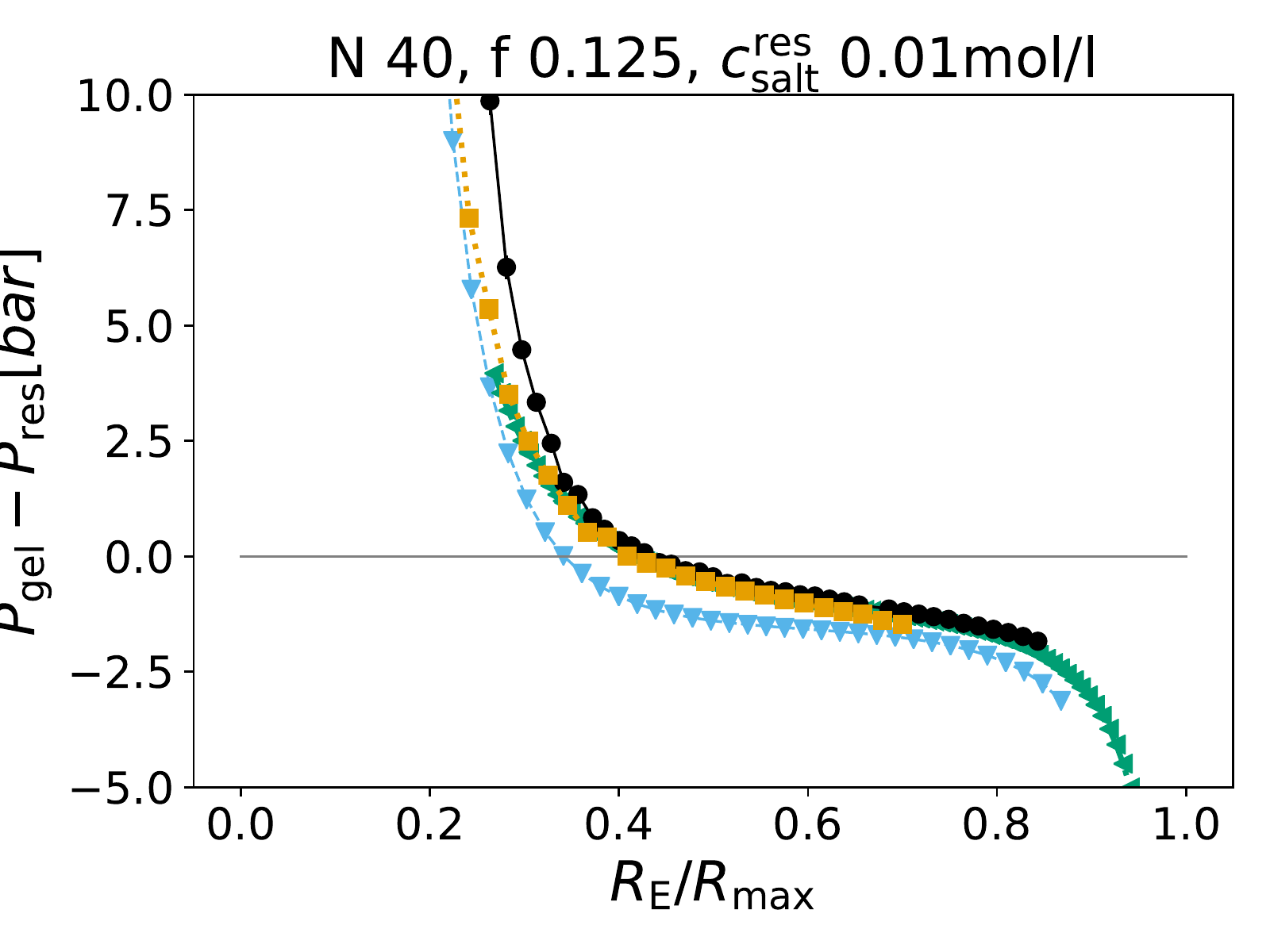}
\includegraphics[width=0.32\linewidth]{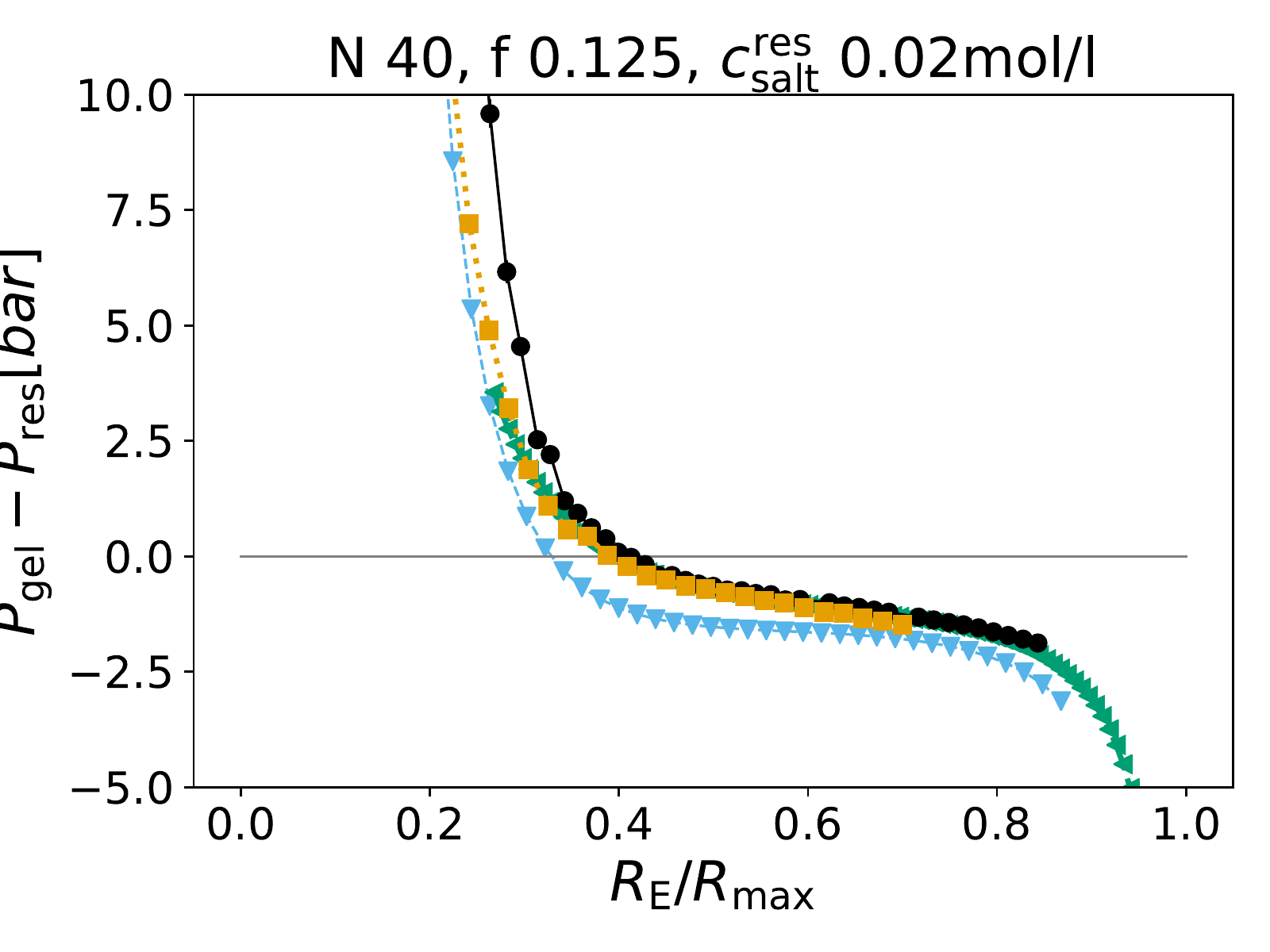}
\includegraphics[width=0.32\linewidth]{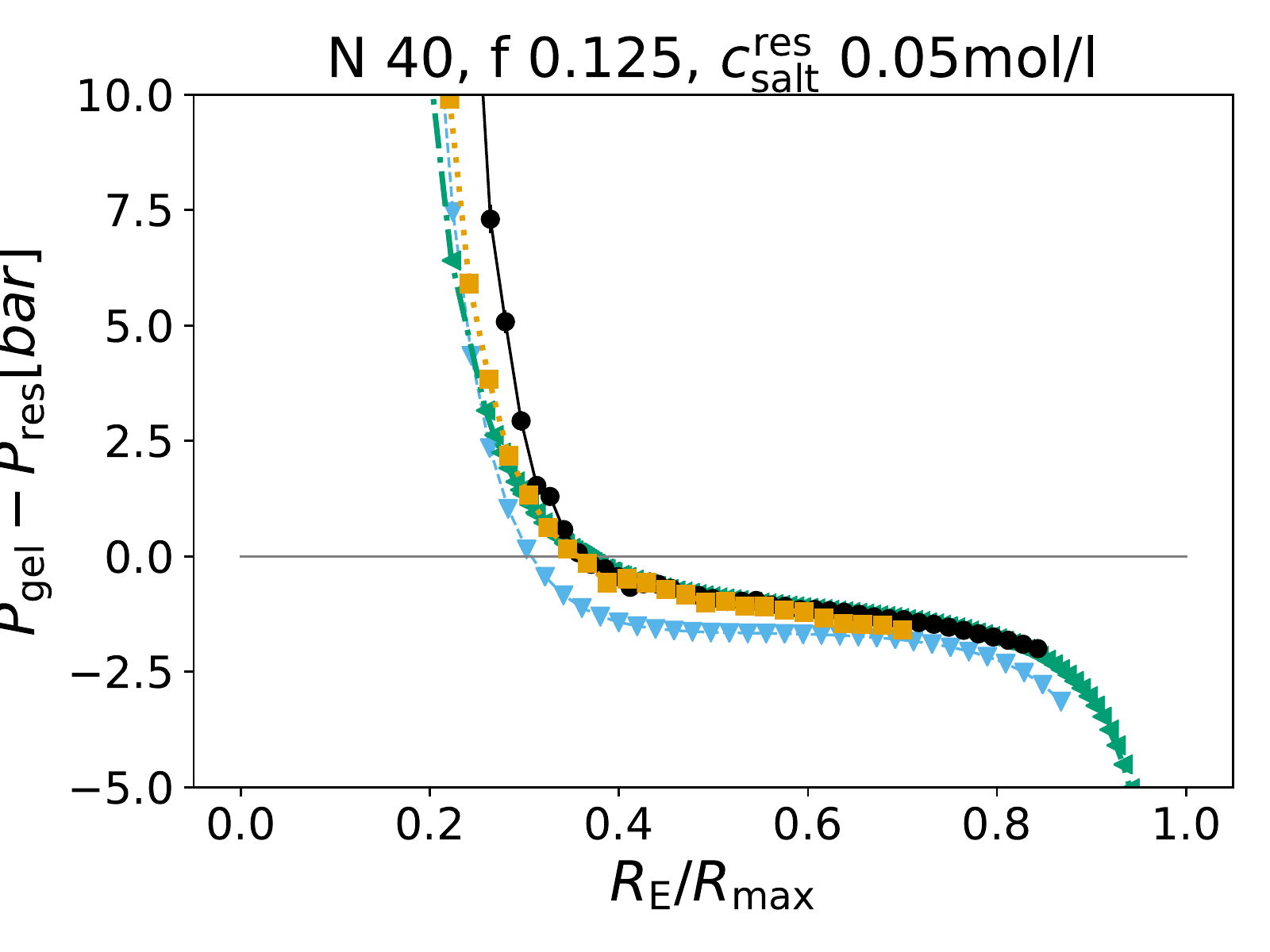}
\includegraphics[width=0.32\linewidth]{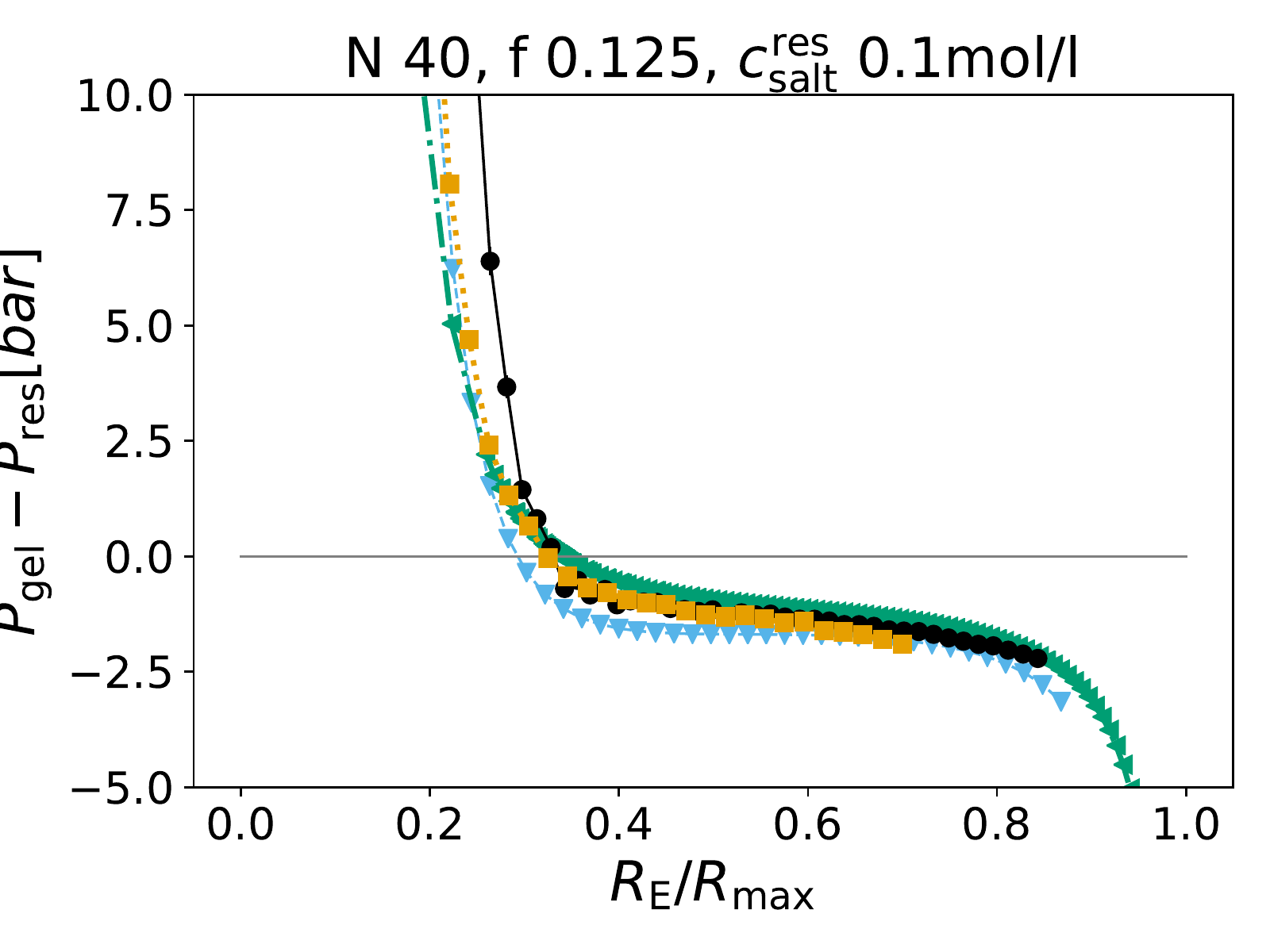}
\includegraphics[width=0.32\linewidth]{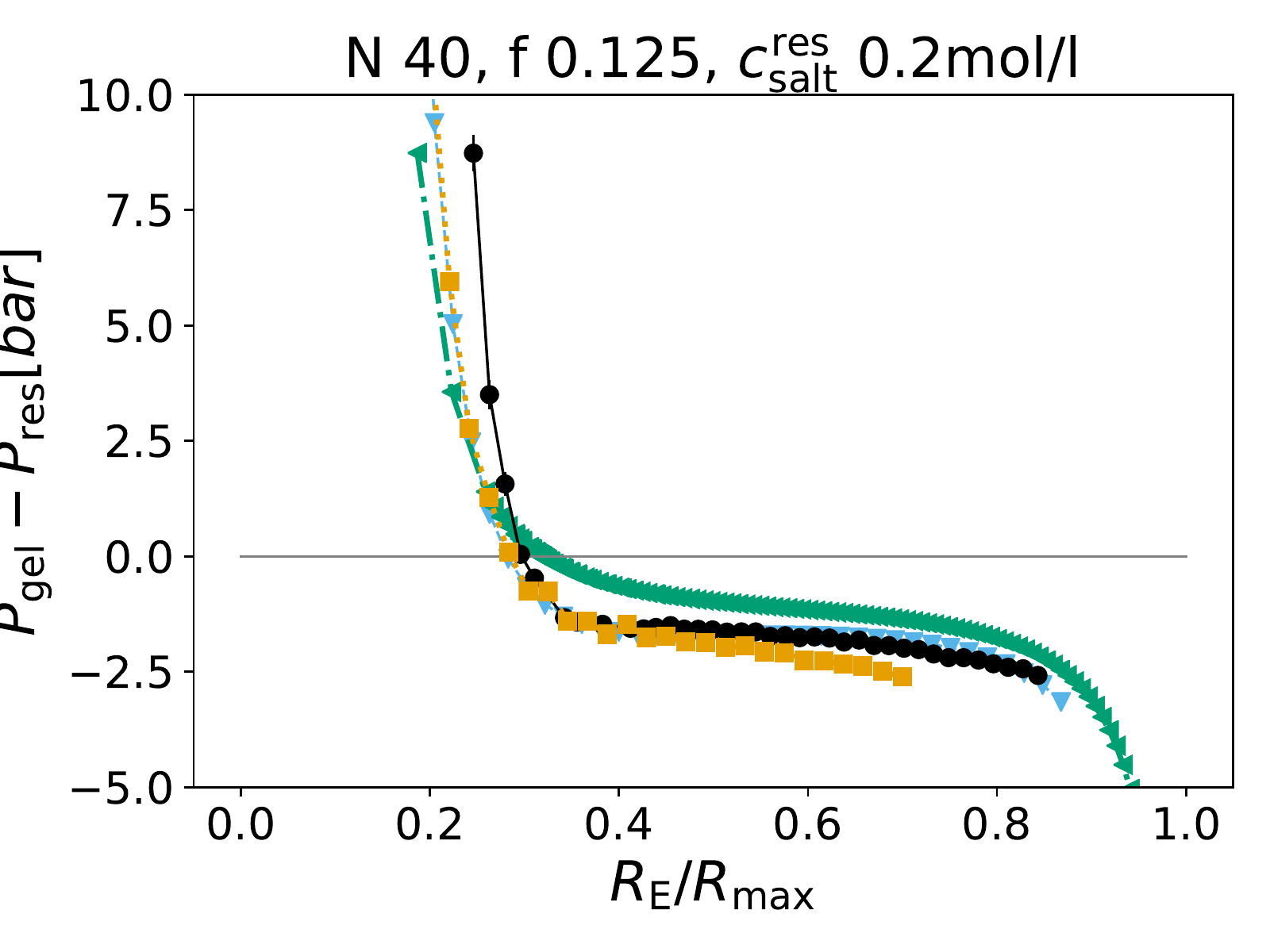}
\includegraphics[width=0.32\linewidth]{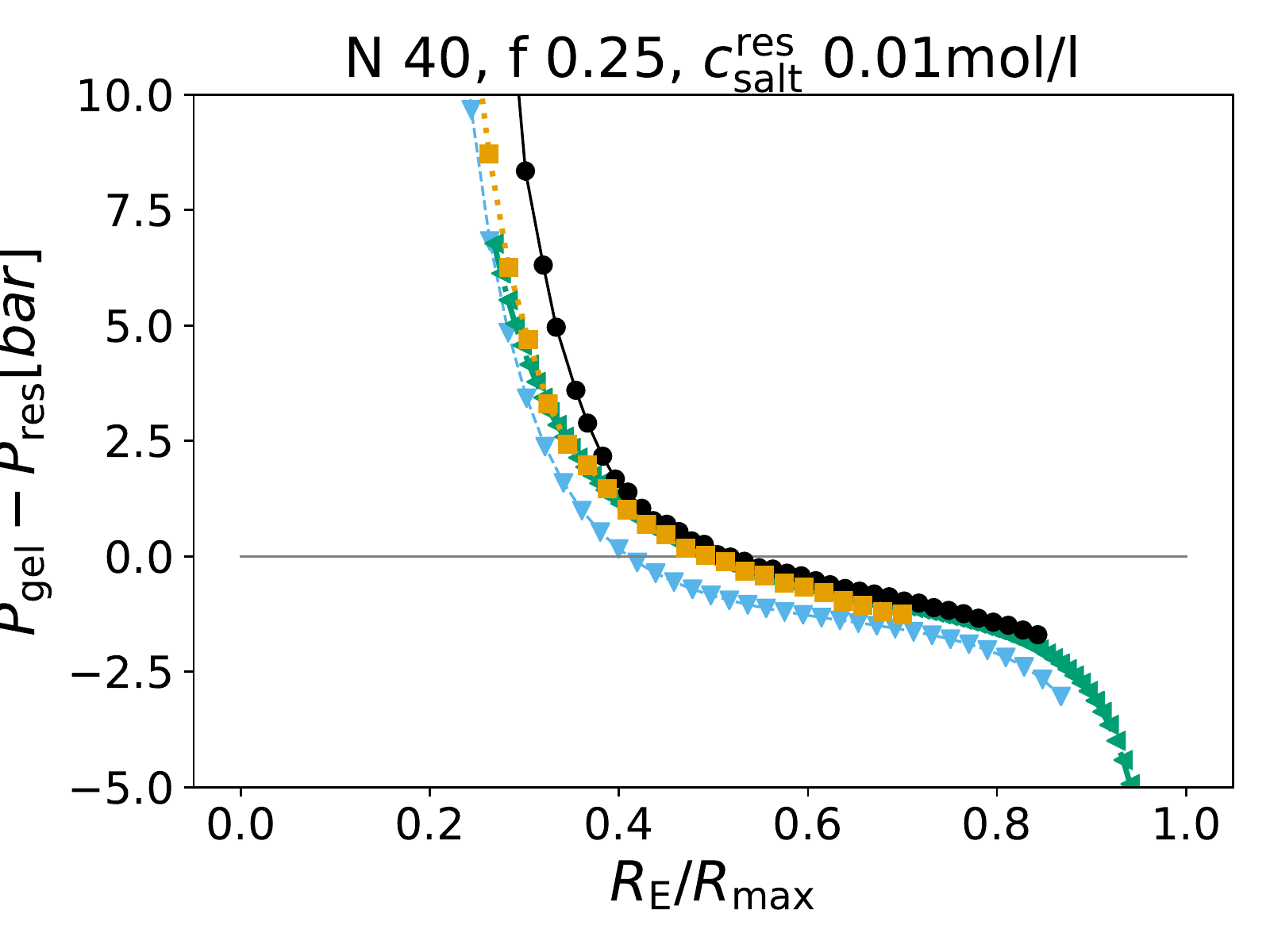}
\includegraphics[width=0.32\linewidth]{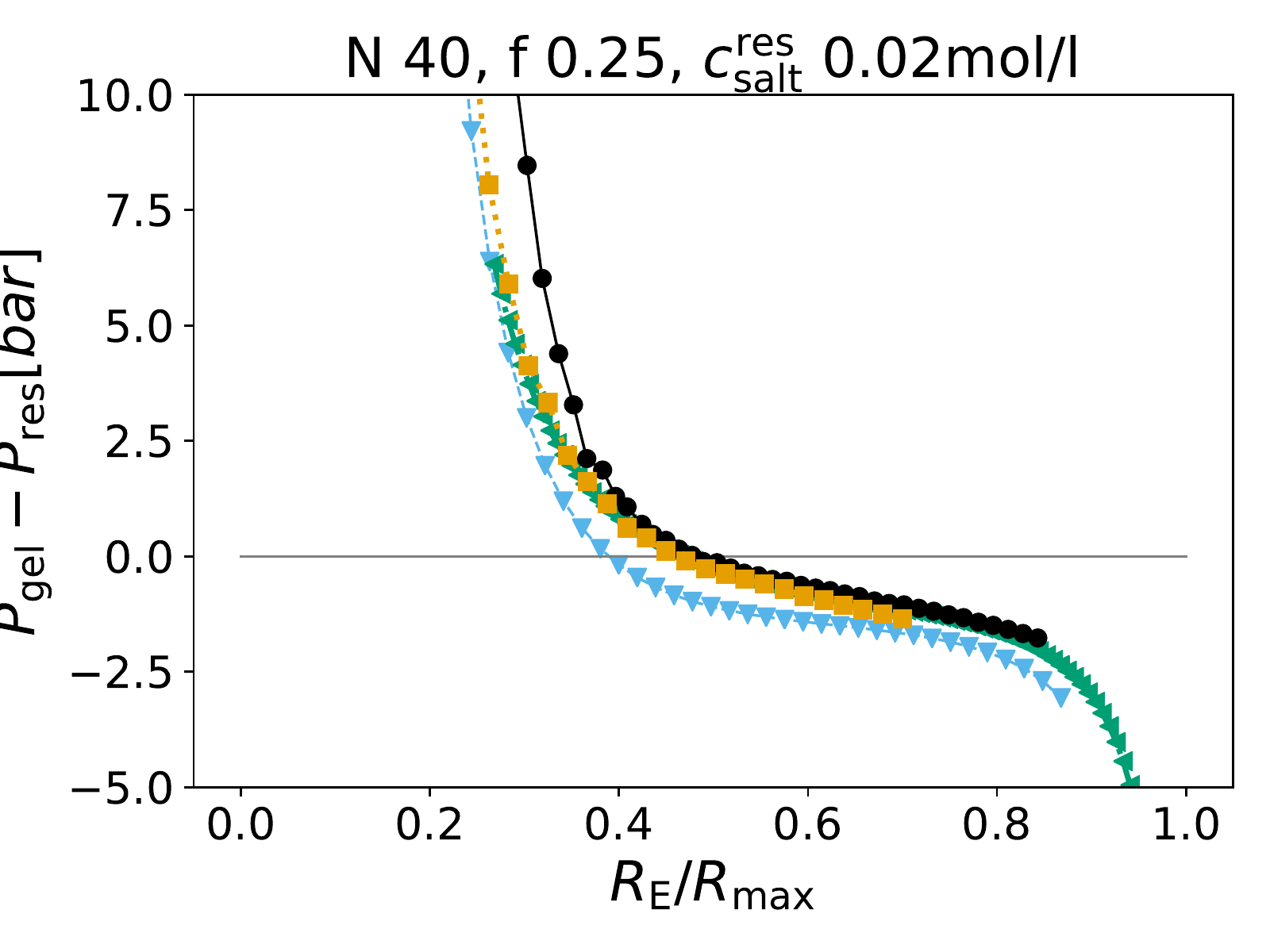}
\includegraphics[width=0.32\linewidth]{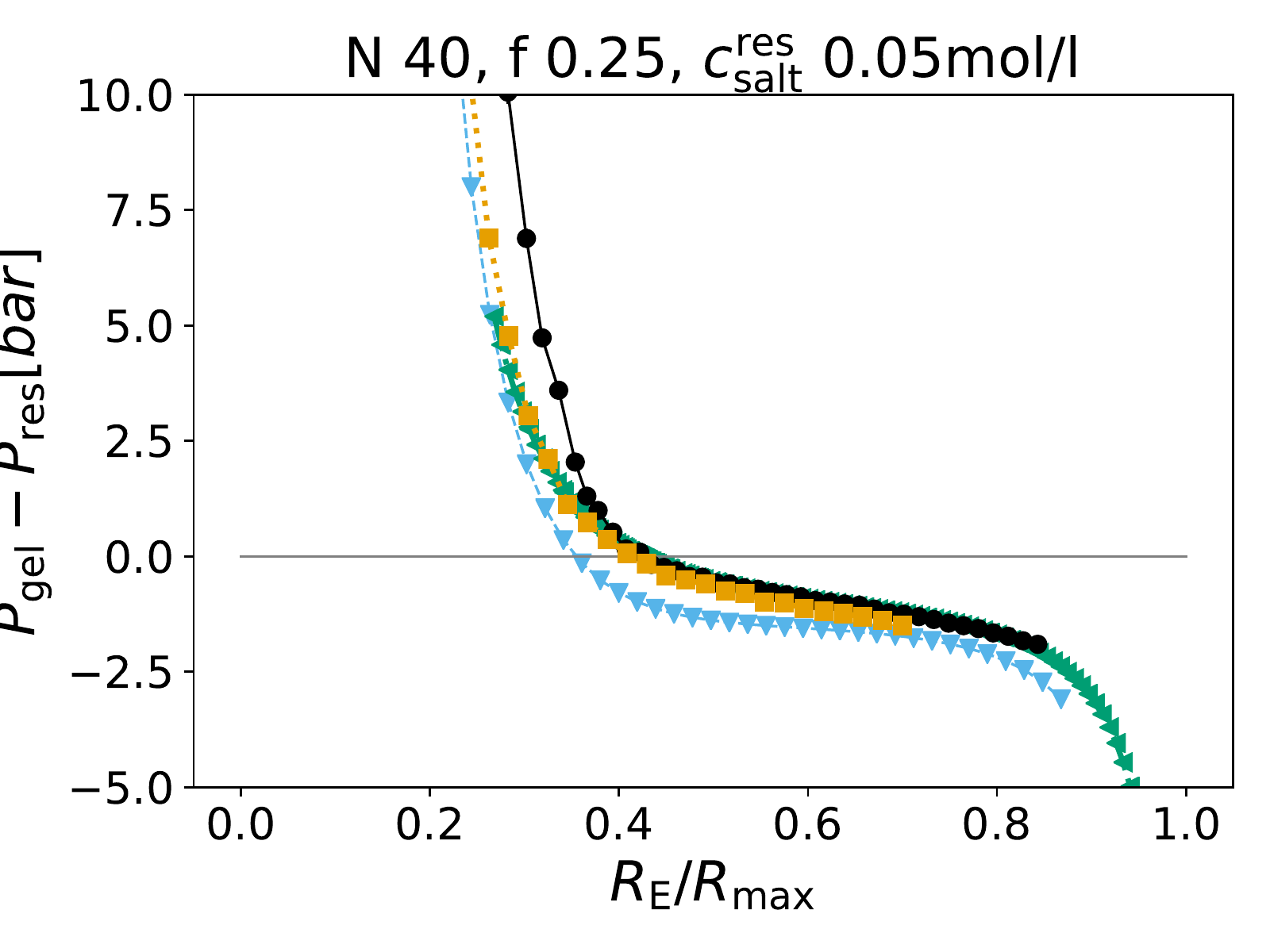}
\includegraphics[width=0.32\linewidth]{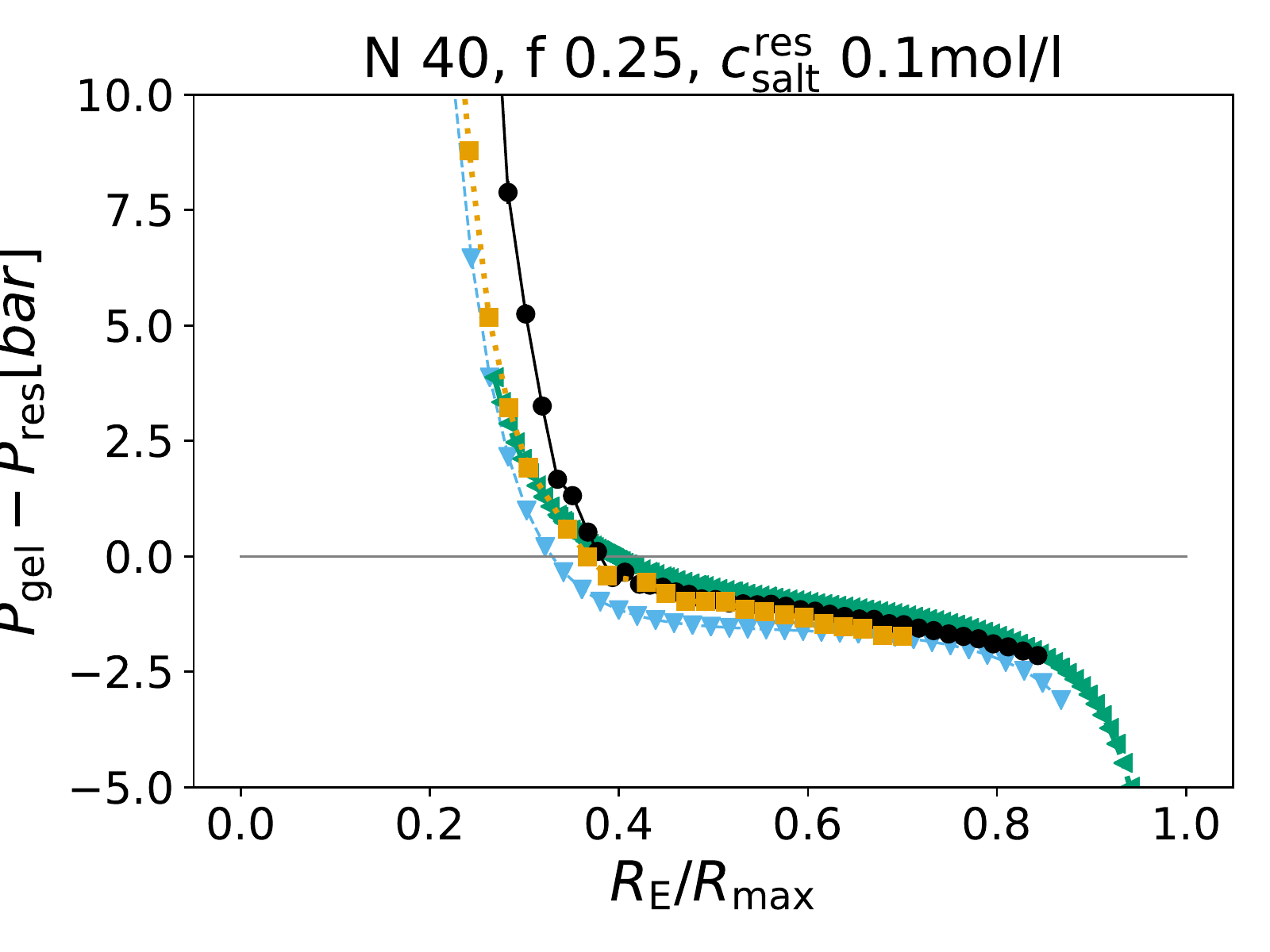}
\includegraphics[width=0.32\linewidth]{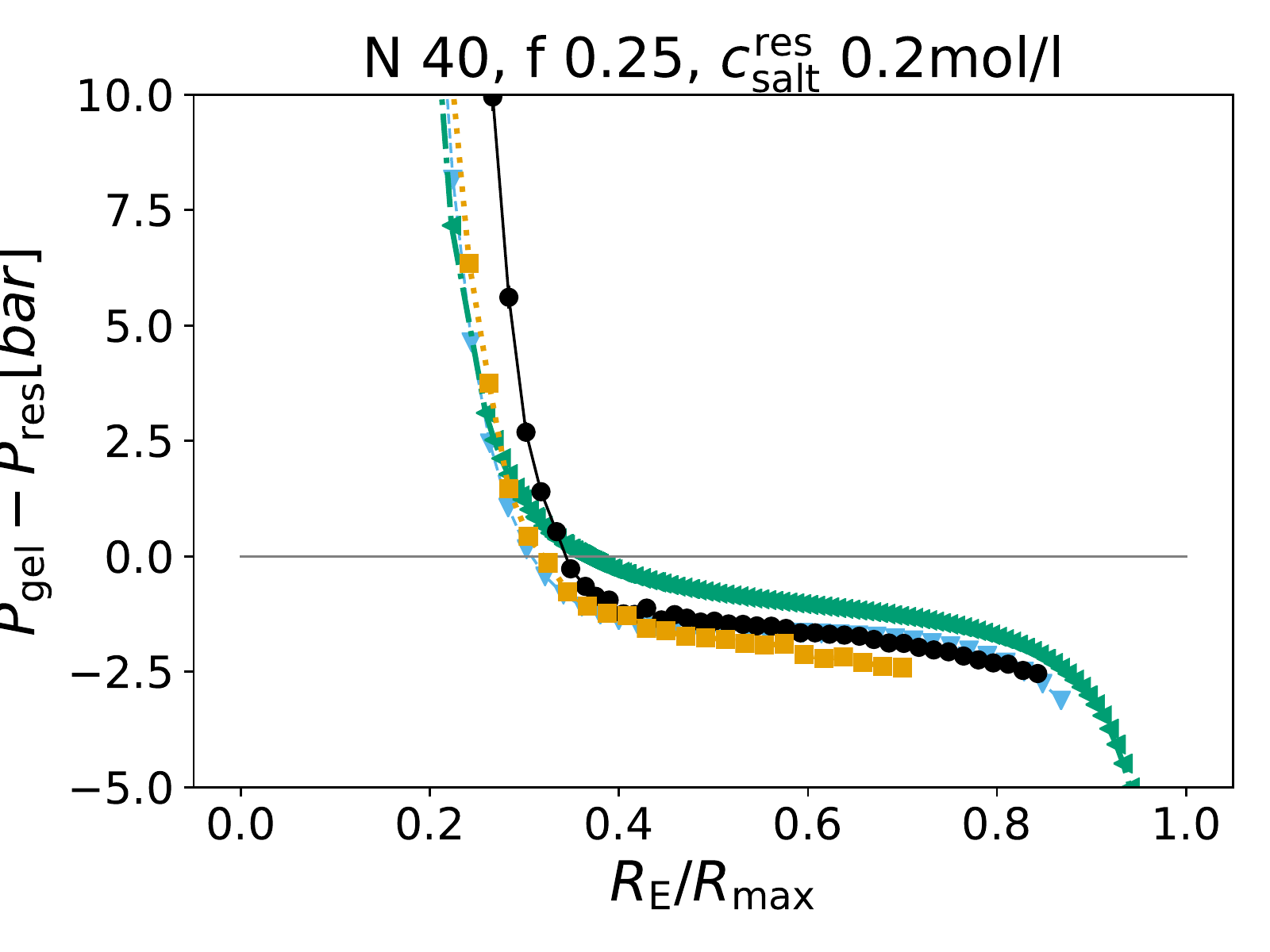}
\includegraphics[width=0.32\linewidth]{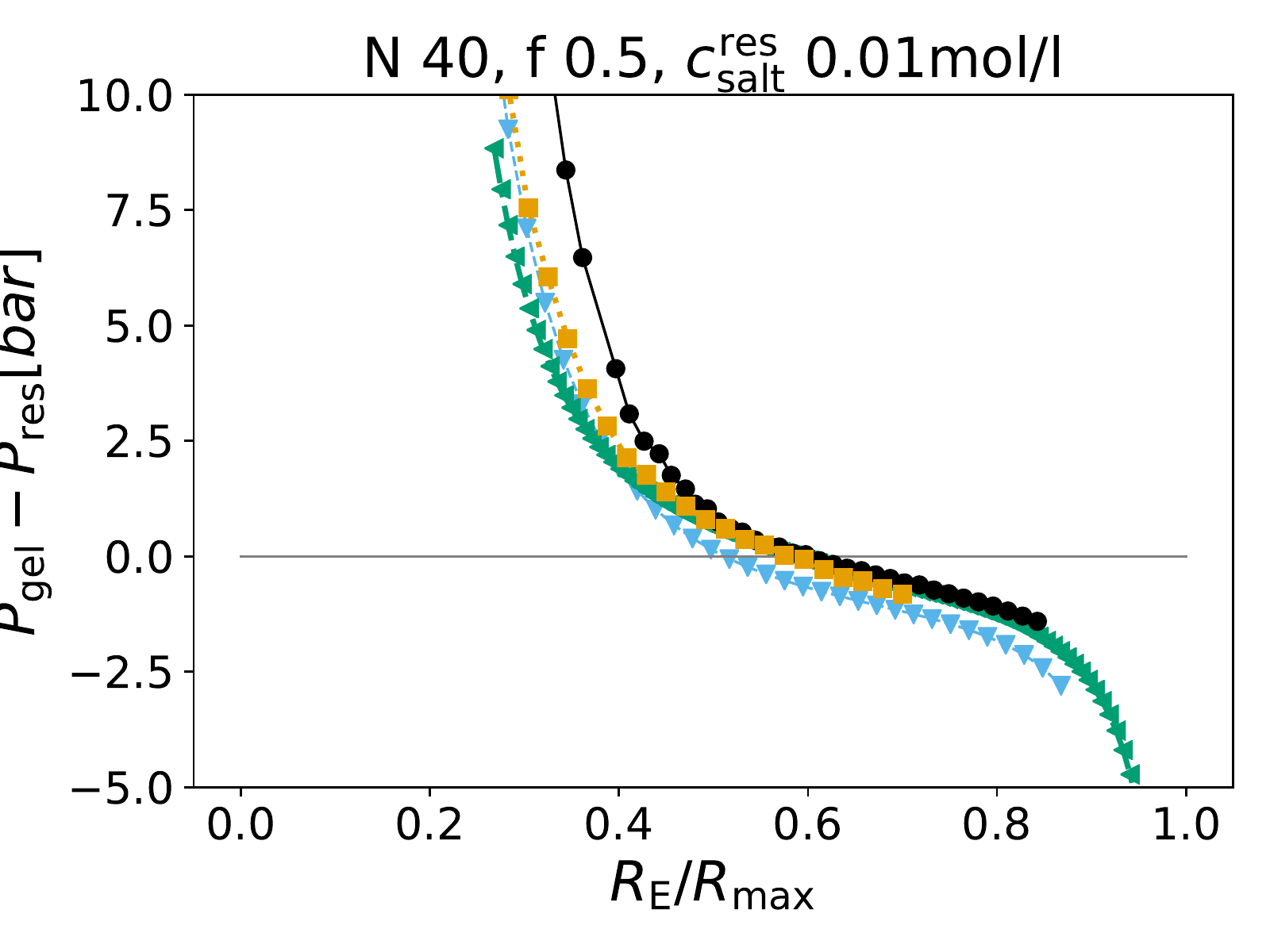}
\includegraphics[width=0.32\linewidth]{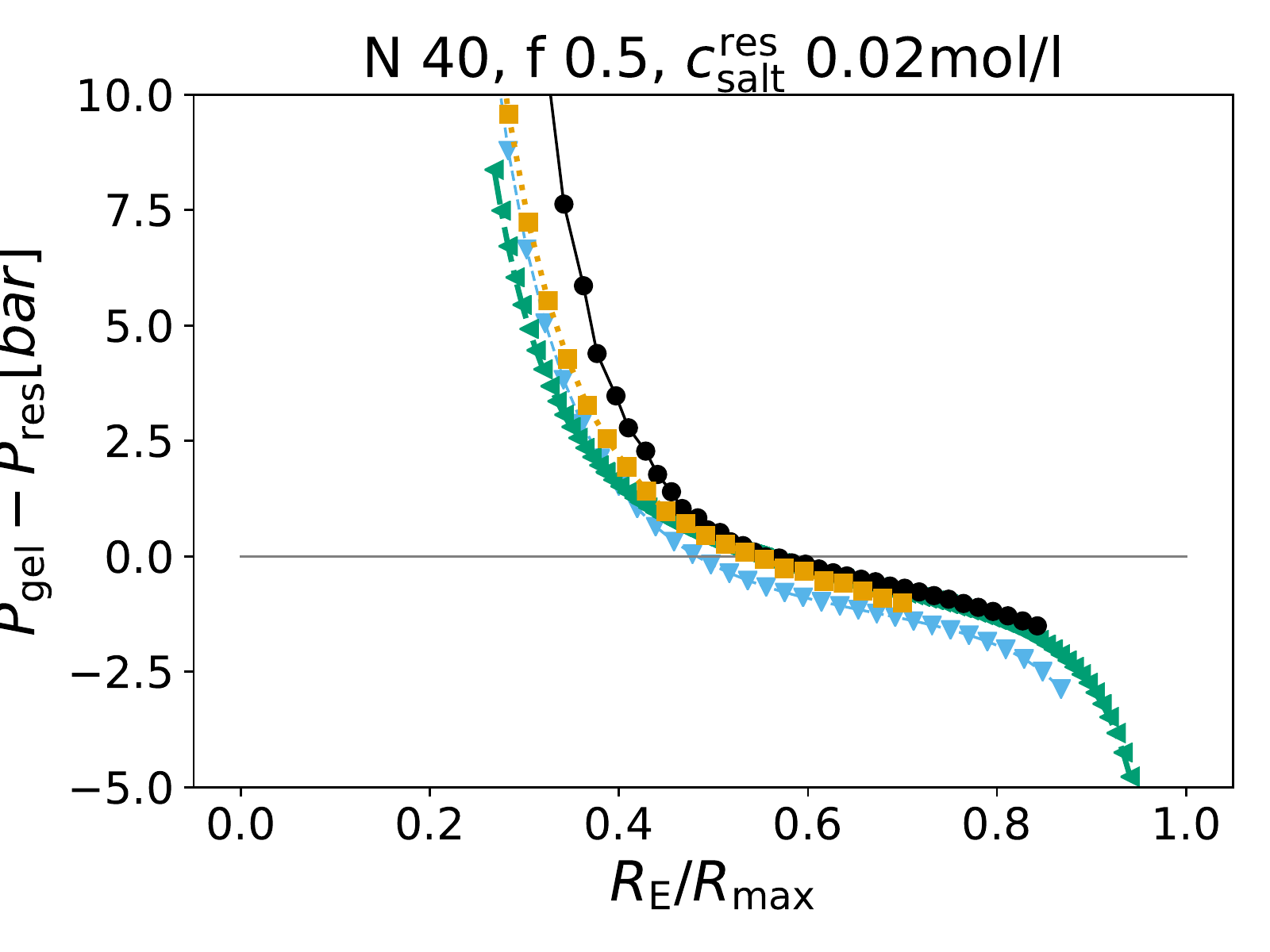}
\includegraphics[width=0.32\linewidth]{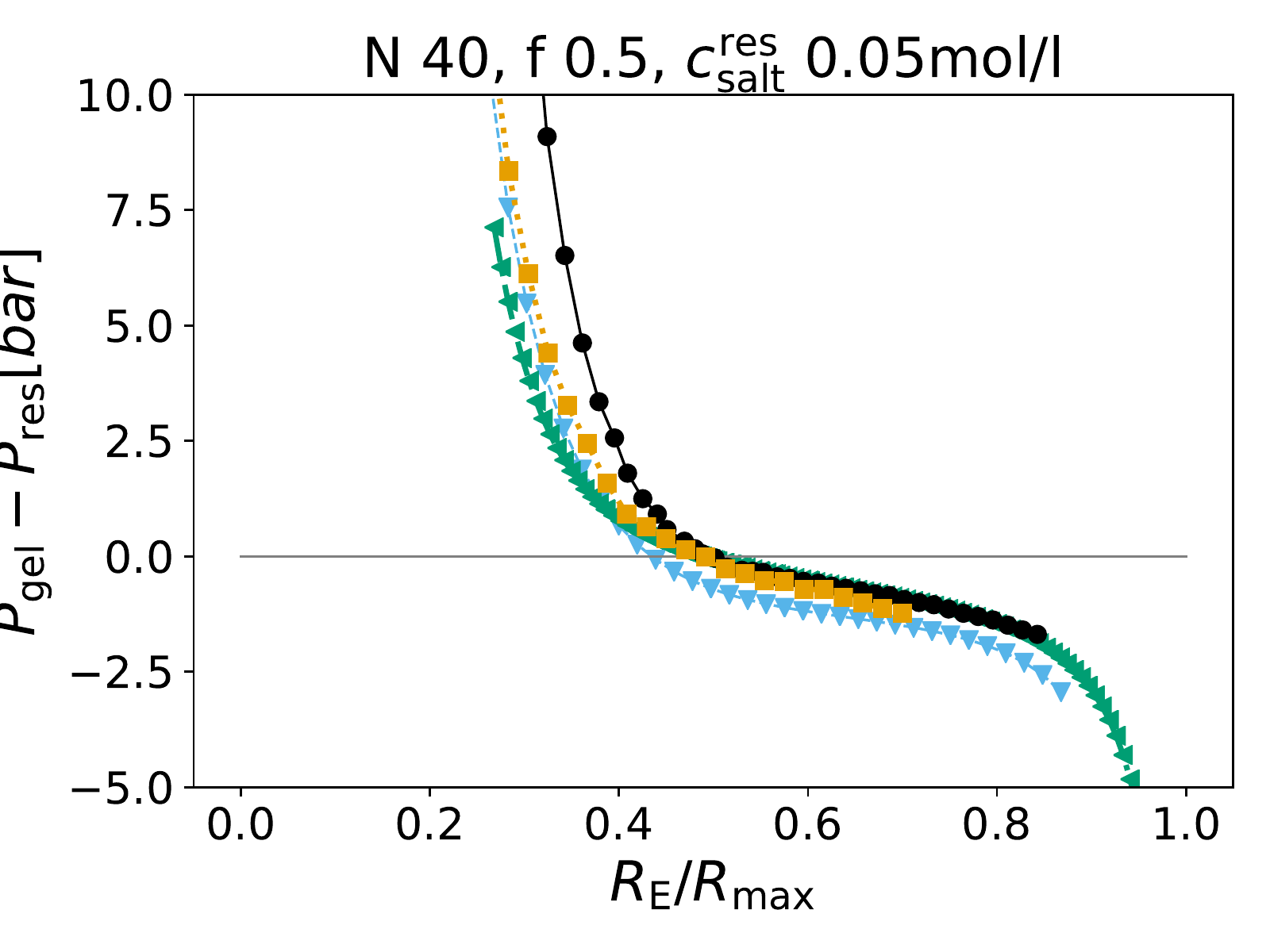}
\includegraphics[width=0.32\linewidth]{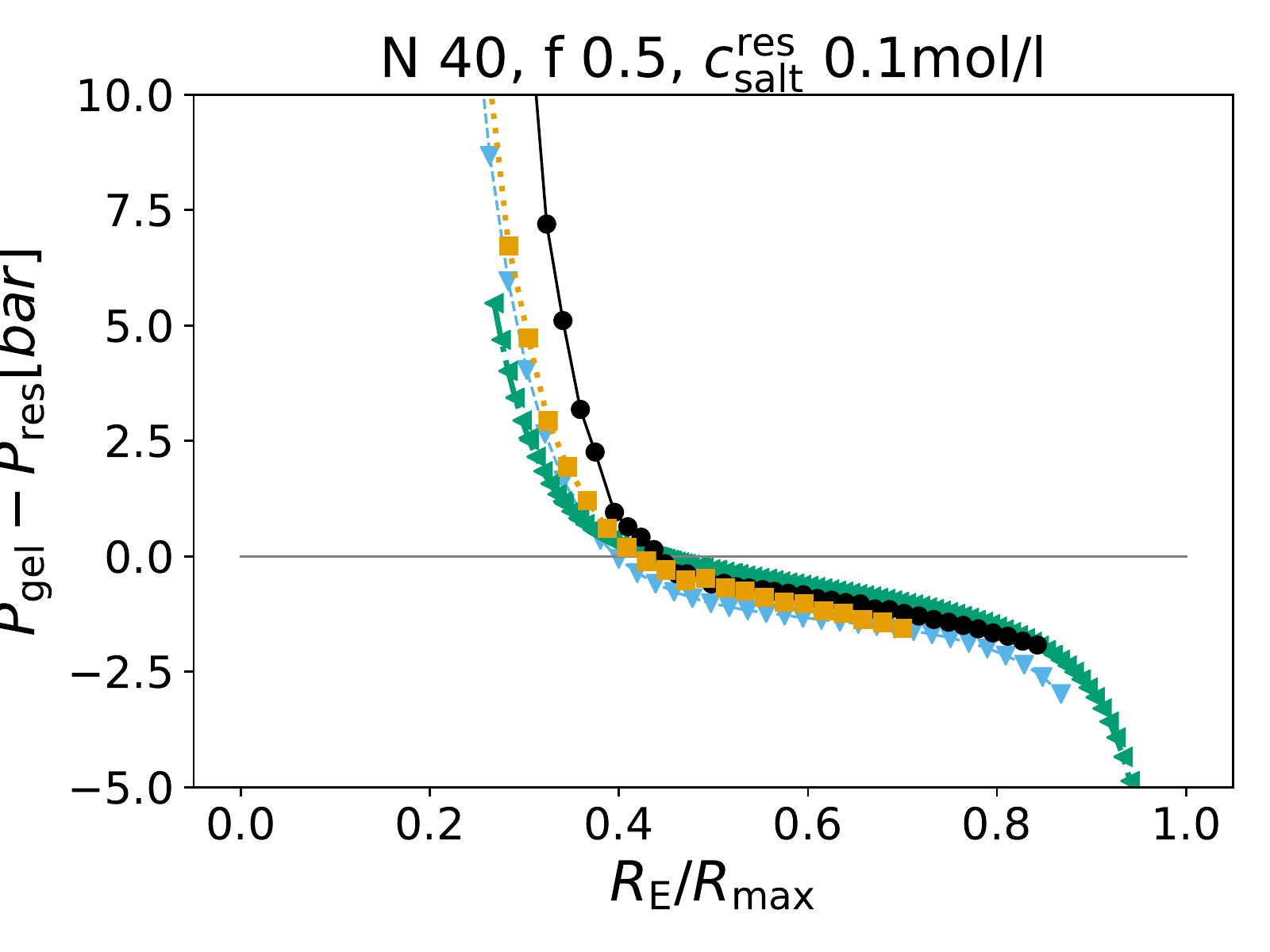}
\includegraphics[width=0.32\linewidth]{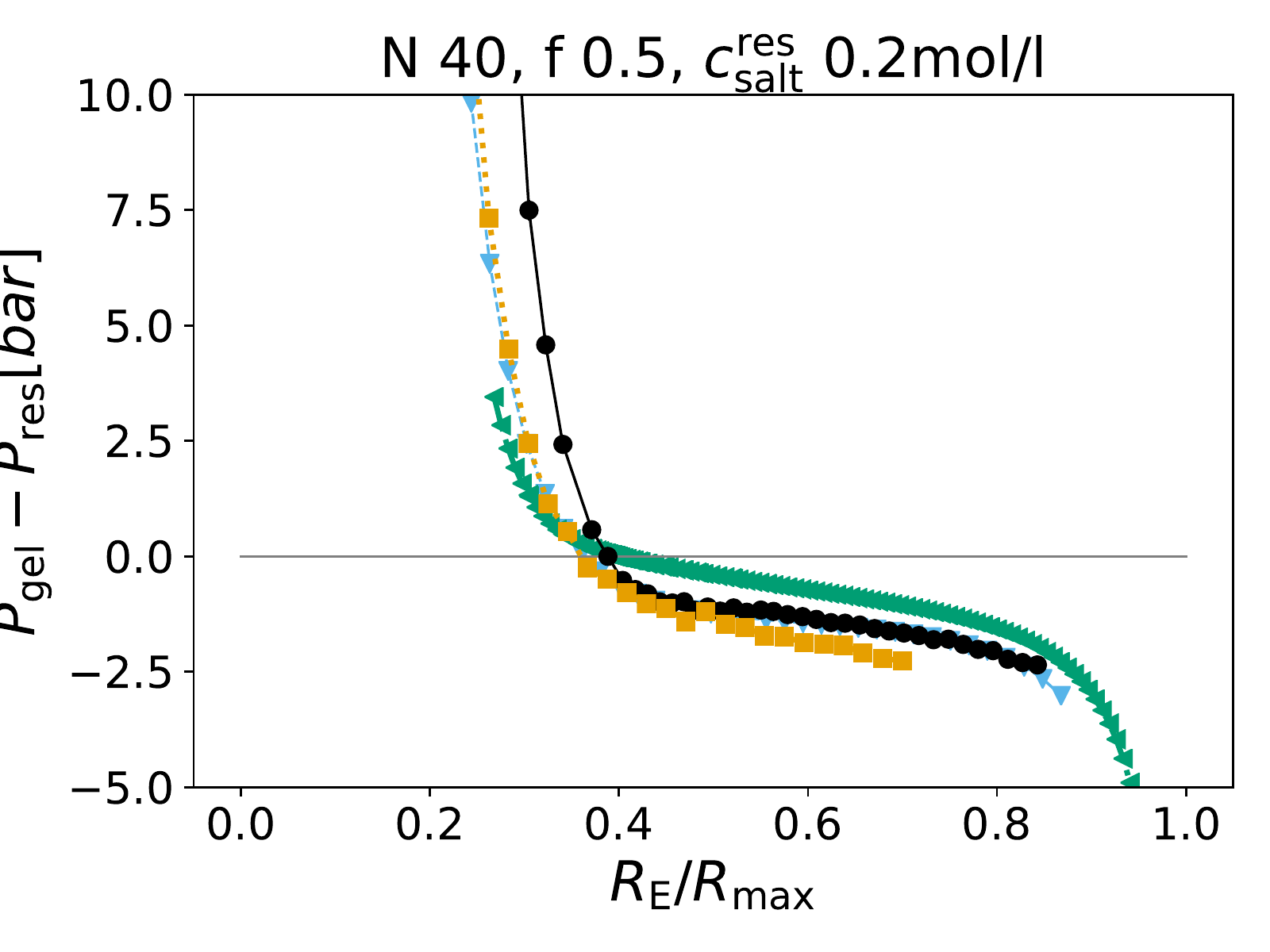}
\end{figure}
\begin{figure}
\includegraphics[width=0.32\linewidth]{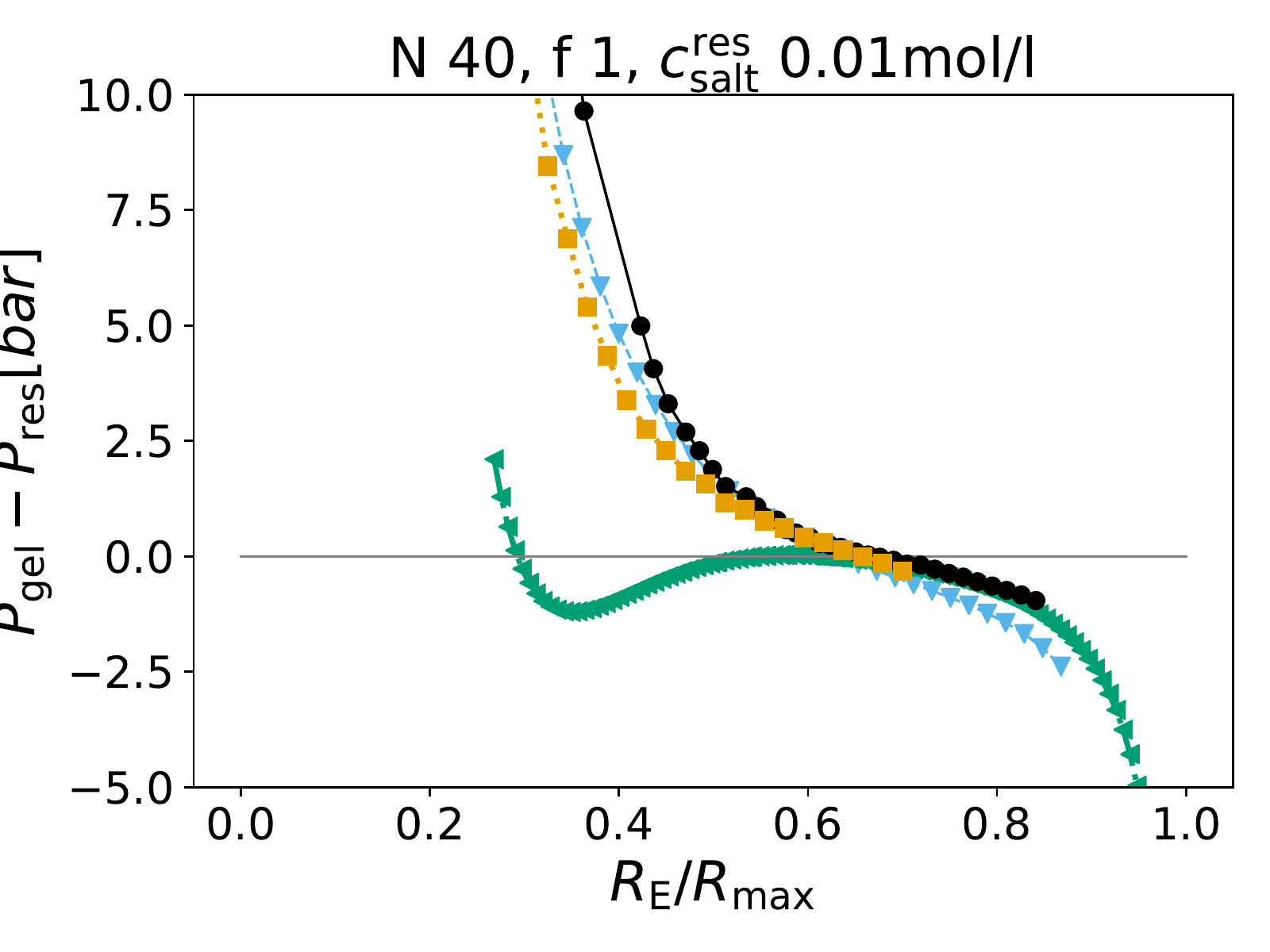}
\includegraphics[width=0.32\linewidth]{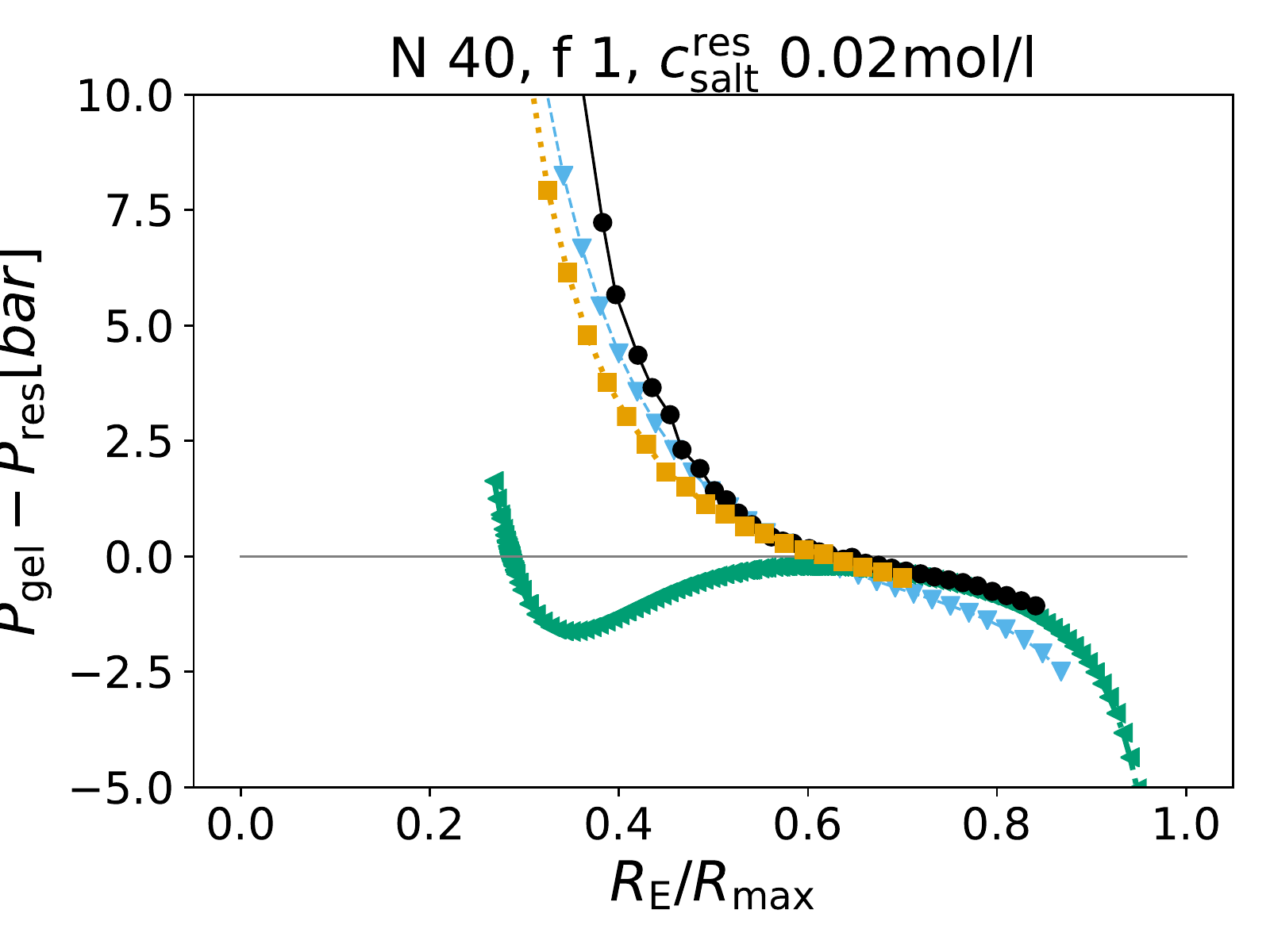}
\includegraphics[width=0.32\linewidth]{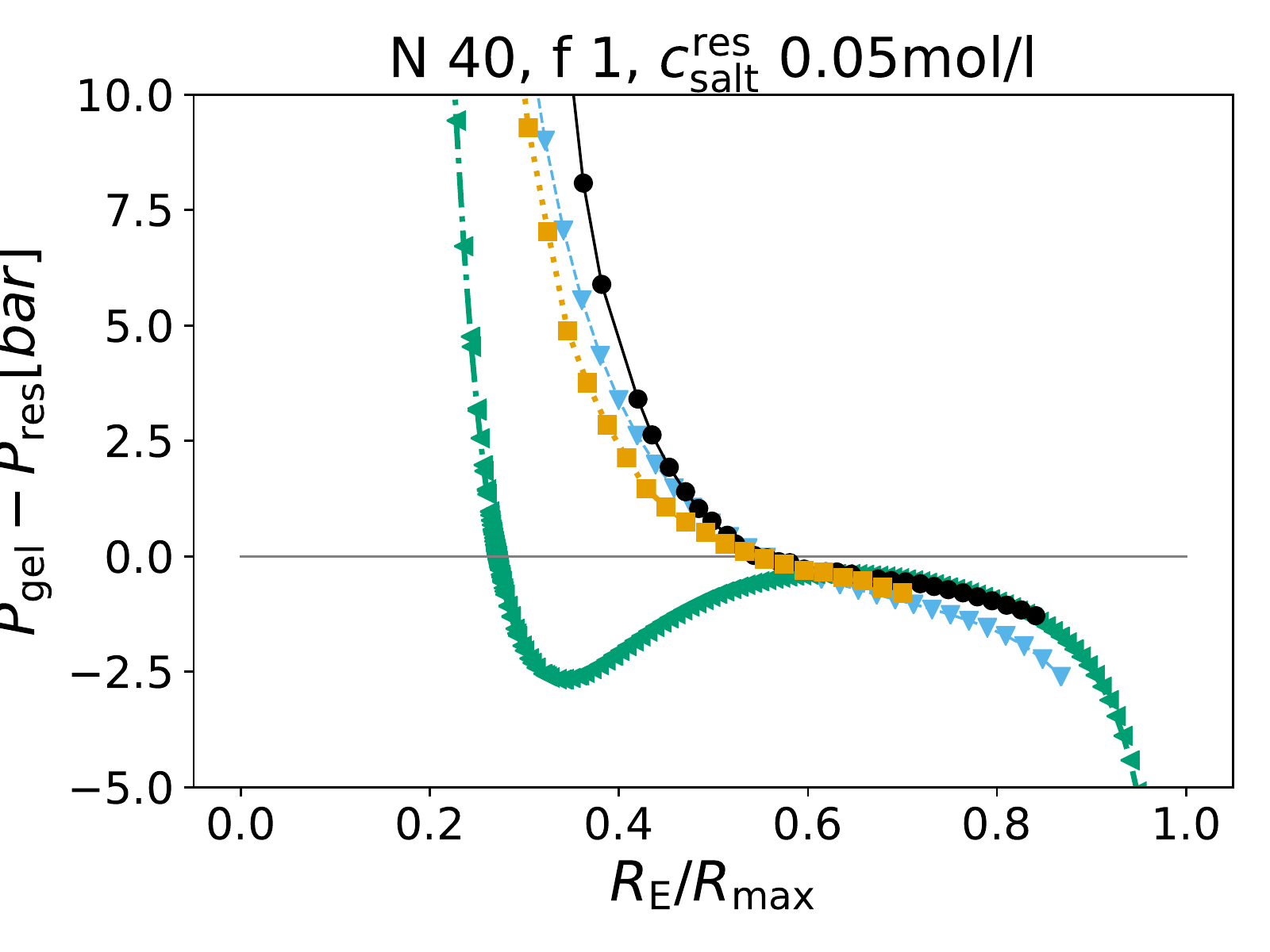}
\includegraphics[width=0.32\linewidth]{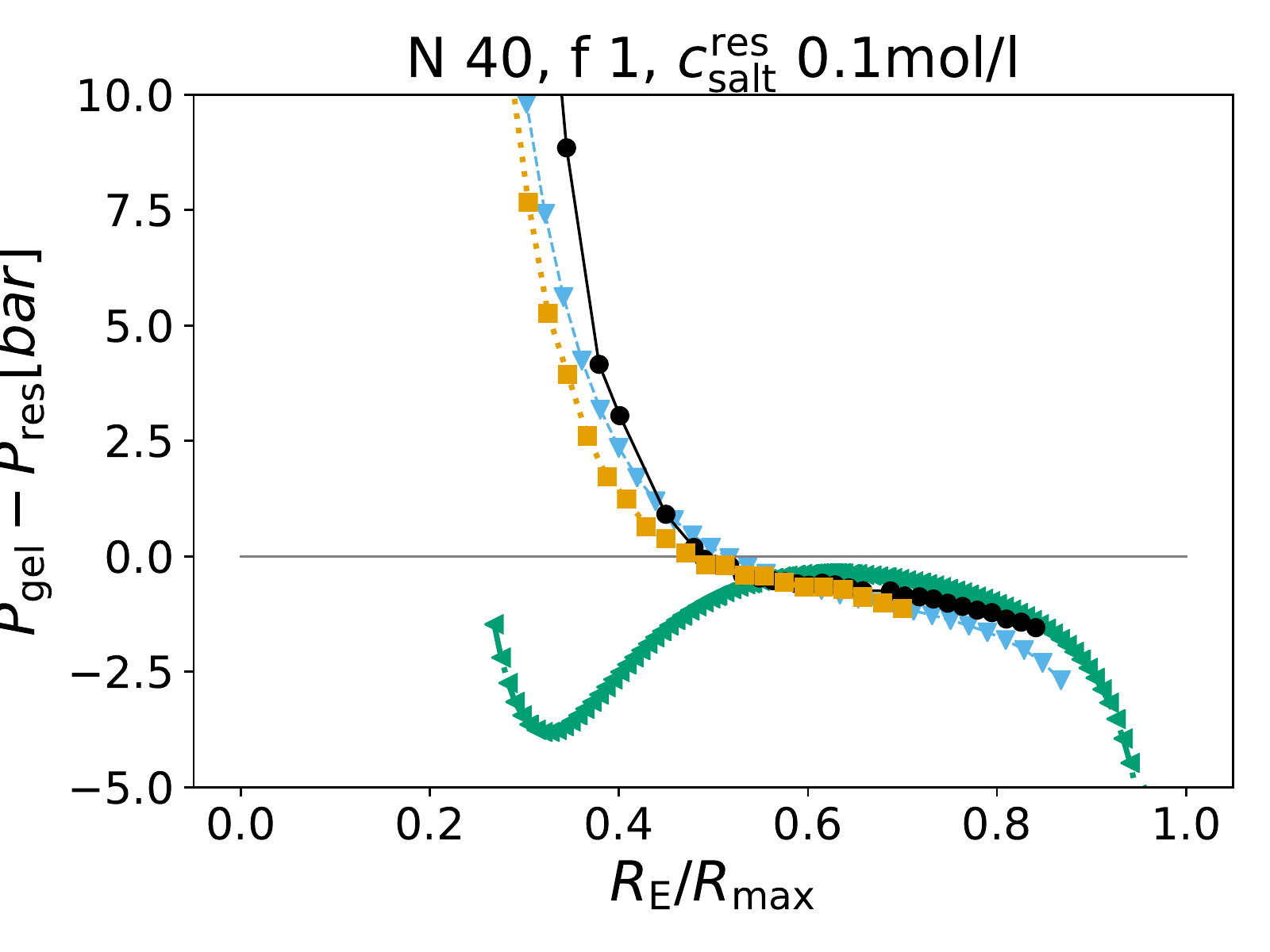}
\includegraphics[width=0.32\linewidth]{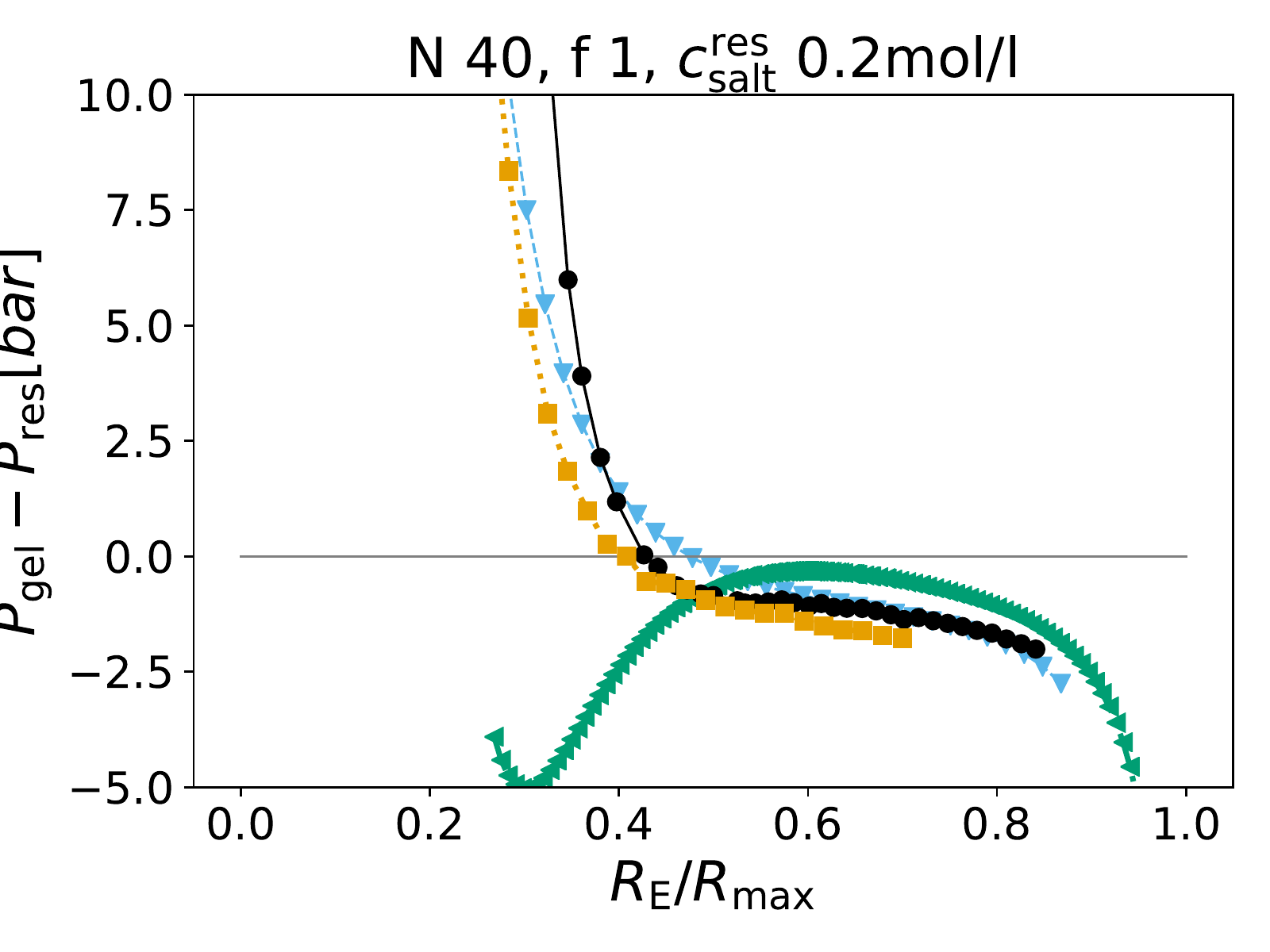}
\includegraphics[width=0.32\linewidth]{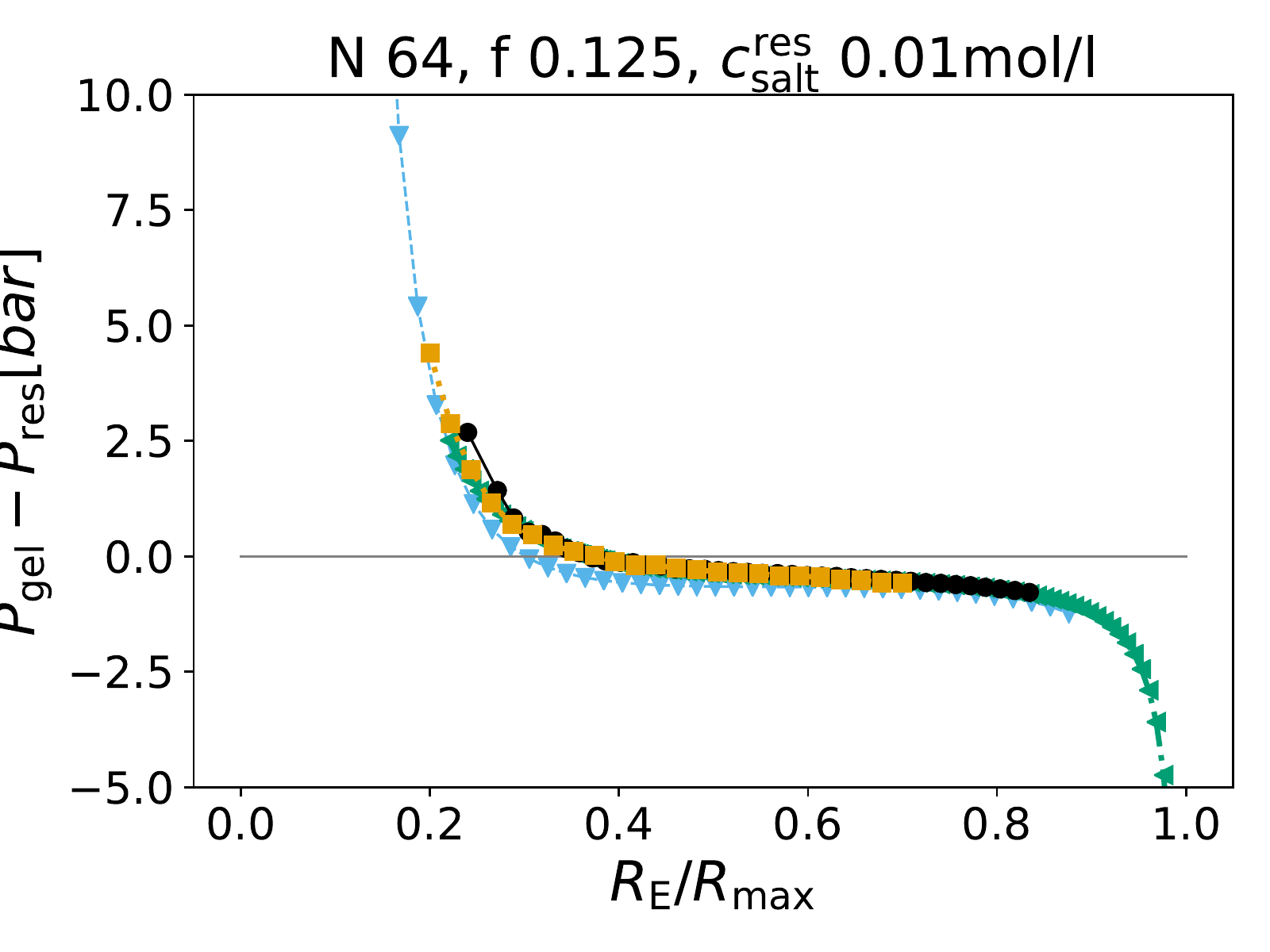}
\includegraphics[width=0.32\linewidth]{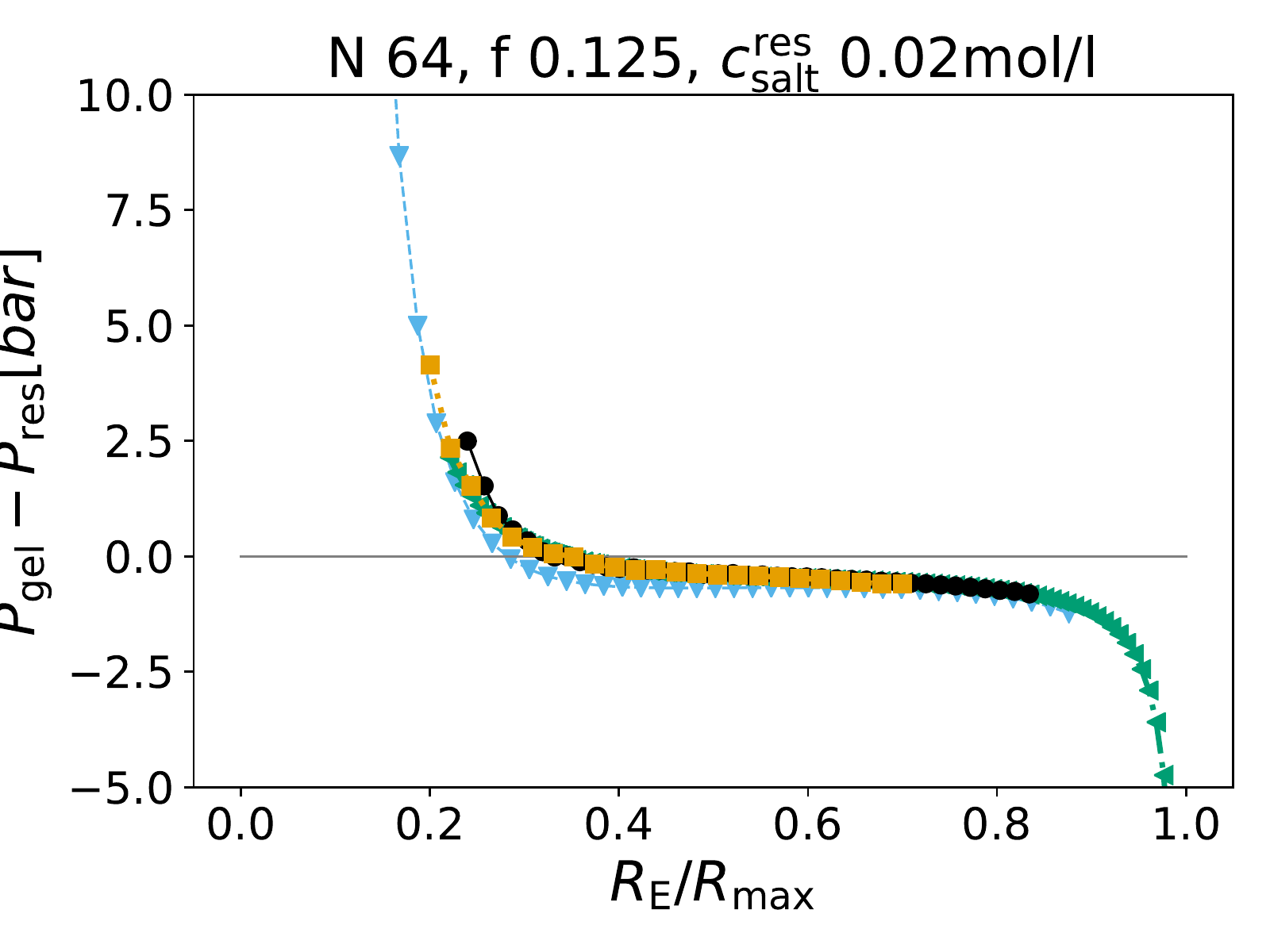}
\includegraphics[width=0.32\linewidth]{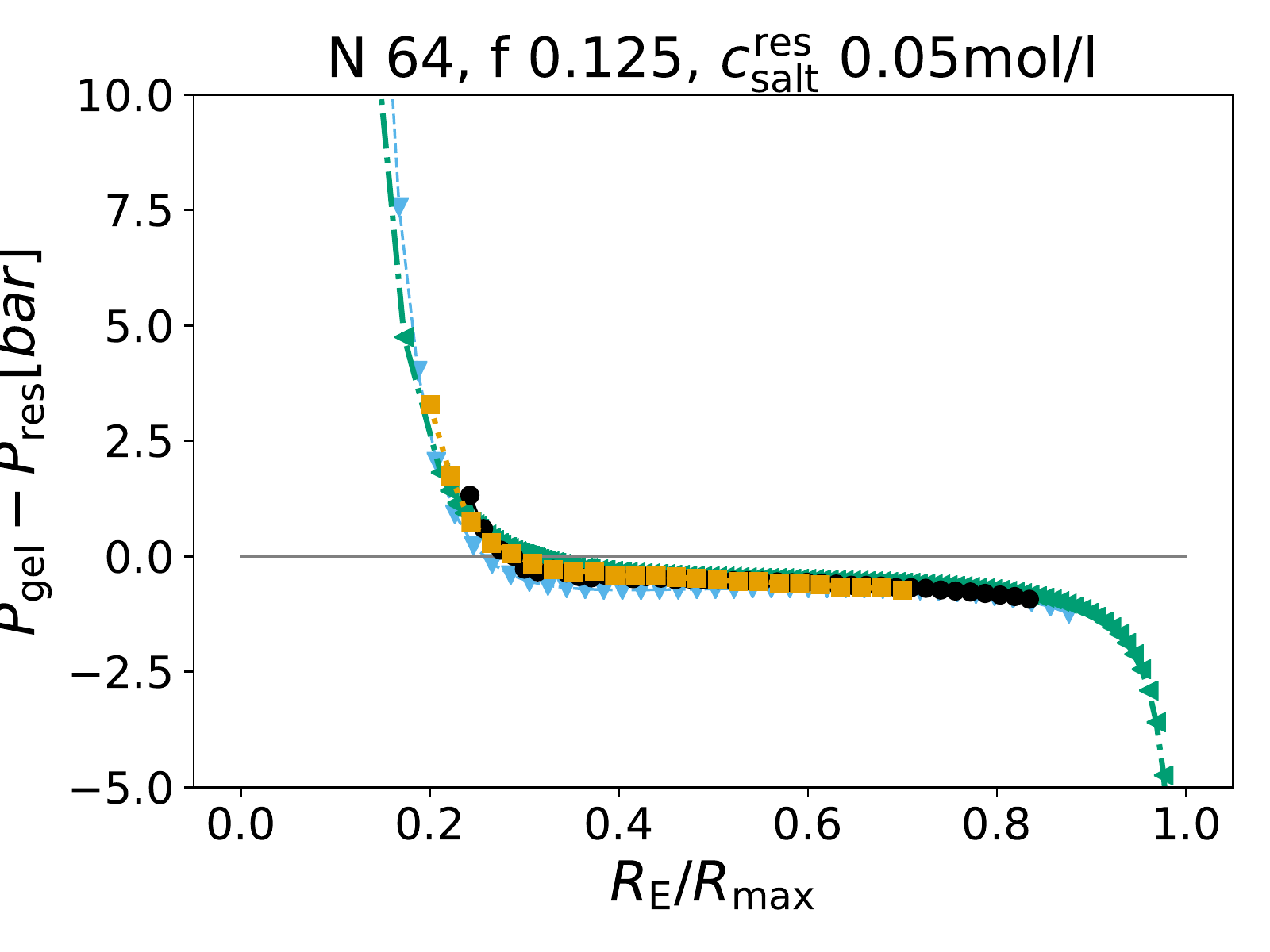}
\includegraphics[width=0.32\linewidth]{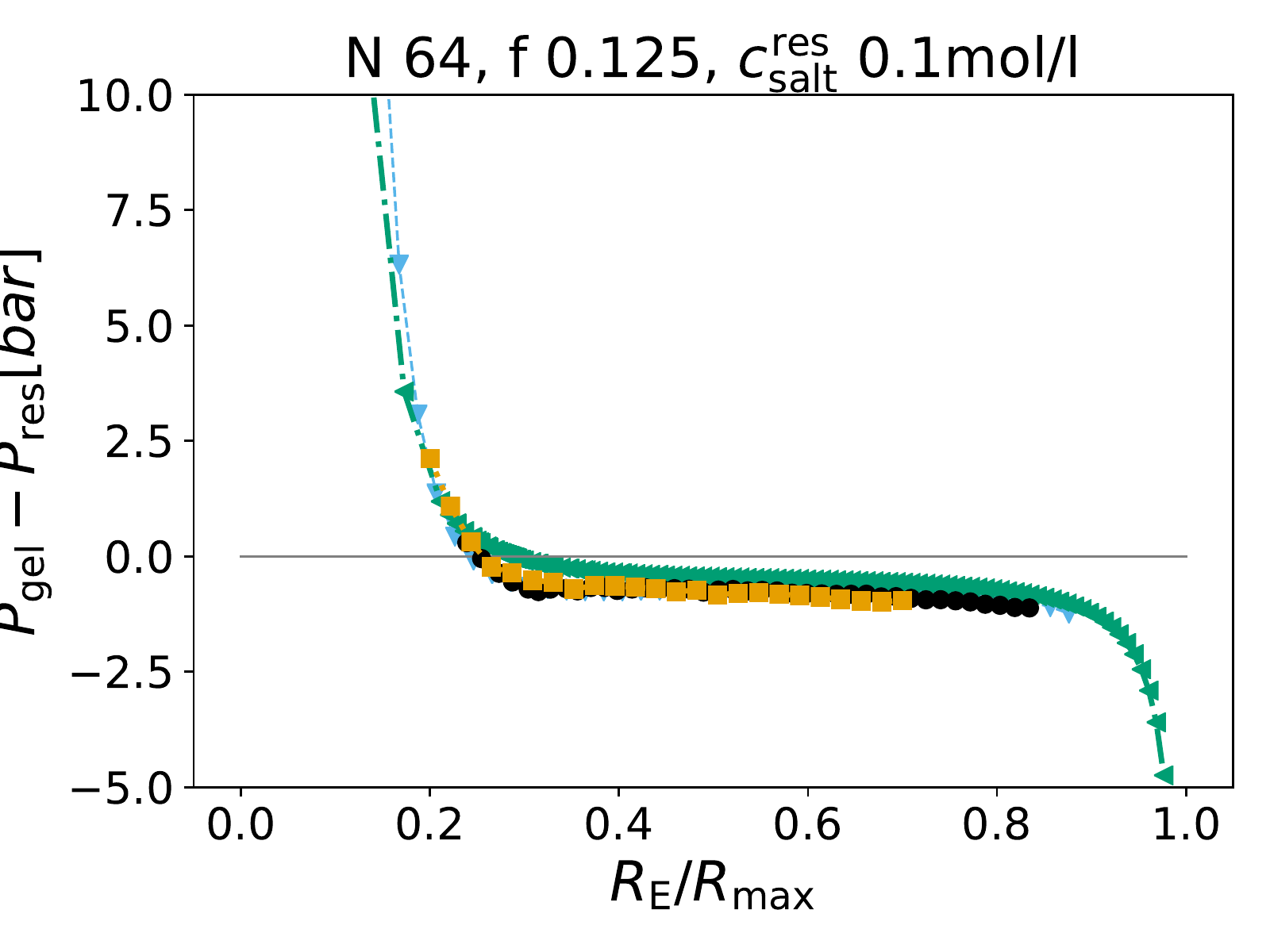}
\includegraphics[width=0.32\linewidth]{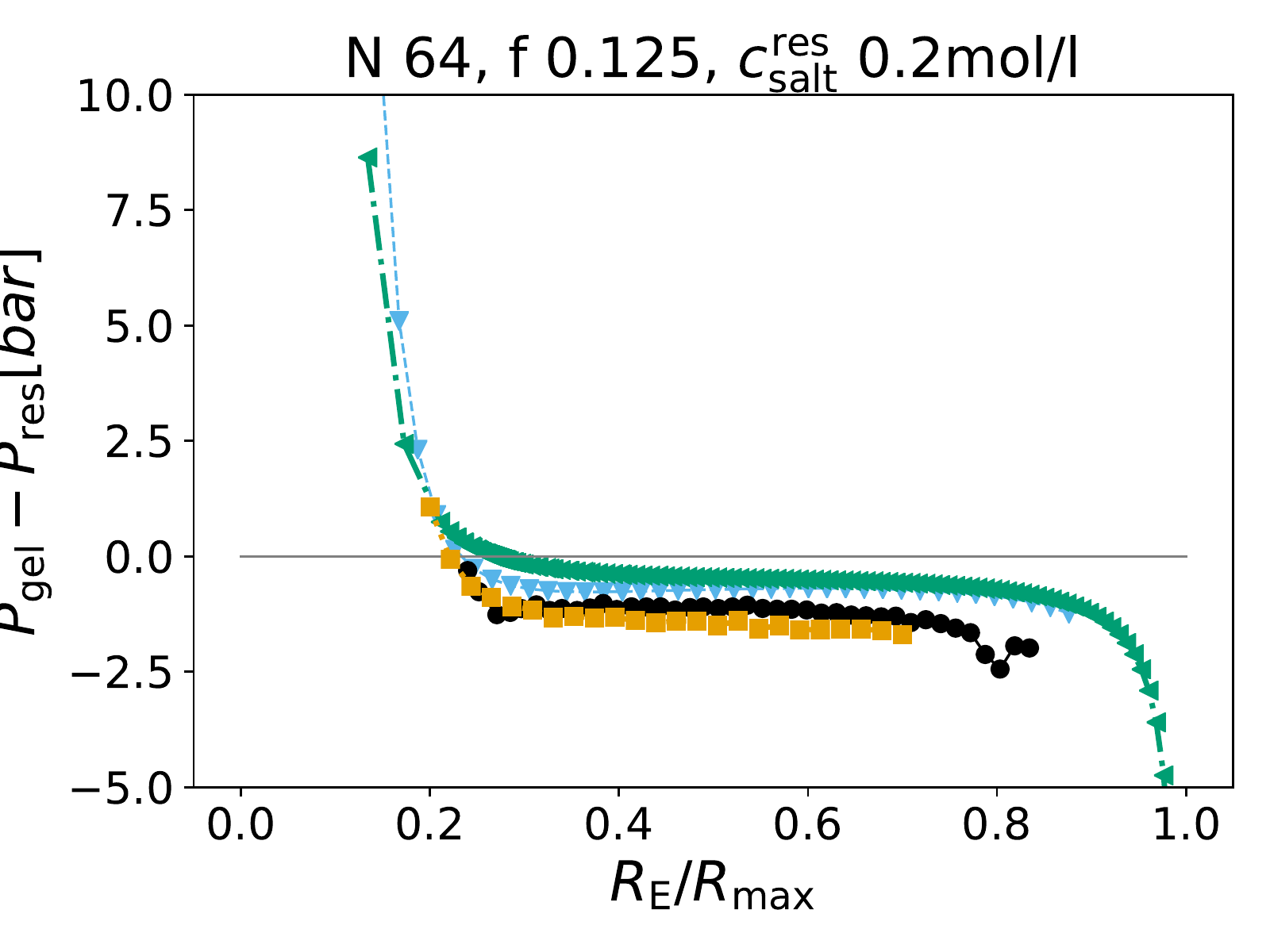}
\includegraphics[width=0.32\linewidth]{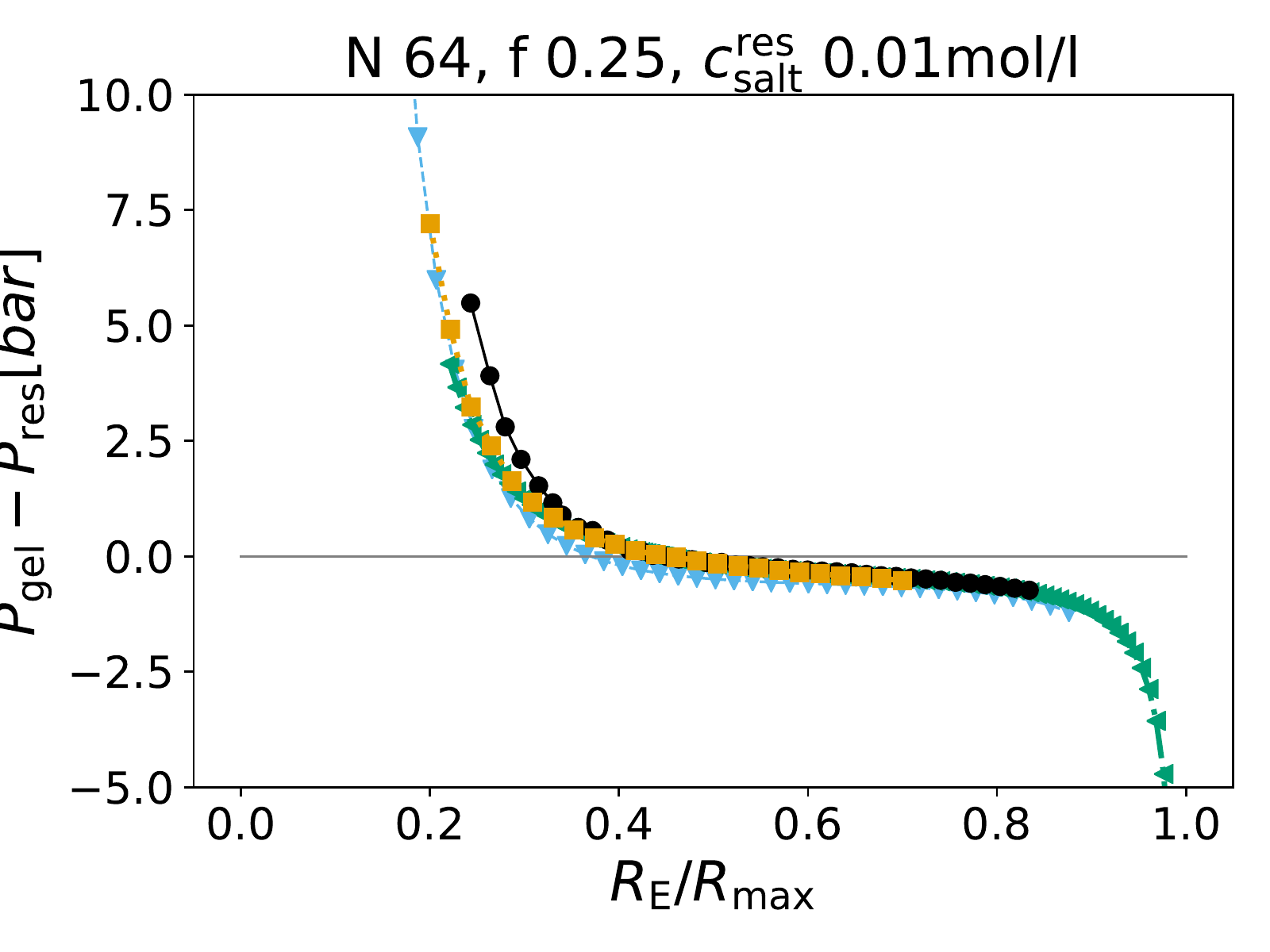}
\includegraphics[width=0.32\linewidth]{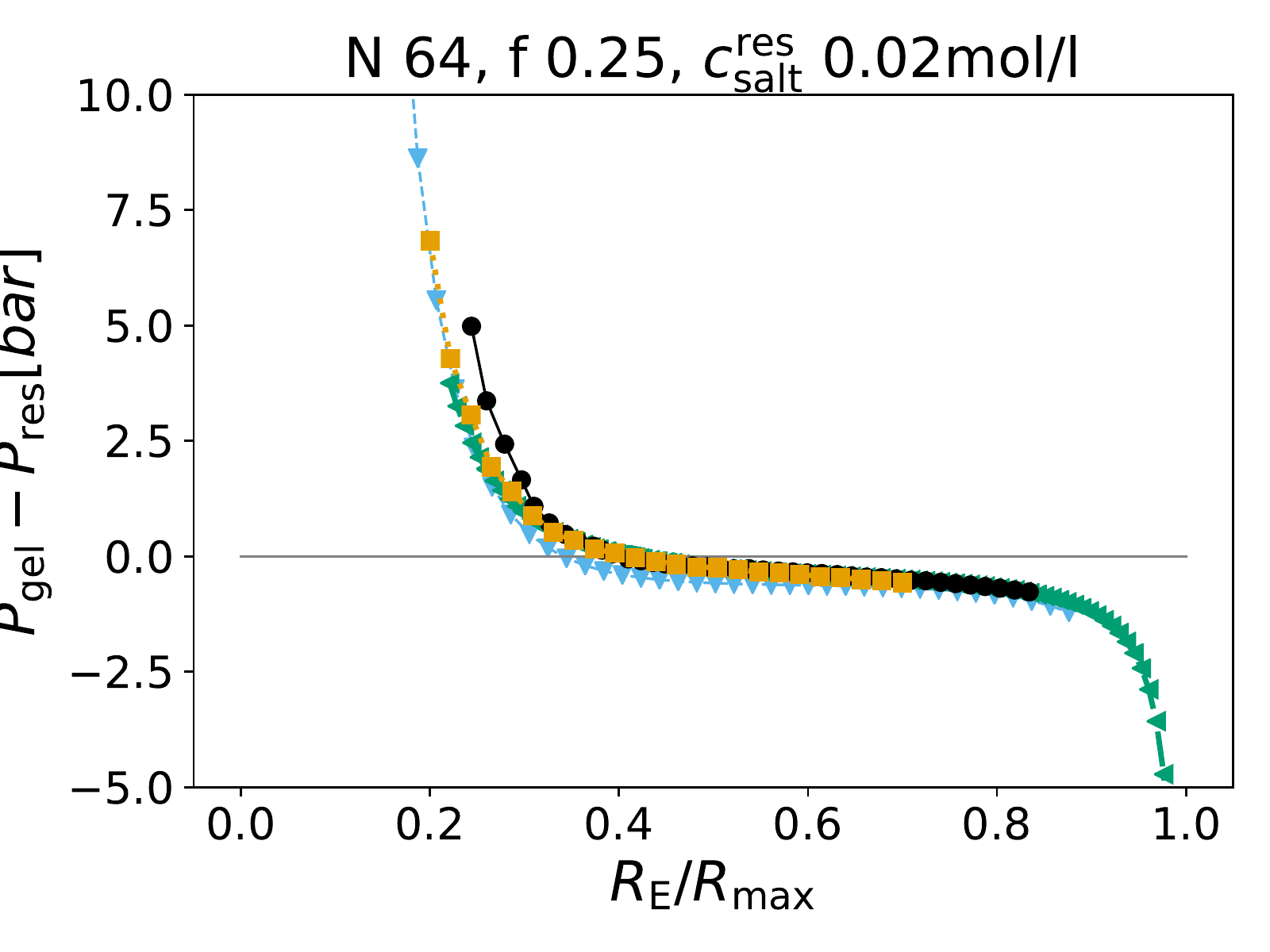}
\includegraphics[width=0.32\linewidth]{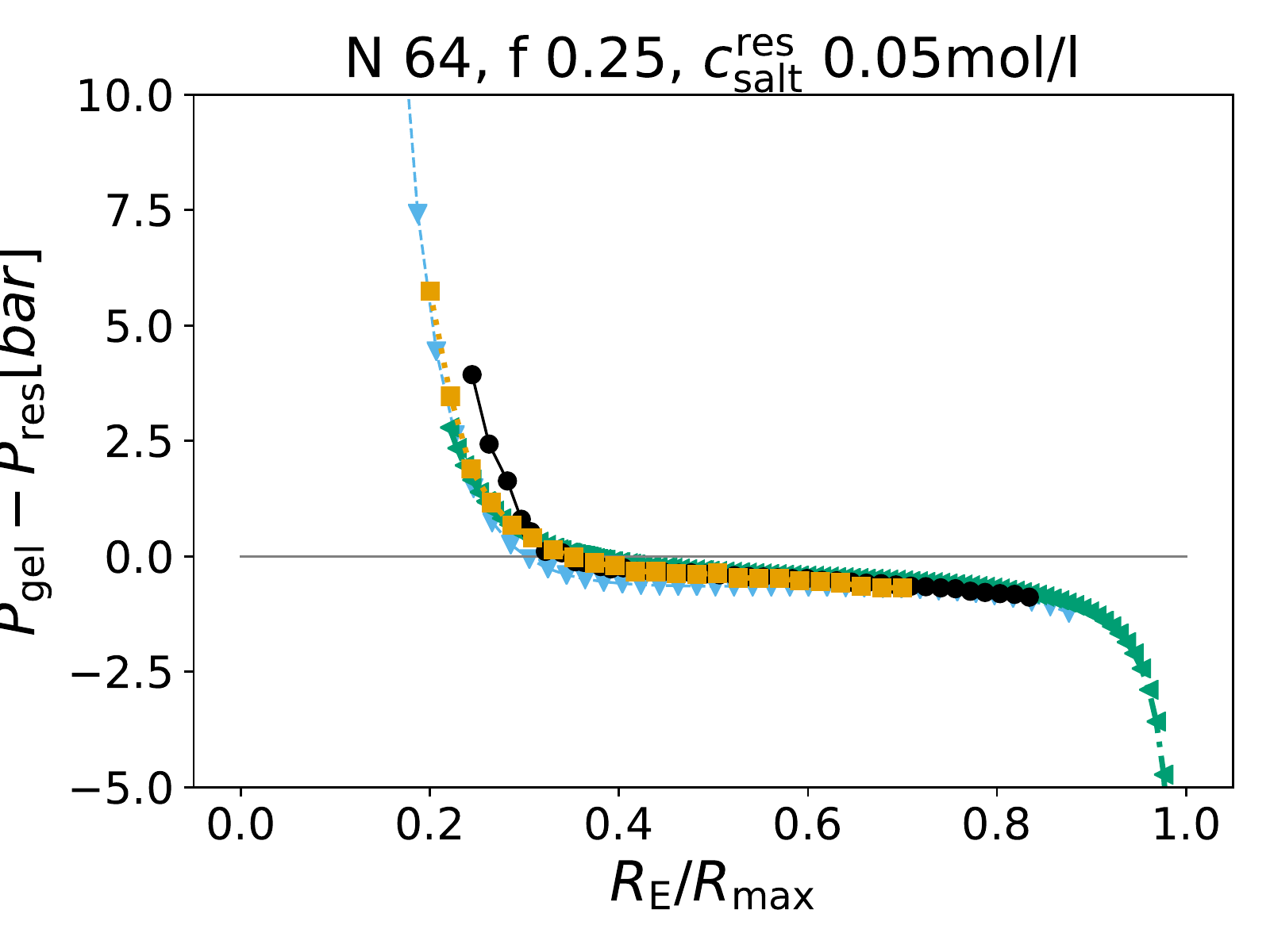}
\includegraphics[width=0.32\linewidth]{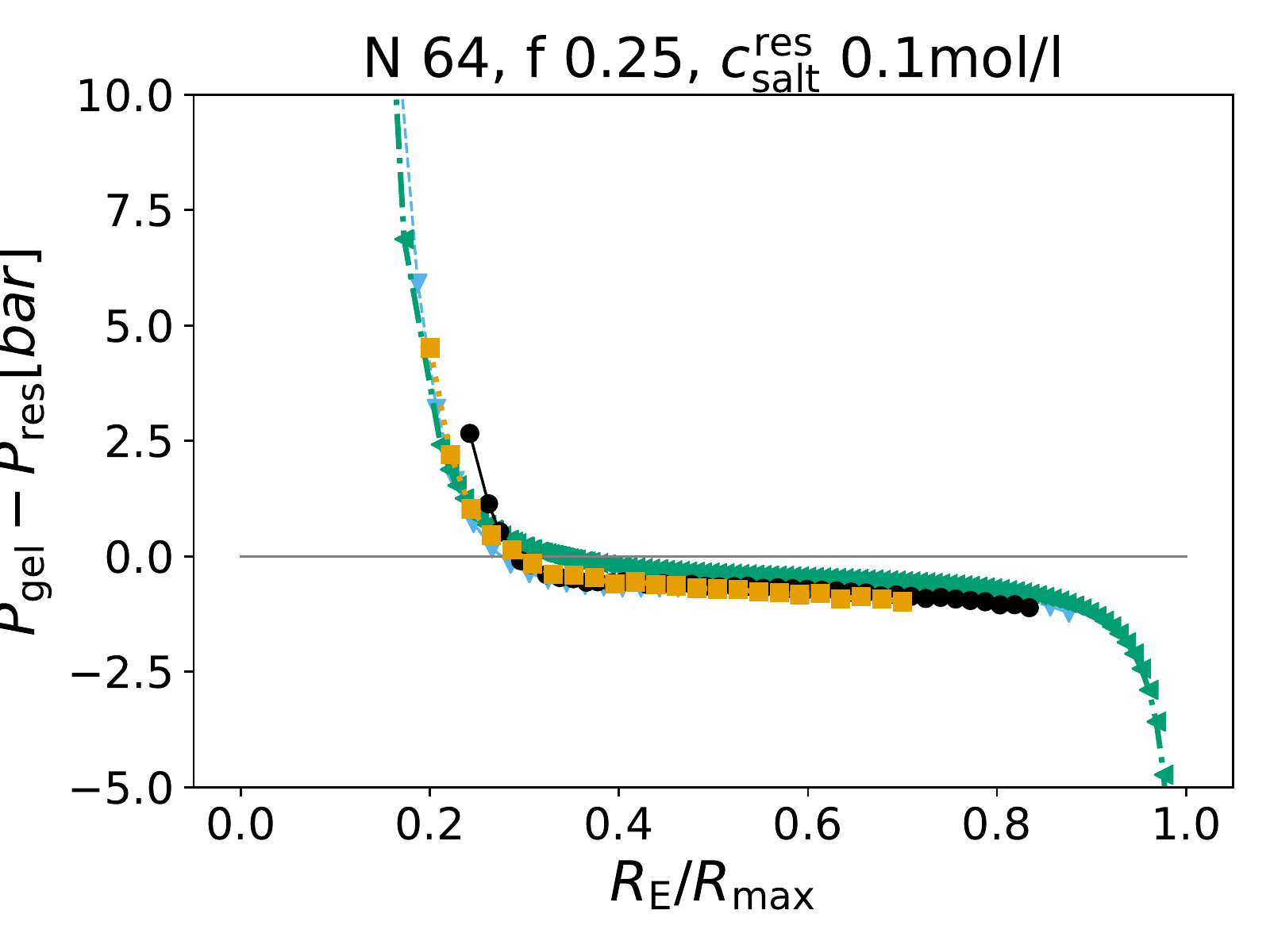}
\includegraphics[width=0.32\linewidth]{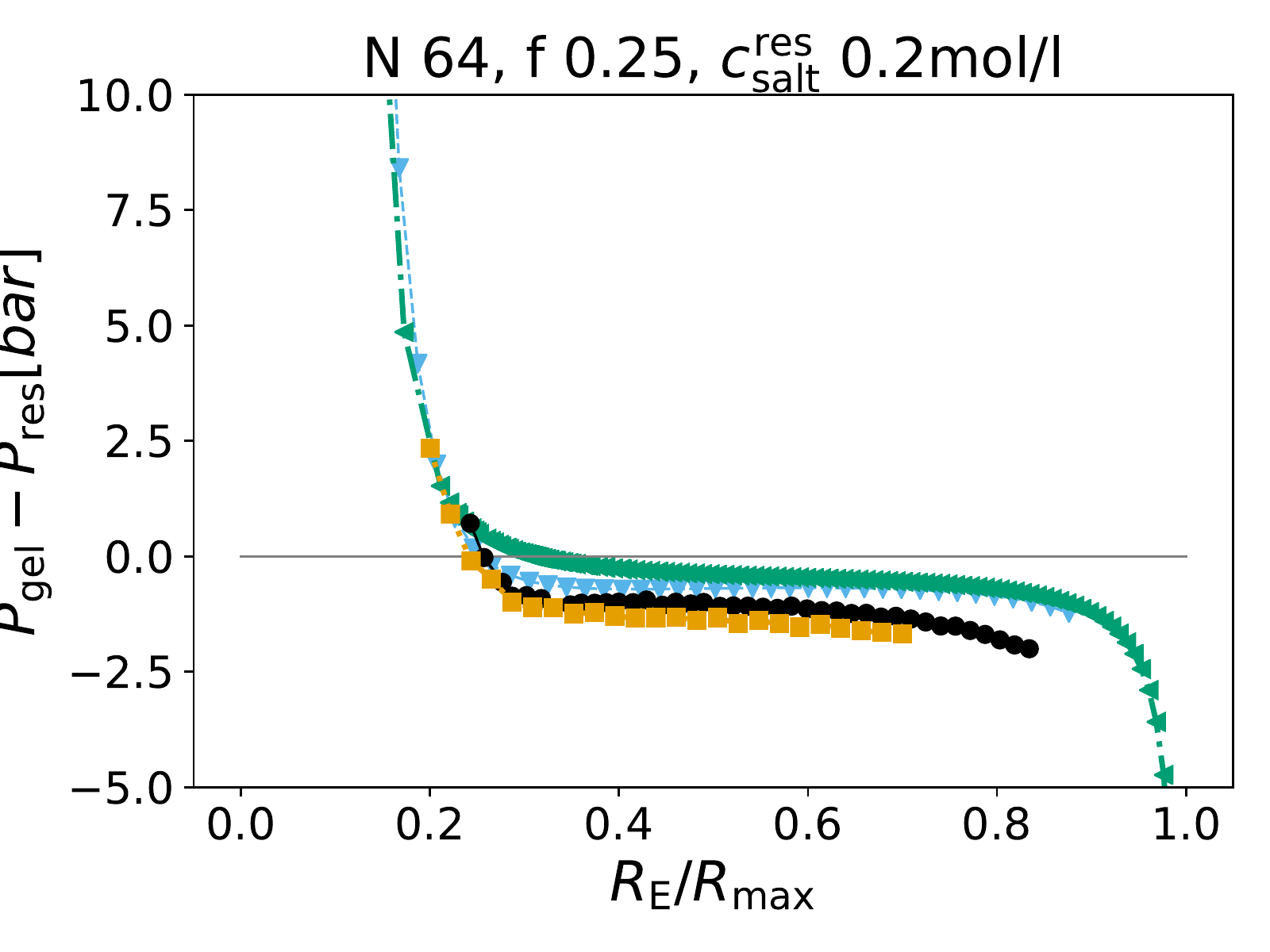}
\end{figure}
\begin{figure}
\includegraphics[width=0.32\linewidth]{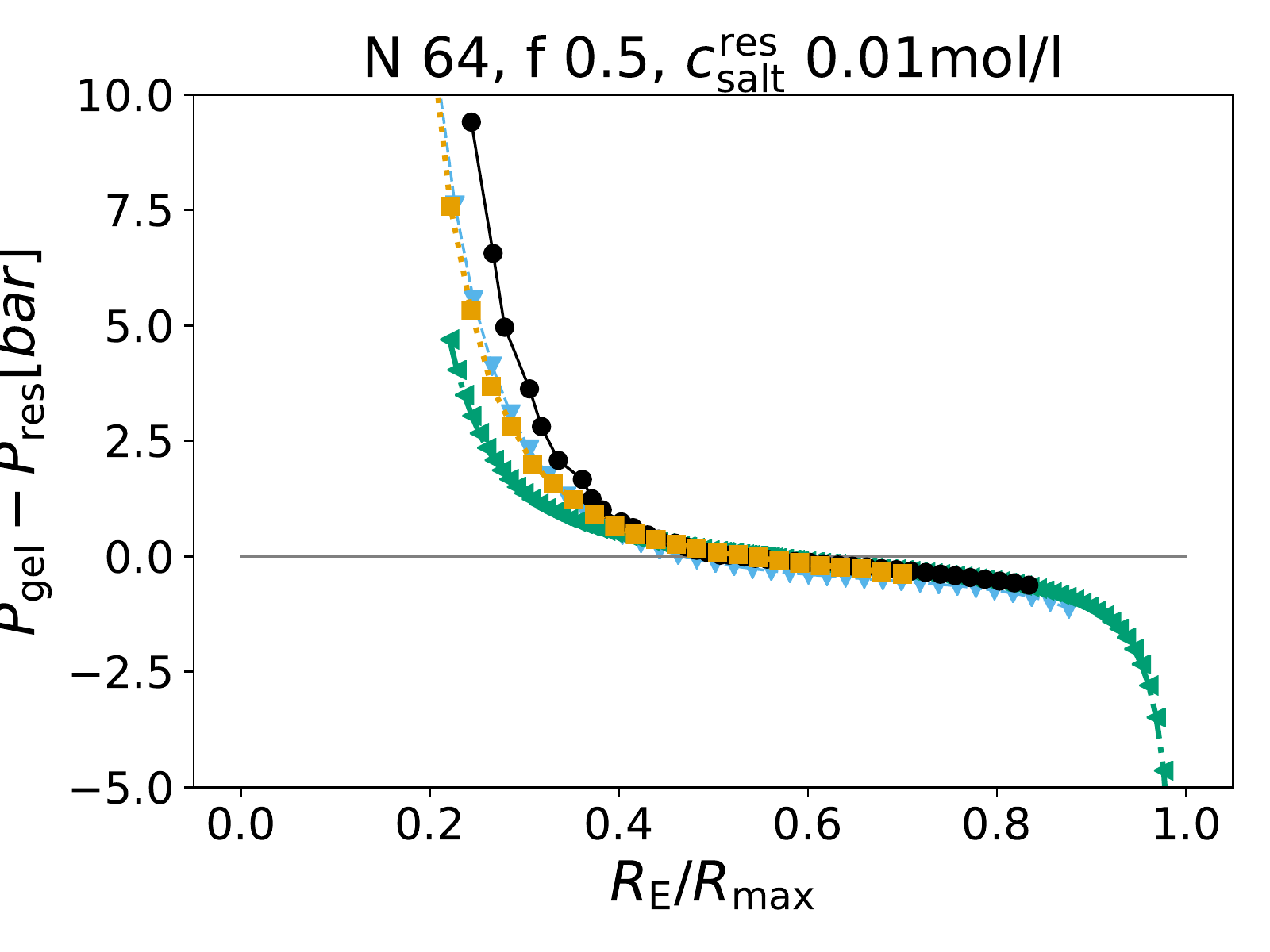}
\includegraphics[width=0.32\linewidth]{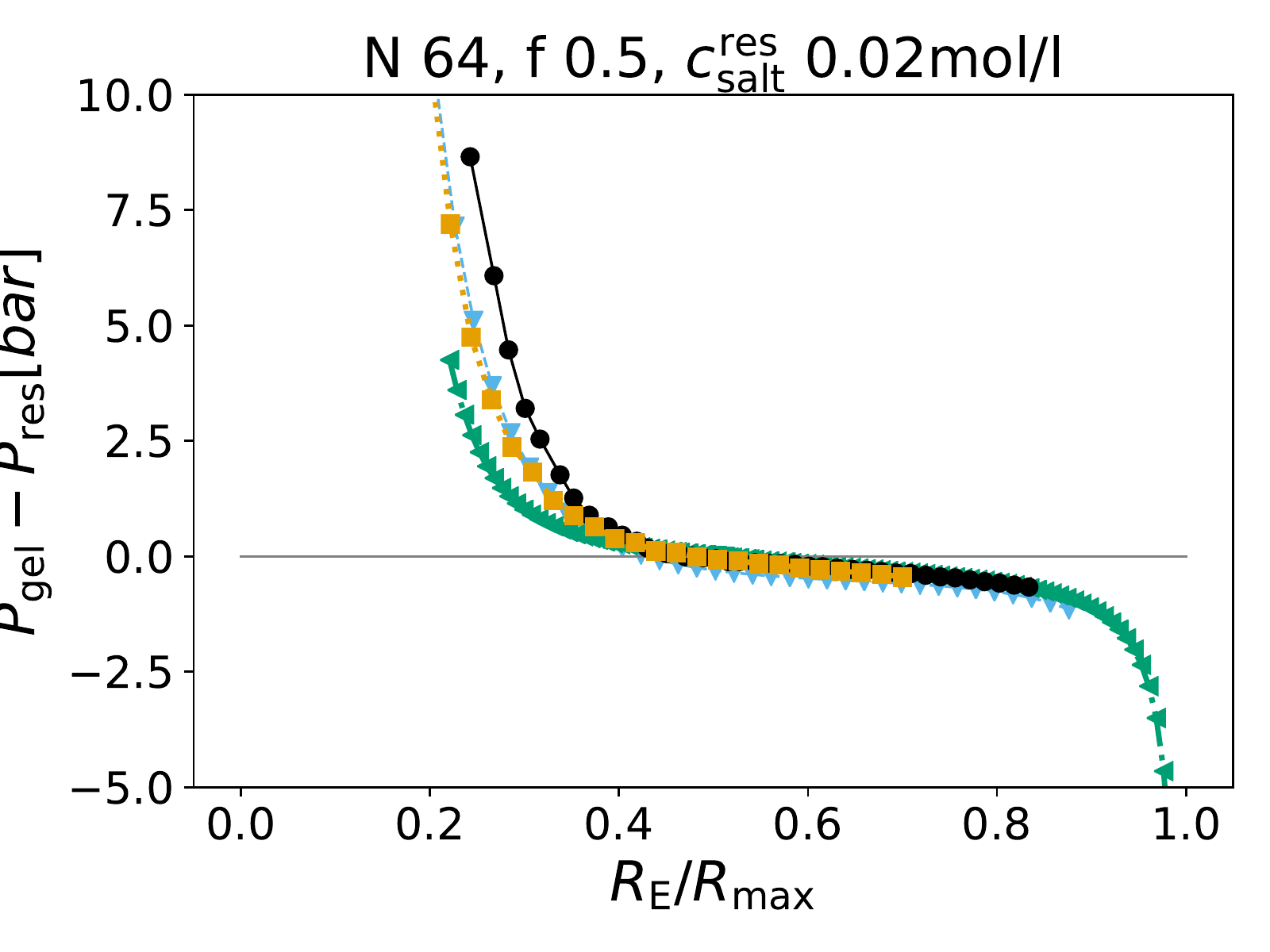}
\includegraphics[width=0.32\linewidth]{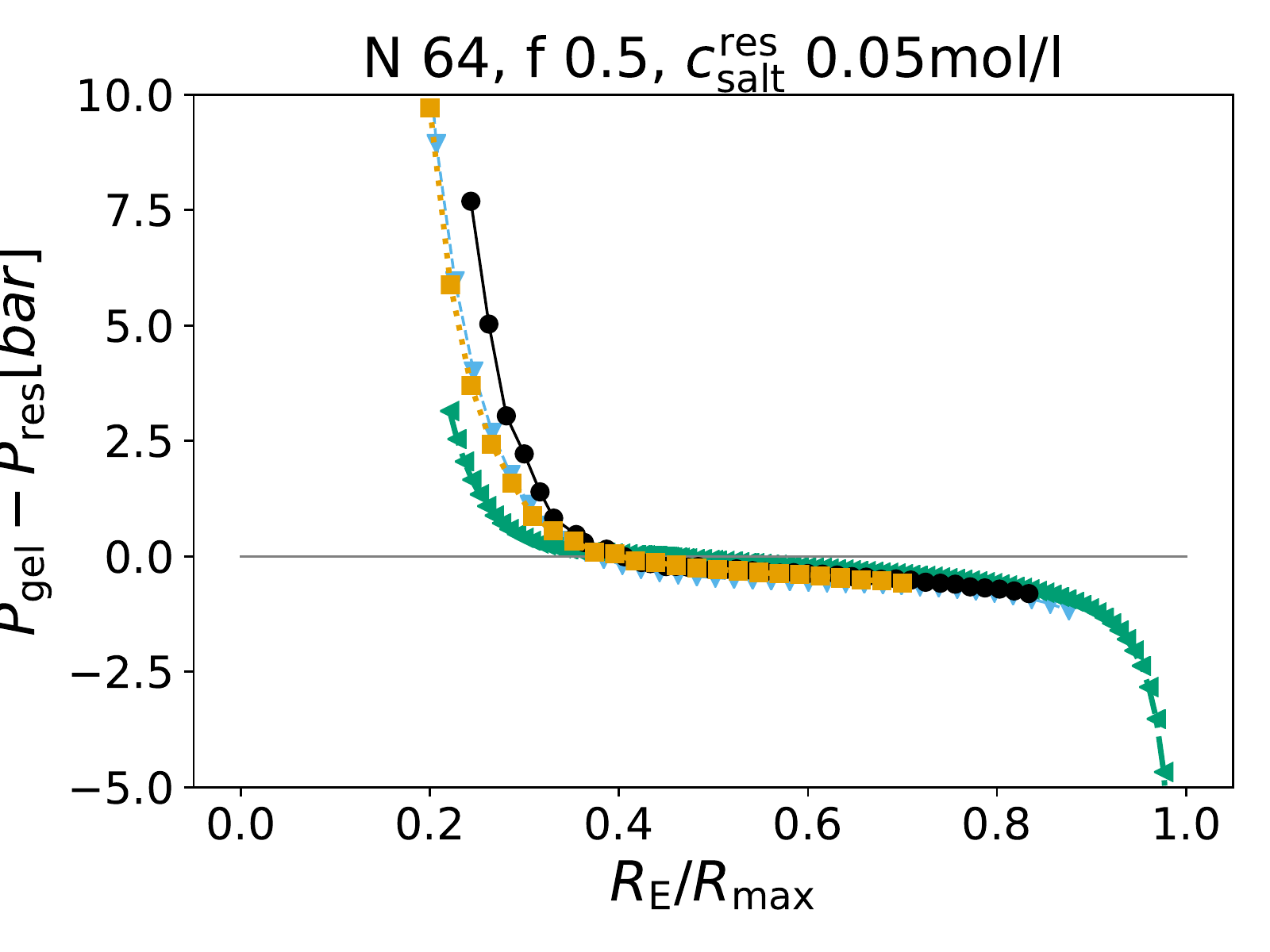}
\includegraphics[width=0.32\linewidth]{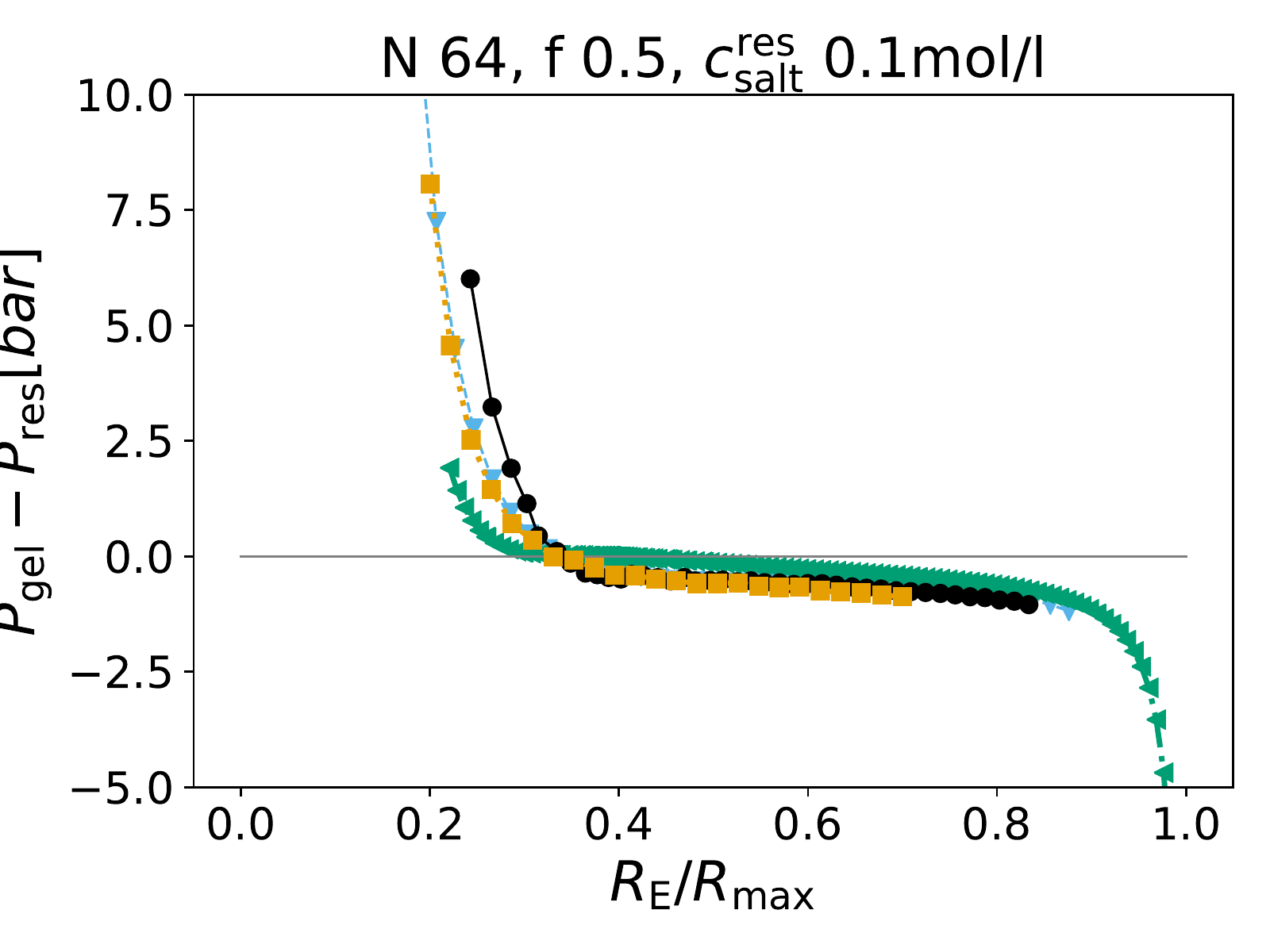}
\includegraphics[width=0.32\linewidth]{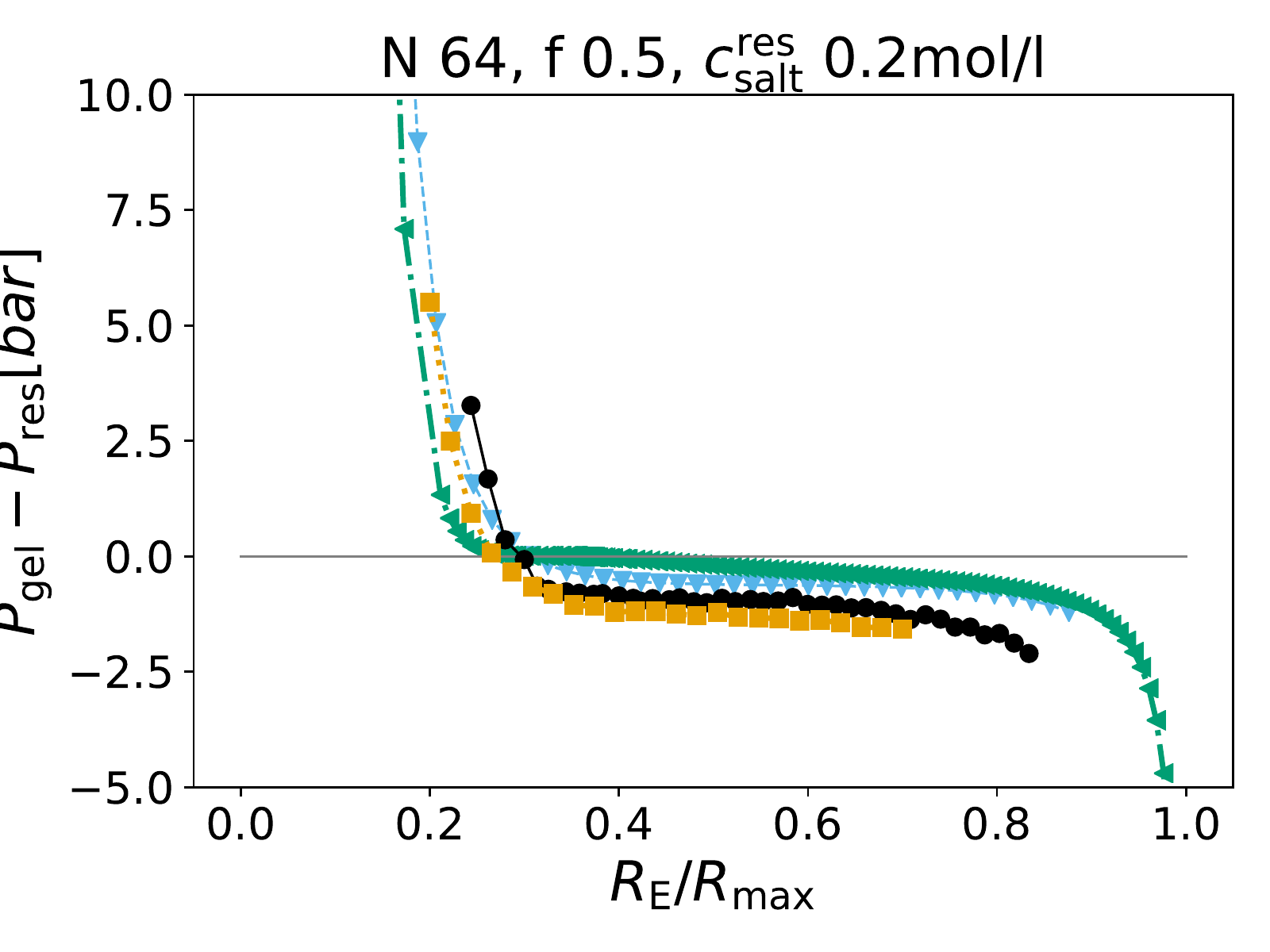}
\includegraphics[width=0.32\linewidth]{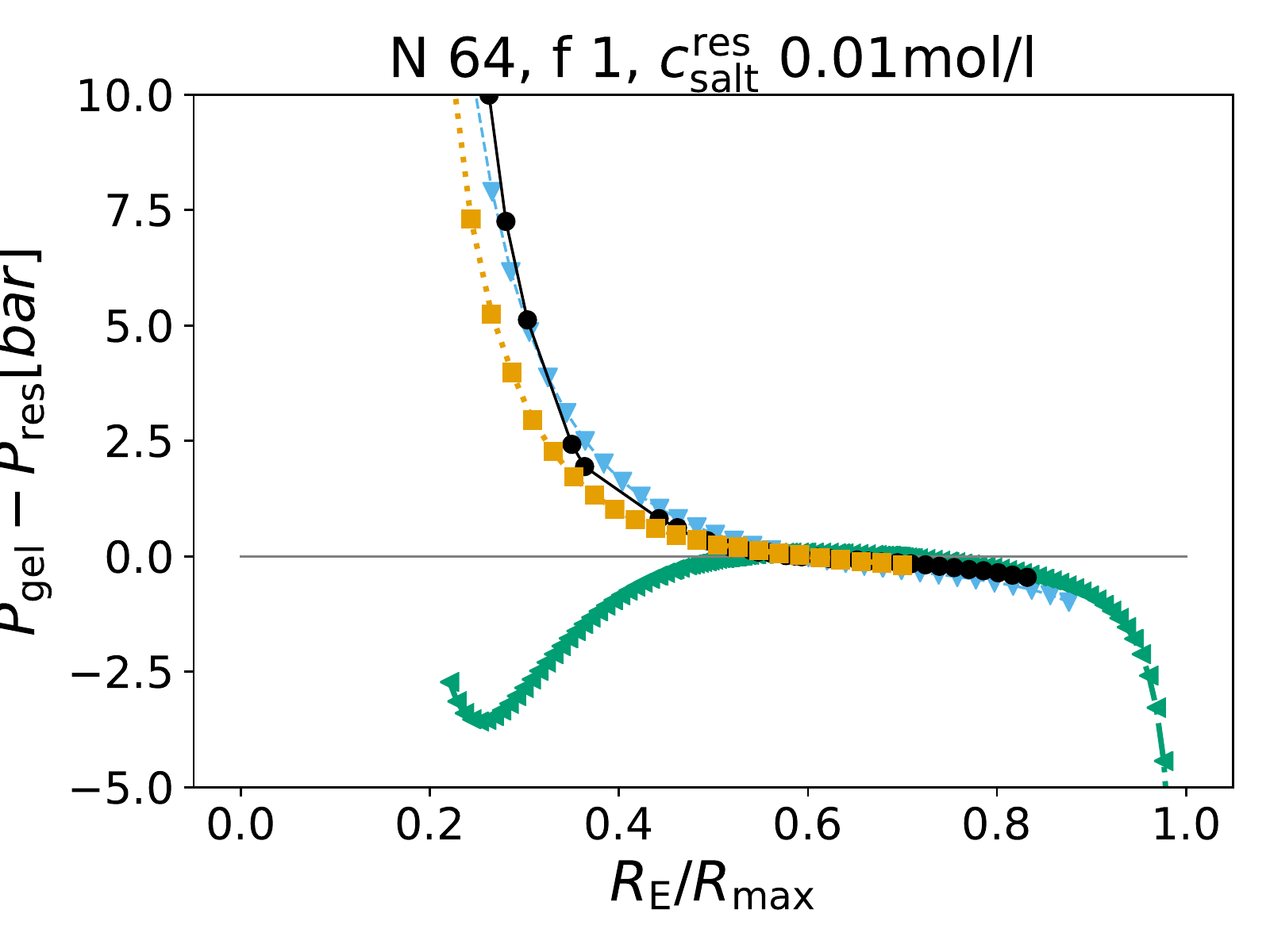}
\includegraphics[width=0.32\linewidth]{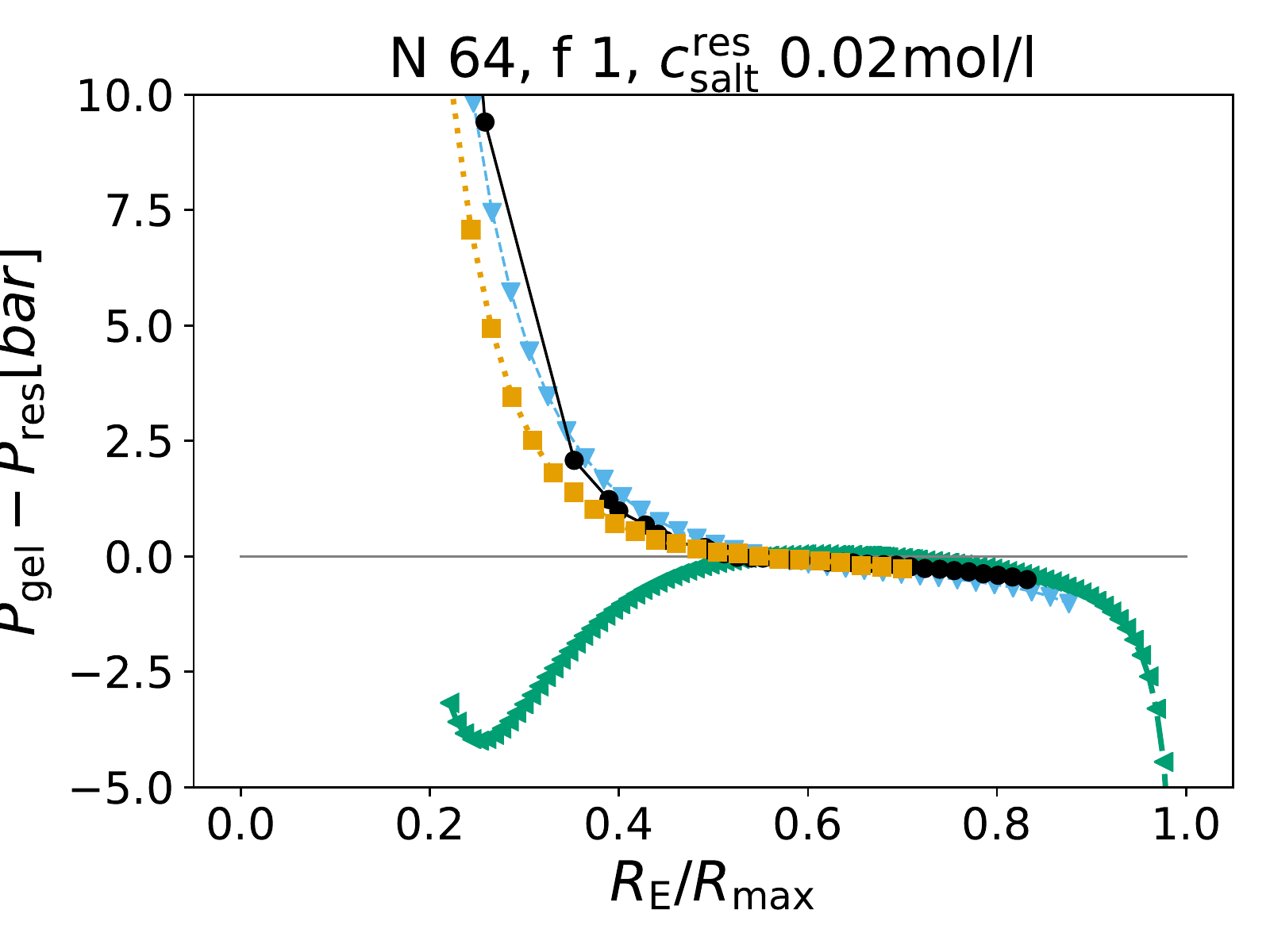}
\includegraphics[width=0.32\linewidth]{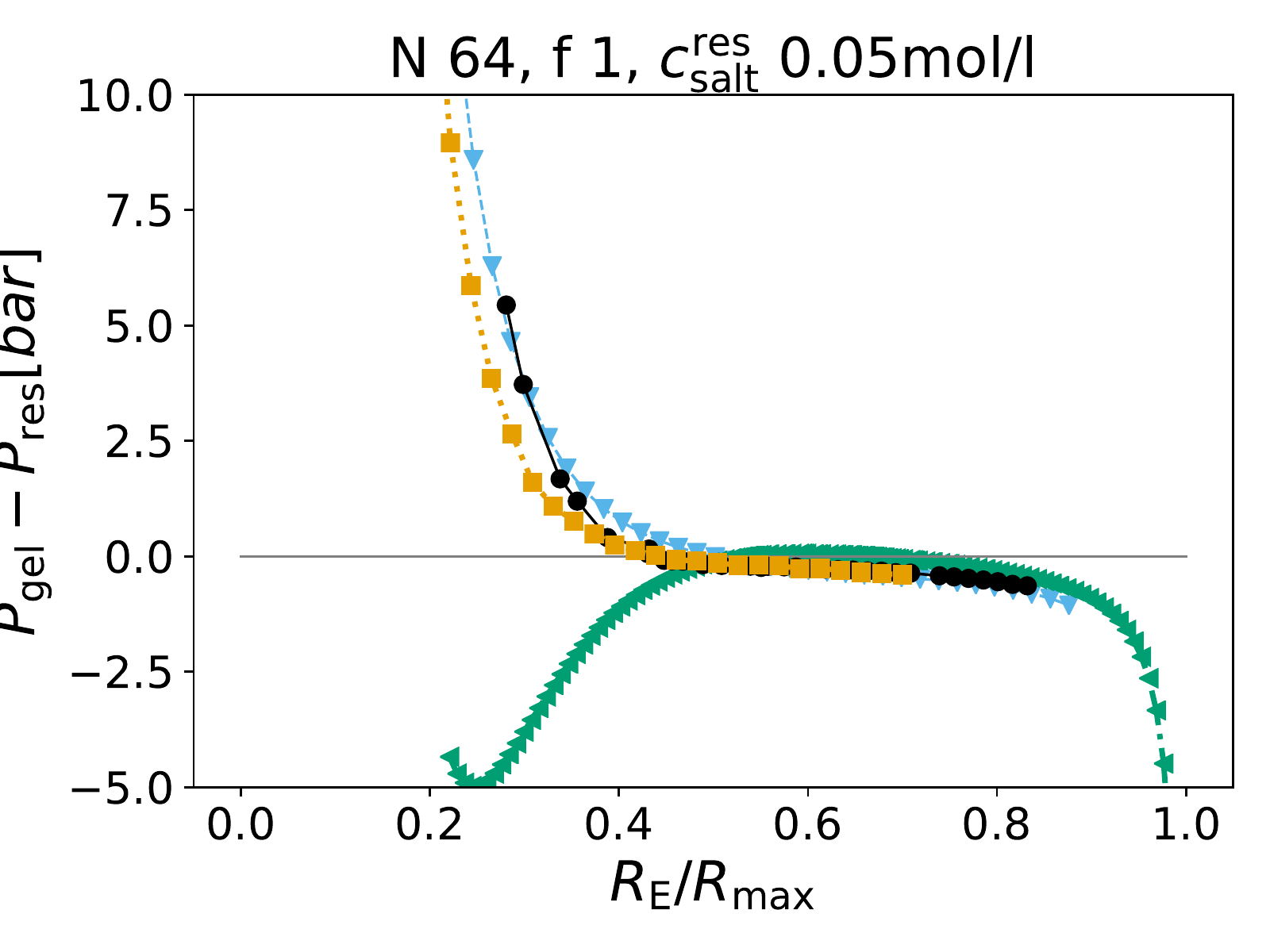}
\includegraphics[width=0.32\linewidth]{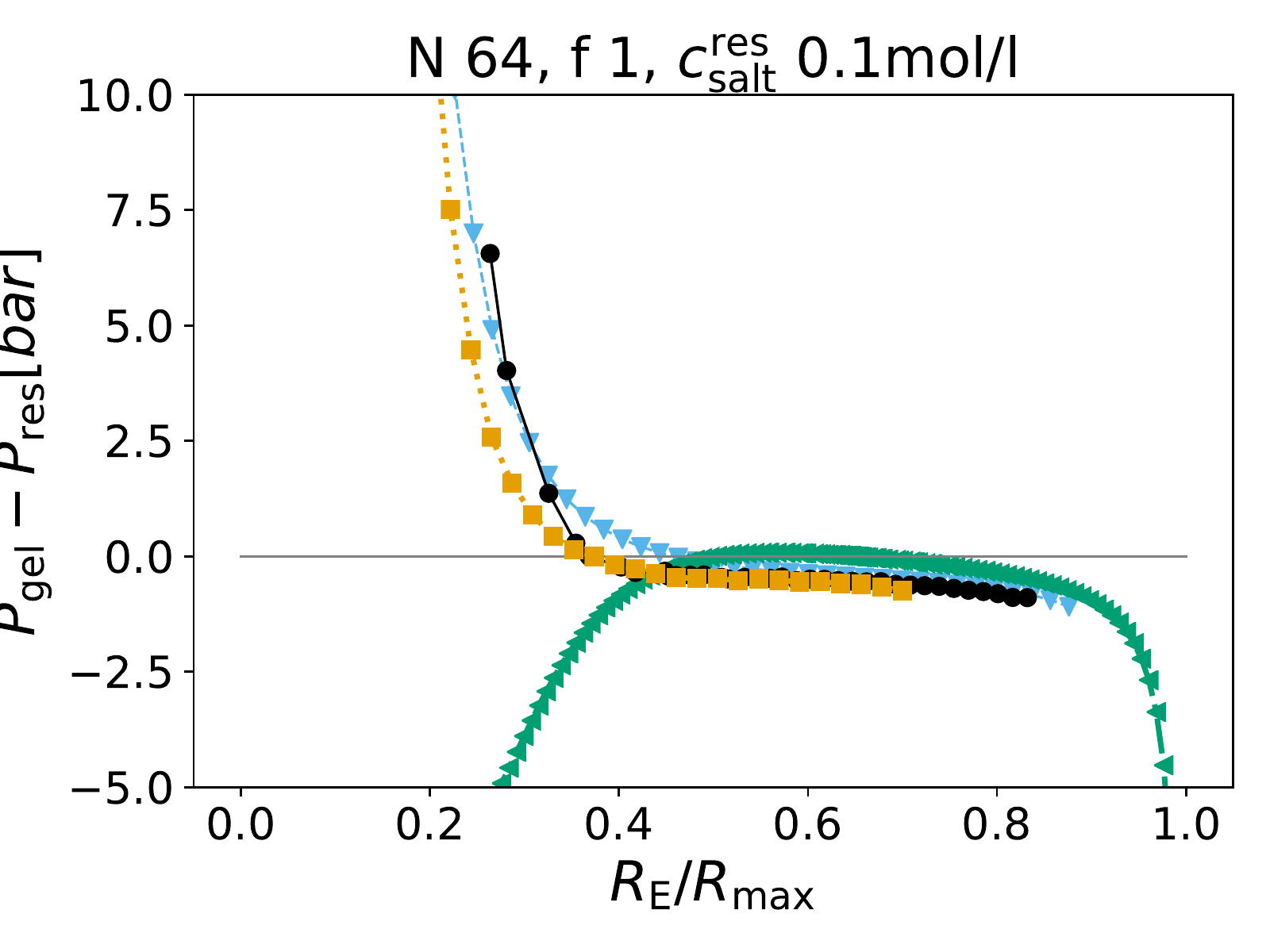}
\includegraphics[width=0.32\linewidth]{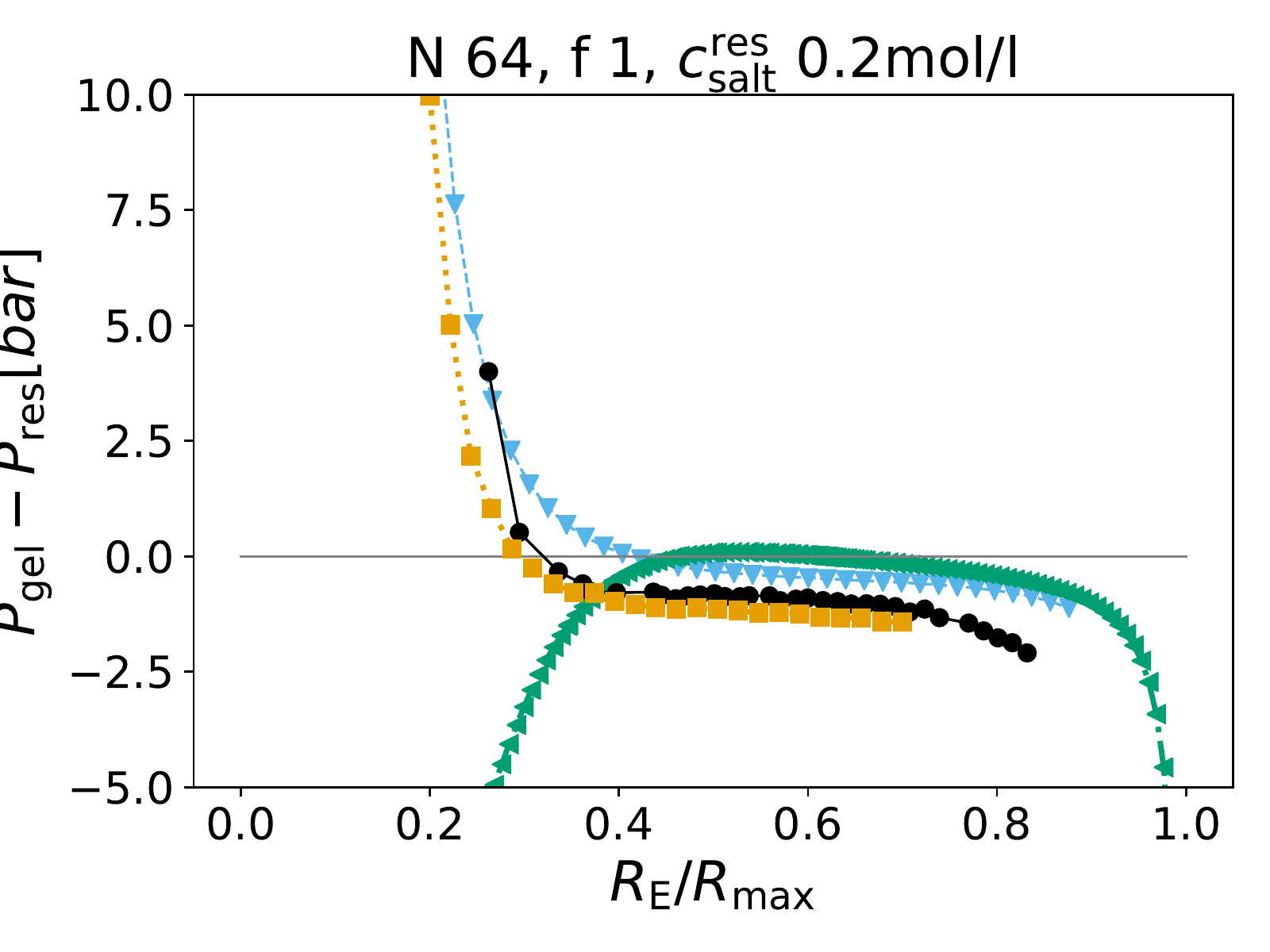}
\includegraphics[width=0.32\linewidth]{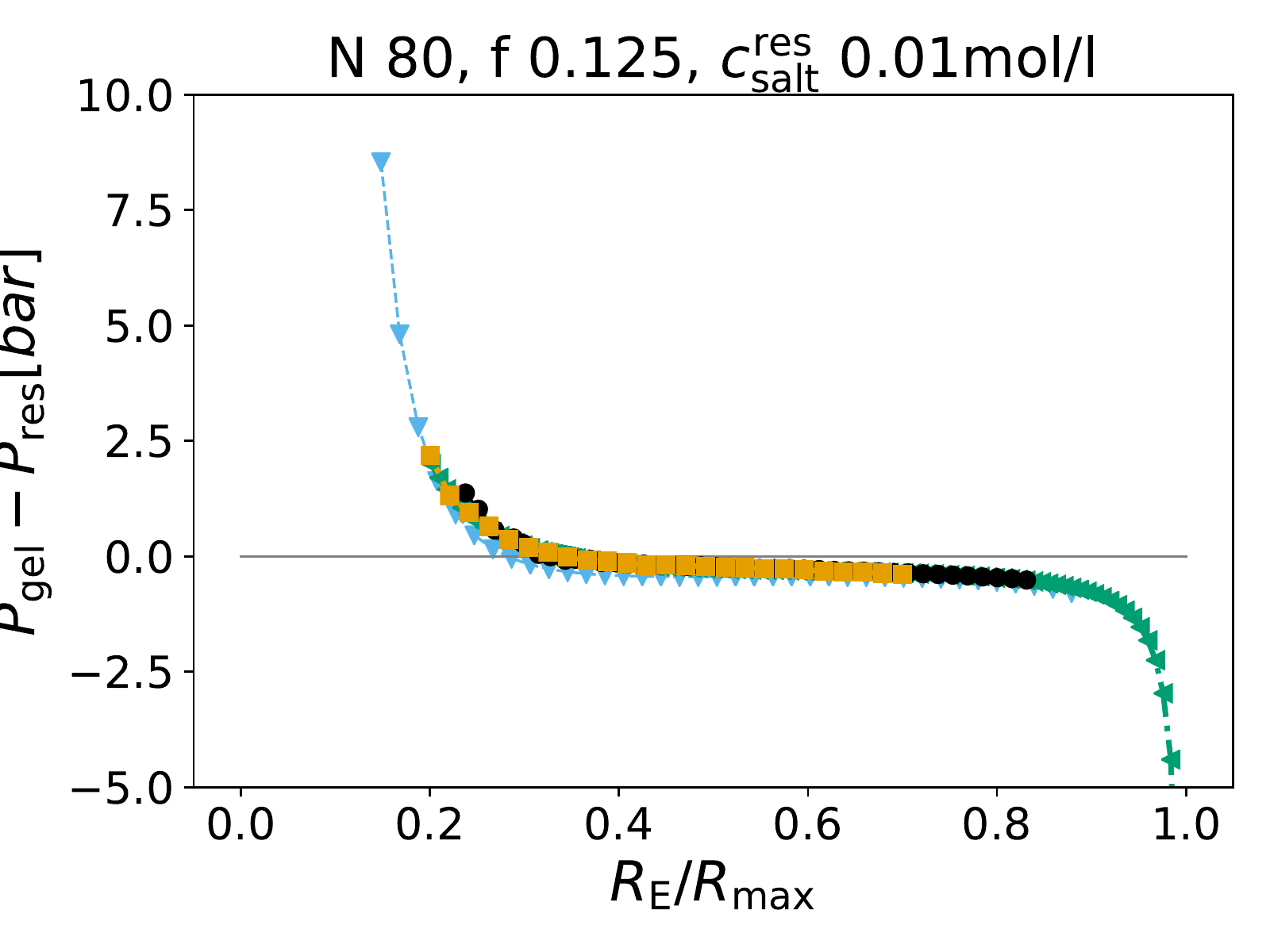}
\includegraphics[width=0.32\linewidth]{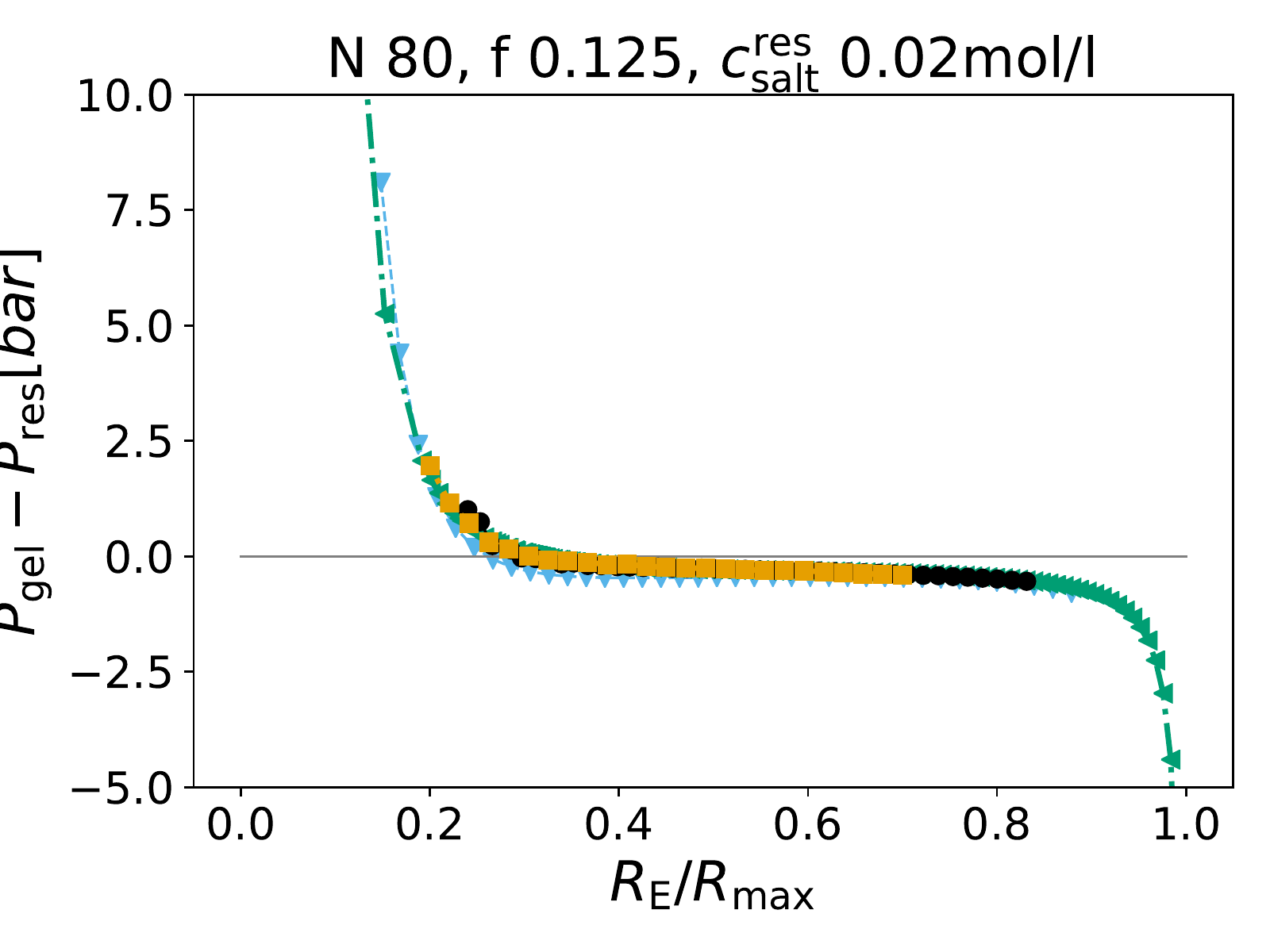}
\includegraphics[width=0.32\linewidth]{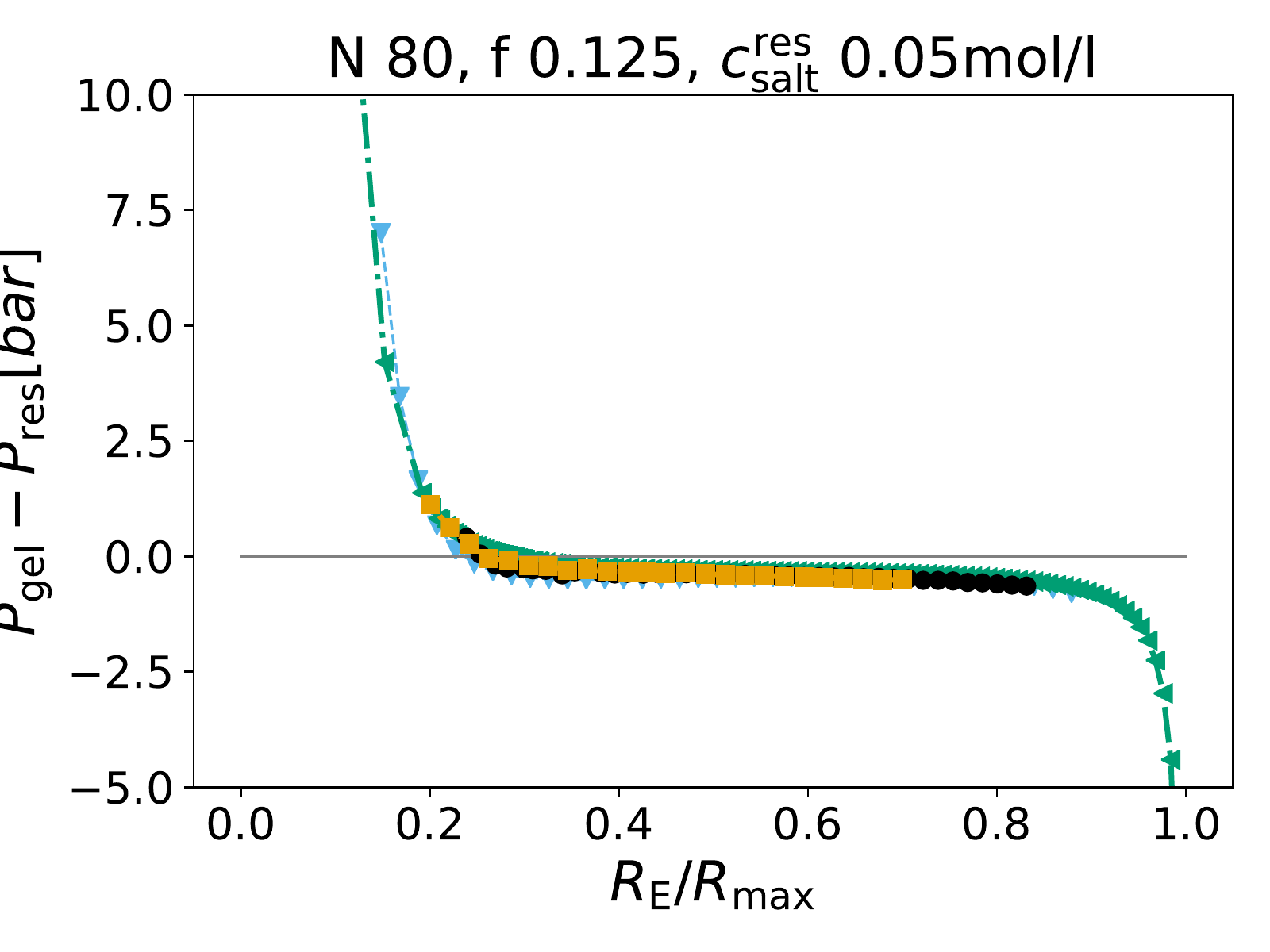}
\includegraphics[width=0.32\linewidth]{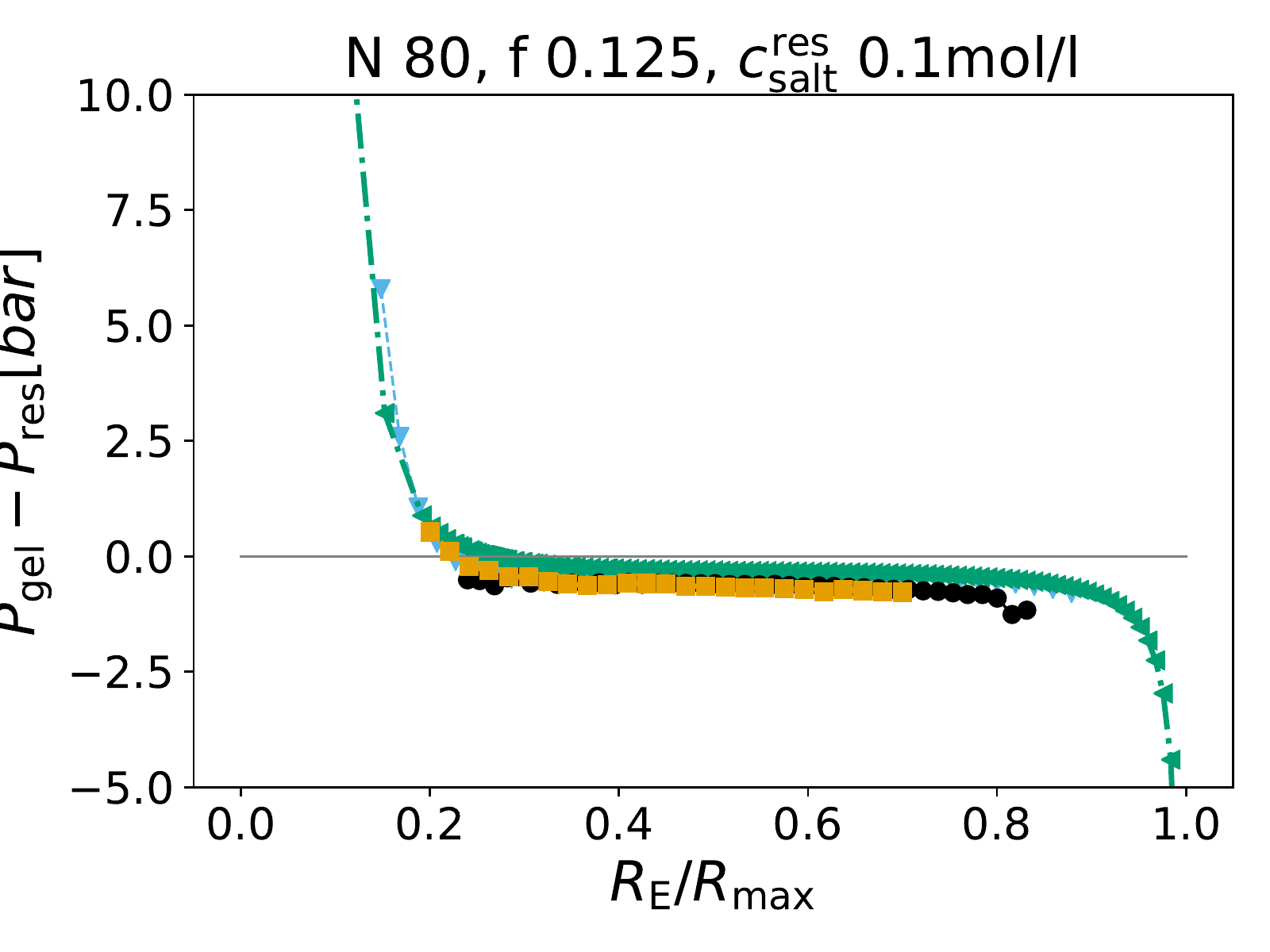}
\includegraphics[width=0.32\linewidth]{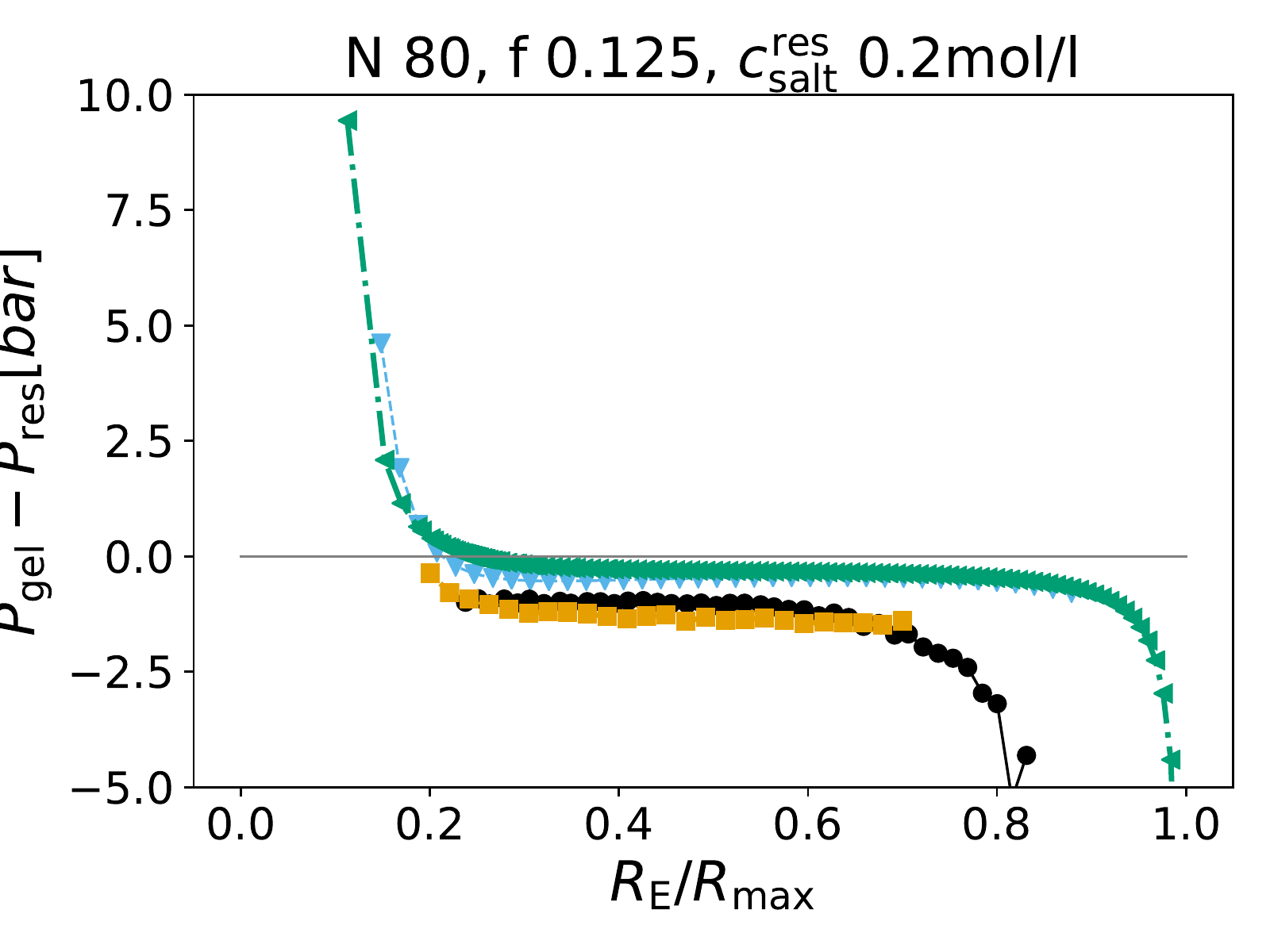}
\end{figure}
\begin{figure}
\includegraphics[width=0.32\linewidth]{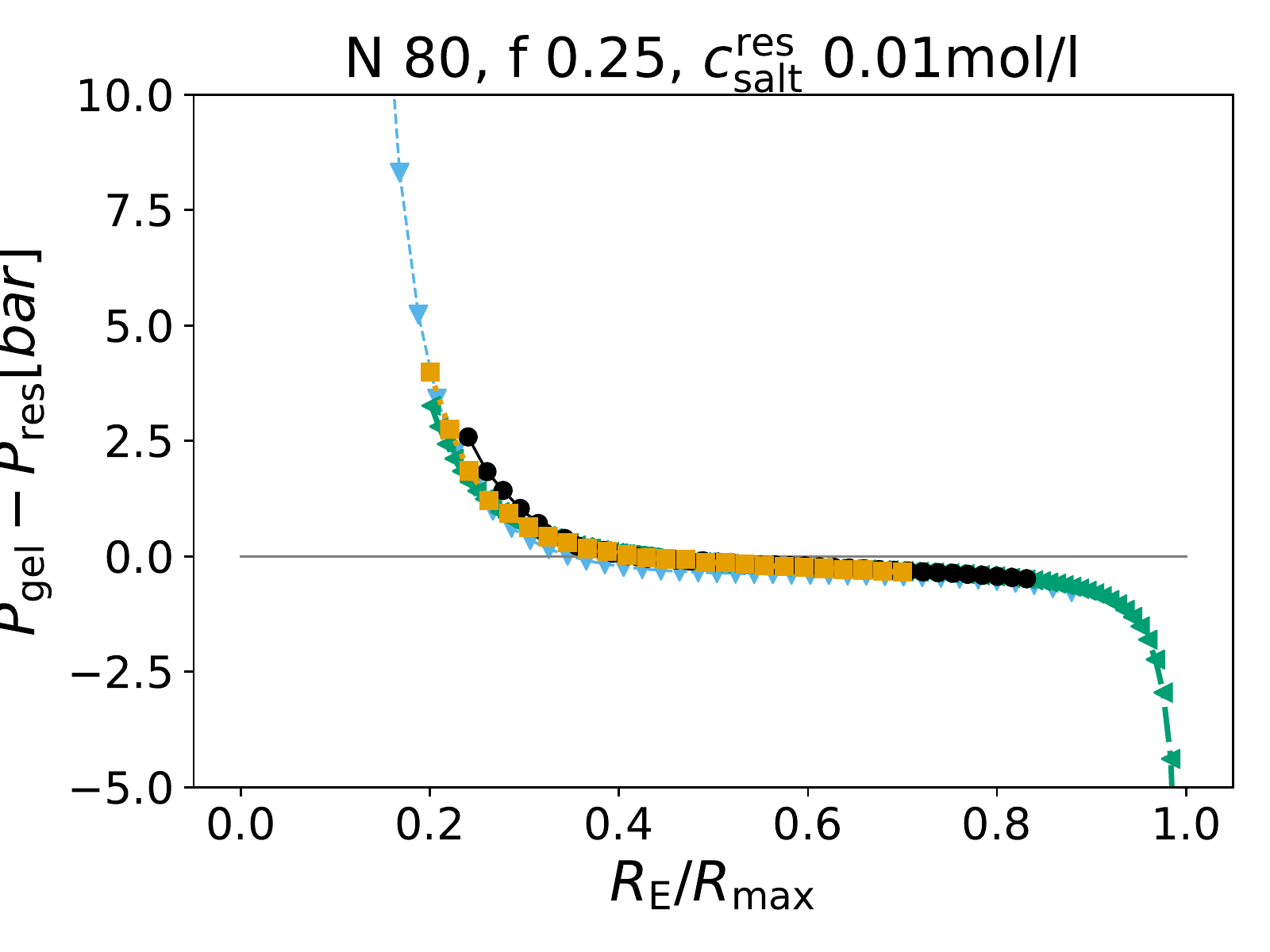}
\includegraphics[width=0.32\linewidth]{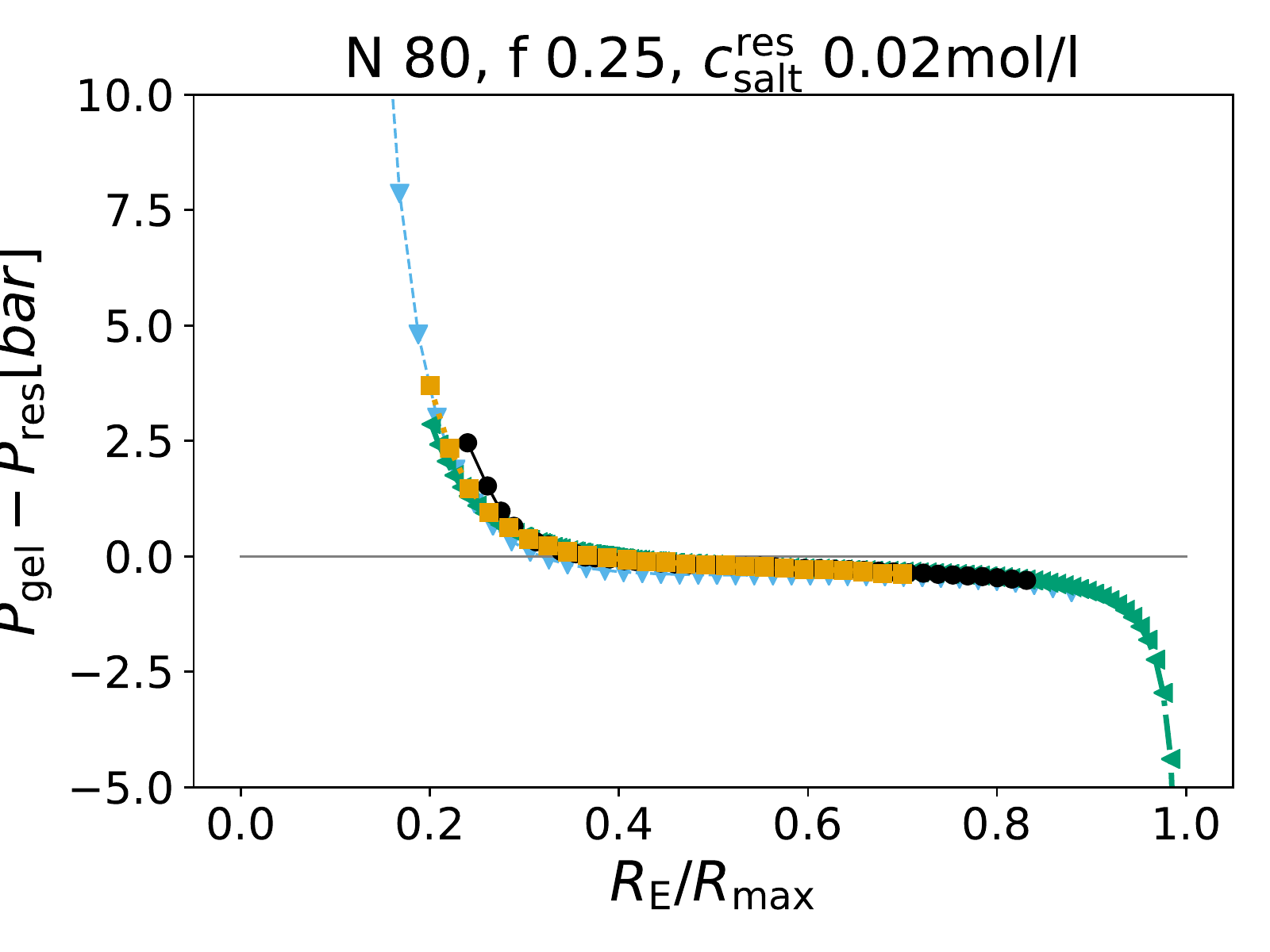}
\includegraphics[width=0.32\linewidth]{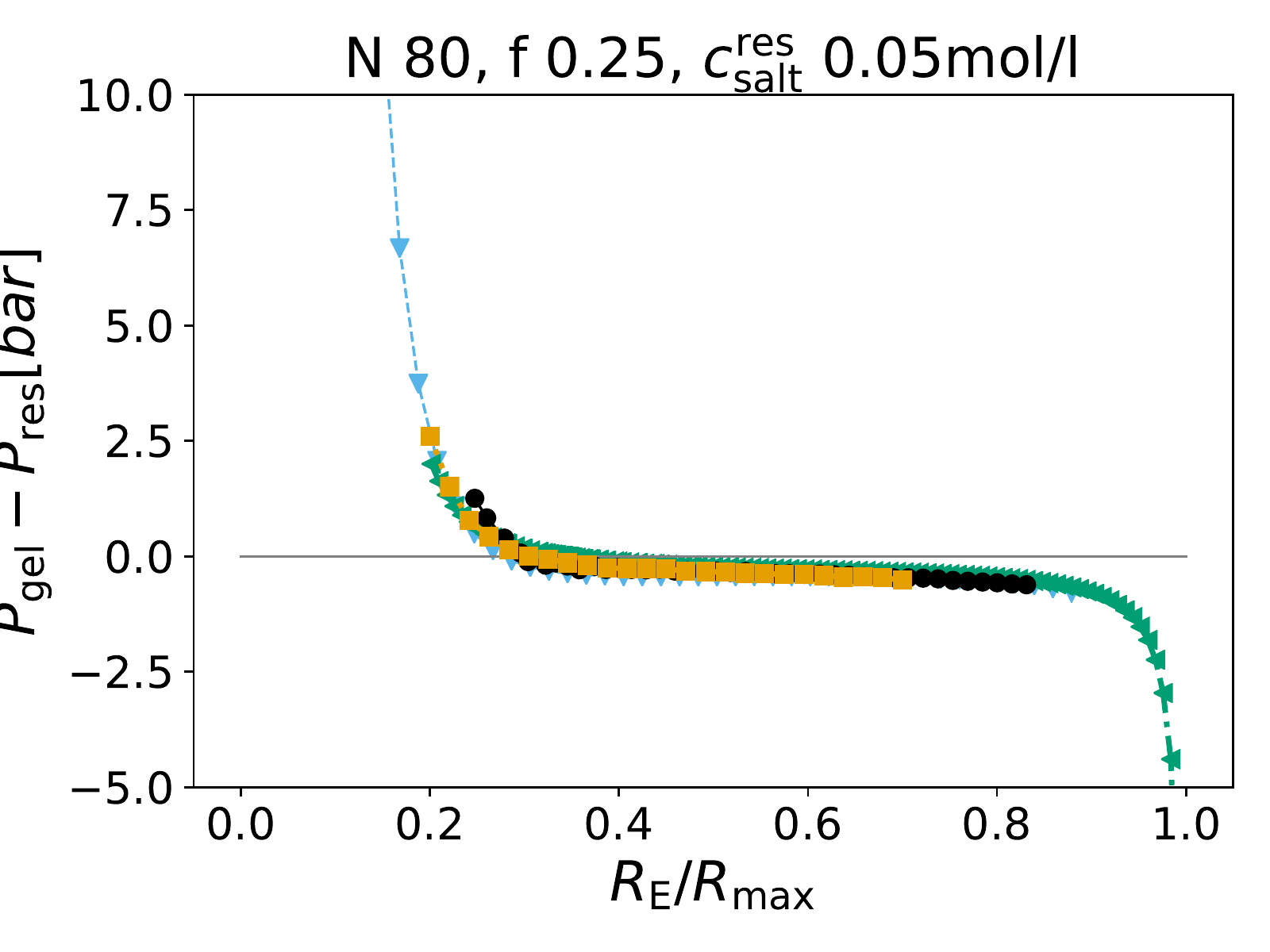}
\includegraphics[width=0.32\linewidth]{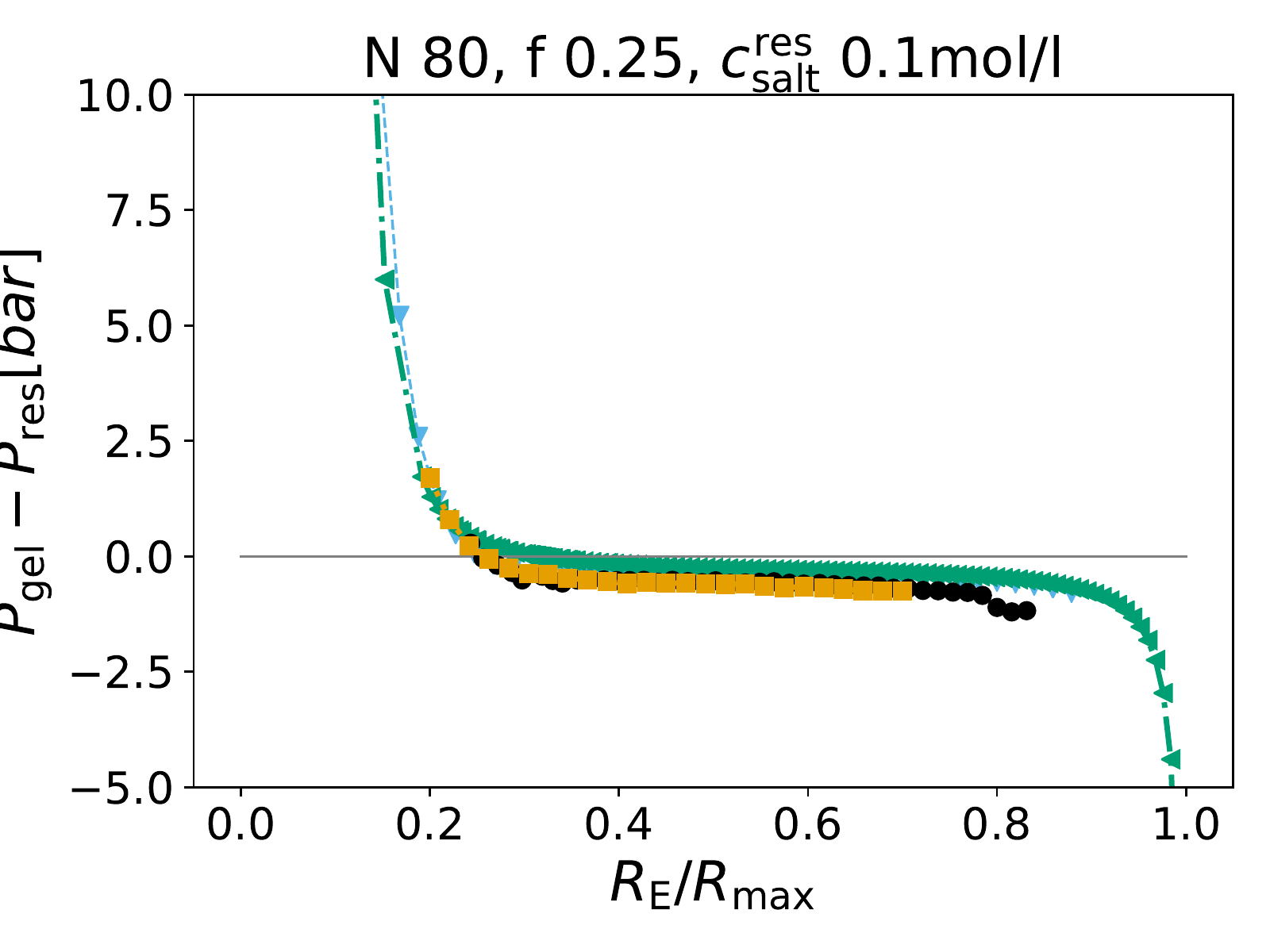}
\includegraphics[width=0.32\linewidth]{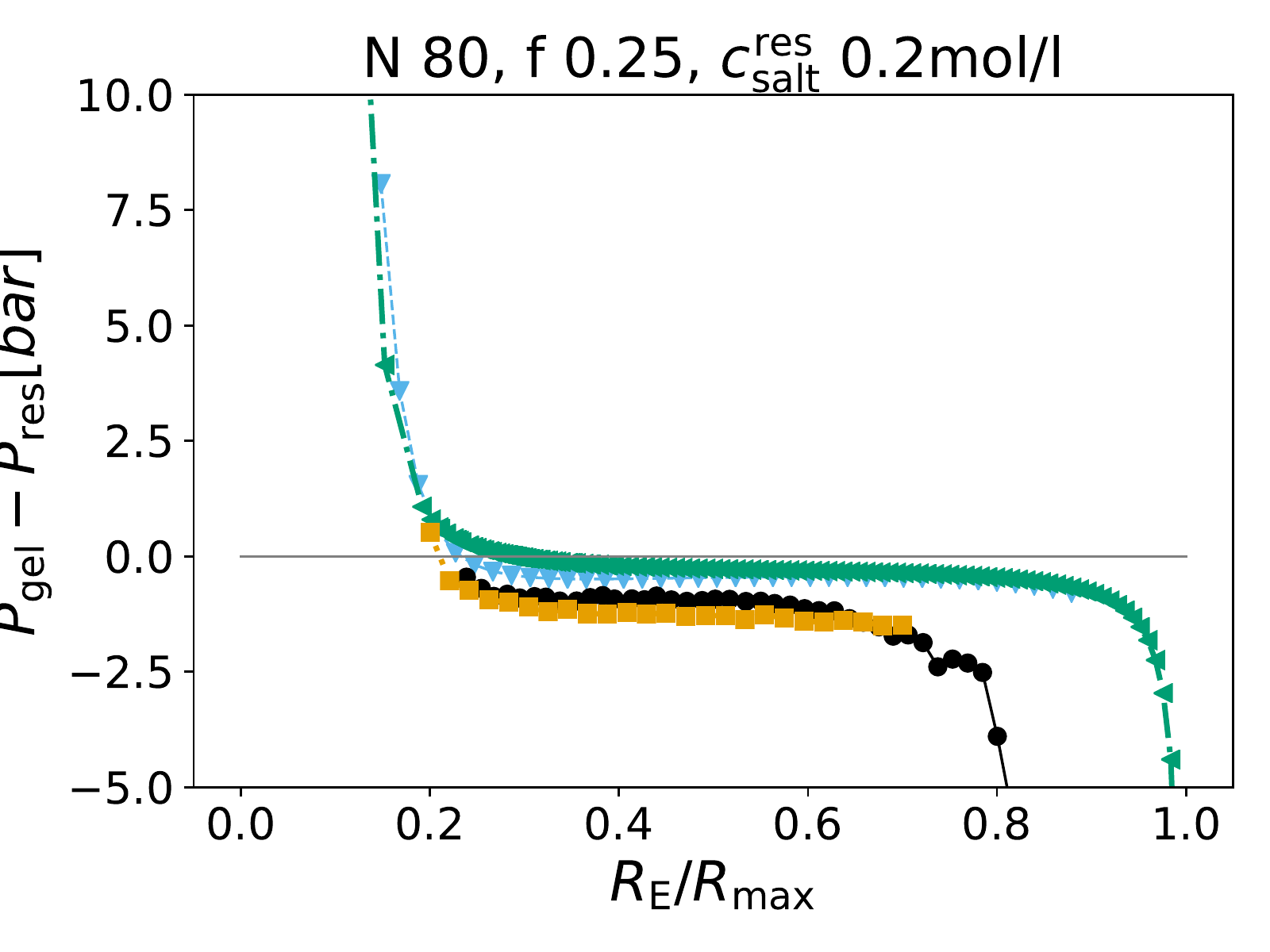}
\includegraphics[width=0.32\linewidth]{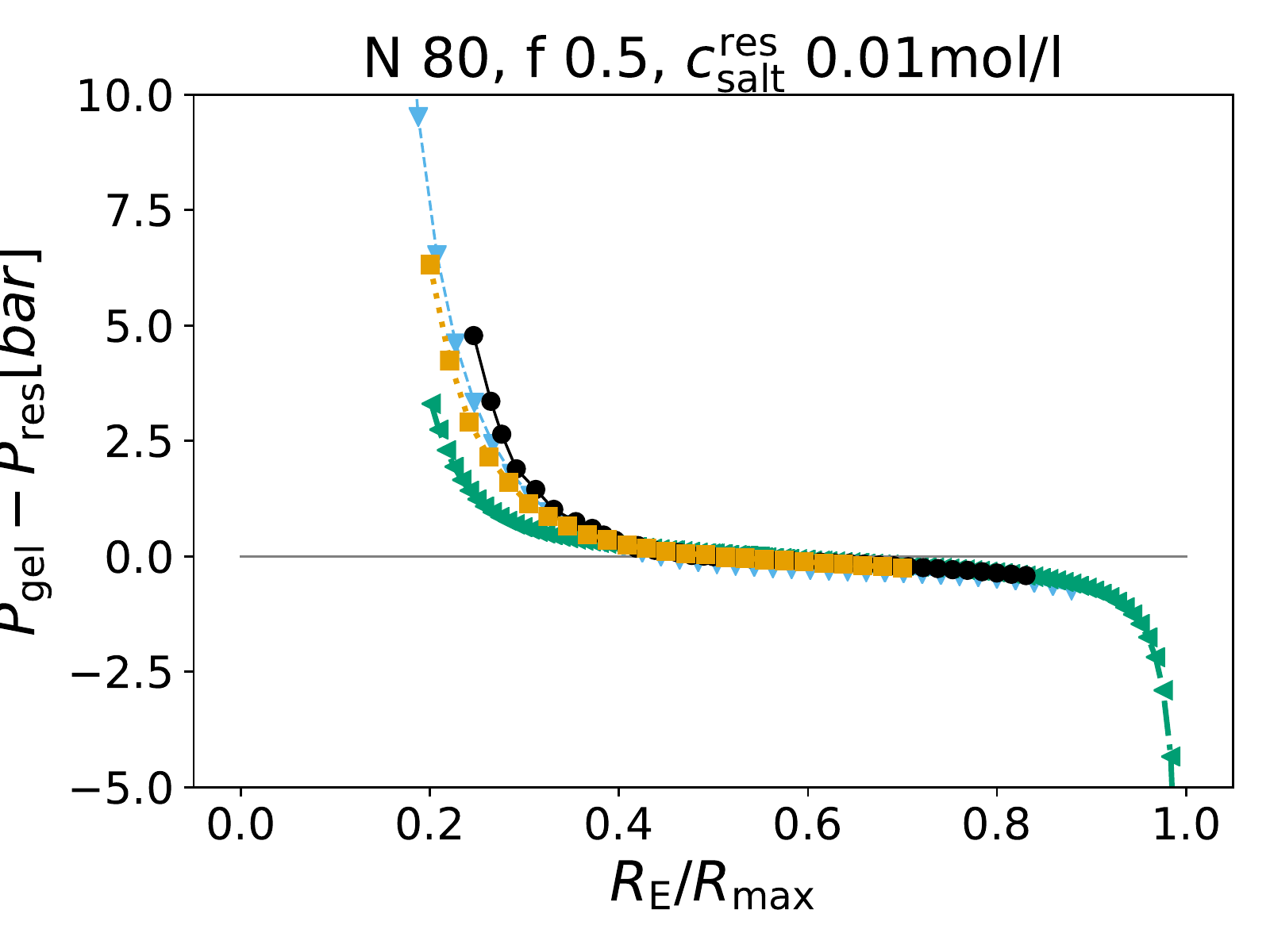}
\includegraphics[width=0.32\linewidth]{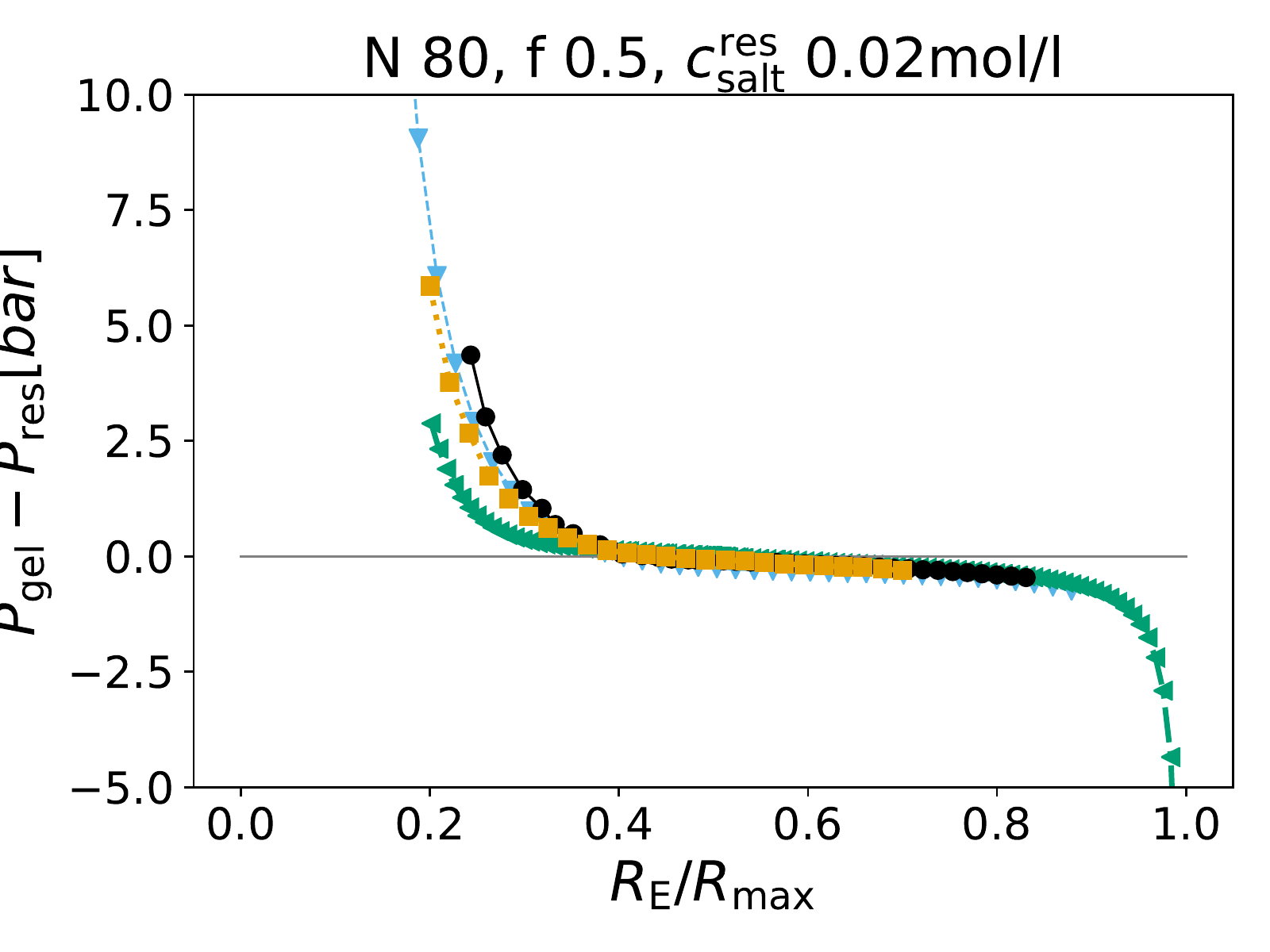}
\includegraphics[width=0.32\linewidth]{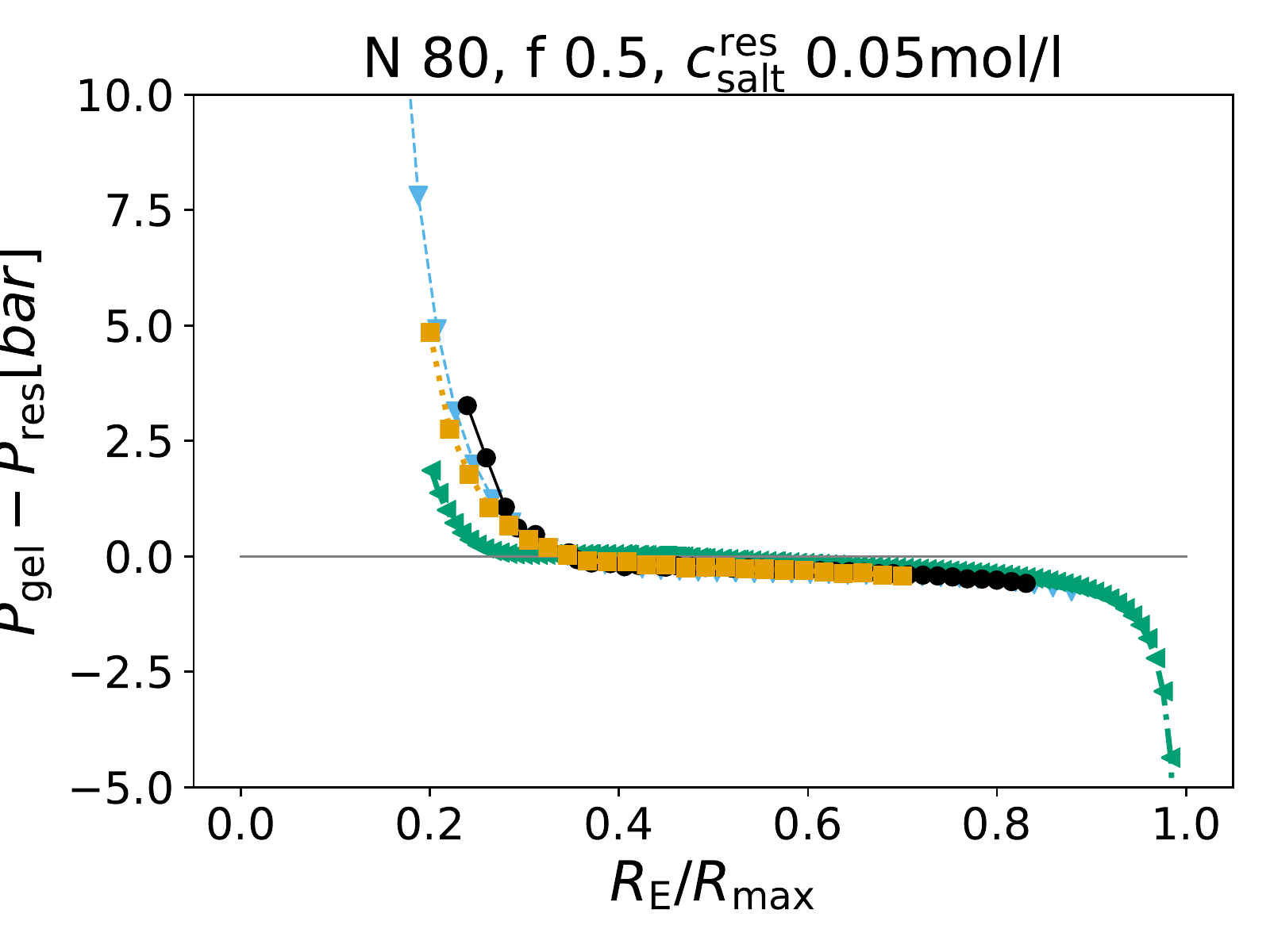}
\includegraphics[width=0.32\linewidth]{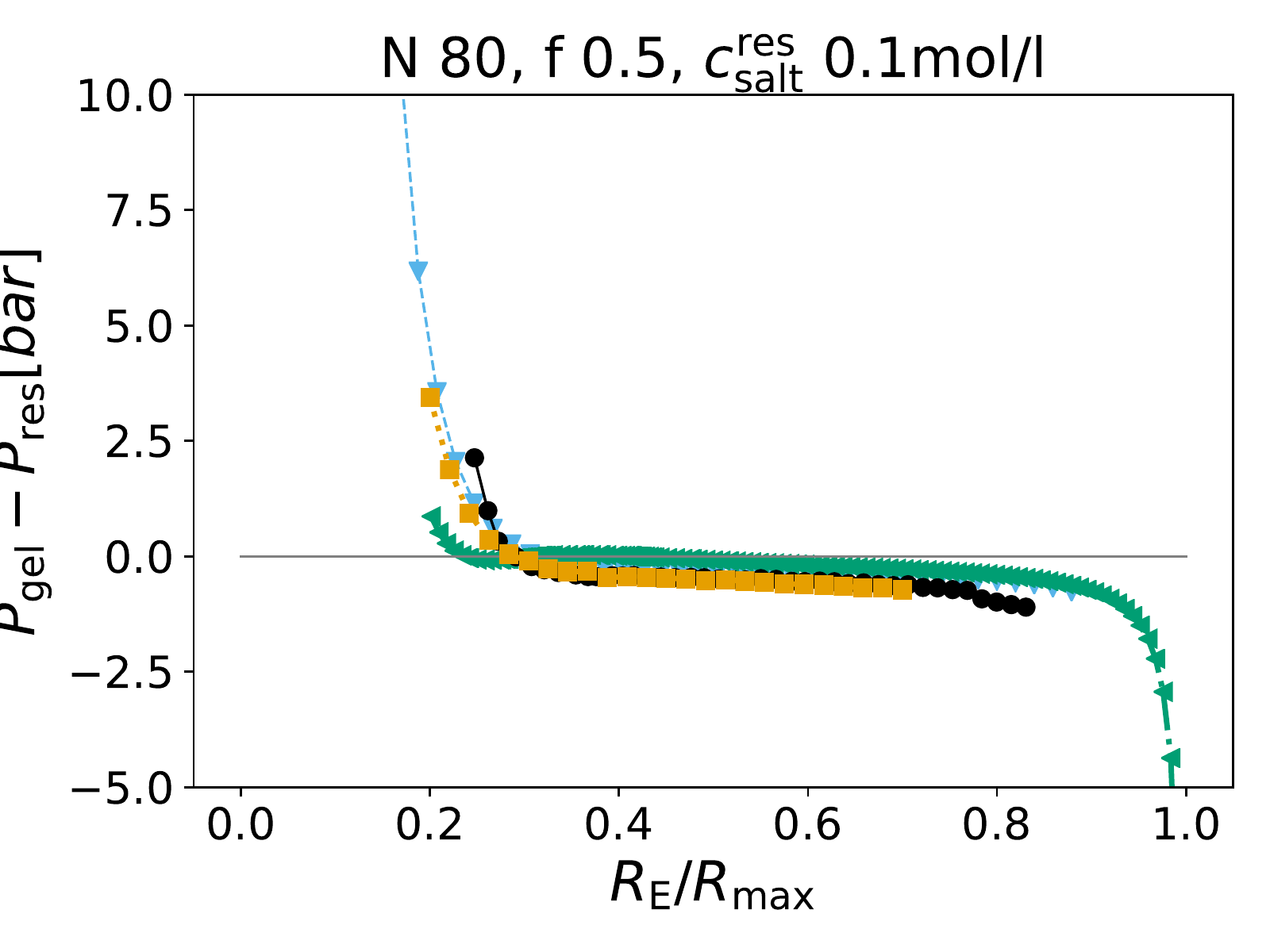}
\includegraphics[width=0.32\linewidth]{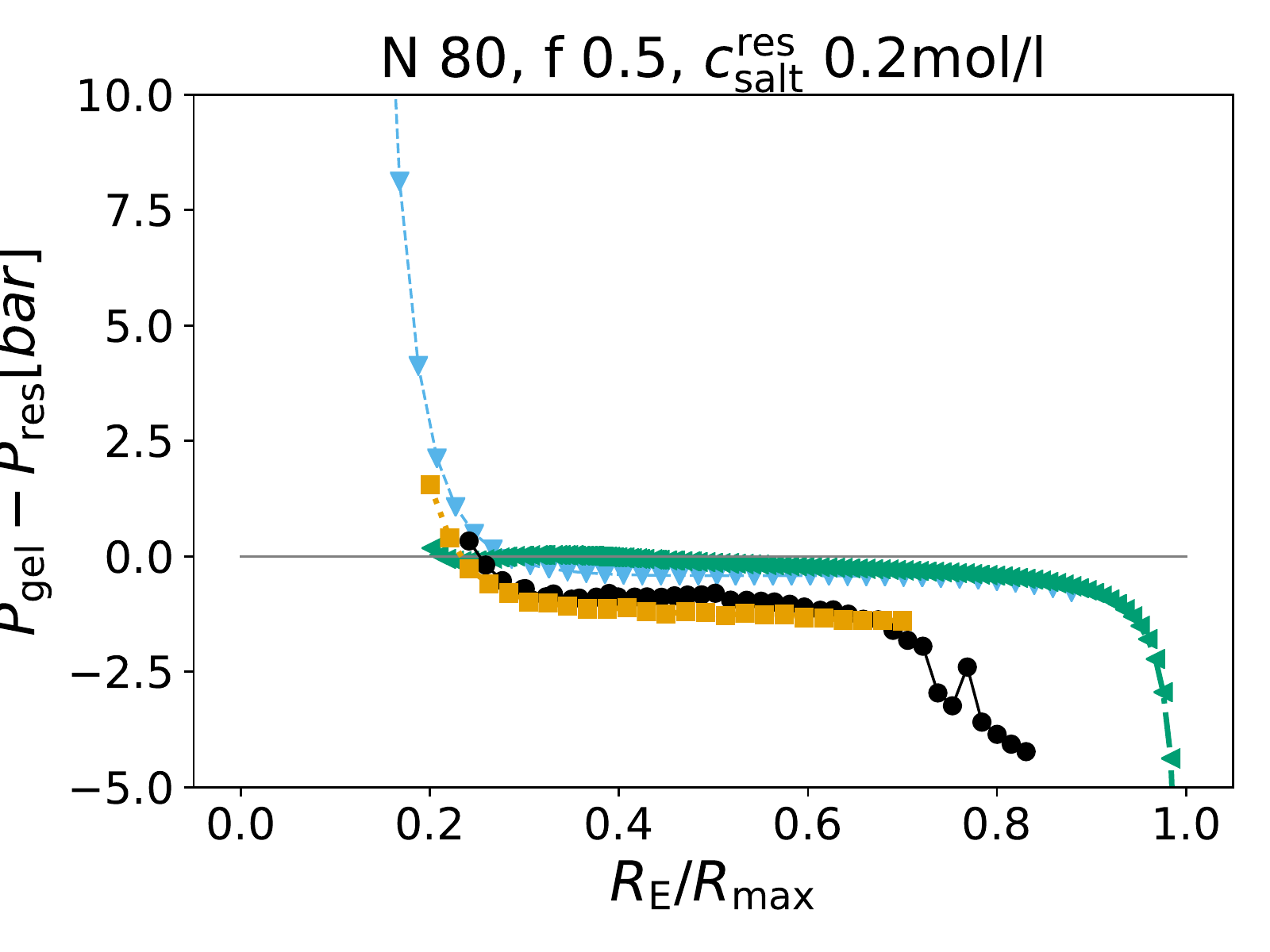}
\includegraphics[width=0.32\linewidth]{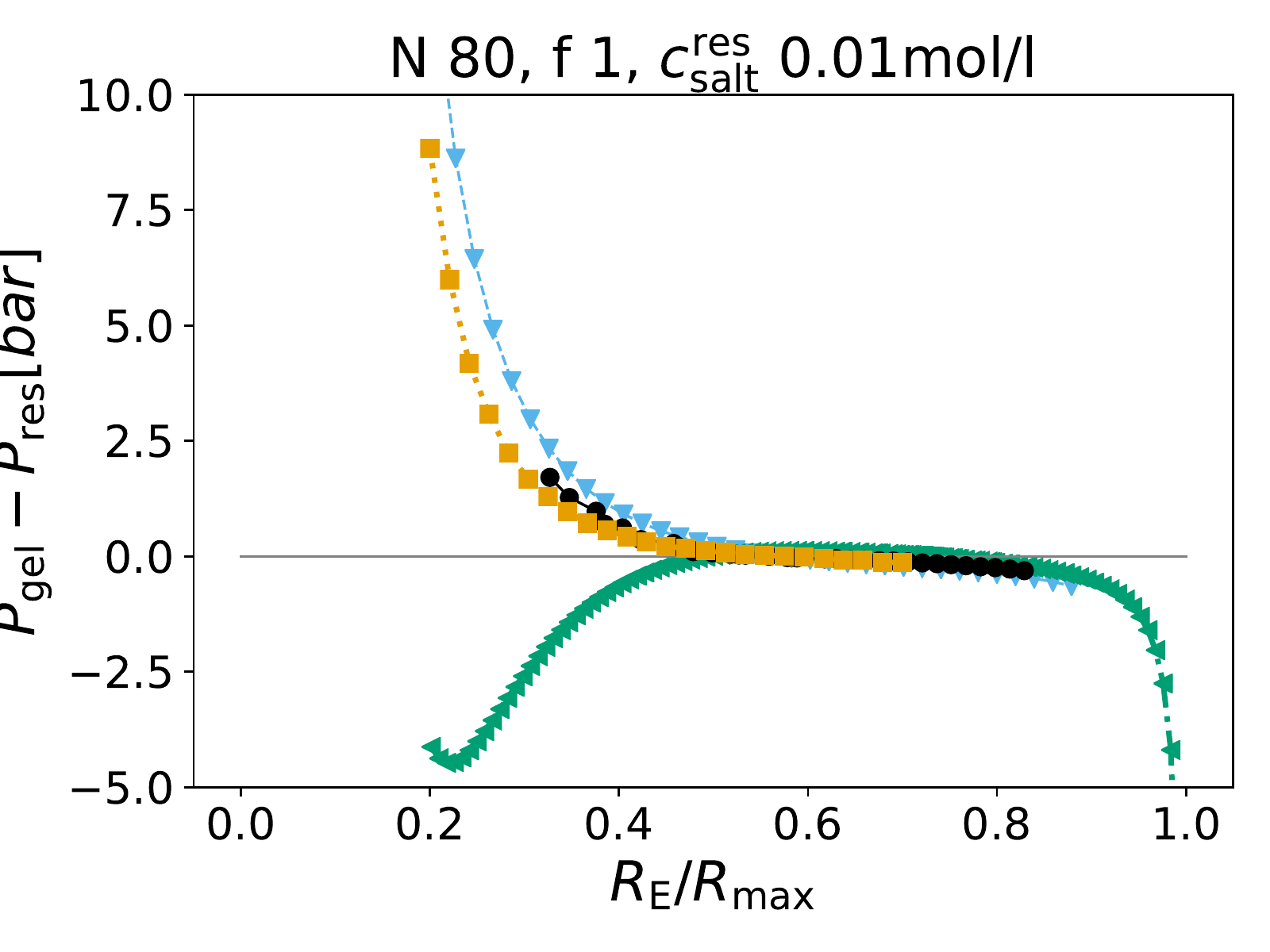}
\includegraphics[width=0.32\linewidth]{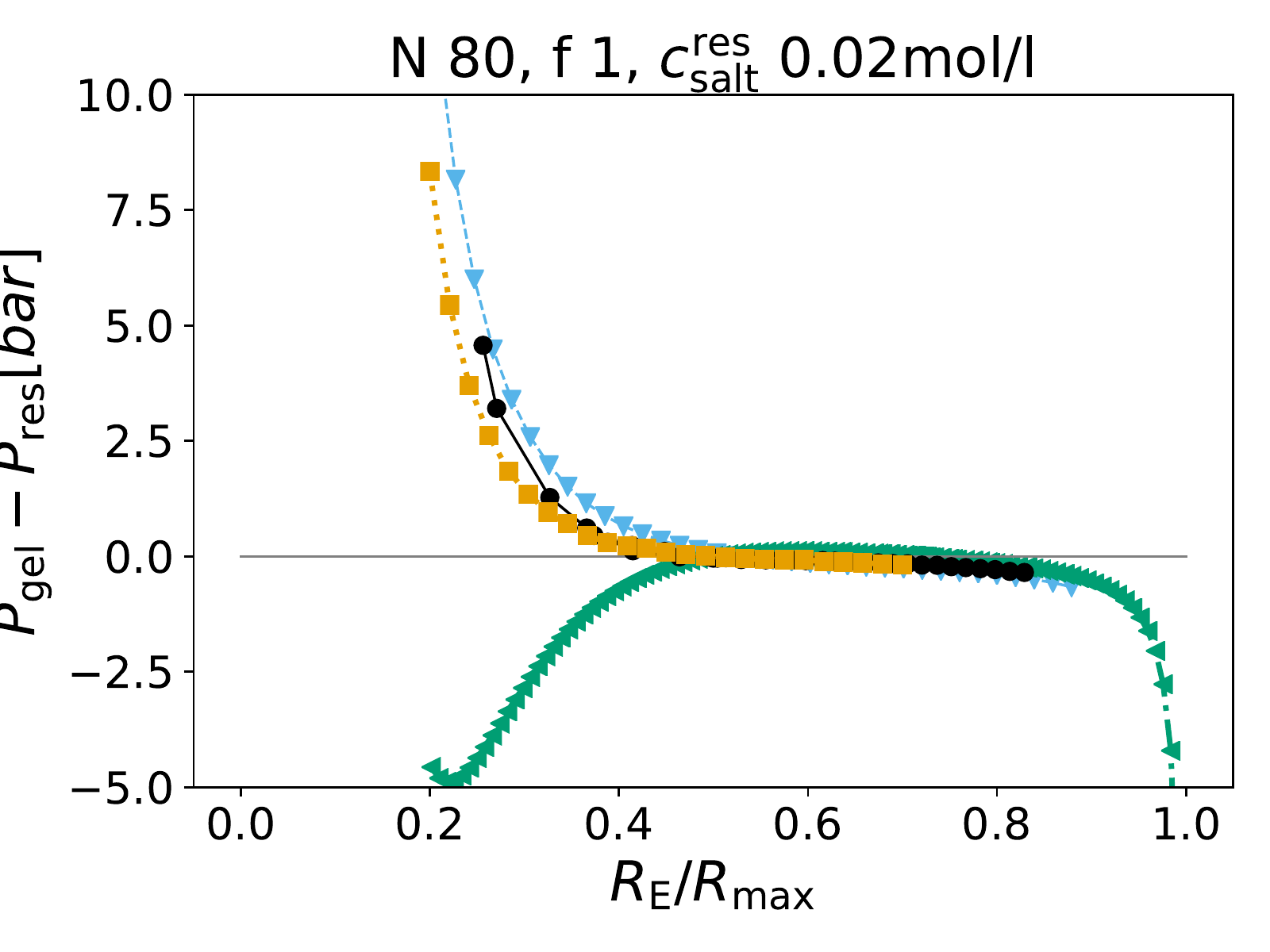}
\includegraphics[width=0.32\linewidth]{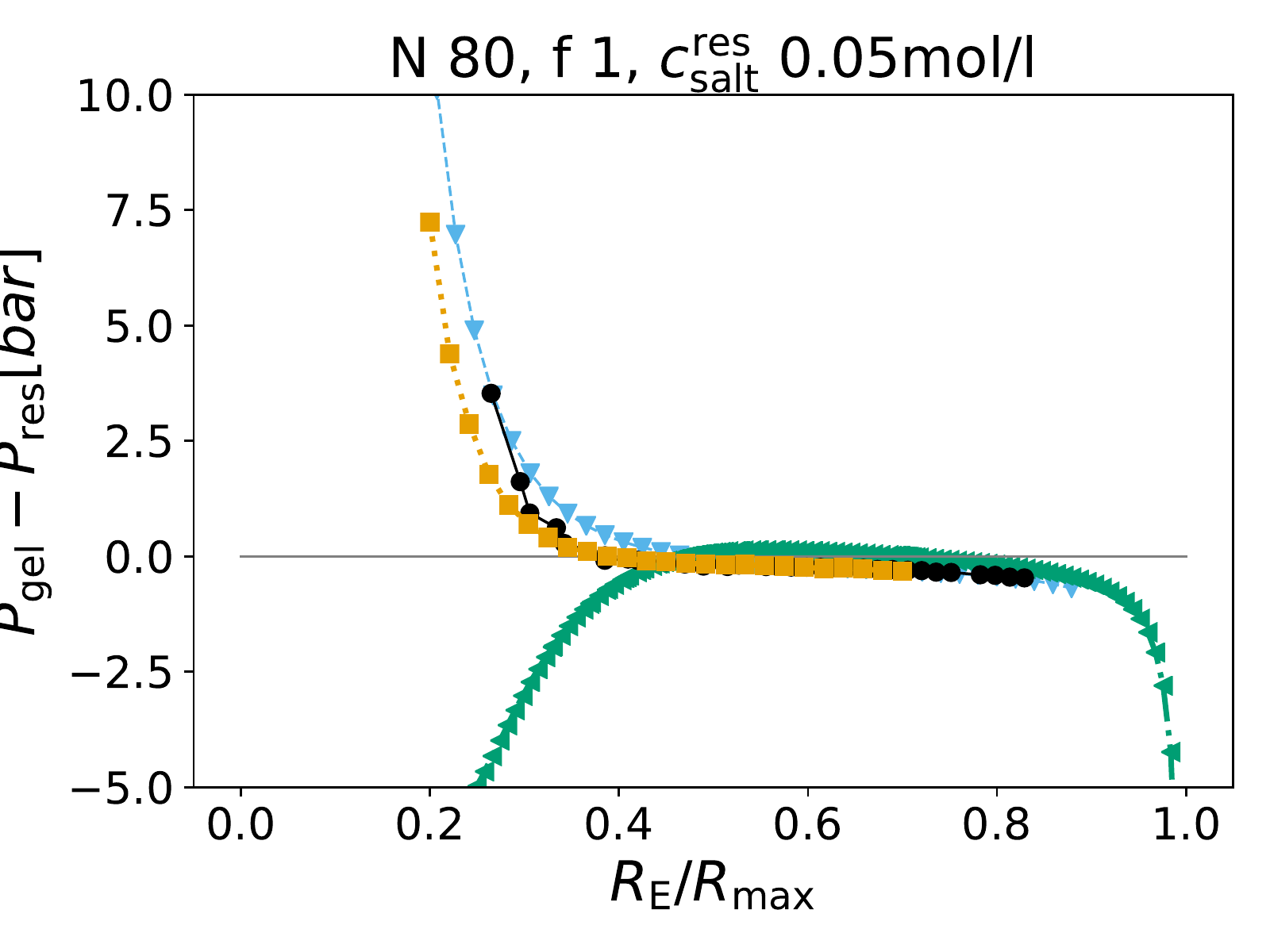}
\includegraphics[width=0.32\linewidth]{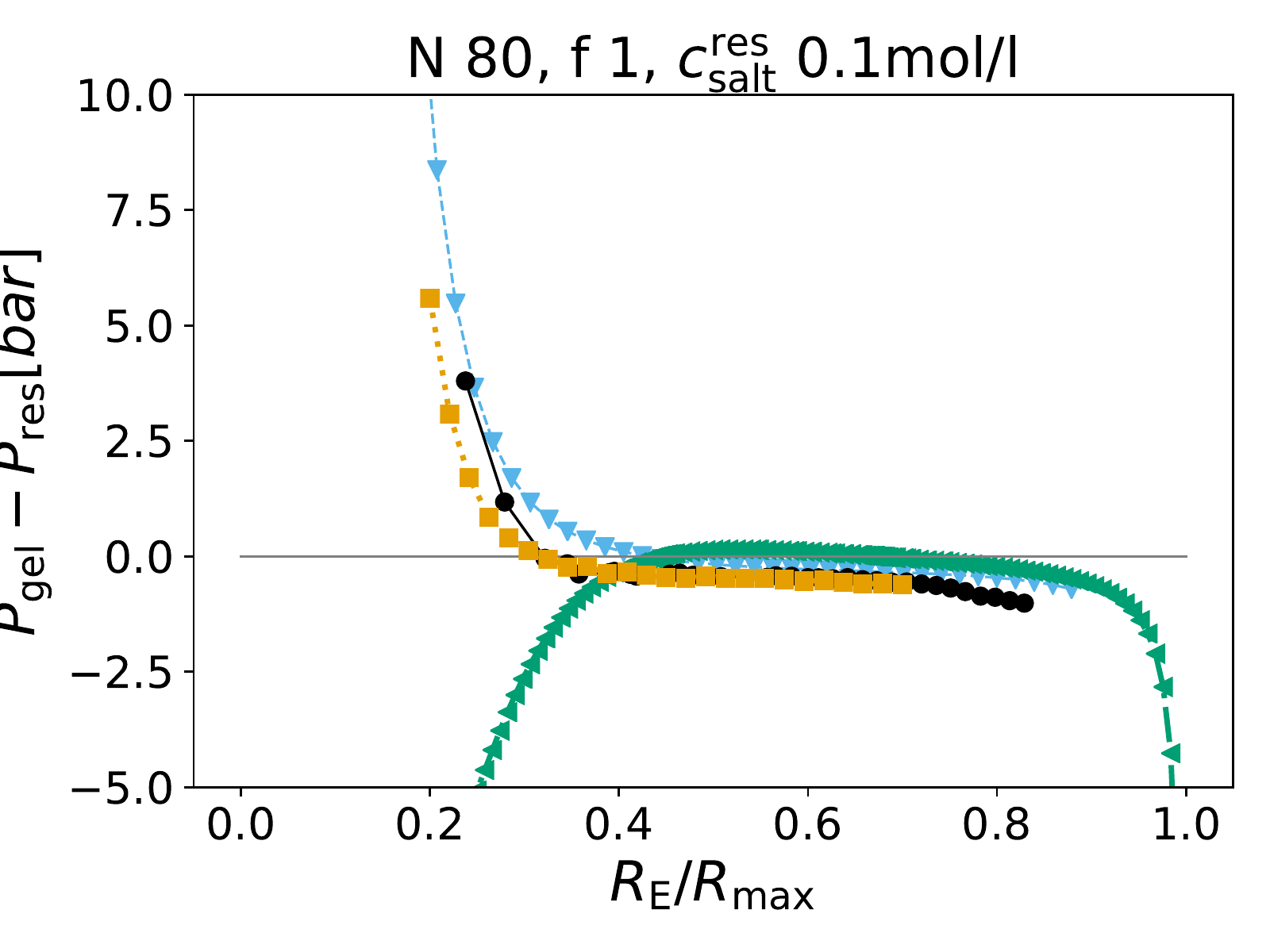}
\includegraphics[width=0.32\linewidth]{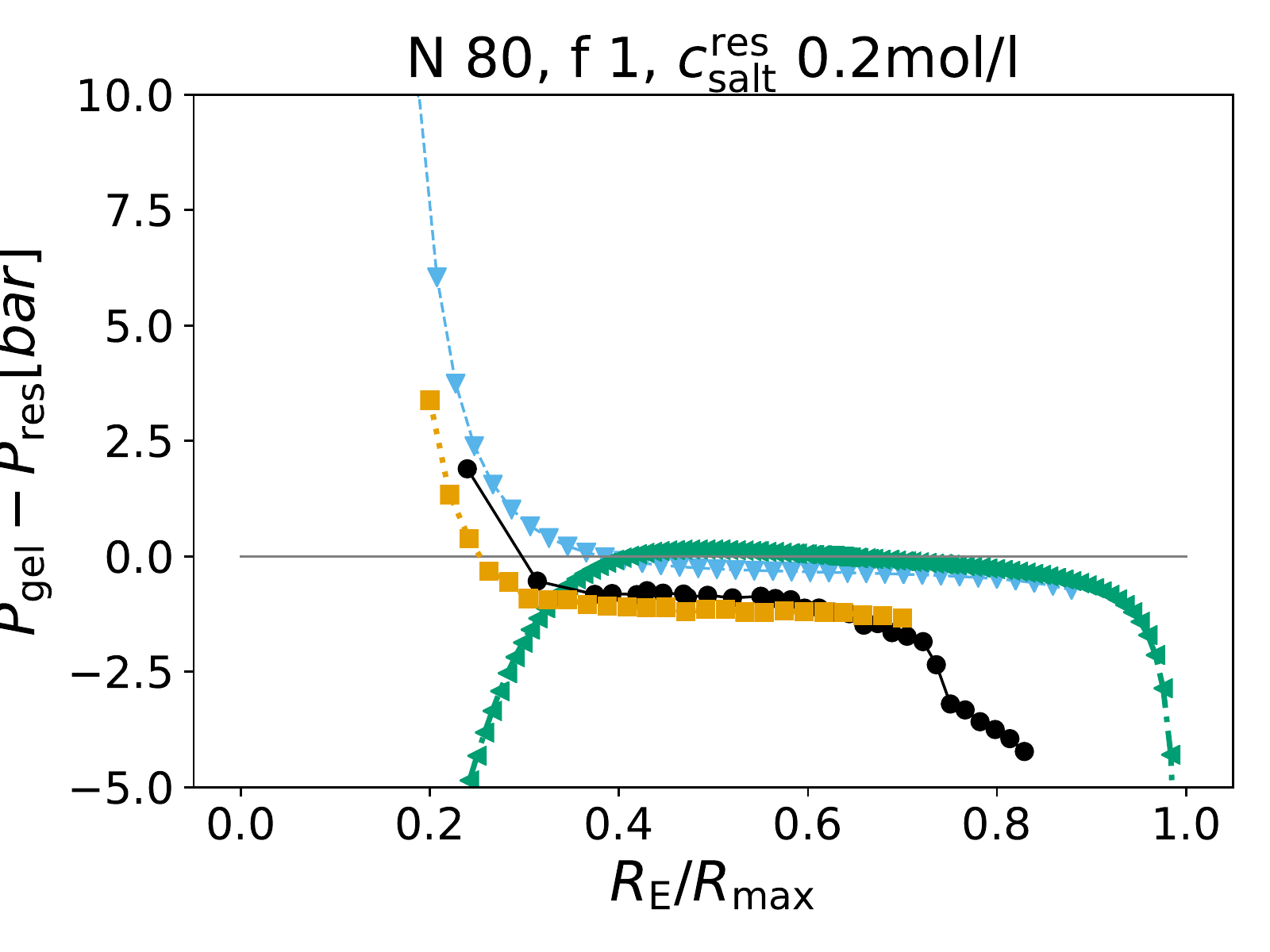}
\end{figure}

\FloatBarrier

\subsection{Swelling equilibria}
For completeness, we show the swelling equilibria as a function salt concentration $\cs$, charge fraction $f$ and chain length $N$. The plots are analogous to the plots in figure 3. For parameters where the Katchalsky model predicts more than one zero crossing in the $PV$ curve, the corresponding value of $\Req$ is set to zero (in order to indicate failure).
\subsubsection{Salt Concentration Dependence}
\begin{figure}
\includegraphics[width=0.32\linewidth, page=1]{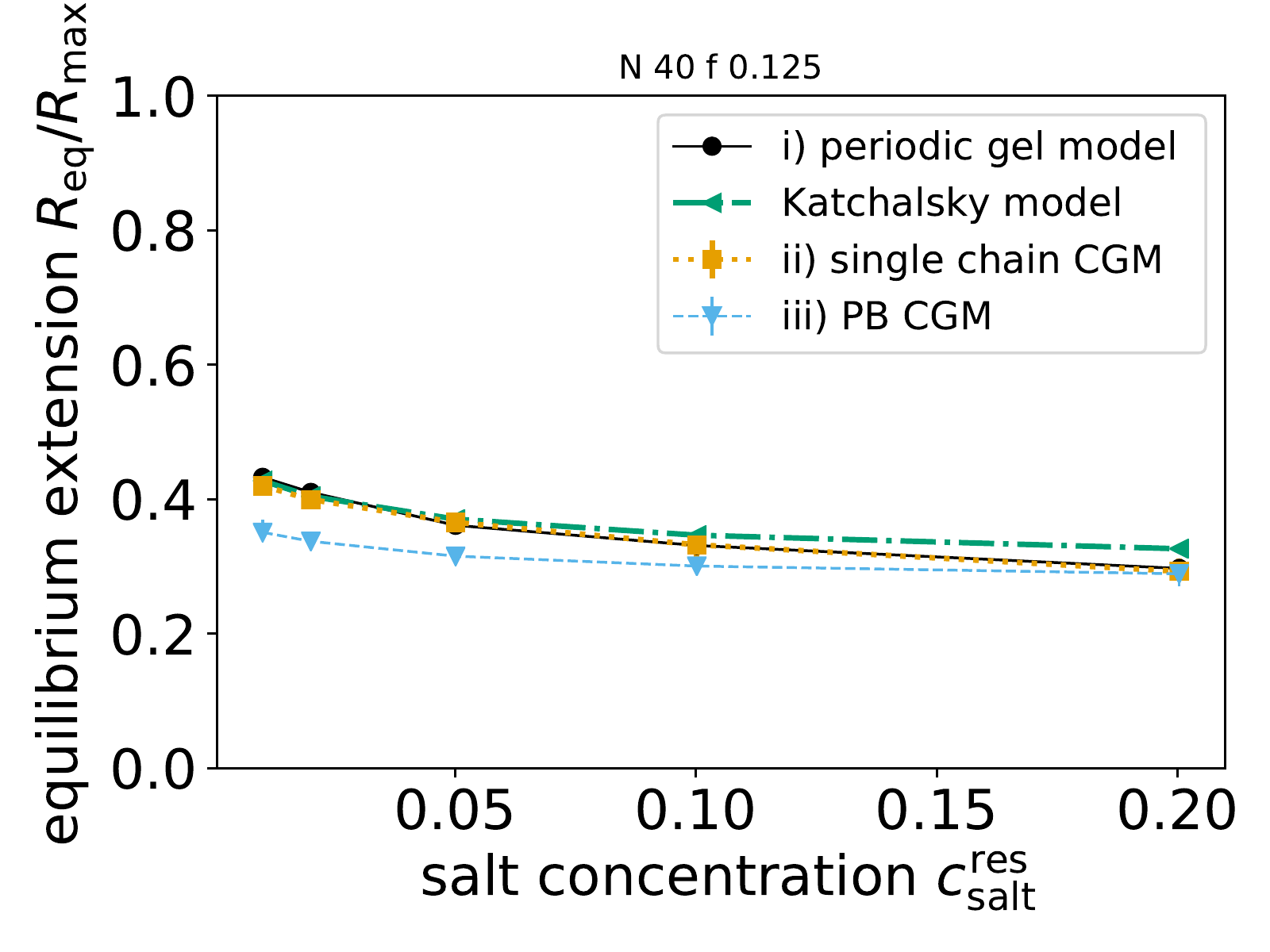}
\includegraphics[width=0.32\linewidth, page=2]{appendix_Q_cs_katchalsky.pdf}
\includegraphics[width=0.32\linewidth, page=3]{appendix_Q_cs_katchalsky.pdf}
\includegraphics[width=0.32\linewidth, page=4]{appendix_Q_cs_katchalsky.pdf}
\includegraphics[width=0.32\linewidth, page=5]{appendix_Q_cs_katchalsky.pdf}
\includegraphics[width=0.32\linewidth, page=6]{appendix_Q_cs_katchalsky.pdf}
\includegraphics[width=0.32\linewidth, page=7]{appendix_Q_cs_katchalsky.pdf}
\includegraphics[width=0.32\linewidth, page=8]{appendix_Q_cs_katchalsky.pdf}
\includegraphics[width=0.32\linewidth, page=9]{appendix_Q_cs_katchalsky.pdf}
\includegraphics[width=0.32\linewidth, page=10]{appendix_Q_cs_katchalsky.pdf}
\includegraphics[width=0.32\linewidth, page=11]{appendix_Q_cs_katchalsky.pdf}
\includegraphics[width=0.32\linewidth, page=12]{appendix_Q_cs_katchalsky.pdf}
\end{figure}
\FloatBarrier
\subsubsection{Charge Fraction Dependence}
\FloatBarrier
\begin{figure}
\includegraphics[width=0.32\linewidth, page=1]{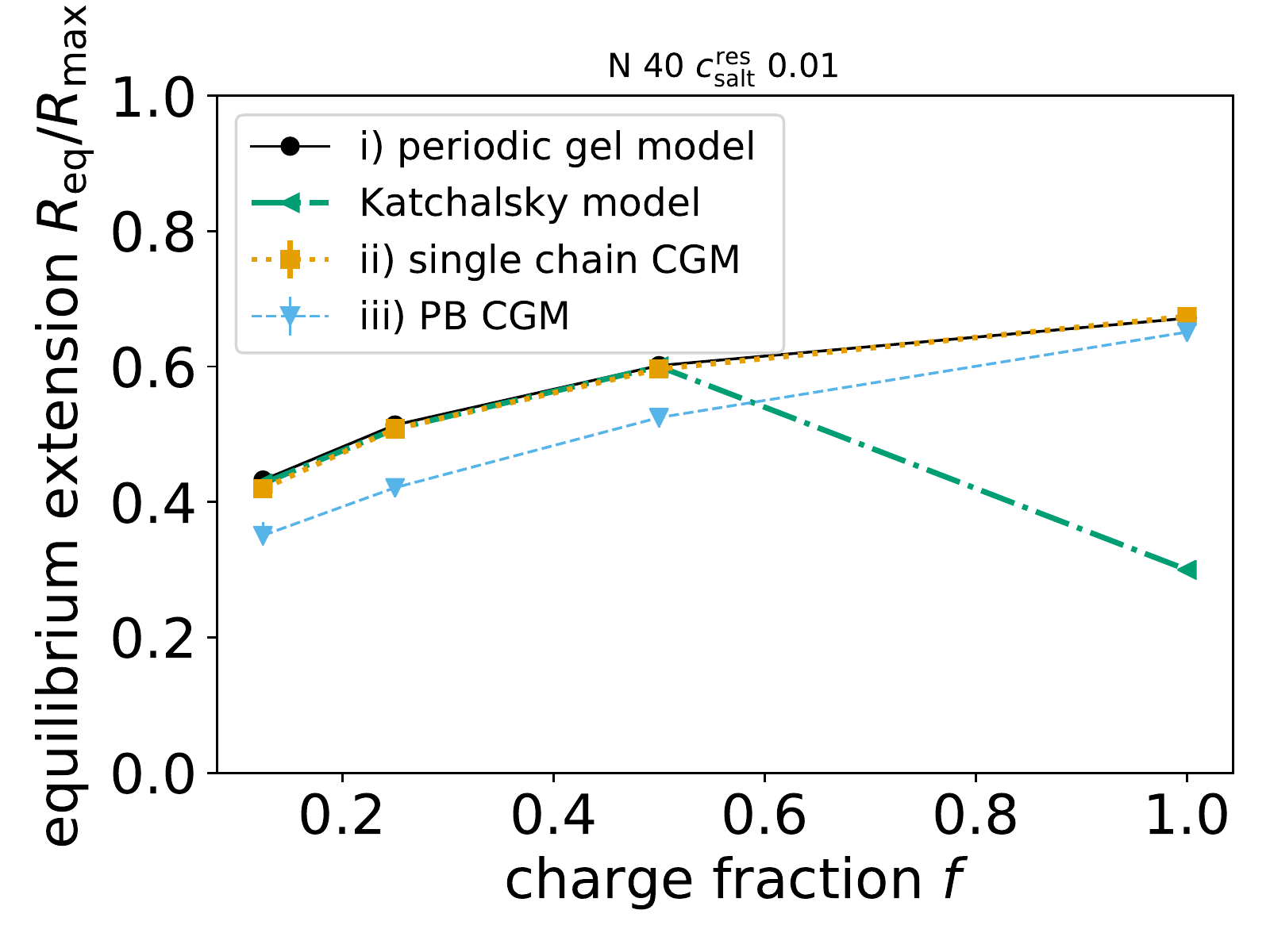}
\includegraphics[width=0.32\linewidth, page=2]{appendix_Q_f_katchalsky.pdf}
\includegraphics[width=0.32\linewidth, page=3]{appendix_Q_f_katchalsky.pdf}
\includegraphics[width=0.32\linewidth, page=4]{appendix_Q_f_katchalsky.pdf}
\includegraphics[width=0.32\linewidth, page=5]{appendix_Q_f_katchalsky.pdf}
\includegraphics[width=0.32\linewidth, page=6]{appendix_Q_f_katchalsky.pdf}
\includegraphics[width=0.32\linewidth, page=7]{appendix_Q_f_katchalsky.pdf}
\includegraphics[width=0.32\linewidth, page=8]{appendix_Q_f_katchalsky.pdf}
\includegraphics[width=0.32\linewidth, page=9]{appendix_Q_f_katchalsky.pdf}
\includegraphics[width=0.32\linewidth, page=10]{appendix_Q_f_katchalsky.pdf}
\includegraphics[width=0.32\linewidth, page=11]{appendix_Q_f_katchalsky.pdf}
\includegraphics[width=0.32\linewidth, page=12]{appendix_Q_f_katchalsky.pdf}
\end{figure}
\begin{figure}
\includegraphics[width=0.32\linewidth, page=13]{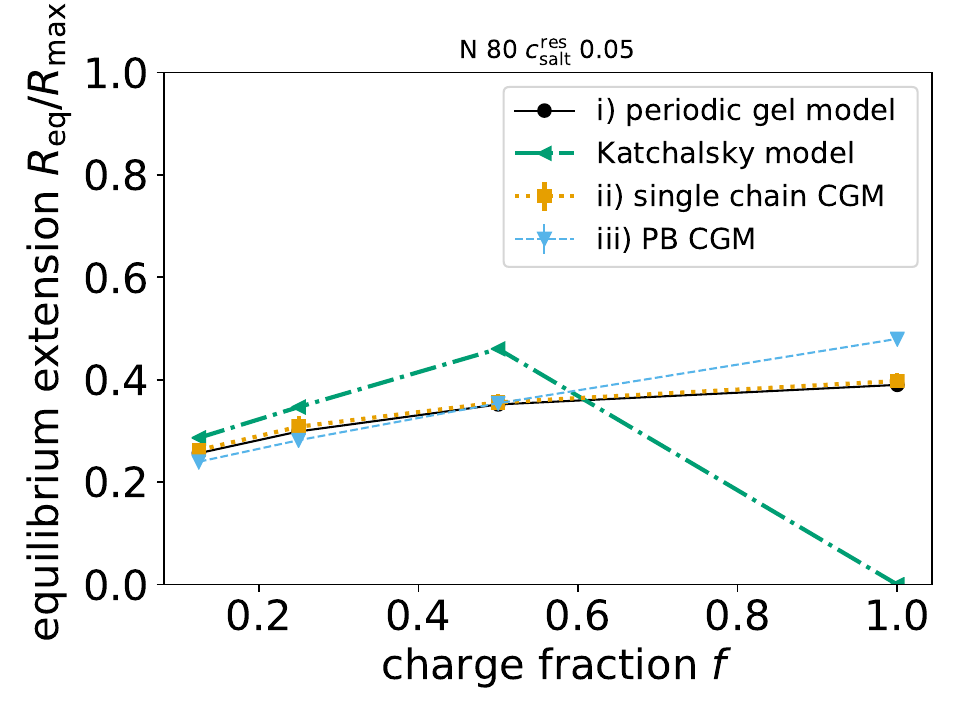}
\includegraphics[width=0.32\linewidth, page=14]{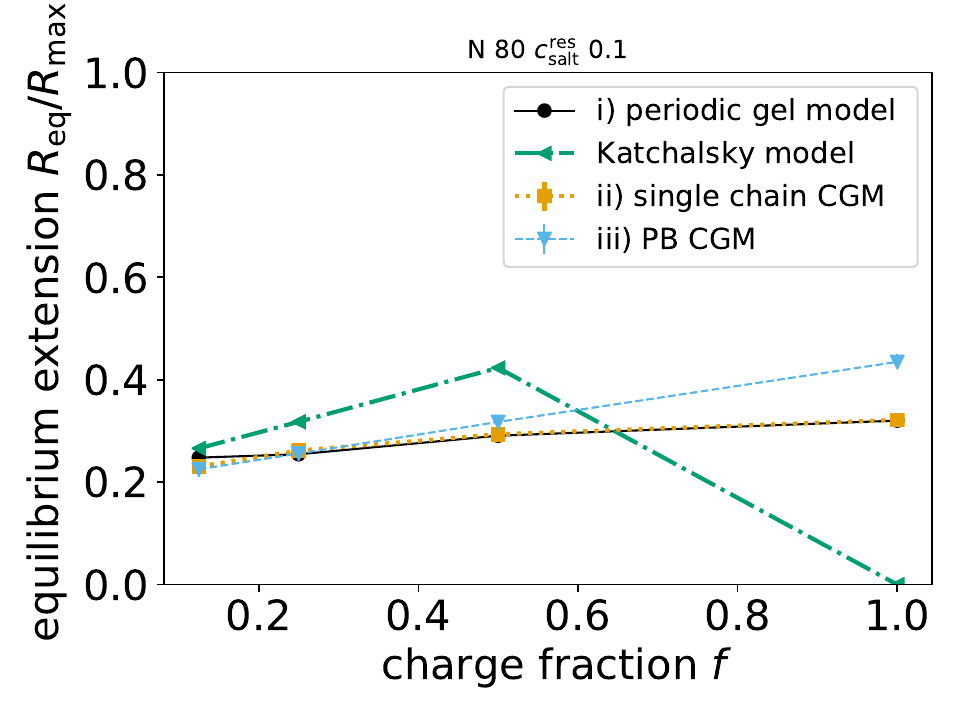}
\includegraphics[width=0.32\linewidth, page=15]{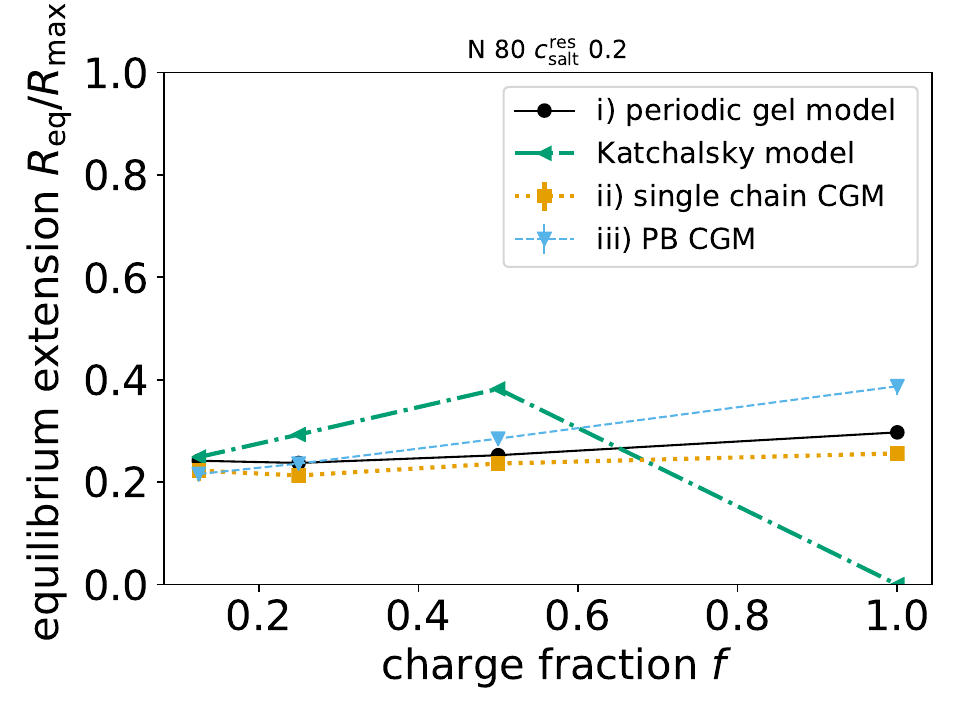}
\end{figure}

\FloatBarrier
\subsubsection{Chain Length Dependence}
\FloatBarrier
\begin{figure}
\includegraphics[width=0.32\linewidth, page=1]{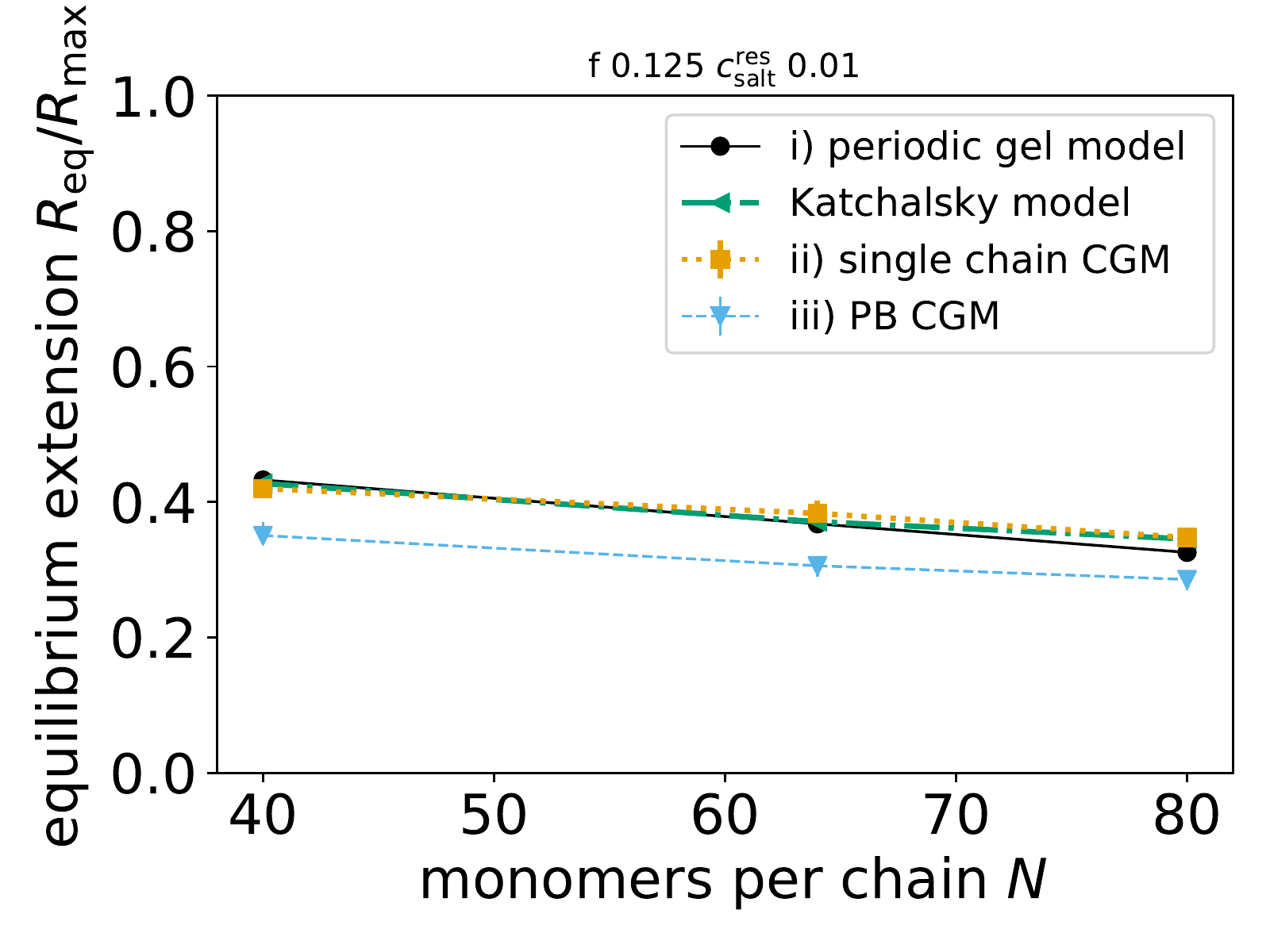}
\includegraphics[width=0.32\linewidth, page=2]{appendix_Q_N.pdf}
\includegraphics[width=0.32\linewidth, page=3]{appendix_Q_N.pdf}
\includegraphics[width=0.32\linewidth, page=4]{appendix_Q_N.pdf}
\includegraphics[width=0.32\linewidth, page=5]{appendix_Q_N.pdf}
\includegraphics[width=0.32\linewidth, page=6]{appendix_Q_N.pdf}
\includegraphics[width=0.32\linewidth, page=7]{appendix_Q_N.pdf}
\includegraphics[width=0.32\linewidth, page=8]{appendix_Q_N.pdf}
\includegraphics[width=0.32\linewidth, page=9]{appendix_Q_N.pdf}
\end{figure}

\begin{figure}
\includegraphics[width=0.32\linewidth, page=10]{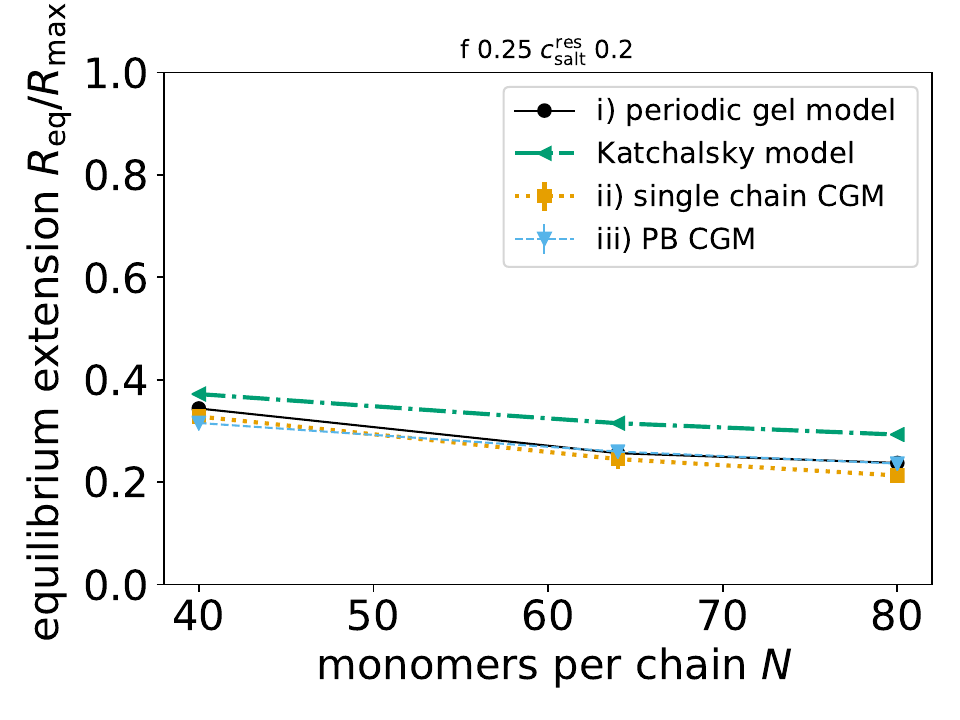}
\includegraphics[width=0.32\linewidth, page=11]{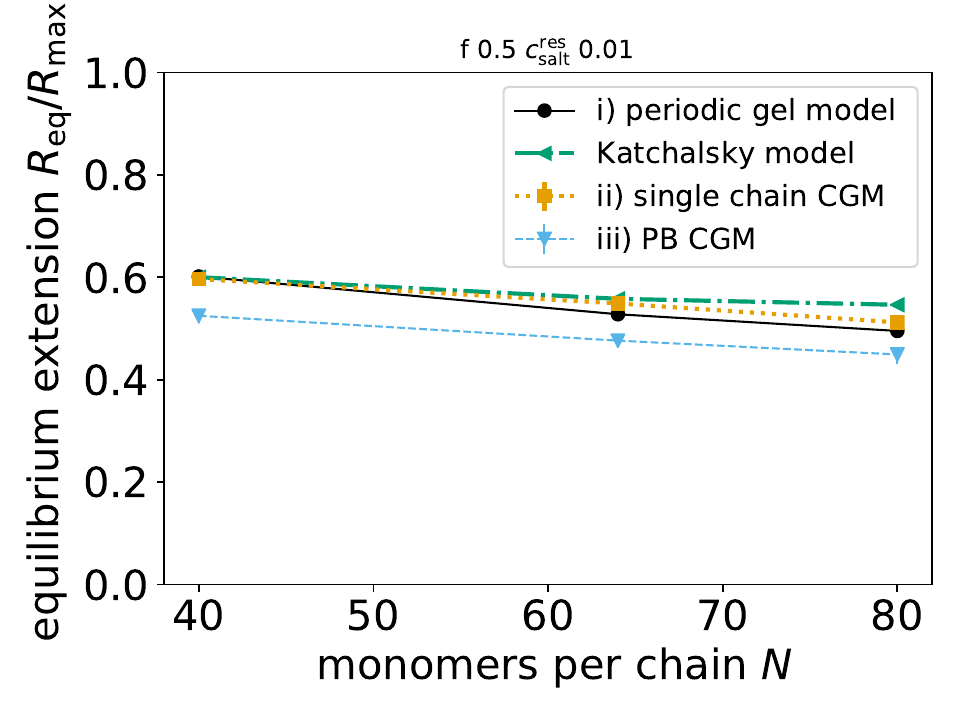}
\includegraphics[width=0.32\linewidth, page=12]{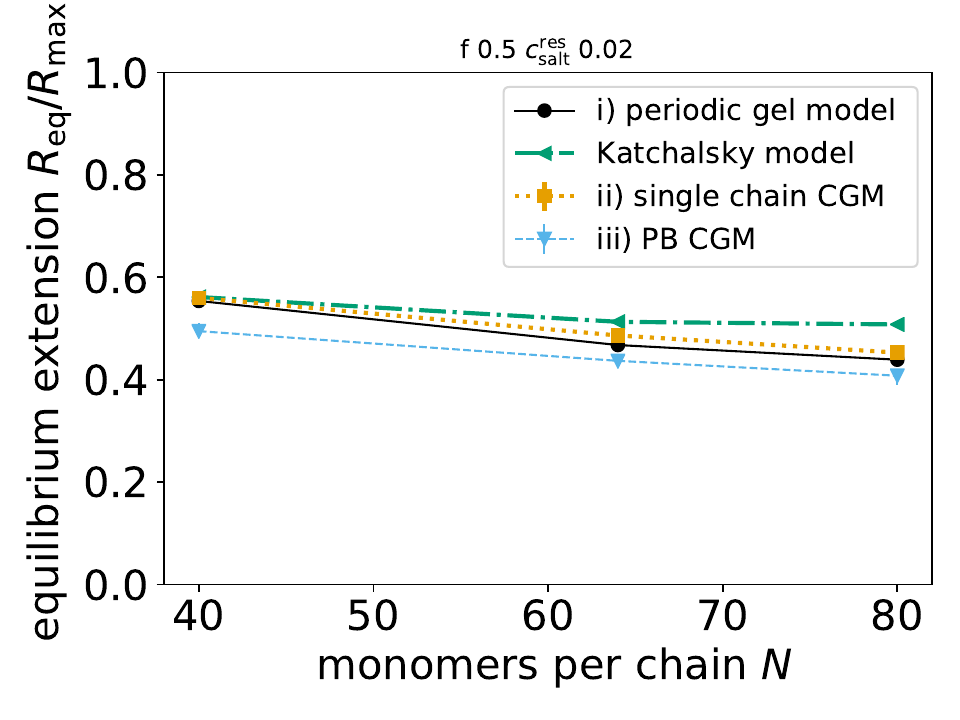}
\includegraphics[width=0.32\linewidth, page=13]{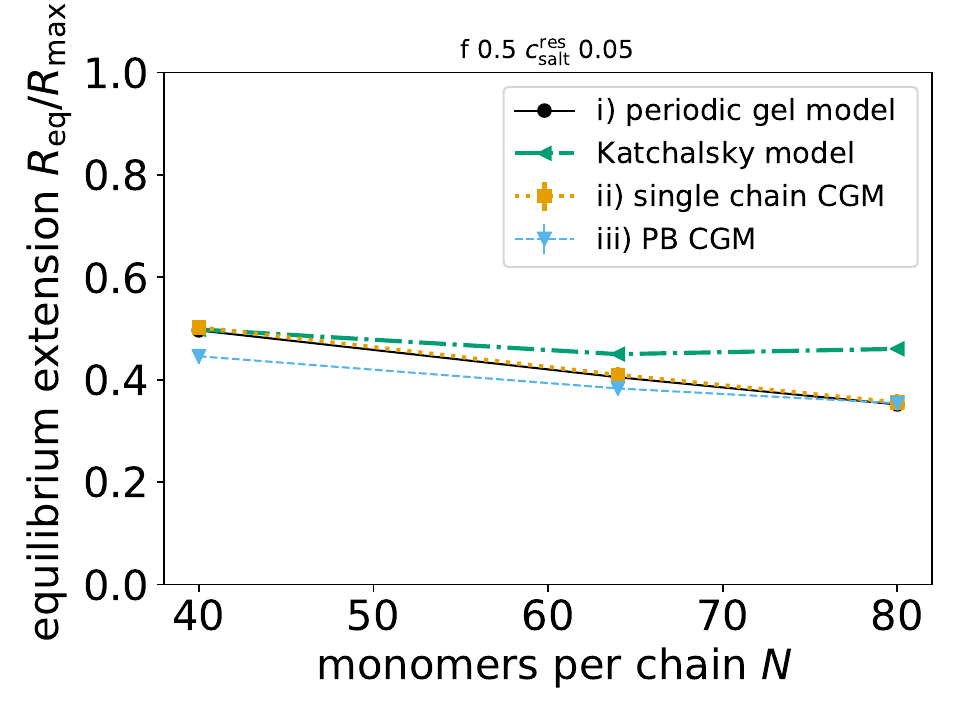}
\includegraphics[width=0.32\linewidth, page=14]{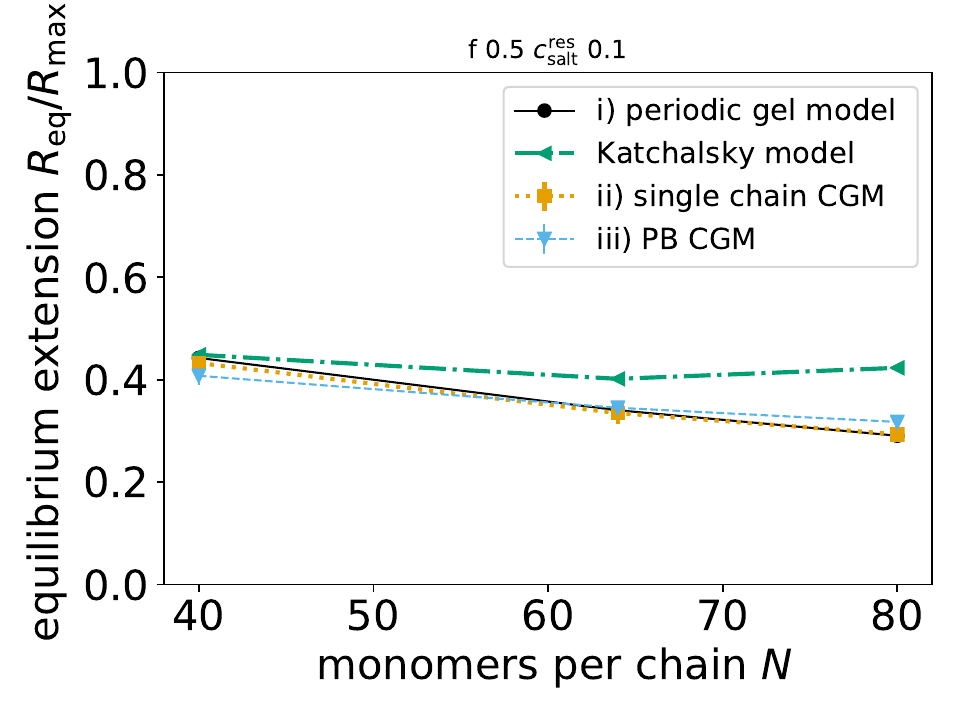}
\includegraphics[width=0.32\linewidth, page=15]{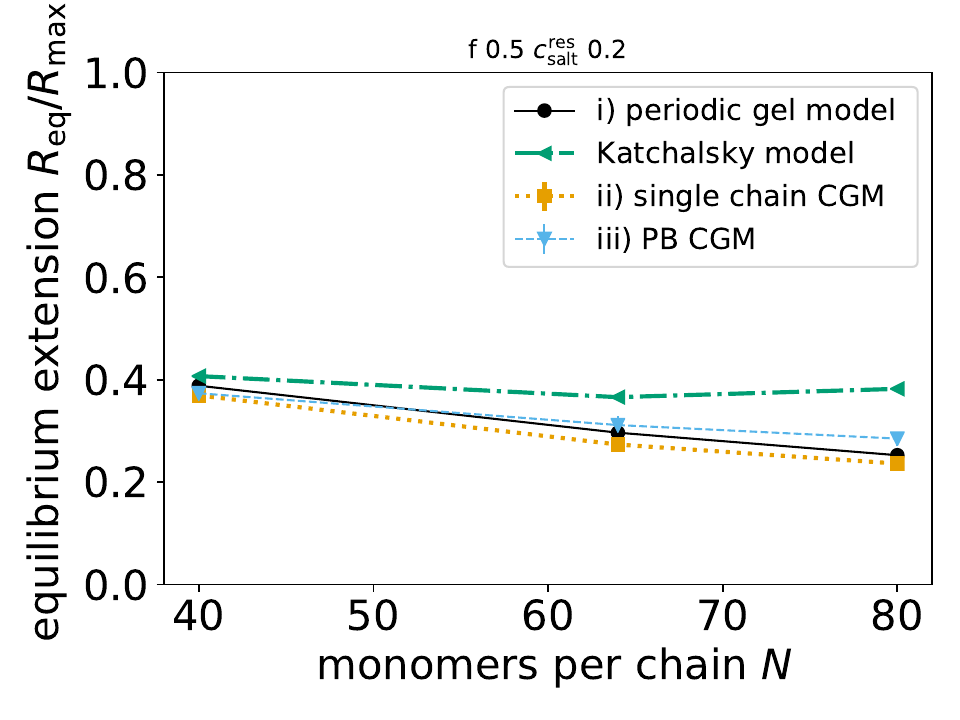}
\includegraphics[width=0.32\linewidth, page=16]{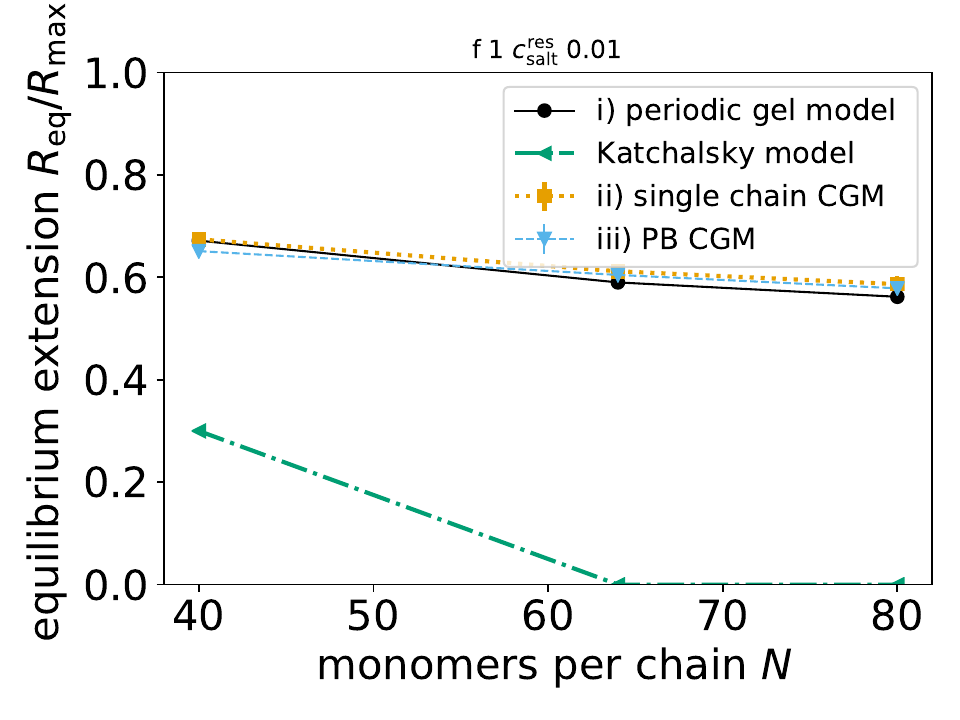}
\includegraphics[width=0.32\linewidth, page=17]{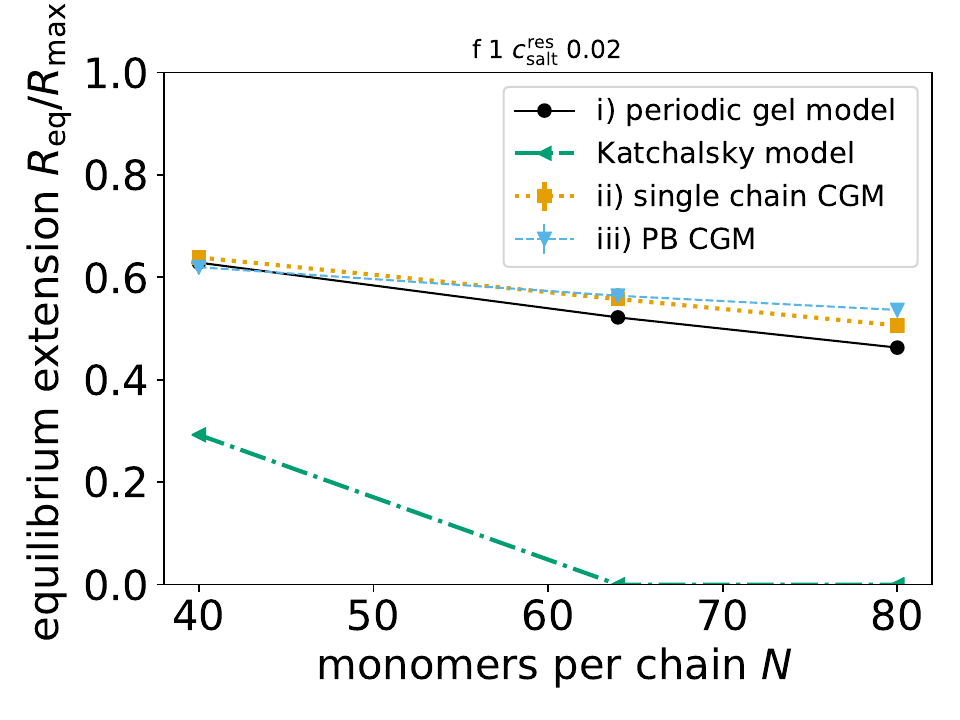}
\includegraphics[width=0.32\linewidth, page=18]{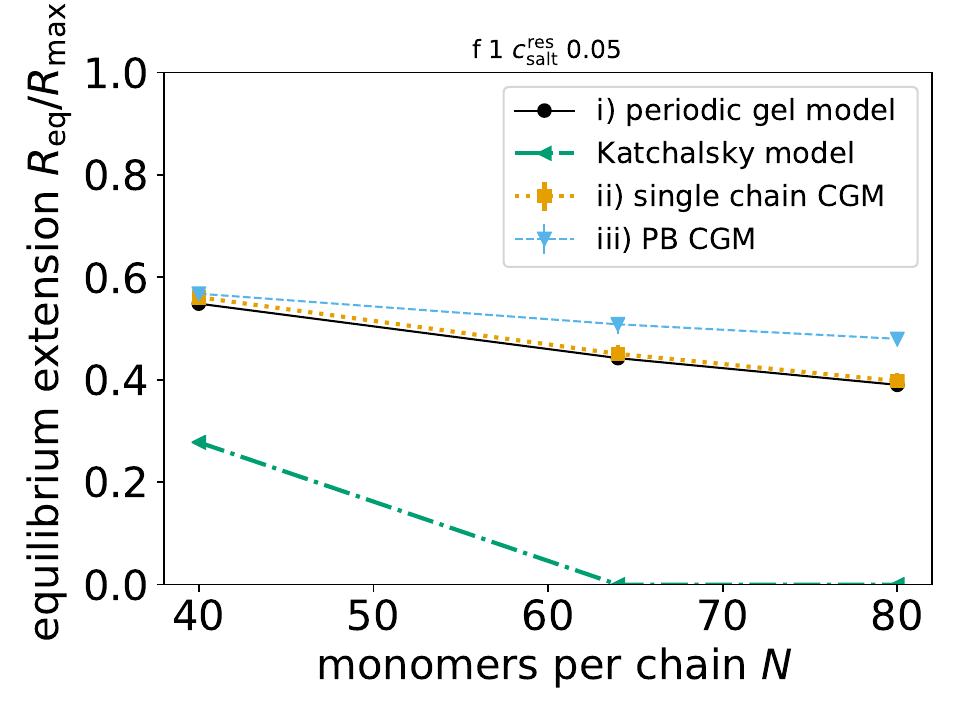}
\includegraphics[width=0.32\linewidth, page=19]{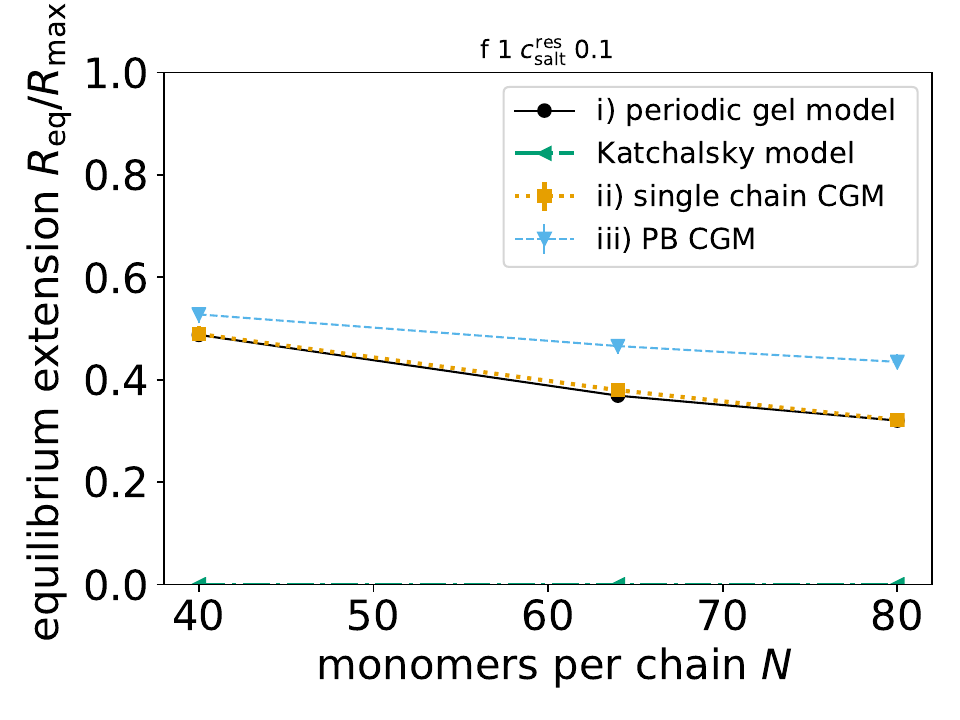}
\includegraphics[width=0.32\linewidth, page=20]{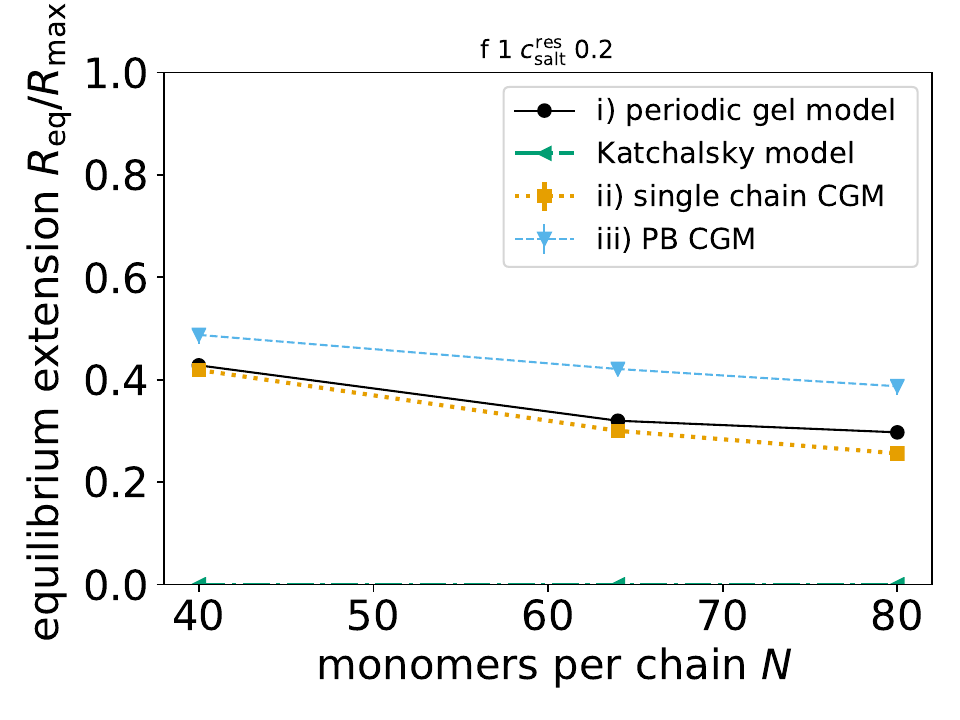}
\end{figure}

\FloatBarrier

\subsection{Comparison of the Models to the Periodic Gel Model}
\begin{figure}
\begin{subfigure}[c]{0.4\textwidth}
\includegraphics[width=0.95\textwidth, page=1]{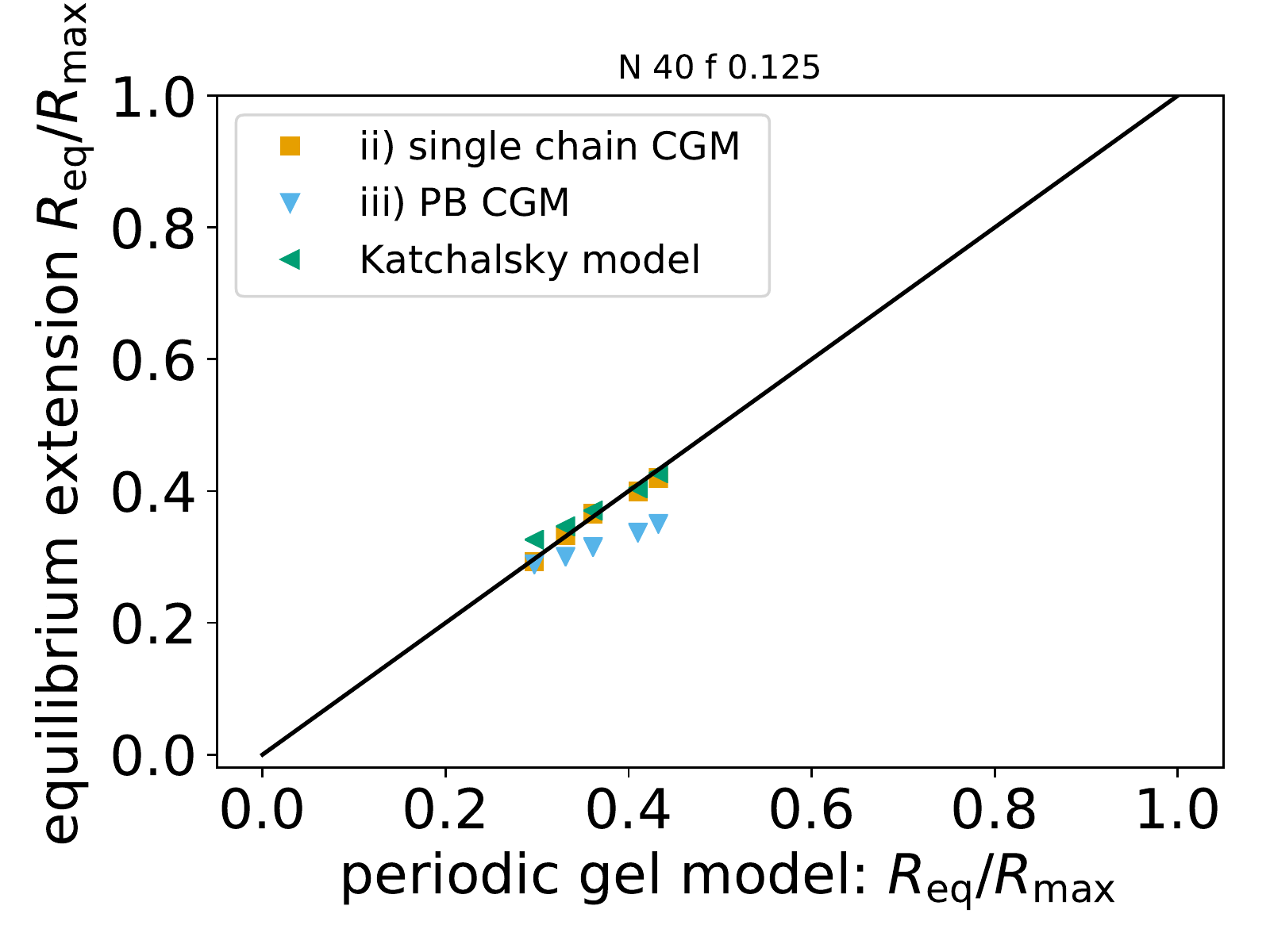}
\subcaption{\label{fig:1}}
\end{subfigure}
\begin{subfigure}[c]{0.4\textwidth}
\includegraphics[width=0.95\textwidth, page=12]{{appendix_correlation_various_cs}.pdf}
\subcaption{\label{fig:2} }
\end{subfigure}
\caption{a) Data set A: N=40, f=0.125, different salt concentrations. b) Data set B: N=80, f=1, different salt concentrations. In each figure the highest salt concentration 0.2 mol/l is associated with the lowest swelling. Reducing the salt concentration increases the swelling. The Katchalsky model fails for f=1. To illustrate this failure, we set the predicted swelling to $\Req/R_\mathrm{max}=0$.}
\end{figure}

We compare the swelling predictions of the two CGMs and the Katchalsky model to the periodic gel model predictions in a parametric plot (similar to figure 5 in the main article).
We seperately show the data for two data sets A: ($N=40, f=0.125$, $\cs \in$ \{0.01, 0.02, 0.05, 0.1, 0.2\} mol/l) and B: ($N=80, f=1$, $\cs \in$ \{0.01, 0.02, 0.05, 0.1, 0.2\} mol/l) in figures \ref{fig:1} and \ref{fig:2}.

Plotting both data sets A and B together, like we do it in figure 5 of the article gives figure \ref{fig:3} below:
\begin{figure}
\includegraphics[width=0.7\textwidth]{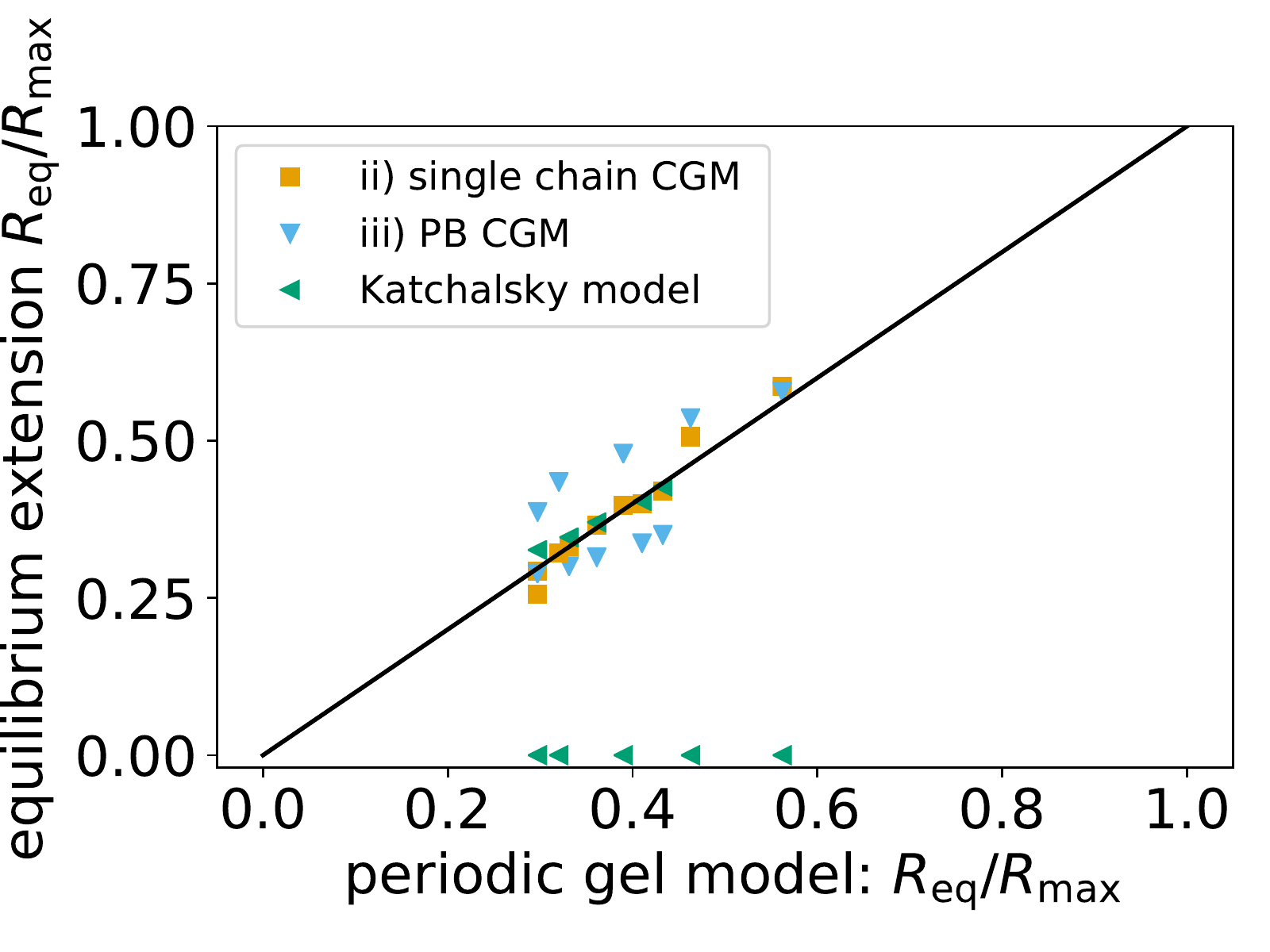}
\caption{Data sets A and B from figures \ref{fig:1} and \ref{fig:2} plotted together. This plot is similar to figure 5 in the manuscript but contains only 1/6 of the data sets.}
\label{fig:3}
\end{figure}

The projection of the predictions from a three-dimensional parameter space on one abscissa results in an apparent \enquote{scattering} of data around the ideal prediction (although the models do not scatter when plotting one data set alone).
This indicates that the different models perform differently well in different parts of the parameter space. 
\FloatBarrier

\end{document}